\documentclass{aa}

\def\mbh{$M_{\rm BH}$\/}

\def\lledd{$L/L_{\rm Edd}$}
\def\nelec{$n_{\mathrm e}$\/}

\def\rfe{$R_\mathrm{FeII}$}

\def\feiiq{\rm Fe{\sc ii}$\lambda$4570\/}
\def\msol{M$_\odot$\/}

\def\chm{$c(\frac{1}{2})$\/}
\def\cqm{$c(\frac{1}{4})$\/}
\def\ltsima{$\; \buildrel < \over \sim \;$}
\def\ltsim{\lower.5ex\hbox{\ltsima}}  
\def\gtsima{$\; \buildrel > \over \sim \;$}

\def\gtsim{\lower.1ex\hbox{\gtsima}}

\def\civ{{\sc{Civ}}$\lambda$1549\/}

\def\cmc{cm$^{-3}$\/}
\def\hb{{\sc{H}}$\beta$\/}

\def\hbbc{{\sc{H}}$\beta_{\rm BC}$\/}

\def\hbnc{{\sc{H}}$\beta_{\rm NC}$\/}

\def\oiiiopt{{\sc{[Oiii]}}$\lambda\lambda$4959,5007\/}
\def\sii{{\sc{[Sii]}}$\lambda\lambda$6731,6717\/}
\def\oii{{\sc{[Oii]}}$\lambda$3727\/}

\def\feiiopt{{Fe \sc{ii}}$_{\rm opt}$\/}
\def\feii{{Fe\sc{ii}}\/}

\def\fe{{\sc{Fe}}\/}

\def\fe76087{{\sc [Fe vii]}$\lambda$6087\/}
\def\oiii{{\sc [Oiii]}$\lambda$5007}

\def\kms{km~s$^{-1}$}

\def\hii{H{\sc ii}\/}

\usepackage{enumerate}

\usepackage{natbib}
\usepackage{graphicx}
\usepackage{lmodern}
\usepackage{rotating}
\usepackage{rotating}
\usepackage{txfonts}
\usepackage{xcolor}
\usepackage{pdflscape}
\usepackage{adjustbox}
\usepackage{subfig}
\usepackage{hyperref}

\usepackage{rotating}
\usepackage{bm}
\def\civ{{\sc{Civ}}$\lambda$1549\AA\ }

\def\hb{{\sc{H}}$\beta$\,}

\def\hbbc{H$\beta_{\rm BC}$\/}
\def\kms{\,km\,s$^{-1}\,$}
\def\k2{$\rm{km/s}$}
\def\feii{{Fe\sc{ii}}\/}
\def\feiiopt{{Fe \sc{ii}}$_{\rm opt}$\/}
\def\feiiq{\rm Fe{\sc ii}$\lambda$4570\/}
\def\feiiq{\rm Fe{\sc ii}$\lambda$4570\/}

\def\rfe{$R_{\rm FeII}$}

\title{Selection of highly-accreting quasars\thanks{}}
\subtitle{Spectral properties of \feiiopt\ emitters not belonging to extreme Population A}

\titlerunning{Narrow emission and Absorption lines among \feiiopt\ strong low-$z$ quasars}

\author{N. Bon\inst{1} \and P. Marziani\inst{2} \and  E. Bon\inst{1}  \and C.A. Negrete\inst{3}  
\and D. Dultzin\inst{3} \and A. del Olmo\inst{4}  \and  M. D'Onofrio\inst{5}  \and    M.L. Mart\'\i nez-Aldama\inst{6}}
\institute{{Astronomical Observatory Belgrade, Volgina 7, 11160 Serbia}
\and{INAF, Osservatorio Astronomico di Padova, IT 35122, Padova, Italy}
\and {Instituto de Astronom\'{\i}a, UNAM, Mexico D.F. 04510, Mexico}
\and{Instituto de Astrofis\'{\i}ca de Andaluc\'{\i}a, IAA-CSIC, E-18008 Granada, Spain}
\and {Dipartimento di Fisica \& Astronomia ``Galileo Galilei'', Universit\`a\ di Padova, Padova,  Italia}
\and {Center for Theoretical Physics, Polish Academy of Sciences, Al. Lotnik\'{o}w 32/46,
02-668 Warsaw, Poland}}
   \date{}

\begin{document} 

\abstract{The quasar class of  extreme Population A (xA)  (also known as super-Eddington accreting massive black holes, 
SEAMBHs) has been hailed as potential distance indicators for cosmology.}
{The aim of this paper is to define tight criteria for their proper identification starting from the main selection 
criterion \rfe $>1 $, and to identify potential intruders not meeting the selection criteria, but nonetheless selected 
as xA because of the coarseness of automatic searches. Inclusion of the spurious xA sources may
dramaticaly increase the dispersion in the Hubble diagram of quasars obtained from virial
luminosity estimates. }
{We studied a sample of 32 low-$z$ quasars originally selected 
from the SDSS DR7 as xA or  SEAMBHs  that have been proved to be almost certainly misclassified sources. 
All of them show moderate-to-strong \feii\ emission and the wide majority 
strong absorption features in their spectra are typical of fairly evolved stellar populations. 
We performed a simultaneous fit of a 
host galaxy spectrum, AGN continuum, FeII template and emission lines to spectra, using the fitting technique based on ULySS, full spectrum fitting package. 
We derive the main accretion parameters (luminosity, black hole mass, and Eddington ratio) and investigate the relation between host galaxy 
properties and AGN. } 
{For sources in our sample (of spectral types corresponding to relatively low Eddington 
ratio), we found an overall consistency between \hbnc, \oiiiopt\ line shifts and the 
mean stellar velocity obtained from the host galaxy fit (within $\lesssim |60|$ \kms).  Non-xA AGN should be distinguished from true xA sources on the basis of several parameters, in addition to the 
ones defining the Main Sequence spectral type: \hb\  asymmetry,   unshifted \oiiiopt, and the intensity ratio between broad and 
narrow component of \hb\ emission line. Only one source in our sample qualify as xA source. }
{Correct classification of spectra contaminated by heavy absorption requires careful determination of the host galaxy 
spectrum. The contamination/misclassification is not usual in the identification of the xAs, nor at low z neither at high z.
We found high fraction of host galaxy spectrum (in half of the sample even higher then 40\%). When absorption lines are prominent, 
and the fraction of the host galaxy
is high, SSP is mimicking FeII, and that may
result in a mistaken identification of FeII spectral features. We have identified several stellar absorption lines that, 
along with the continuum shape, may lead to an 
overestimate of \rfe, and therefore to the misclassification of sources as xA sources. 
 } 
\maketitle
   
%
\defcitealias{marzianietal09}{M09}
\defcitealias{negreteetal18}{Paper I}
\defcitealias{sulenticetal14}{S14}
\defcitealias{sulenticetal07}{S07}
\defcitealias{marzianisulentic14}{MS14}


\section{Introduction}

Quasars show properties that make them potential cosmological probes: they are plentiful, very luminous, and 
detected at very early cosmic epochs (currently out to redshift 7). However, they have never been successfully exploited 
as distance indicators in the past decades. Their luminosity is spread over six orders of magnitude, making them antithetical 
to conventional standard candles. Attempts at providing one or more parameters tightly correlated with luminosity were 
largely unsuccessful in the past decades (i.e., the ``Baldwin effect'' did not live up to its cosmological expectations 
(\citealt[e.g.,][]{popovickovacevic11,bianetal12,geetal16}, see also \citealt{sulenticetal00a} for a synopsis up to 1999). 
Even the next generations of supernova surveys are unlikely to overcome the redshift limit at $\sim 1.5$ \citep{hook13}. 
At the time of writing there is no established distance indicator in the range of redshift $1.5 \lesssim z \lesssim 4$, 
where important information could be gained on the cosmic density of matter and on the dynamic nature of the dark 
energy \citep[e.g.,][ and references therein]{donofrioburigana09}.  

Realistic expectations are now kindled by isolating a class of quasars with some constant property from which the quasar 
luminosity can be estimated independently of its redshift. For instance, the non-linear relation between UV and X ray 
luminosity has been used to build the Hubble diagram up to redshift $\approx$ 5.5 \citep{risalitilusso15,risalitilusso19}. 
Other approaches are being tested as well \citep[e.g., ][see also \citealt{czernyetal18} for a recent review]{watsonetal11}.  
A promising possibility is provided by quasars that are accreting at high (possibly super-Eddington) rates \citep{wangetal14}. 
Physically, in a super-Eddington accretion regime, a geometrically and optically thick structure known as a  ``thick disk'' 
is expected to develop \citep{abramowiczetal88}. The accretion flow remains optically thick   so that radiation pressure  
``fattens'' it.  When the mass accretion rate becomes super-Eddington, the emitted radiation is advected toward the black hole,
so that the source luminosity increases only with the logarithm of accretion rate. In other words, the radiative efficiency 
of the accretion process is expected to decrease, yielding an asymptotic behavior of the luminosity as a function of the 
mass accretion rate \citep{abramowiczetal88,mineshigeetal00,wataraietal00,sadowski11}. In observational terms, the 
luminosity-to-black hole mass ratio ($L$/\mbh $\propto$ \lledd) should tend  toward a well-defined value. 
As the accretion rate increases above $\approx 0.1$, the disk may become first ``slim'' and then  ``thick'' 
in supercritical regime   \citep[][and references therein]{abramowiczstaub14}. The resulting ``thick'' accretion disk 
is expected to emit a steep soft and hard X-ray spectrum,  with hard X-ray photon index (computed between 2 and 20 KeV) 
converging toward $\Gamma_\mathrm{hard} \approx 2.5$ \citep{wangetal13}. Observationally, results are less clear. 
There is a broad consensus that the soft X-ray slope and the index $\alpha_{\mathrm{oX}}$  depend on Eddington ratio 
and can be steep at high accretion rate \citep{bolleretal96,wangetal96,sulenticetal00b,dewanganetal02,grupeetal10,benschetal15}.
In the hard X-ray domain data on weak-lined quasars which are believed to be all xAs \citep{martinez-aldamaetal18}  
suggest weakness, but not necessarily with a steep slope \citep[][]{shemmeretal10,nietal18}, possible because the X-ray 
emission is seen through a dense outflow. More powerful X-ray instrumentation than presently available is needed for 
accurate derivation of the hard-X continuum shape of sources that are anyway X-ray weak compared to the general population 
of quasars \citep{brightmanetal19}. Quasars hosting thick disks should radiate at a well-defined limit because their luminosity
is expected to saturate close to the Eddington luminosity (hence the attribution of ``extremely radiating quasars'') 
even if the mass accretion rate becomes highly super-Eddington \citep{mineshigeetal00}. Their physical and observational 
properties are only summarily known. However, our ability to distinguish sources in different accretion states has greatly 
improved thanks to the exploitation of an empirical correlation set known as the ``main sequence'' (MS) of quasars  
\citep{borosongreen92,sulenticetal00a,sulenticetal00b}.



The MS concept originates from a principal component analysis carried out on the spectra of $\approx$ 80 Palomar-Green (PG) 
quasars by \citet{borosongreen92}. These authors identified a first eigenvector  dominated by an anticorrelation  
between  the \oiii\ peak intensity and the strength of optical \feii\ emission.  Along E1 FWHM of \hb\ and \feii\ prominence 
are also correlated (Fig. 9 of \citealt{borosongreen92}), and define a sequence based on optical parameters which are 
easily measurable on single-epoch spectra of large samples of quasars. 

The Eigenvector 1 (E1) in a parameter space of four dimensions 
(4DE1, \citealt{sulenticetal00a, sulenticetal00b, sulenticetal07})  is especially useful  to isolate different spectral 
types and, among them, spectral type that may be associated with  extreme phenomena.  4DE1 involves optical, UV and X-ray. 
Its dimension (1) the full width at half maximum of  \hb, FWHM (\hb), 2) the ratio of the equivalent widths of \feii\ 
emission at 4570 \AA\ and \hb, \rfe = W (\feiiq) / W (\hb) $\approx F$ (\feiiq) / $F$ (\hb). (1) and (2) define what has 
come to be known as the optical plane of E1 quasars main sequence (MS);  3) the photon index in the soft X-rays domain, 
$\Gamma_\mathrm{soft}$, and 4) the blueshift of the high ionization line \civ. \citet{sulenticetal00a} proposed two main 
populations on the basis of the quasar systematic trends in the optical plane  (FWHM(\hb) vs \rfe) of the 4DE1 parameter 
space: Population A for quasars with FWHM (\hb) $<$ 4000 \kms\ and Population B for those with FWHM (\hb)$>$ 4000 \kms.  
The two populations are not homogenous and they show trends in spectral properties, especially within Pop. A 
\citep{sulenticetal02}. For this reason, the optical plane of E1 was divided into $\Delta$FWHM (\hb) = 4000 \kms\ and 
$\Delta$ \rfe = 0.5. This defined the A1, A2, A3, A4  bins as  \rfe\ increases, and the B1, B1 +, B1 ++   
defined as FWHM (\hb) increases (see Figure 1 of \citealt{sulenticetal02}).  Similarly,   B2, B2+ and so on for each 
interval of the 2 strip, with \rfe\ in the range 0.5 -- 1.   Thus, spectra belonging to the same bin are expected to 
have fairly similar characteristics concerning line profiles and optical and UV line ratios 
\citep{sulenticetal07,zamfiretal10}. 
The MS may be driven by  Eddington ratio \lledd\ convolved with the effect of orientation 
\citep[e.g.][]{boroson02,ferlandetal09,marzianietal01, shenho14,sunshen15}, although this view is not void of challenges.  
Physically, quasars may be distinguished by differences in Eddington ratio (mainly the horizontal axis as for A1,A2,A3, etc.) or by orientation (mainly the vertical axis for a fixed black hole mass). 

Quasars are considered high accretors (hereafter xA quasars, for extreme Population A quasars) following the work of  \citet[][hereafter MS14]{marzianisulentic14}, if they satisfy the selection criterion:


\begin{equation}
R_\mathrm{FeII} = \frac{EW(\mathrm{FeII} \lambda4570)}{EW(H\beta)} > 1.0
\end{equation}

At low redshift, we can identify  xA quasars following the method described in  (MS14), i.e. by isolating sources that have \rfe $\ge 1$\ i.e., belonging to spectral types A3 and A4, or to bins B3 and B4\ if FWHM (\hb)$>$ 4000 \kms. Super-Eddington accretors  can be identified from   $\Gamma_\mathrm{soft}$ and from the $\Gamma_\mathrm{hard}$ (2-20 keV) as well \citep{wangetal13,wangetal14}. This method requires deep spectral observations from space-borne instrumentation, and at present, cannot be applied to large samples. The MS of quasars offers the simplest selection criterion \rfe $> 1$. A similar selection criterion has been defined through the fundamental plane of accreting black holes \citep{duetal16a}, a relation between Eddington ratio (or dimensionless accretion rate), 
and \rfe\ and the $D$ parameter defined as the ratio between the FWHM and the dispersion $\sigma$ of the \hb\ line profile ($FWHM/\sigma(H\beta)$) \citep{duetal16a}.  The fundamental plane can  be written as two linear relations   between $\log  \dot{\cal M}$\ and $ L_{\rm bol}/L_{\rm Edd}$ versus  $\approx \alpha  + \beta  \frac{\rm FWHM}{\sigma}+\gamma R_{\rm FeII},$ where $\alpha (>0), \beta (<0), \gamma\ (>0)$ are  reported by \citet{duetal16a}.  The values of  Eddington ratio and $\dot{{\mathcal M}}$ derived from the fundamental plane equation are large enough to qualify the xA sources satisfying \rfe$> 1$ as SEAMBHs. The converse may not be true, 
since some SEAMBHs have been identified that corresponds to spectral types A2 and even A1 (e.g. Mrk 110). The point is that A1 and A2 show the minimum value of $D$ as their \hb\ most closely resemble Lorentzian functions while  in A3 and A4 a blue-shifted excess leads to an increase in $D$. In the following we will consider \rfe$> 1$\  (or 1.2 if doubtful borderline cases have to be excluded, following \citealt{negreteetal18}, hereafter \citetalias{negreteetal18}) as a necessary condition to consider a source xA or SEAMBH, with the two terms. 

As mentioned, accretion theory supports the empirical finding of \citetalias{marzianisulentic14} on xA sources.  First,  $L/L_\mathrm{Edd} \sim {\cal O} (1)$ (up to a few times the Eddington luminosity)  is a  physically motivated condition.   The ability to obtain a redshift independent distance then stems from the knowledge of the \lledd\ with a small dispersion around a characteristic value, and from the ability to estimate the black hole mass (\lledd $\propto $ L/\mbh).    The preliminary analysis carried out in the last two years (e.g. Paper I) emphasize the need to avoid the inclusion of ``intruders'' in the Hubble diagram build from xA, as they can significantly increase the dispersion in the distance modulus.


In this paper, we take advantage of  the sample of quasars in \citetalias{negreteetal18}  that were selected from an automatic analysis, and we focus on the sources that were affected by strong contamination of the host galaxy and that turned out not to be xA sources. The identification of large samples of xA sources needed for their cosmological exploitation is and  will remain based on surveys collected from fixed apertures or, at best, diffraction limited PSFs $\approx 0.1$ arcsec, as in the case of Euclid \citep{euclid11}.  Therefore, the broad line emitting regions will be always unresolved, and contaminated by emission from regions more distant from the central continuum source. Specifically, a major role is played by the host galaxy stellar spectrum. We will therefore devote the paper to a detailed  study of the emission properties and of the host spectrum of the ``intruders'' in order to better define exclusion criteria. 
 
Section \ref{sample} describes the method  followed for the sample selection. 
The merit of the sample is to provide sources covering a relatively wide range of \rfe\ with typical low-luminosity 
type-1 properties, for which  several intriguing properties of the host galaxy and of the AGN can be measured for the same object:  
age and chemical compositions as well as radial velocity shifts of narrow emission lines associated with the 
AGN narrow line region (NLR). In addition, the host galaxy spectrum effect on the appearance of the AGN spectrum can be 
thoroughly analysed.  We then describe several approaches aimed at obtaining the spectroscopic components associated 
with the AGN continuum and the emission spectrum (Section \ref{anal}). 
Section \ref{results} provides measurements and results on the host spectrum, internal line shifts 
(analysing in detail the use of the \oii\ doublet whose rest frame wavelength  is dependent on electron density), 
and narrow and broad emission line parameters. Section \ref{disc} discusses the results in the  context of the quasar 
main sequence, trying to assess the main factors affecting the \mbh\ and \lledd\ estimates in small samples. In 
Section \ref{sec:conclusions} are reported main conclusions and a summary of the paper.

\section{Sample selection}
\label{sample}

The quasar sample presented by \citet{shenetal11}  consists of 105,783 quasar spectra of the SDSS DR7 and was vetted 
following several filters: $z  <0.8$ to cover the range around H$\beta$ and include the \feiiq\  
and 5260 \AA\ blends; (2) S/N$>$ 20.  Only  2,734 spectra satisfy these criteria, reduced to 468 
with (3) \rfe $\ge$ 1.  S/N and \rfe\ were estimated through the automatic measurements after continuum 
normalization at  5100 \AA. Then we measured the EW of FeII and \hb\ in the ranges 4435-4686 and 4776-4946 respectively 
\citep{borosongreen92} to estimate  \rfe.  Among the 468 sources \cite{negreteetal18} found 134 of 
them whose spectra  are either noisy or are of the intermediate type (Sy $\sim$ 1.5), 
that is, the emission of the broad component of \hb\ is very weak compared to its narrow component, 
which is usually intense. These authors excluded  them to a have a final sample of  334 sources properly 
classified as type 1 with \rfe $\gtrsim 1$. Thirty-two of  334 sources showed  strong contamination from the host galaxy. 
It turned out that the host-galaxy contamination mimicked the \feii\ emission features customarily  found around \hb, leading to an overestimate of \rfe\ from the automated measurement (see \ref{section_mimicing}). The study of this sample  (hereafter HG) is presented in this paper, 
while the rest of the sample (which we found out is in part suited for our cosmological project) 
has been in an independent paper devoted to the exploitation of xA quasars for cosmological parameter estimates 
(Paper I).  

Table \ref{basic} reports the identification, the redshift, the $g$ magnitude, the color index $g-r$, 
the specific flux at  6cm in mJy (FIRST), the log of the specific flux at 2500\AA, and the radio-loudness parameter  
$R = f_{\nu}(6 \mathrm{cm})/f_{\nu}$(2500\AA) \citep{jiangetal07,kellermannetal89}. According to \citet{jiangetal07} 
radio sources are classified as radio-quiet for $R \le 10$, and radio-loud for $R > 10$. Data reported in Table \ref{basic} 
are taken from the table of  \cite{shenetal11}, where radio properties are included by matching  SDSS DR7 quasar with the 
FIRST \cite{whiteetal97}. The radio fluxes densities are subject to a considerable uncertainty (to be a factor of 
$\approx$2 from a coarse analysis on the FIRST maps): the sources are faint, the continuum are extrapolated from 20cm 
to 6cm using a power-law with index 0.5, and affected by reduction residuals in the maps. The radio power is actually 
modest; in the case of \object{SDSSJ151600.39+572415.7} at $z\approx 0.2$, $\log P_{\nu}[W/Hz] \approx 30.4$ , 
which is typical of radio detected sources in spectral type A2 \citep{gancietal19} to which this source belongs. 
Similar considerations apply to the other two sources. On the basis of the results of \citet{gancietal19}, 
the three sources may not be even truly RL source in the sense of having a relativistic  jet \citep{padovani17}.

\begin{table}[tp!]
\tabcolsep=2pt
	\caption{Basic  properties of the HG sample}\label{basic}
		{\scriptsize
		\begin{tabular}{ccccccccr}
		
			\hline\hline\			
			\hspace{5mm}SDSS ID	&	$z$   	&	$g$     	&	$g-r$     	&	$f$(6cm)	&	$\log f$(2500\AA)	&	R	\\
(1) & (2) & (3) & (4) & (5) & (6) & (7)  \\
\hline
J003657.17-100810.6	&	0.19	&	17.84	$\pm$	0.02	&	0.34	$\pm$	0.03	&		&	-26.89	&		\\
J010933.91+152559.0	&	0.23	&	18.97	$\pm$	0.02	&	0.44	$\pm$	0.04	&		&	-27.24	&		\\
J011807.98+150512.9	&	0.32	&	19.16	$\pm$	0.02	&	0.47	$\pm$	0.04	&		&	-27.19	&		\\
J031715.10-073822.3	&	0.27	&	19.08	$\pm$	0.03	&	0.56	$\pm$	0.05	&		&	-26.93	&		\\
J075059.82+352005.2	&	0.41	&	19.37	$\pm$	0.02	&	0.24	$\pm$	0.03	&		&	-27.14	&		\\
J082205.19+584058.3	&	0.31	&	19.48	$\pm$	0.02	&	0.46	$\pm$	0.04	&		&	-27.21	&		\\
J082205.24+455349.1	&	0.30	&	18.38	$\pm$	0.02	&	0.44	$\pm$	0.04	&		&	-27.02	&		\\
J091017.07+060238.6	&	0.30	&	19.13	$\pm$	0.04	&	0.68	$\pm$	0.05	&		&	-27.06	&		\\
J091020.11+312417.8	&	0.26	&	18.73	$\pm$	0.01	&	0.42	$\pm$	0.03	&		&	-27.07	&		\\
J092620.62+101734.8	&	0.27	&	19.25	$\pm$	0.02	&	0.55	$\pm$	0.03	&		&	-27.53	&		\\
J094249.40+593206.4	&	0.24	&	18.88	$\pm$	0.02	&	0.46	$\pm$	0.04	&		&	-27.19	&		\\
J094305.88+535048.4	&	0.32	&	19.18	$\pm$	0.03	&	0.48	$\pm$	0.05	&		&	-27.16	&		\\
J103021.24+170825.4	&	0.25	&	18.63	$\pm$	0.02	&	0.39	$\pm$	0.03	&		&	-27.19	&		\\
J105530.40+132117.7	&	0.18	&	17.61	$\pm$	0.02	&	0.22	$\pm$	0.04	&		&	-26.64	&		\\
J105705.40+580437.4	&	0.14	&	17.66	$\pm$	0.05	&	0.37	$\pm$	0.07	&		&	-26.87	&		\\
J112930.76+431017.3	&	0.19	&	18.46	$\pm$	0.03	&	0.60	$\pm$	0.04	&		&	-27.37	&		\\
J113630.11+621902.4	&	0.21	&	18.72	$\pm$	0.02	&	0.49	$\pm$	0.05	&		&	-27.06	&		\\
J113651.66+445016.4	&	0.12	&	17.71	$\pm$	0.04	&	0.59	$\pm$	0.05	&	1.95	&	-26.97	&	18.3	\\
J123431.08+515629.2	&	0.30	&	19.05	$\pm$	0.03	&	0.54	$\pm$	0.05	&	1.07	&	-27.42	&	27.9	\\
J124533.87+534838.3	&	0.33	&	18.51	$\pm$	0.03	&	0.36	$\pm$	0.05	&		&	-27.08	&		\\
J125219.55+182036.0	&	0.20	&	18.98	$\pm$	0.02	&	0.79	$\pm$	0.03	&		&	-27.18	&		\\
J133612.29+094746.8	&	0.25	&	19.10	$\pm$	0.02	&	0.49	$\pm$	0.04	&		&	-27.33	&		\\
J134748.06+404632.6	&	0.27	&	19.17	$\pm$	0.02	&	0.82	$\pm$	0.03	&		&	-27.28	&		\\
J134938.08+033543.8	&	0.20	&	18.70	$\pm$	0.03	&	0.53	$\pm$	0.05	&		&	-27.16	&		\\
J135008.55+233146.0	&	0.27	&	18.14	$\pm$	0.02	&	0.23	$\pm$	0.05	&		&	-27.00	&		\\
J141131.86+442001.0	&	0.26	&	18.90	$\pm$	0.03	&	0.61	$\pm$	0.04	&		&	-27.18	&		\\
J143651.50+343602.4	&	0.30	&	19.22	$\pm$	0.01	&	0.35	$\pm$	0.04	&		&	-27.44	&		\\
J151600.39+572415.7	&	0.20	&	18.41	$\pm$	0.02	&	0.58	$\pm$	0.04	&	2.27	&	-26.96	&	20.4	\\
J155950.79+512504.1	&	0.24	&	18.82	$\pm$	0.02	&	0.39	$\pm$	0.04	&		&	-27.26	&		\\
J161002.70+202108.5	&	0.22	&	18.80	$\pm$	0.02	&	0.61	$\pm$	0.03	&		&	-27.25	&		\\
J162612.16+143029.0	&	0.26	&	19.71	$\pm$	0.02	&	0.88	$\pm$	0.03	&		&	-27.62	&		\\
J170250.46+334409.6	&	0.20	&	18.14	$\pm$	0.01	&	0.41	$\pm$	0.02	&		&	-27.10	&		\\
\hline
\end{tabular}
 {Notes: (1) SDSS name, (2) redshift, (3) $g$ magnitude, (4) color index $g-r$,
 (5) specific flux at 6cm in mJy (= $10^{-26}$ erg s$^{-1}$ cm$^{-2}$ Hz$^{-1}$), (6) log of the specific flux per unit frequency at 2500\AA, in erg s$^{-1}$ cm$^{-2}$Hz$^{-1}$, 
 (7) radio-loudness parameter $R \equiv f_{\nu}(6 \mathrm{cm})/f_{\nu}$(2500\AA) \cite{jiangetal07}.}
}
\end{table}

\section{Data analysis}
\label{anal}

In Paper I we have found subsample of 32 sources with strong contamination by host galaxy. 
That analysis was done using {\em  specfit} \citep{kriss94}. Here we made data analysis using technique based on {\em  ULySS} \citep{Kol09}. 
Results are compared with two separate techiques. One based on {\em  specfit} and {\em  STARLIGHT}\footnote{
Fitting procedure with {\em  specfit} was done as described in Paper I. 
Since we found prominent galactic absorption lines in residuals, and \hb\ profile appeared as double peaked in some cases, 
we considered an additional {\em  specfit}  component - the spectrum of NGC 3779, 
a quiescent giant elliptical galaxy with an evolved stellar population, 
as a reference template. As a second approach we used STARLIGHT \citep{cid-fernandesetal05} to subtract the host galaxy contribution before 
running the \textit{specfit} analysis.}, and another 
one based on {\it DASpec}\footnote{Written by P. Du (private communication). The code is used in e.g., \cite{duetal16a} and \cite{zhangetal19}). 
{\it DASpec} is based on Levenberg-Marquardt minimization and can perform multi-component spectral fitting including AGN continuum, 
emission lines, Fe II template, and host contribution simultaneously.}. 
Since we obtained fairly consistent result with all techiques, 
we made more detailed analysis only with {\em  ULySS},  
such as Monte Carlo simulations and $\chi^2$ maps.
Therefore, all results presented in tables and on figures are done with {\em  ULySS}.

\subsection{{\em  ULySS - Full Spectrum Fitting}}

The analysis was performed using {\em  ULySS}\footnote{{The ULySS full
spectrum fitting package is available at: http://ulyss.univ-lyon1.fr/}}, a full spectrum fitting software
package, that we adopted for fitting Sy1 spectra with models representing a
linear combination of non-linear model components {--} AGN continuum, host
galaxy, \feii\ template and emission lines. The detailed description is given in
\citep{bonetal16}, where {\em  ULySS} has been used for the first time for fitting Sy 1
spectra.\\
\indent Before running the fitting procedure we converted
vacuum wavelengths into air, using the IAU definition \cite{morton91}, since the wavelength calibration of the SDSS spectra 
is in Heliocentric vacuum wavelength, while the components of the model are in air wavelengths. 
Therefore, all analysis were done in air wavelengths.\\

\textbf{Line and continuum fittings}\\

\indent We adjusted {\em  ULySS} to analyze simultaneously all components
that contribute to the flux in the wavelength region $\lambda\lambda=[3700,6800] \AA$. The model that we used for the fit
represents bounded linear
combination of non-linear components - (\textit{i}) stellar population spectrum, 
convolved with a line-of-sight velocity broadening
function, (\textit{ii}) an AGN continuum model represented with a power law
function, (\textit{iii}) a sum of Gaussian functions
accounting for AGN emission lines in analysed spectral domain,
and (\textit{iv})
{\ion{Fe}{II}} template. \\
\indent In order to eliminate overall shape differences between the observed stellar and galactic spectra, 
the model is multiplied by a polynomial function that is
a linear combination of Legendre polynomials. The introduction of this
polynomial in the fit ensures that results are
insensitive to the Galactic extinction, normalization and the flux
calibration of a galaxy and stellar
template spectra \citep{Kol08}.  The polynomial is replacing 
the prior normalization to the pseudo-continuum
that other methods need. We have used a third order of the polynomial
in the fit, in order to model at best the extinction function, and at
the same time to prevent that the higher order terms of the polynomial
affect the fit of broad emission lines and AGN continuum. \\
\indent The single stellar population spectra (SSP) used for the fit of the host
galaxy are spline interpolated over an age-metallicity grid of stellar population models from
the library of SSPs computed by \cite{Vazdekis99} with the Miles library
\citep{SanchezBlazquez2006}. \\
\indent Emission lines are fitted with the sum of Gaussians in the following way:\\
(\textit{i}) all Balmer lines, as well as HeII are fitted with four components -
narrow, two broad components that fit the wings of the lines and very broad
component; \\
(\textit{ii}) to tie widths, shifts and intensities of the [\ion{O}{III}]
lines, we defined two 
separate components of the model - a narrow component and a semi-broad
component. Intensity ratio was kept to 3:1 \citep{dimitrijevicetal07};\\
(\textit{iii}) the rest of emission lines are mainly fitted with two Gaussians -
for the fit of narrow and semi-broad component. Eventhough in some cases the fit was possible with smaller number of 
Gaussian components, in order to stay consistent and perform the analysis in the same way for all spectra, we used for the whole 
sample the same number of components. \\
\indent We used the semi-empirical FeII template by \citet{marzianietal09}, obtained from a high 
resolution spectrum of I Zw 1 
starting from 4000 \AA. In the range underlying the  \hb\ profile the FeII emission was modeled with the help  of FeII emission from the photoionization code CLOUDY version 07.01 
\citep{ferlandetal98}. 

The AGN model is generated with the same sampling and at the
same resolution as the observations, and the fit is performed in the pixel space. The fitting method consists of a non-linear minimization procedure for
minimizing the $\chi^2$ between an observed
spectrum and a model. The fitting procedure applies
the Levenberg-Marquardt minimization technique \citep{Marquardt63}.
The coefficients of the multiplicative polynomial
are determined by least-squares method at each evaluation of
the function minimized by the Marquardt-Levenberg routine. As well, at each iteration the weight of each
component is determined using
a bounding value least-square method \citep{lawsonhanson95}. \\
\indent The simultaneous fit of all components in the model, that implies as well the
simultaneous analysis of kinematic and evolutionary parameters of the stellar population, 
minimizes in a most efficient way many degeneracies between AGN model components reported in the literature, such as: 
(\textit{i}) degeneracy between fractions of AGN
continuum and the host galaxy \citep{bonetal14,moultaka05},  (\textit{ii}) SSP age-metallicity degeneracy \citep{Kol09}, and 
(\textit{iii}) degeneracy between stellar velocity dispersion 
and SSP metallicity \citep{Kol08a}.


\section{Results}
\label{results}

\subsection{Immediate SSP and spectral classification results}

The results of the host galaxy single population analysis with ULySS are reported in Table \ref{longtable}.  The table lists, after 
the SDSS ID (Col. 1) the specific flux measured at 5100 $\AA$ (as proposed by \citealt{richardsetal06}), the light fraction of power law continuum, and the power law spectral index (Cols. 3-4). Cols. 5-8 report information on the SSP 
analysis: the light fraction of the host galaxy, the SSP age, 
and the SSP shift with respect to the rest frame defined by the SDSS-provided redshift value.  The shift of the H$\beta$ narrow component (\hbnc) and of narrow [OIII]$\lambda$5007 line are reported in Cols. 9--10. 
The shift and width  (the Gaussian dispersion $\sigma$) of FeII lines are listed in Col. 11 -- 12. Cols. 13--16 list the 
flux and $\sigma$\ for \hbnc\ and [OIII]$\lambda$5007. Fig. \ref{fits_plots} in Appendix \ref{sap} shows a 
spectral atlas with the main components. 

Classification  concerning spectral type assignment along the E1 MS optical diagram and AGN classification according to  \cite{veron-cettyveron06} are presented in Table \ref{classification}.  The Table lists: the FWHM and Flux of \hbbc\ (Cols. 2-3), \rfe\ and the main sequence spectral type (Cols. 4-5), 
along with the classification of the Catalogue of \citet{veron-cettyveron06}. Sources for which no classification is given in the 
Catalogue are recognizable as type-1 AGN in the SDSS (from S1.0 to 1.8). However, the classification of some of them 
(for example J162612.16) might not have been easy on old spectroscopic data, right because of the strong host contamination. Cols. 7 to 8 list the FWHM   
\hbbc\ and \rfe\ following \citet{shenetal11}. The corresponding spectral type is listed in Col. 9. 
The last columns report, in this order, the FW at 1/4, 1/2, 3/4 and 0.9 \hbbc\ peak intensity, and 
the \hbbc\ centroid at quarter and half maximum, \cqm\ and \chm. These parameters are useful in the asymmetry  and the 
shift analysis, especially at 1/4 of maximum intensity. Both \hb\ and \oiiiopt\ are often affected by asymmetries  close 
to the line base. The 1/4 maximum intensity provides a suitable level to detect and quantify these asymmetries.

\subsection{Spectral Type Classification along the quasar MS: not xA sources in almost all cases}

\begin{figure}
	\begin{center}
	\vspace{-0.cm}
	
	\hspace{-.5cm}
		\includegraphics[width=0.45\textwidth]{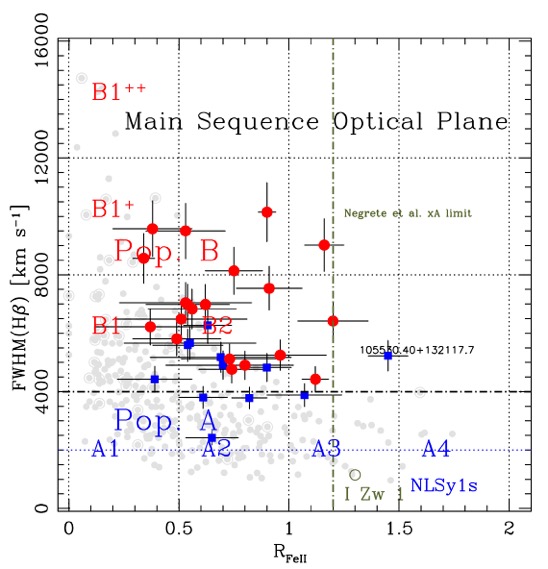}\\
\caption{The optical plane of the E1 MS, FWHM \hbbc\ vs \rfe. Sources from our sample are represented as red and blue circles; 
the grey symbols represent the MS from the sample of \citet{zamfiretal10}, with RL sources identified by an outer circle. 
The red circles are sources with the $D$\ parameter larger than 1.5; for the blue squares $D \lesssim 1.5$. The horizontal 
dot-dashed line marks the limit between Population A and at FWHM $=$ 4000 \kms. Dotted lines separate spectral types and 
NLSy1s. The vertical dot-dashed olive line identifies the \rfe=1.2 limit for xA ``safe'' identification according to 
\citetalias{negreteetal18}.  Only one source qualifies as a true xA candidates. The position of the prototypical xA source 
I Zw 1 is marked. }\label{fig:e1fl}
	\end{center}
\end{figure}

The HG sample sources remain by all measurements relatively strong \feii\ emitters, with \rfe$\gtrsim 0.3$. 
Fig. \ref{fig:e1fl} shows the location of the 32 sources in the optical plane of the E1 MS 
(represented with red and blue circles). 
The \rfe\ and FWHM \hb\ place the sources predominantly into the B2 and A2 spectral bins; 
only one source can be considered genuine xA candidate.

 There is good agreement between our measurements of \rfe\ and those of \citet{shenetal11}: 
 from the measurements reported in Table \ref{longtable},  
 \rfe\ (\citealt{shenetal11}) $ \approx (1.07 \pm 0.08)$\rfe +  ($0.18 \pm 0.06$), 
 implying that \citet{shenetal11} values are systematically higher by 18\%. The reason for this 
 disagreement could be that \citet{shenetal11} did not take into account the host galaxy contribution \citep{Sniegowska2018}.
 This analysis would imply that 5/32 sources could be classified as xA with \rfe $>$1, following \citet{shenetal11}.  The number reduces to only 1 out of 32 if the most restrictive criterion \rfe$\ge 1.2$\ is applied. 
 Parameter $D=FWHM_{H\beta}/\sigma_{H\beta}$ \ 
 distinguishes sources on the Fig. \ref{fig:e1fl} in two groups. Sources with $D> 1.5$ show a more Gaussian-like H$\beta$ 
 profile. $D < 1.5$ implies a more Lorenzian-like profile. 
 
 \indent As mentioned above, only one source (SDSS J105530.40+132117.7) is confirmed as xA in the full HG sample
 after SSP analysis, applying the selection criterion \rfe\ $\gtrsim 1.2$. 
This source will be individually discussed in \S \ref{disc}. The restriction to  \rfe$\ge 1.2$ is operational, 
to avoid contamination from a fraction of borderline objects that may not be really xA: 
since typical uncertainties are $\delta$\rfe $\approx$0.1 at 1$\sigma$\ confidence level, the presence of ``imitators'' 
should be reduced  by 95\%\ the number expected with the limit at \rfe =1. Therefore, source SDSS J105530.40+132117.7
should be considered true xA and analysed as such at a confidence level $\gtrsim 4\sigma$.

	\begin{sidewaystable*}
\vspace{17cm}
\caption{Measurements of the host galaxy, AGN continuum, and emission lines parameters with ULySS.}\label{longtable}	
		\fontsize{8.2}{8} \selectfont\tabcolsep=3pt
			\begin{tabular}{ccccccccccccccccc}			
				\hline\hline \\[0.5ex]
				 SDSS ID &	$f_{\lambda}$(5100) & AGN $f$	&	sp. index	&
SSP $f$	&	SSP age	&	SSP $cz$	&	SSP $\sigma_{\star}$    &  H$\beta_\mathrm{NC}$ z	&	 	OIII $z$	&
 	FeII $z$	 &	FeII $\sigma$	&	H$\beta_\mathrm{NC}$  $\sigma$	&	H$\beta_\mathrm{NC}$ flux	&
[OIII]  $\sigma$	&	[OIII] flux	\\
		&	[$10^{-17}$erg/s/cm$^{2}/$\AA]		 &	[\%]	&		&	[\%]	&	[Myr]	&	[\kms]	&  [\kms]	&	[\kms]	&		[\kms]	&	[\kms]	&	[\kms]	&	[\kms]	&  [$\star\star$]	&  [\kms]	& [$\star\star$]	\\
					 (1) &	(2)	&	(3)	&	(4)	&	(5)	&
(6)	&	(7)	&	(8)	&	(9)	&	 	(10) 	&	(11)  &	*(12)	&
(13)	&	(14) & (15) & (16)	\\
				\hline \\[0.5ex]
J003657.17-100810.6	&	29.74	&	46.8	$\pm$	0.3	&	-0.66	$\pm$	0.23	&	44.0	$\pm$	0.2	&	5398	$\pm$	469	&	60	$\pm$	9	&	196	+/-	10	&	-8	$\pm$	20	&	38	$\pm$	6	&	288	$\pm$	173	&	1635	$\pm$	143	&	189	$\pm$	26		&	2.80	$\pm$	0.01	&	155	$\pm$	10	&	6.21	$\pm$	0.15	\\
J010933.91+152559.0	&	10.87	&	50.9	$\pm$	0.7	&	-0.04	$\pm$	0.15	&	38.6	$\pm$	0.7	&	4445	$\pm$	905	&	119	$\pm$	15	&	138	+/-	17	&	4	$\pm$	26	&	95	$\pm$	29	&	30	$\pm$	174	&	1492	$\pm$	227	&	138	$\pm$	31		&	1.86	$\pm$	0.36	&	129	$\pm$	40	&	1.45	$\pm$	0.58	\\
J011807.98+150512.9	&	10.02	&	55.8	$\pm$	0.6	&	-0.25	$\pm$	0.05	&	34.5	$\pm$	0.6	&	6649	$\pm$	245	&	-40	$\pm$	20	&	170	+/-	22	&	-60	$\pm$	17	&	-43	$\pm$	7	&	252	$\pm$	62	&	1375	$\pm$	252	&	119	$\pm$	19		&	1.87	$\pm$	0.04	&	139	$\pm$	11	&	5.01	$\pm$	0.66	\\
J031715.10-073822.3	&	12.87	&	64.3	$\pm$	0.6	&	-0.81	$\pm$	0.04	&	27.3	$\pm$	0.6	&	1595	$\pm$	329	&	170	$\pm$	16	&	104	+/-	20	&	132	$\pm$	11	&	145	$\pm$	11	&	88	$\pm$	97	&	1953	$\pm$	799	&	145	$\pm$	12		&	3.84	$\pm$	0.28	&	102	$\pm$	15	&	2.55	$\pm$	0.17	\\
J075059.82+352005.2	&	6.91	&	59.1	$\pm$	0.4	&	-1.25	$\pm$	0.06	&	33.5	$\pm$	0.5	&	1431	$\pm$	89	&	66	$\pm$	10	&	95	+/-	14	&	107	$\pm$	18	&	104	$\pm$	5	&	232	$\pm$	138	&	1302	$\pm$	155	&	129	$\pm$	21		&	1.75	$\pm$	0.23	&	153	$\pm$	6	&	7.86	$\pm$	0.22	\\
082205.19+584058.3	&	8.00	&	69.2	$\pm$	0.6	&	-0.25	$\pm$	0.20	&	22.5	$\pm$	0.6	&	6252	$\pm$	477	&	-7	$\pm$	26	&	127	+/-	31	&	-45	$\pm$	24	&	68	$\pm$	13	&	568	$\pm$	65	&	2256	$\pm$	497	&	47	$\pm$	24		&	0.82	$\pm$	0.06	&	134	$\pm$	19	&	3.74	$\pm$	0.50	\\
082205.24+455349.1	&	17.47	&	77.4	$\pm$	0.4	&	-0.79	$\pm$	0.01	&	13.7	$\pm$	0.3	&	3493	$\pm$	194	&	60	$\pm$	21	&	100	+/-	27	&	-5	$\pm$	24	&	-3	$\pm$	14	&	252	$\pm$	106	&	1273	$\pm$	89	&	76	$\pm$	29		&	0.52	$\pm$	0.52	&	186	$\pm$	16	&	0.00	$\pm$	0.44	\\
091017.07+060238.6	&	12.22	&	50.3	$\pm$	0.3	&	-2.00	$\pm$	0.00	&	42.5	$\pm$	0.3	&	3037	$\pm$	74	&	154	$\pm$	11	&	116	+/-	13	&	99	$\pm$	56	&	66	$\pm$	12	&	176	$\pm$	208	&	1408	$\pm$	248	&	170	$\pm$	70		&	0.86	$\pm$	0.11	&	161	$\pm$	20	&	4.59	$\pm$	0.13	\\
091020.11+312417.8	&	13.89	&	60.0	$\pm$	0.5	&	-0.48	$\pm$	0.11	&	34.2	$\pm$	0.5	&	3434	$\pm$	52	&	71	$\pm$	14	&	151	+/-	17	&	73	$\pm$	15	&	45	$\pm$	14	&	387	$\pm$	145	&	1676	$\pm$	623	&	147	$\pm$	19		&	2.55	$\pm$	0.10	&	124	$\pm$	17	&	1.91	$\pm$	0.21	\\
092620.62+101734.8	&	9.46	&	47.7	$\pm$	0.6	&	-0.12	$\pm$	0.17	&	45.5	$\pm$	0.6	&	5623	$\pm$	206	&	66	$\pm$	12	&	142	+/-	15	&	61	$\pm$	31	&	22	$\pm$	23	&	872	$\pm$	105	&	1310	$\pm$	251	&	148	$\pm$	36		&	1.70	$\pm$	0.21	&	95	$\pm$	37	&	1.16	$\pm$	1.01	\\
094249.40+593206.4	&	14.30	&	57.6	$\pm$	0.5	&	-0.21	$\pm$	0.03	&	35.6	$\pm$	0.5	&	5724	$\pm$	417	&	84	$\pm$	12	&	137	+/-	14	&	95	$\pm$	29	&	39	$\pm$	33	&	318	$\pm$	158	&	1454	$\pm$	206	&	173	$\pm$	35		&	1.76	$\pm$	0.06	&	135	$\pm$	35	&	0.84	$\pm$	1.96	\\
094305.88+535048.4	&	9.69	&	51.4	$\pm$	0.3	&	-1.85	$\pm$	0.04	&	40.8	$\pm$	0.3	&	4428	$\pm$	352	&	14	$\pm$	15	&	149	+/-	17	&	35	$\pm$	32	&	41	$\pm$	12	&	-128	$\pm$	80	&	1288	$\pm$	284	&	176	$\pm$	37		&	1.58	$\pm$	0.37	&	154	$\pm$	24	&	4.87	$\pm$	0.64	\\
103021.24+170825.4	&	16.30	&	64.1	$\pm$	0.3	&	-0.90	$\pm$	0.09	&	21.3	$\pm$	0.3	&	9363	$\pm$	290	&	134	$\pm$	13	&	86	+/-	16	&	173	$\pm$	38	&	76	$\pm$	25	&	-88	$\pm$	278	&	1881	$\pm$	118	&	47	$\pm$	41		&	0.29	$\pm$	0.04	&	212	$\pm$	68	&	2.31	$\pm$	0.17	\\
105530.40+132117.7	&	38.72	&	70.4	$\pm$	0.3	&	-1.20	$\pm$	0.03	&	23.9	$\pm$	0.2	&	3072	$\pm$	246	&	41	$\pm$	11	&	106	+/-	14	&	-81	$\pm$	32	&	93	$\pm$	17	&	-215	$\pm$	213	&	1701	$\pm$	200	&	131	$\pm$	37		&	0.72	$\pm$	0.00	&	132	$\pm$	0	&	2.26	$\pm$	0.57	\\
105705.40+580437.4	&	40.27	&	36.8	$\pm$	0.5	&	-0.63	$\pm$	0.09	&	55.9	$\pm$	0.5	&	1601	$\pm$	177	&	51	$\pm$	9	&	144	+/-	10	&	30	$\pm$	15	&	-6	$\pm$	27	&	253	$\pm$	143	&	829	$\pm$	77	&	81	$\pm$	18		&	1.24	$\pm$	2.31	&	130	$\pm$	40	&	1.59	$\pm$	0.98	\\
112930.76+431017.3	&	20.58	&	35.2	$\pm$	0.4	&	-1.10	$\pm$	0.16	&	58.6	$\pm$	0.3	&	3333	$\pm$	159	&	100	$\pm$	6	&	110	+/-	8	&	87	$\pm$	26	&	94	$\pm$	23	&	364	$\pm$	232	&	1378	$\pm$	214	&	139	$\pm$	31		&	1.47	$\pm$	0.24	&	103	$\pm$	35	&	1.19	$\pm$	0.63	\\
113630.11+621902.4	&	19.49	&	55.6	$\pm$	0.5	&	-0.89	$\pm$	0.10	&	36.4	$\pm$	0.4	&	1933	$\pm$	74	&	130	$\pm$	14	&	143	+/-	16	&	51	$\pm$	26	&	-35	$\pm$	10	&	327	$\pm$	139	&	2206	$\pm$	456	&	171	$\pm$	33		&	3.00	$\pm$	0.01	&	179	$\pm$	14	&	7.16	$\pm$	0.38	\\
113651.66+445016.4	&	42.49	&	36.6	$\pm$	0.3	&	-1.13	$\pm$	0.22	&	57.1	$\pm$	0.3	&	2535	$\pm$	191	&	-8	$\pm$	8	&	147	+/-	9	&	-79	$\pm$	12	&	67	$\pm$	14	&	70	$\pm$	294	&	1633	$\pm$	340	&	165	$\pm$	14		&	3.28	$\pm$	0.21	&	84	$\pm$	22	&	1.85	$\pm$	0.38	\\
123431.08+515629.2	&	11.37	&	40.4	$\pm$	0.8	&	-0.17	$\pm$	0.29	&	52.1	$\pm$	0.8	&	2838	$\pm$	408	&	40	$\pm$	11	&	117	+/-	13	&	44	$\pm$	12	&	0	$\pm$	14	&	368	$\pm$	113	&	1391	$\pm$	384	&	31	$\pm$	0		&	1.97	$\pm$	0.37	&	164	$\pm$	19	&	4.48	$\pm$	0.20	\\
124533.87+534838.3	&	10.65	&	53.4	$\pm$	0.6	&	-0.71	$\pm$	0.13	&	38.3	$\pm$	0.7	&	1355	$\pm$	66	&	-8	$\pm$	14	&	112	+/-	18	&	34	$\pm$	22	&	102	$\pm$	34	&	385	$\pm$	241	&	1464	$\pm$	314	&	107	$\pm$	28		&	1.27	$\pm$	0.09	&	56	$\pm$	64	&	0.28	$\pm$	0.75	\\
125219.55+182036.0	&	18.69	&	61.1	$\pm$	0.5	&	-0.14	$\pm$	0.07	&	30.9	$\pm$	0.5	&	3182	$\pm$	250	&	106	$\pm$	14	&	142	+/-	16	&	76	$\pm$	32	&	77	$\pm$	7	&	204	$\pm$	156	&	1457	$\pm$	208	&	166	$\pm$	51		&	1.77	$\pm$	0.20	&	73	$\pm$	13	&	2.51	$\pm$	0.13	\\
133612.29+094746.8	&	12.96	&	48.4	$\pm$	0.6	&	-0.19	$\pm$	0.09	&	43.5	$\pm$	0.5	&	6914	$\pm$	709	&	79	$\pm$	15	&	178	+/-	16	&	-46	$\pm$	42	&	54	$\pm$	23	&	337	$\pm$	106	&	1324	$\pm$	179	&	153	$\pm$	56		&	1.16	$\pm$	0.58	&	84	$\pm$	36	&	1.27	$\pm$	0.13	\\
134748.06+404632.6	&	13.72	&	44.5	$\pm$	0.4	&	-0.84	$\pm$	0.08	&	49.3	$\pm$	0.4	&	4761	$\pm$	427	&	122	$\pm$	15	&	201	+/-	16	&	-36	$\pm$	122	&	14	$\pm$	23	&	926	$\pm$	88	&	1430	$\pm$	386	&	289	$\pm$	183		&	1.14	$\pm$	0.14	&	88	$\pm$	33	&	0.87	$\pm$	0.11	\\
134938.08+033543.8	&	18.60	&	50.0	$\pm$	0.3	&	-1.81	$\pm$	0.07	&	44.0	$\pm$	0.2	&	5606	$\pm$	738	&	90	$\pm$	13	&	169	+/-	14	&	35	$\pm$	28	&	73	$\pm$	10	&	-71	$\pm$	250	&	1656	$\pm$	339	&	159	$\pm$	35		&	1.91	$\pm$	0.43	&	190	$\pm$	17	&	6.39	$\pm$	0.06	\\
135008.55+233146.0	&	20.82	&	66.1	$\pm$	0.3	&	-0.73	$\pm$	0.18	&	22.9	$\pm$	0.3	&	3766	$\pm$	229	&	80	$\pm$	15	&	150	+/-	18	&	85	$\pm$	30	&	13	$\pm$	15	&	193	$\pm$	106	&	2500	$\pm$	176	&	71	$\pm$	34		&	0.34	$\pm$	0.24	&	214	$\pm$	16	&	0.00	$\pm$	0.03	\\
141131.86+442001.0	&	15.64	&	38.1	$\pm$	0.4	&	-1.15	$\pm$	0.15	&	55.3	$\pm$	0.4	&	4704	$\pm$	719	&	58	$\pm$	15	&	221	+/-	16	&	21	$\pm$	44	&	-14	$\pm$	5	&	1321	$\pm$	148	&	995	$\pm$	324	&	170	$\pm$	56		&	1.18	$\pm$	0.45	&	139	$\pm$	6	&	5.52	$\pm$	0.08	\\
143651.50+343602.4	&	8.43	&	44.4	$\pm$	0.4	&	-2.00	$\pm$	0.00	&	47.3	$\pm$	0.3	&	9489	$\pm$	1150	&	4	$\pm$	21	&	198	+/-	22	&	-88	$\pm$	54	&	-51	$\pm$	12	&	192	$\pm$	134	&	1755	$\pm$	498	&	203	$\pm$	62		&	2.26	$\pm$	0.68	&	229	$\pm$	13	&	0.00	$\pm$	1.34	\\
151600.39+572415.7	&	24.42	&	35.8	$\pm$	0.6	&	-0.22	$\pm$	0.05	&	57.6	$\pm$	0.6	&	1595	$\pm$	785	&	96	$\pm$	9	&	172	+/-	11	&	48	$\pm$	17	&	-21	$\pm$	13	&	313	$\pm$	162	&	1286	$\pm$	331	&	170	$\pm$	22		&	3.06	$\pm$	0.21	&	156	$\pm$	18	&	3.13	$\pm$	0.45	\\
155950.79+512504.1	&	11.79	&	63.3	$\pm$	0.6	&	-0.19	$\pm$	0.11	&	27.8	$\pm$	0.6	&	4366	$\pm$	506	&	126	$\pm$	21	&	152	+/-	24	&	171	$\pm$	24	&	48	$\pm$	12	&	224	$\pm$	179	&	1488	$\pm$	235	&	144	$\pm$	32		&	2.00	$\pm$	0.32	&	120	$\pm$	18	&	4.27	$\pm$	0.07	\\
161002.70+202108.5	&	18.09	&	40.7	$\pm$	0.6	&	-0.07	$\pm$	0.16	&	50.9	$\pm$	0.6	&	3439	$\pm$	397	&	79	$\pm$	9	&	141	+/-	11	&	24	$\pm$	16	&	-34	$\pm$	6	&	291	$\pm$	126	&	1316	$\pm$	178	&	155	$\pm$	18		&	3.24	$\pm$	0.20	&	153	$\pm$	13	&	9.03	$\pm$	0.16	\\
162612.16+143029.0	&	8.57	&	40.7	$\pm$	0.6	&	-0.29	$\pm$	0.16	&	48.1	$\pm$	0.5	&	11220	$\pm$	645	&	112	$\pm$	16	&	198	+/-	17	&	40	$\pm$	27	&	435	$\pm$	21	&	98	$\pm$	152	&	2252	$\pm$	302	&	253	$\pm$	31		&	3.74	$\pm$	0.30	&	138	$\pm$	30	&	3.26	$\pm$	0.17	\\
170250.46+334409.6	&	20.89	&	50.8	$\pm$	0.4	&	-0.30	$\pm$	0.30	&	39.5	$\pm$	0.3	&	9704	$\pm$	392	&	50	$\pm$	11	&	156	+/-	12	&	108	$\pm$	36	&	92	$\pm$	12	&	279	$\pm$	102	&	1324	$\pm$	161	&	161	$\pm$	48		&	1.54	$\pm$	0.05	&	185	$\pm$	18	&	5.65	$\pm$	0.17	\\
\hline
				
				\end{tabular}\\
		 {Notes:} {(1) SDSS ID of the object ; (2) flux measured at 5100 $\AA$ in $10^{-17}$erg/s/cm$^{2}/$\AA 
		 (hereafter in the table [$\star\star$]) \citep{richardsetal06}; 
		 (3) fraction of power law continuum; 
		 (4) power law spectral index (errors present 1$\sigma$ dispersion obtained from Monte Carlo simulations); (5) contribution of the host galaxy; 
		 (6) SSP age (errors present a dispersion obtained from Monte Carlo simulations); (7) mean stellar velocity; (8) stellar velocity dispersion} 
		 {	(9) shift of the H$\beta$ narrow component; 
		 (10)  shift of narrow [OIII]$\lambda$5007 line;}
		 {(11)  shift of FeII lines; (12) width of FeII lines; 
		 (13) width of H$\beta$ narrow component; (14) flux of H$\beta$ narrow component; } 
          {(15) width of narrow [OIII]$\lambda$5007 component; (16) flux of narrow [OIII]$\lambda$5007 component.}
	\end{sidewaystable*}

\begin{figure*}
	\centering
	\includegraphics[width=18cm]{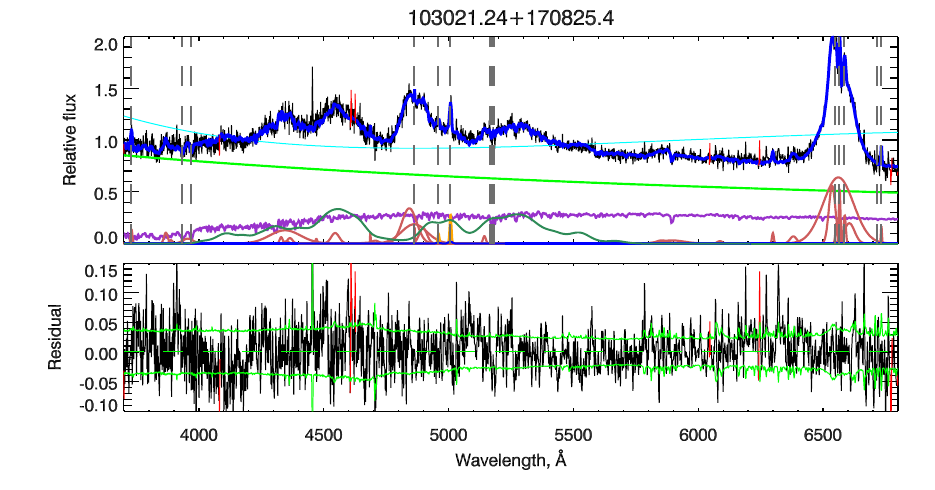} 
	\caption{The example of the spectra with strong FeII emission. 
         In the upper panel the black line represents the observed
		spectrum, blue line the best fit model, and the cyan line represents the multiplicative polynomial,
		while the green, light red, and violet lines represent components of the ULySS best
		fit model: violet -- stellar population, red -- emission lines, and green -- AGN
		continuum. Grey vertical lines mark the wavelengths in the air of the next lines: [OII]3727.5,
		CaII H \& K, H$\beta$ narrow component, [OIII]$\lambda\lambda$4959, 5007 narrow components and MgI b lines, 
		narrow components of H$\alpha$, [NII]$\lambda\lambda$46548,6583 and [SII]$\lambda\lambda$46716,6731, respectively.
		Residuals from the best fit (black line) are shown on the bottom panel .
		The dashed line is the zero-axis, and the green solid line shows the level of the noise. 
		Red lines in both panels correspond to outliers of the fit.
	}\label{strong_FeII}
\end{figure*}

\begin{figure*}%
		\centering
	\subfloat[]{{\includegraphics[width=0.95\columnwidth]{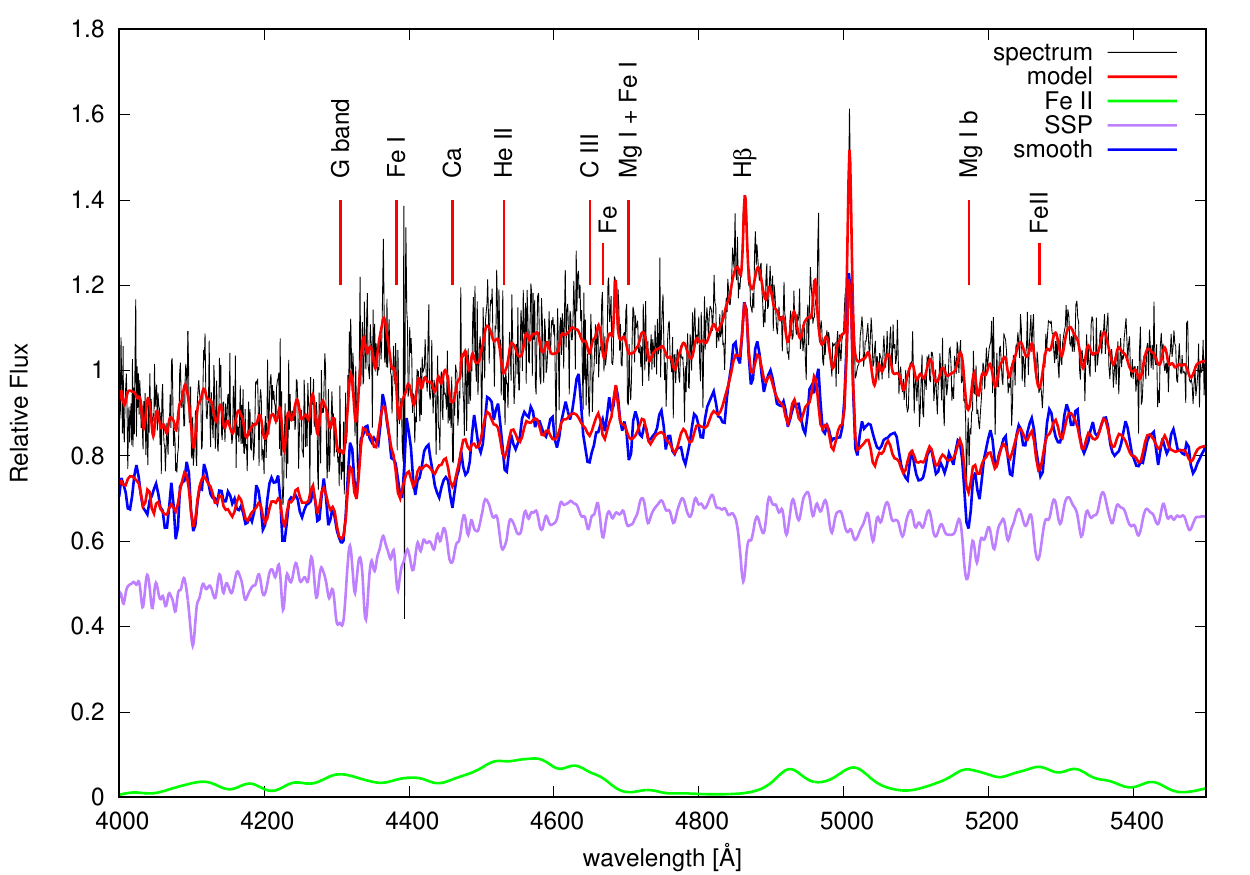} }}%
		\qquad
		\subfloat[]{{\includegraphics[width=0.95\columnwidth]{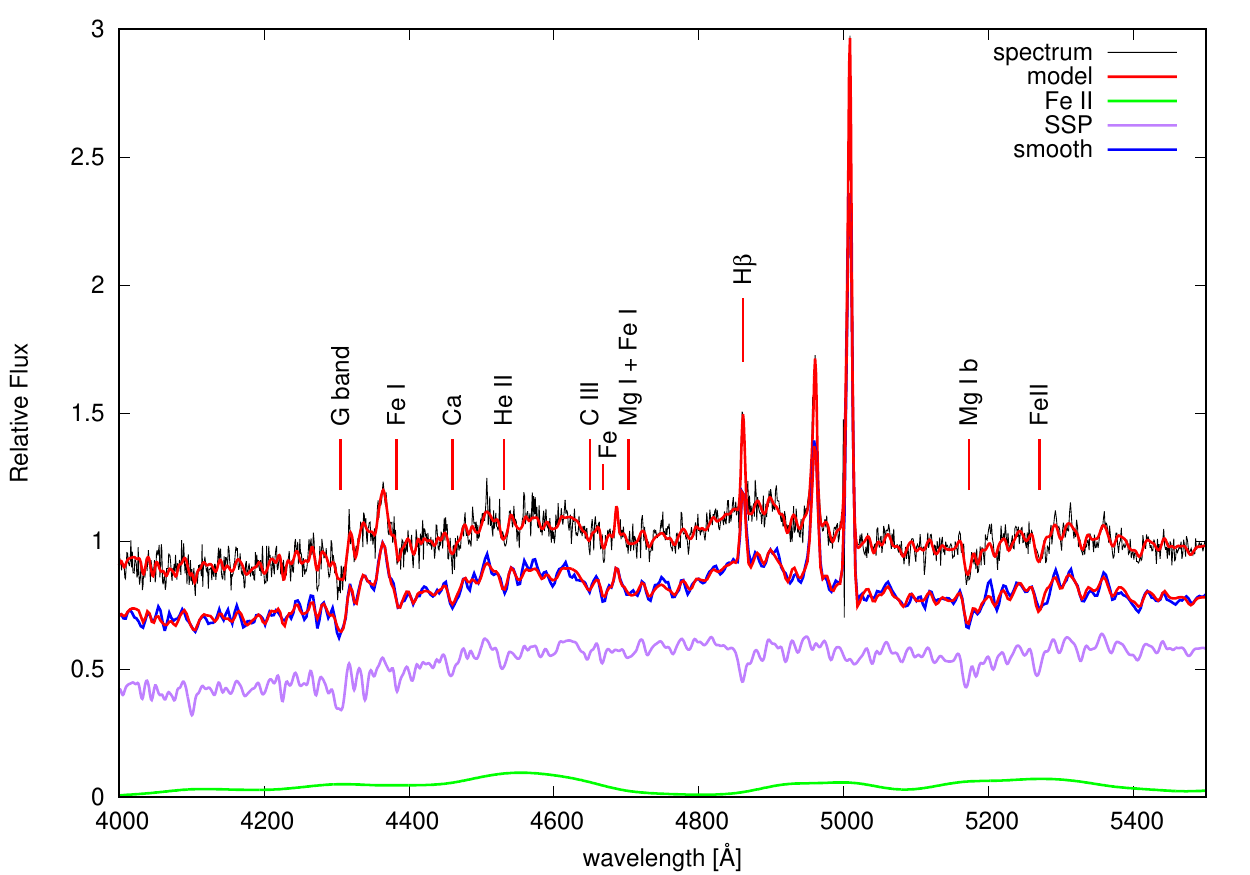} }}%
		\caption{Examples of spectra (J092620.62+101734.8 and J113651.66+445016.4, respectively) from our sample
		where the host galaxy mimic strong \feii\ emission leading to a mistaken identification of strong FeII emitters. 
			Spectra on the plots (a) and (b) have widely different SNR. The panels show (from top to bottom):  
			the real spectra and the best fitting model; the smoothed spectra overlapped with the best fitting model;
			single stellar population spectra that were used in the best fitting model; and
	the \feii\ template used in the fit. Some prominent absorption lines are
	marked on the plots. }%
		\label{mimics}%
	\end{figure*}

\subsection{The reason of the xA misclassification: contamination by host galaxy absorptions}\label{section_mimicing}

A main issue is why the HG sources were misclassified in the first place. 
An example of HG spectrum with the various  fit components is shown in Fig. \ref{strong_FeII}.  
We   notice  the high contribution of the  host galaxy spectrum which is a general feature of the sample.  
Only one source has SSP fraction between 10\% and 20\% to the total flux. 
In all other spectra we find very high fraction of the SSP component (in 16 object even higher then 40\%).  
The feature that can be used as an indication of strong contamination from the host galaxy is primarily 
the MgIb feature  that is almost always observed along with \hb\ and \oiiiopt. 
When absorption lines are prominent, and the fraction of the host galaxy is high,  
we detected that SSP is mimicking FeII, and that this may lead  to mistaken identification 
of FeII  spectral features (see Fig. \ref{mimics}). As one can see on the Fig. 
\ref{mimics}, the superposition of high fraction of the SSP on the FeII template, mimics   
FeII emission lines. This effect is more noticeable in the case of high SNR, as shown on the right hand side of the 
Fig. \ref{mimics}. 
The combined effect of the G band at about 4220 \AA\ and the Ca absorption at 4455 \AA\  creates
the impression of an excess emission around 4300 \AA, as expected from multiplets m27  and m28 (\feii\ 
multiplet wavelengths and information on spectral terms were taken from the \citealt{moore45} multiplet tables). 
The CaI absorption  apparently delimits the blue side of the $\lambda4570$ blend, due mostly to the \feii\ m38 and m37 lines. 
At the red end of the blend, the CIII 4650 \\ \AA\ ,  Fe 4668 \AA\ , and FeI absorptions at $4600 -  4650$ \AA\ help again 
create illusion of a bump around  
$\lambda4570$. The stellar continuum  remains relatively flat down to $\approx $ 4400 - 4500 \AA, and steepens short-ward; 
this behavior also contributes to the visual impression of a FeII $\lambda 4570$ emission blend. 

Similar considerations apply on the red side of \hb:  the MgIb ``green triplet'' 
cuts the continuum between the line of m42 at the blue edge of the blend (at 5169 \AA), and the shortest wavelength line of m49 at 5197 \AA. 
The FeI absorption at 5270 \AA\ corresponds roughly to a 5295 \AA\ dip between two pairs of lines of m48 and m49 (5265 \AA\ and 5316 \AA,  
corresponding to the transitions  z$^{4}$D$_{1\frac{1}{2}}^\mathrm{o}$ \, $\rightarrow$ a$^{4}$G$_{2\frac{1}{2}}$\ and 
z$^{4}$F$_{4\frac{1}{2}}^\mathrm{o}$ \, $\rightarrow$ a$^{4}$G$_{5\frac{1}{2}}$: \citealt{moore45}). 
Again, the Fe I absorption at 5335 \AA\ finds a rough correspondence in the dip  at $\approx$5349 \AA\ between two lines  of m48 and m49 (at 5316 \AA\ and 5363 \AA\ of m48). Last, at the red end of the $\lambda 5130$\ blend, FeI triplet at $\lambda$ 5406 with the  possible contribution of the HeII 5412 absorptions, contributes  the illusion of significant emission also on the red side of \hb.  This explains the misidentification of the xA sources by the automatic procedure or by a superficial inspection.   

\begin{figure}%
        \centering
    {{\includegraphics[width=0.95\columnwidth]{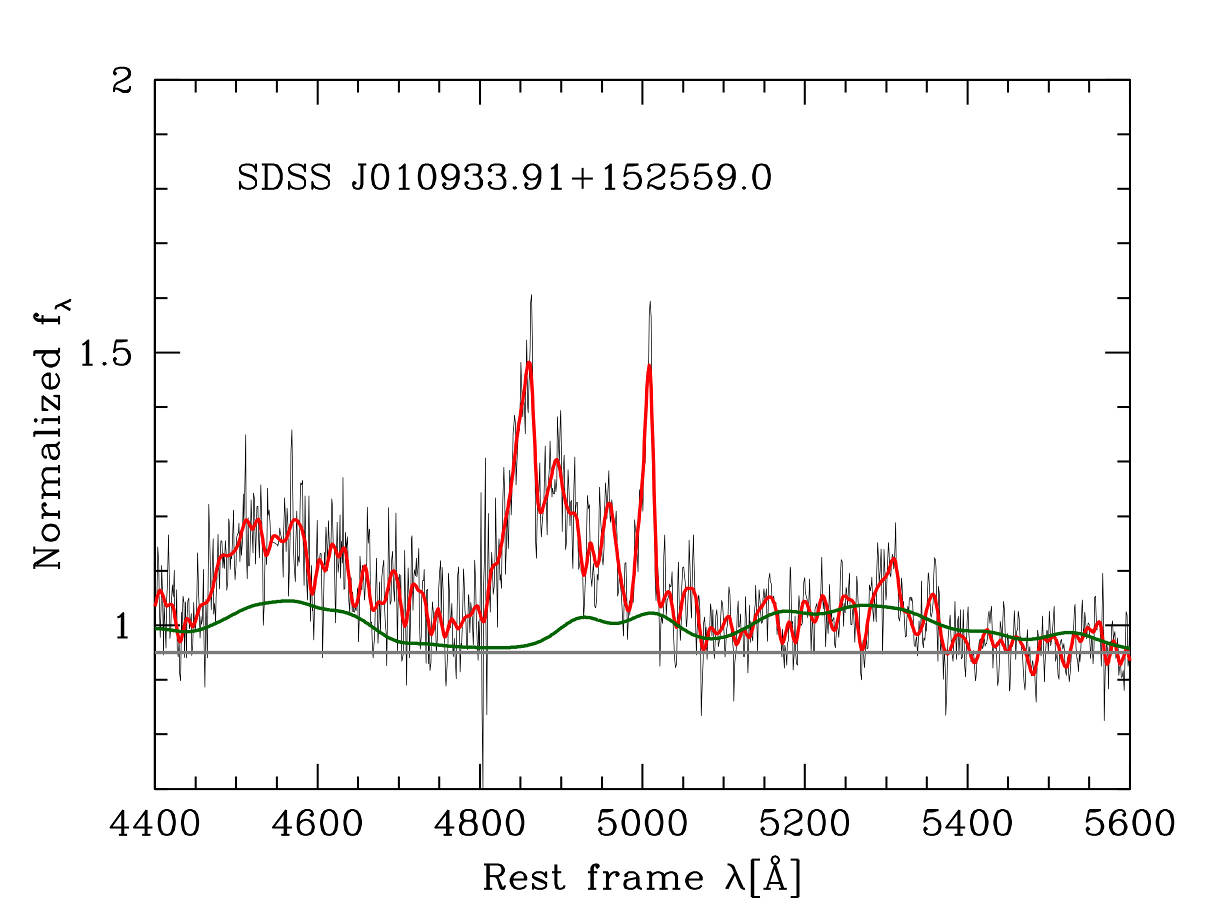} }}%
        \caption{Example of rest-frame spectrum whose features are misinterpreted because of heavy 
        contamination by the absorption spectrum of the host galaxy and because of insufficient spectral coverage. 
        The spectrum is shown with the original SDSS resolution (thin black line) and after rebinning to model 
        data with significantly lower resolution (thick red line). The flat continuum (grey line) erroneously 
        suggests a significant deviation from the \feii\ template (dark green line).  }%
        \label{fig:disp}%
    \end{figure}

In case of a significantly lower dispersion, the strong host contamination creates the appearance of a blue \feii\ 
emission blend at $\lambda$4570 much stronger than the red one at $\lambda 5300$, even if the S/N ratio is high. 
This phenomenon -- that can be misinterpreted if the spectral coverage does not extend below 4000 \AA\ in the rest frame -- 
might be responsible for early claims of a different blue-to-red \feii\ intensity ratio. Fig. \ref{fig:disp} 
illustrates how a spectrum heavily contaminated by the host galaxy, with insufficient spectral coverage and/or dispersion may 
lead to an incorrect placement of the continuum that in turns implies an anomalous ratio between the \feii\ blends on the 
blue and red side. Independently from resolution, little can be said on the spectrum if S/N$\lesssim$10: the G and MgIb 
bands are lost in noise if they have $W \sim 1 $ \AA. If the resolution is high, noise can be reduced by filtering, but 
little can be done in the case of low resolution ($\ll$ 1000).  An
accurate \feii\ measurements necessitates of S/N$\sim$30 in the
continuum, and of inverse spectral resolution R$\gtrsim$1000 in the
case of a significant contamination  by the host galaxy spectrum.

\begin{figure*}
	\begin{center}
		\includegraphics[width=0.33\textwidth]{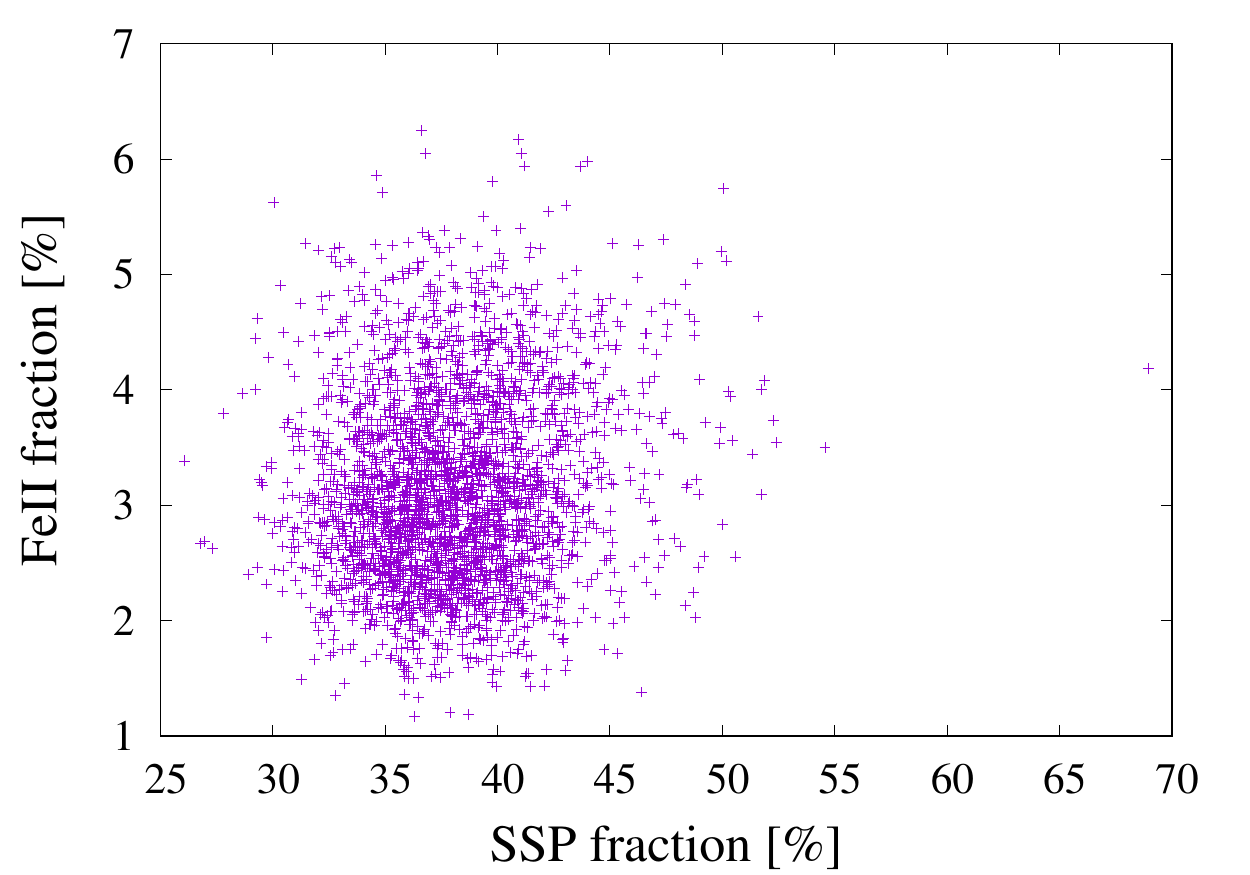}
		\includegraphics[width=0.33\textwidth]{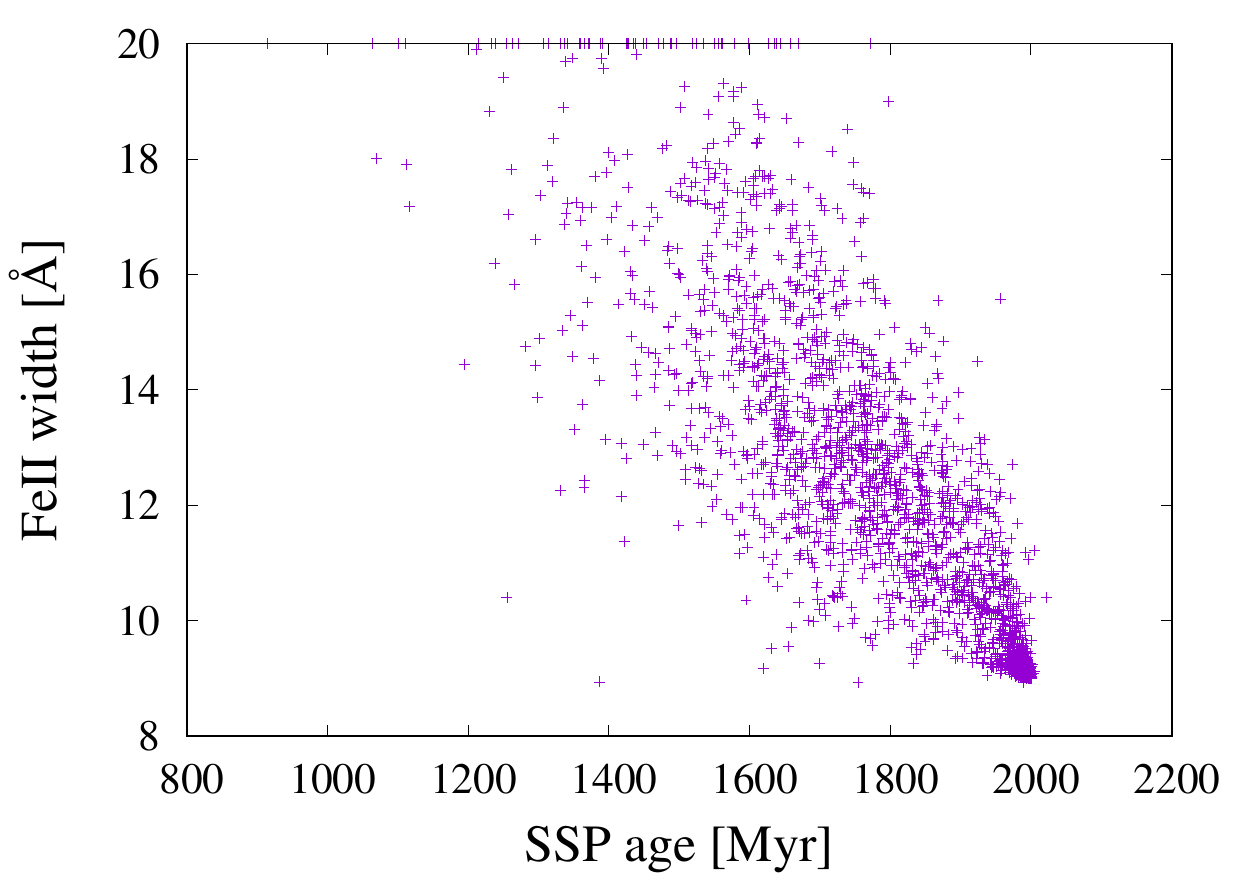}

		\caption{Results of 3000 Monte Carlo simulations for the object SDSSJ124533.87+534838.3
		reveal no dependency between SSP and FeII fraction 
		(left), and a degeneracy between SSP age and \feii\ width (right). 
		}\label{MonteCarlo}
	\end{center}
\end{figure*}

Monte Carlo simulations (for more information how one can use Monte Carlo simulation with ULySS see \citep{Kol09}) 
showed independence between prominence of \feii\ and SSP fraction, even if we might have expected to find correlation between 
these two parameters (see Fig. \ref{MonteCarlo}, left). Cross correlation also showed lack of dependence between these 
two parameters (for example, for the case of SDSS J124533.87+534838.3 we found $r=0.13$, $P=7.12E-12$). Besides, we 
found degeneracies between the age of the dominant stellar population on one hand and the fraction of \feii\ template and  
the width of emission lines that make up \feii\ template on the other, in the sense that we find older stellar populations 
when we have lower fraction of FeII, and narrower \feii\ lines (this confirms also  Pearson's cross-correlation 
which for example for SDSSJ 124533.87+534838.3 gives $r=-0.87$, $P=8E-15$). The Fig. \ref{MonteCarlo} shows just an example;
different ages are involved with different objects. Relative age inferences should not be affected, although the simulations 
show that the actual uncertainty is larger ($\sim 1 Gyr$) than the ones reported in Table \ref{longtable}, 
which are formal uncertainties from the MC simulations. Found degeneracies are not necessarily due to a physical reason, and could be due to 
the technical fitting problem. In order to decrease degeneracies between parameters of stellar population and FeII template to the minimum, it is advisable to perform the 
simultaneous fit of these components of the model, as done. This implies that the degeneracies could be even higher in the 
non-simultaneous fit of the components.

 \begin{table*} 
 	\caption{Spectral classification.}\label{classification}	
 		{\fontsize{8.2}{8} \selectfont\tabcolsep=3pt
 			\begin{tabular}{cccccccccccccr}
 				\hline\hline\\[0.5ex]
 				\hspace{5mm}  SDSS ID	&	FWHM H$\beta_\mathrm{BC}$ 	&	Flux H$\beta_\mathrm{BC} $  	&	$R_{FeII}$	&	$EV1_{class}$	&		$AGN_{class}$	&	FWHM(H$\beta_\mathrm{BC})^{\ast}$	&	$R_\mathrm{FeII}^{\ast}$	&	$EV1_{class}^{\ast}$	&	FW 1/4	&	FW 3/4	&	FW 9/10	&	C 1/4	&	C 1/2	\\
 				\hspace{7mm}(1)	&	(2)	&	(3)	&	(4)	&	(5)	&	(6)	&	(7)	&	(8)	&	(9)	&	(10)	&	(11)	&	(13)	&	(13)	&	(14)	\\
 				\hline\\[0.5ex]
J003657.17-100810.6	&	4827	&	27.1	$\pm$	2.3	&	0.9	$\pm$	0.1	&	B2	&	S1	&	5860	&	1.1	&	B3	&	8634	&	3104	&	1933	&	239	&	-300	\\
J010933.91+152559.0	&	4903	&	36.4	$\pm$	6.7	&	0.8	$\pm$	0.2	&	B2	&	S1	&	4629	&	0.9	&	B2	&	7675	&	3038	&	1795	&	526	&	166	\\
J011807.98+150512.9	&	9501	&	36.7	$\pm$	5.6	&	0.5	$\pm$	0.2	&	B2+	&	S1	&	11091	&	0.9	&	B2+	&	13381	&	6103	&	3675	&	1390	&	1423	\\
J031715.10-073822.3	&	8564	&	28.1	$\pm$	1.2	&	0.3	$\pm$	0.1	&	B1+	&	S1	&	9024	&	0.8	&	B2+	&	12605	&	4473	&	2475	&	1063	&	198	\\
J075059.82+352005.2	&	6830	&	38.8	$\pm$	4.4	&	0.6	$\pm$	0.2	&	B2	&	S1	&	8008	&	0.8	&	B2+	&	10371	&	5103	&	1718	&	541	&	-141	\\
J082205.19+584058.3	&	8136	&	36.7	$\pm$	4.1	&	0.8	$\pm$	0.1	&	B2+	&	--	&	7729	&	0.9	&	B2	&	11652	&	5171	&	3102	&	-316	&	-311	\\
J082205.24+455349.1	&	7538	&	32.8	$\pm$	4.6	&	0.9	$\pm$	0.1	&	B2	&	S1	&	9677	&	1.2	&	B3+	&	11626	&	5044	&	3591	&	800	&	595	\\
J091017.07+060238.6	&	7042	&	32.5	$\pm$	4.5	&	0.5	$\pm$	0.3	&	B2	&	AGN	&	7947	&	0.7	&	B2	&	9888	&	5592	&	4695	&	562	&	93	\\
J091020.11+312417.8	&	5807	&	16.6	$\pm$	1.2	&	0.5	$\pm$	0.2	&	B1	&	--	&	6469	&	0.7	&	B2	&	7120	&	4770	&	1170	&	447	&	474	\\
J092620.62+101734.8	&	4906	&	28.4	$\pm$	7.0	&	0.7	$\pm$	0.3	&	B2	&	S1	&	8350	&	0.9	&	B2+	&	8287	&	2972	&	1728	&	166	&	320	\\
J094249.40+593206.4	&	3880	&	19.9	$\pm$	0.7	&	1.1	$\pm$	0.2	&	A3	&	S1	&	3982	&	1.2	&	A3	&	7252	&	2219	&	1318	&	209	&	1127	\\
J094305.88+535048.4	&	9578	&	34.5	$\pm$	3.2	&	0.4	$\pm$	0.2	&	B1+	&	AGN	&	9457	&	0.5	&	B2+	&	13643	&	6343	&	4626	&	-478	&	-483	\\
J103021.24+170825.4	&	6416	&	50.4	$\pm$	6.4	&	1.2	$\pm$	0.2	&	B3	&	--	&	6285	&	1.7	&	B4	&	8414	&	4275	&	2065	&	-251	&	-160	\\
J105530.40+132117.7	&	5228	&	12.6	$\pm$	0.7	&	1.5	$\pm$	0.1	&	B3	&	X	&	8013	&	1.8	&	B4+	&	7985	&	3233	&	1926	&	-791	&	-1012	\\
J105705.40+580437.4	&	2418	&	26.8	$\pm$	2.5	&	0.6	$\pm$	0.1	&	A2	&	--	&	2826	&	0.8	&	A2	&	3868	&	1520	&	898	&	208	&	239	\\
J112930.76+431017.3	&	5176	&	22.8	$\pm$	5.2	&	0.7	$\pm$	0.3	&	B2	&	S1	&	9154	&	1.1	&	B3+	&	8909	&	3036	&	1794	&	197	&	-32	\\
J113630.11+621902.4	&	6982	&	31.0	$\pm$	2.1	&	0.6	$\pm$	0.1	&	B2	&	AGN	&	7445	&	0.9	&	B2	&	11417	&	4140	&	1929	&	757	&	447	\\
J113651.66+445016.4	&	5118	&	16.6	$\pm$	2.4	&	0.7	$\pm$	0.2	&	B2	&	S1	&	7214	&	1.1	&	B3	&	8099	&	3181	&	1867	&	876	&	617	\\
J123431.08+515629.2	&	5662	&	26.0	$\pm$	4.9	&	0.6	$\pm$	0.3	&	B2	&	AGN	&	8896	&	0.7	&	B2+	&	9765	&	4005	&	2970	&	1019	&	129	\\
J124533.87+534838.3	&	6484	&	33.6	$\pm$	7.5	&	0.5	$\pm$	0.3	&	B2	&	AGN	&	6956	&	0.7	&	B2	&	10050	&	4353	&	3043	&	-802	&	-207	\\
J125219.55+182036.0	&	6272	&	29.5	$\pm$	2.7	&	0.6	$\pm$	0.1	&	B2	&	--	&	7675	&	0.8	&	B2	&	10749	&	3038	&	1242	&	-497	&	-426	\\
J133612.29+094746.8	&	5247	&	26.5	$\pm$	5.0	&	1.0	$\pm$	0.2	&	B2	&	--	&	6556	&	1.3	&	B3	&	7310	&	3664	&	1248	&	-255	&	80	\\
J134748.06+404632.6	&	4427	&	29.3	$\pm$	3.0	&	0.4	$\pm$	0.2	&	B1	&	S1	&	5208	&	0.5	&	B1	&	6638	&	2837	&	1730	&	521	&	650	\\
J134938.08+033543.8	&	3781	&	19.1	$\pm$	1.1	&	0.8	$\pm$	0.1	&	A2	&	AGN	&	3464	&	1.1	&	A3	&	6402	&	2337	&	1375	&	-784	&	-1208	\\
J135008.55+233146.0	&	10145	&	46.0	$\pm$	1.6	&	0.9	$\pm$	0.0	&	B2+	&	X	&	11098	&	1.2	&	B3+	&	14354	&	6556	&	4003	&	-11	&	-19	\\
J141131.86+442001.0	&	6219	&	20.9	$\pm$	2.0	&	0.4	$\pm$	0.2	&	B1	&	S1	&	6779	&	0.8	&	B2	&	7736	&	4977	&	3941	&	207	&	336	\\
J143651.50+343602.4	&	6985	&	32.5	$\pm$	6.3	&	0.5	$\pm$	0.2	&	B2	&	S1	&	7102	&	0.7	&	B2	&	9204	&	5533	&	4565	&	806	&	583	\\
J151600.39+572415.7	&	3798	&	16.6	$\pm$	0.1	&	0.6	$\pm$	0.1	&	A2	&	S1	&	4444	&	0.6	&	B2	&	5666	&	2555	&	1519	&	326	&	145	\\
J155950.79+512504.1	&	4765	&	31.4	$\pm$	4.0	&	0.7	$\pm$	0.2	&	B2	&	S1	&	5593	&	0.9	&	B2	&	7385	&	1790	&	964	&	-29	&	165	\\
J161002.70+202108.5	&	4424	&	19.5	$\pm$	1.2	&	1.1	$\pm$	0.1	&	B3	&	--	&	5080	&	1.2	&	B3	&	6428	&	2834	&	1728	&	416	&	441	\\
J162612.16+143029.0	&	9017	&	31.8	$\pm$	2.6	&	1.2	$\pm$	0.1	&	B3+	&	--	&	9620	&	1.4	&	B3+	&	12665	&	5783	&	3511	&	-805	&	-803	\\
J170250.46+334409.6	&	5588	&	42.7	$\pm$	4.6	&	0.5	$\pm$	0.2	&	B2	&	AGN	&	6528	&	0.7	&	B2	&	10023	&	3935	&	2902	&	444	&	-137	\\
 				 		  \hline		\end{tabular}}
  {Notes:} (1) SDSS ID of the object; (2) FWHM H$\beta$; (3) F H$\beta$ BC - flux of broad H$\beta$ line component; 
  (4) $R_{FeII}$; (5) classification of the spectra using EV1 diagram;  (6) AGN classification according to \citet{veron-cettyveron06}--Sy 1 - Seyfert 1 galaxy; -- AGN - unclassified AGN;  (7) FWHM of broad H$\beta$ from \citep{shenho14};  (8) calculated $R_{FeII}$ using data from \citep{shenho14}; (9) classification of spectra on EV1 diagram, calculated from \citep{shenho14} data. (10) FW 1/4 - full width of H$\beta$ line; (11) FW3/4 of H$\beta$; (12) FW9/10 H$\beta$; (13) C 1/4 - centroid of H$\beta$ line measured at 1/4 of maximum intensity;	(7) C 1/2 - centroid of H$\beta$ line measured at 1/2 of maximum intensity. 
 \end{table*}

\subsection{Consistency between AGN emission and host galaxy absorption
spectrum}\label{consistency_cz_hb_oiii}

\begin{figure}[h!]
 \centering
 \includegraphics[width=\columnwidth]{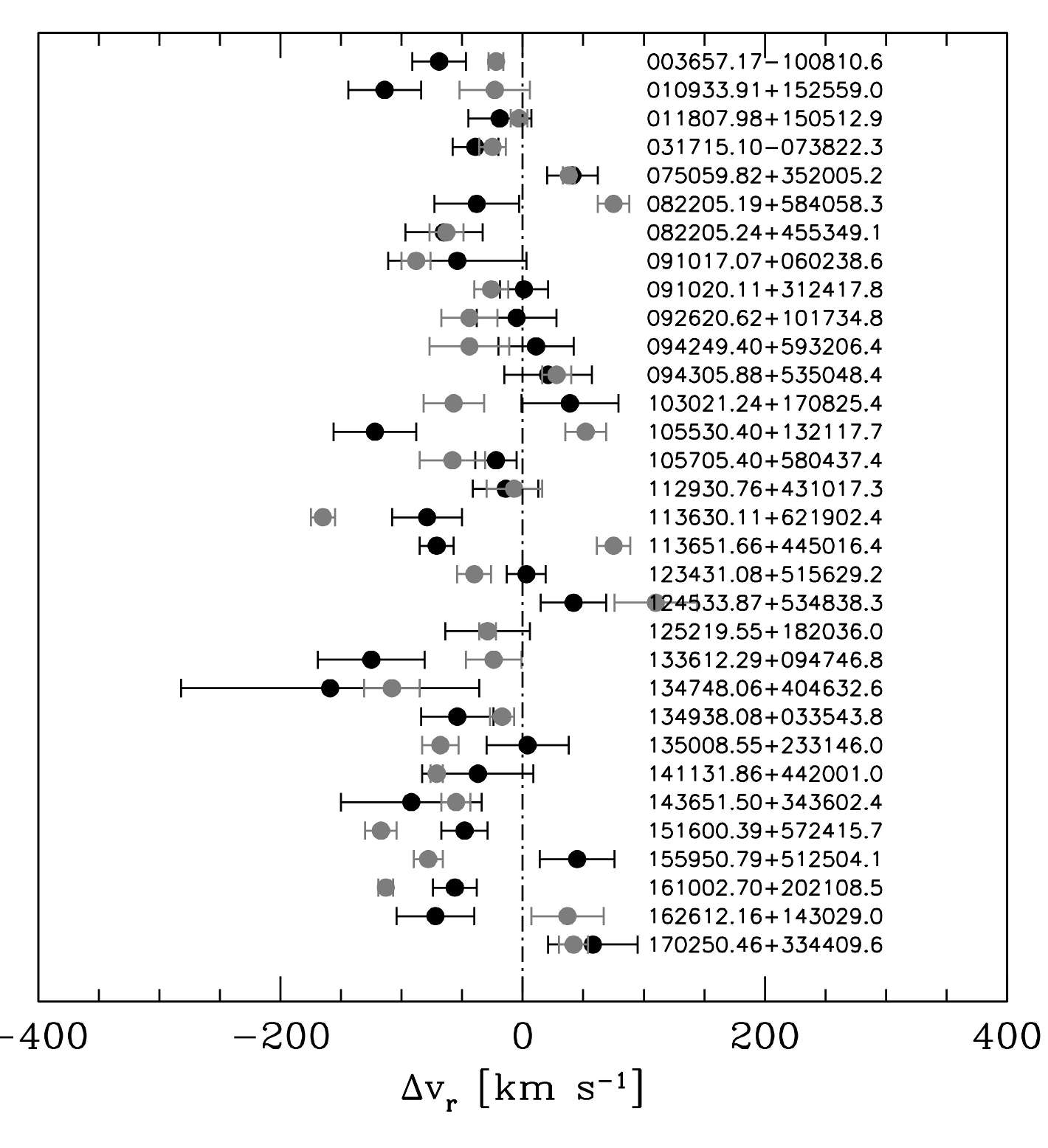} 
 \caption{Radial velocity difference between \hbnc\ (grey) and \oiiiopt\ (black) with respect to the HG reference frame. }\label{fig:z}
\end{figure}

Generally speaking, there is a good consistency between estimators of the systemic redshift of the host galaxy and low 
ionization narrow emission lines \citep[a fact known since the early study of][]{condonetal85}. The systemic redshift of the 
host may be estimated using the atomic 21 cm hydrogen lines or emission from molecular CO which usually give results in 
close agreement \citep{mirabelsanders88}. A third method is provided by the absorption features of the old stellar 
population of the host galaxies. The tips of the narrow emission line \hb\ and H$\alpha$\ can be considered the best 
estimator the system redshift of the host galaxy \citep{letaweetal07}. Significant  differences are found mostly for the 
high ionization lines such as \oiiiopt. The agreement between narrow low-ionization lines and the systemic redshift 
estimators has the important implication that any shift with respect to them can be considered also a shift with respect 
to the host. This is an advantage as the inter-line shifts between low- and high- narrow ionization lines are easy to 
measure. The amplitude of the relative shifts is known to depend on the location along the main sequence. In extreme Pop. B 
shifts between Balmer lines and \oiii\ are generally modest $\lesssim 100$ \kms\ \citep{eracleoushalpern04}. In Pop. B 
\oiiiopt\ are often blueward asymmetric close to the line base, but the peak shift is roughly consistent with systemic 
redshift \citep[see the diagram of average \oiii\ shift along the MS in][]{marzianietal18}. In Pop. A and especially among 
xAs the \oiii\ shifts become larger, and may reach several hundred \kms\ in the case of the so-called blue outliers 
\citep{zamanovetal02,komossaetal08,zhangetal11,craccoetal16,marzianietal16a}, believed to be relatively frequent at high 
Eddington ratio or high luminosity. 

Figure \ref{fig:z}  shows the radial velocity difference between \hbnc\ (grey) and \oiiiopt\ (black) with respect to the mean stellar  velocity reference frame (HG). The comparison shows that 1) both \hb\ and \oiiiopt\ 
shifts are consistent with HG with some scatter (54 \kms for the case of H$\beta$ and 61 \kms for [OIII]).

The Pearson's cross-correlation between parameters pointed out the high cross-correlation coefficient (r=0.62, P-value=4.83E-05) between 
the shift of narrow component of H$\beta$ line and cz. 
On the other hand we did not find the correlation between the 
SSP cz and the shift of narrow component of [OIII]4959,5007 lines (Pearson's cross-correlation coefficient is just 
r=0.27, and P-value=0.12).

\begin{figure*}
	\begin{center}
		\includegraphics[width=0.9\textwidth]{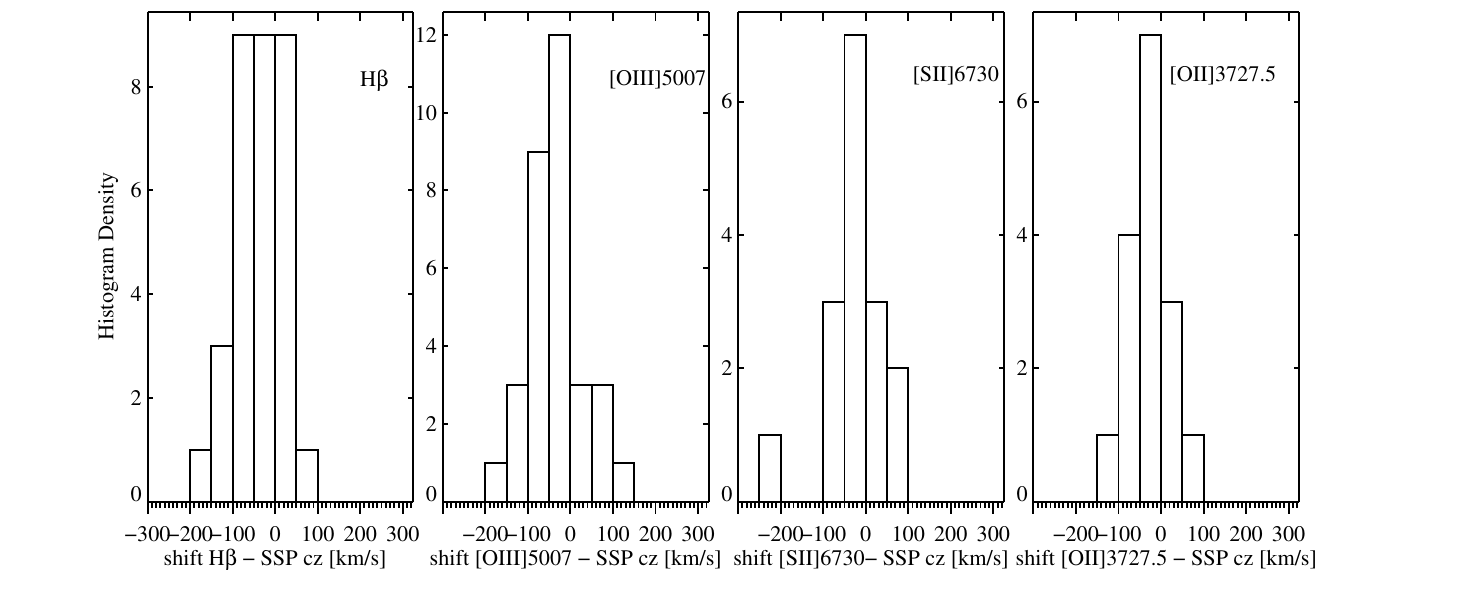}
		\caption{Distribution of the difference in values of z
			derived from mean stellar velocity and different emission lines. From left to
			right: H$\beta$, [OIII]$\lambda$5007, [SII]$\lambda$6731, and [OII]$\lambda$3728.5.
		}\label{shifts_histograms}
	\end{center}
\end{figure*}

In Figure \ref{shifts_histograms}  we compare measurements of shifts, derived from narrow components of different lines: H$\beta$, 
\oiii, [SII]$\lambda$6731 and [OII]$\lambda$3727.5, in respect to the mean stellar velocity (cz). 
We notice a small systematic effects of blueshift of the NLR with respect to the host (-34 $\pm$ 54 \kms, -31 $\pm$ 61 \kms, -22  $\pm$ 72, 
and -25 $\pm$ 50, for H$\beta$, 
\oiii, [SII]$\lambda$6731 and [OII]$\lambda$3727.5, respectively.)

\subsubsection{\oii}

The \oii\ doublet deserves special attention. The ratio of the two components of the doublet  $\mathcal{R} =I(^{2}$D$_{5/2} \rightarrow ^{4}$S$_{\frac{3}{2}})/I(^{2}$D$_{3/2} \rightarrow ^{4}$S$_{\frac{3}{2}}$) = $I$($\lambda$3729)/$I$($\lambda$3727) is sensitive to electron density \nelec\ \citep{osterbrockferland06} with an extremely weak dependence on electron temperature \citep{cantoetal80}. The wavelengths are 3726.04 and 3728.80 \AA\ in air and 3727.10 and 3729.86 \AA \ in vacuum.  When the doublet is resolved, the measurement of the two component is straightforward. However, the spectral resolution of the SDSS and the intrinsic width of the \oii\ doublet in AGNs make the doublet most often unresolved. In this case, the peak wavelength of the \oii\ doublet is sensitive to the ratio and hence to \nelec\ (see Appendix \ref{consistency_cz_hb_oii} for a discussion on the issue). 

\begin{figure}
 \begin{center}
\includegraphics[width=0.4\textwidth]{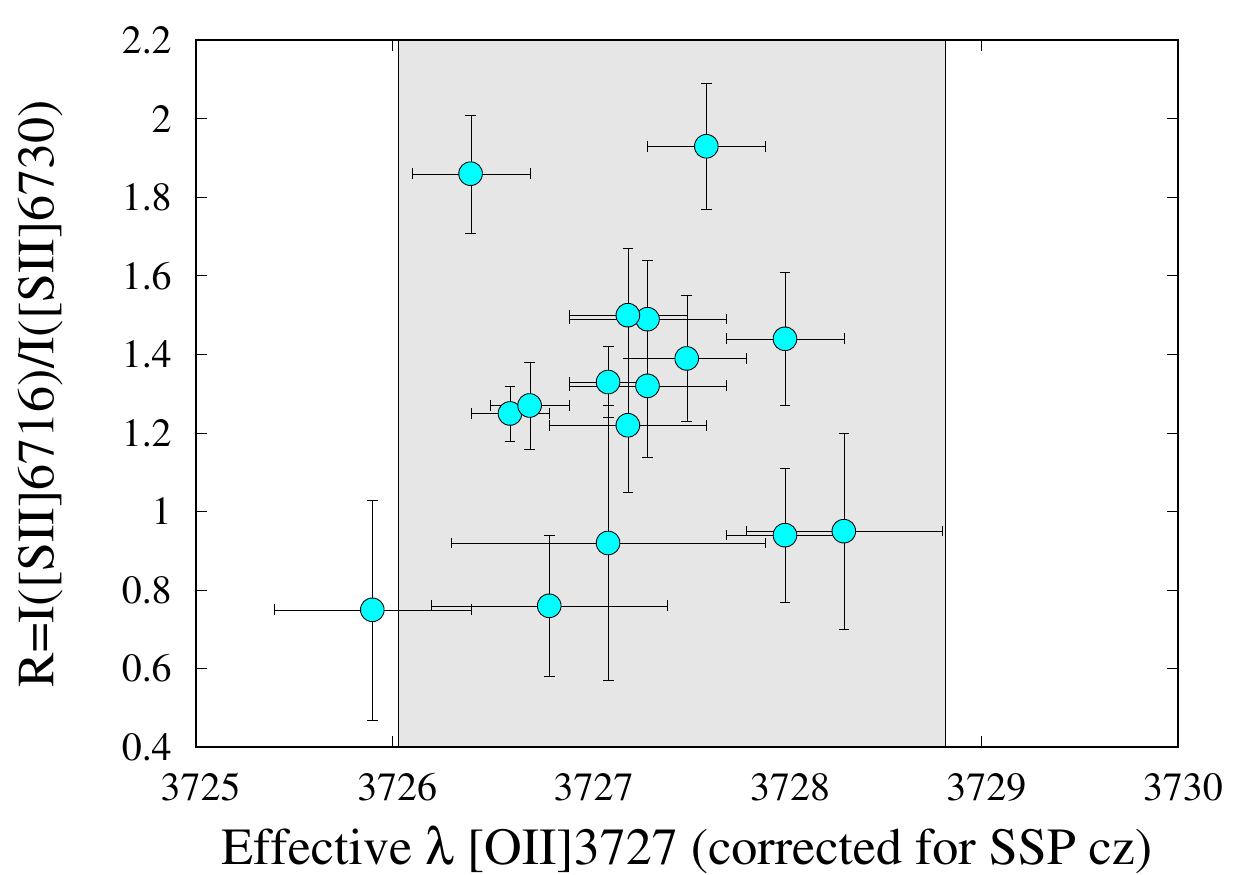}
\caption{ {The intensity ratio [SII]$\lambda$6717/$\lambda$6731 as a function of effective wavelength of 
unresolved [OII]$\lambda\lambda$3726,3729 doublet corrected for SSP cz. 
Region inside the physical limits of the effective wavelengths of
\oii \ is shaded on the plot. }}\label{R_vs_[OII]}
\end{center}
\end{figure}

Since [OII]$\lambda\lambda$3726,3729 lines in our sample are not resolved, we fitted the \oii \ doublet 
with a single Gaussian. We used the ratio between [SII]$\lambda\lambda$6717,6731 lines to test a correspondence between 
the wavelength peak and an independent density estimator (the procedure works relatively well for \hii\ spectra, 
as described in the Appendix \ref{consistency_cz_hb_oii}).  Only in the case of 16 objects we succeeded to fit 
[SII]$\lambda\lambda$6717,6731 lines, and therefore to calculate their intensity ratio. Table \ref{Table_OIIeff_R} list the measured 
effective wavelength of [OII]$\lambda\lambda$3726,3729 doublet, effective [OII] wavelength corrected for the SSP shift, and the 
ratio between the intensities of 
[SII] lines. Figure \ref{R_vs_[OII]}  represents the $R$[SII]=[SII]$\lambda$6717/$\lambda$6731 as a function of 
[OII]$\lambda\lambda$3726,3729 doublet effective wavelength, for unresolved doublets, corrected for the shift of SSP $cz$. 
We emphasizes the importance of the de-shifting the spectra for SSP $cz$, since yet after the de-shifting the spectra 
for SSP $cz$, the correlation between effective 
wavelength of [OII]  and [SII]$\lambda\lambda$6717,6731 intensity comes into agreement with theoretical predictions. 
There is an overall consistency between the prediction of the  $R$[SII] and the effective wavelength 
$\lambda_\mathrm{eff}$ of the \oii\ doublet. Only three sources deviate from a clear trend, in one case $R$[SII] suggesting low density and $\lambda_\mathrm{eff}$\ high density, while in two cases the $\lambda_\mathrm{eff}$\ around 3728 \AA\ in air is suggesting low density and    $R$[SII]\ high density.  
Accepted at face values, the first condition may be associated with \oii\ being predominantly emitted in the AGN 
narrow-line region, while the second case may imply dominance by \hii\ regions in the \oii, and perhaps by a 
denser shock-heated region for the\ \sii\ emission. However, these inferences remain highly speculative, 
given the possibility that blueshifted emission associated with a wind may contaminate at a low-level the 
\oii\ profile \citep{kauffmannmaraston19}. A larger sample with higher S/N is needed to ascertain whether 
these discrepancies are seen statistically, and  may hint at a particular physical scenario.

Great care should be used in assuming a reference wavelength for \oii. \hii\ regions may be dominated by relatively-low density emission, yielding $\mathcal{R}$[SII]$\approx 1.5$. On the converse, emission within the NLR may be weighted in favor of much higher density gas (\nelec\  $\gtrsim 10^{3}$ \cmc), which implies ${R} \approx 0.4$. It cannot be given for granted that the spectra of our sample are dominated by NLR emission. The \oii/\oiii\ ratio is larger in \hii\ than in AGN.  The SDSS aperture at the typical $z \approx 0.25$, the scale is 3.943 kpc arcsec$^{-1}$; within the 3 arcsec aperture, $\lesssim 12$ kpc, most of the light of the host galaxy should be also included. AGN show complex density behavior in their circumnuclear regions, depending on the presence of nuclear outflows   \citep{maddox18,kakkadetal18}, and some mixing between high-ionization narrow line region gas and low-ionization \hii\ regions is found for fixed size apertures \citep{thomasetal18}. Electron density is also dependent on star formation rate \citep{kaasinenetal17}.  We might therefore expect a dependence on physical condition as well as on aperture size.  

The dependence on \nelec\ implies a wavelength shift that is $\lesssim$ 1.5\AA\ (Appendix \ref{consistency_cz_hb_oii}), and therefore much larger than the accuracy of the wavelength scale of SDSS spectra. One should never forget that neglecting the dependence on density, and using a fixed wavelength as a reference, may bias redshift estimates and at least introduce a significant scatter if \hbnc\ and \oii\ redshift are averaged together, even if in most cases is not possible to do otherwise. The average wavelength of the present sample  is $\bar{\lambda}_\mathrm{eff} \approx 3728.3$ \AA\ (vacuum) and 3727.2 \AA\ (air), which corresponds to $R$[OII] %
around unity, and \nelec\  $\sim 10^{2.7}$ \cmc\ \citep[Fig. 8.6 of ][]{pradhannahar15}. The value is not far from the expectation for the lower density limit typical of the NLR \citep{netzer90}. This results may be a direct consequence of the location of the sources along the MS. For xA sources,  \nelec\ might be higher reflecting a compact NLR with a larger density \citep{zamanovetal02}. On the other hand, if the aperture is large enough, circumnuclear and nuclear star formation may be dominating the \oii\ emission. Ascertain the  systematic trends of $\mathcal{R}$ would require an extensive work whose scope is much beyond that of the present work.

\begin{table} 
 	\caption{[OII]$\lambda\lambda$3726,3729 effective wavelengths}\label{Table_OIIeff_R}	
 		{\fontsize{8.2}{8} \selectfont\tabcolsep=3pt
 			\begin{tabular}{cccr}
 				\hline\hline\\[0.5ex]
 				\hspace{5mm}  SDSS ID	& [OII]$\lambda\lambda$3726,3729 	&	$\rm[OII]^{*}\lambda\lambda$3726,3729	&	R	\\
                                \hspace{7mm} & [\AA] & [\AA] & \\
 \hspace{7mm}(1)	&	(2)	&	(3)	&	(4)	\\
 				\hline\\[0.5ex]
J003657.17-100810.6	&	3728.2	$\pm$	0.3	&	3727.5	&	1.39	$\pm$	0.16	\\
J010933.91+152559.0	&	3728.7	$\pm$	0.4	&	3727.2	&	1.22	$\pm$	0.17	\\
J091020.11+312417.8	&	3728.5	$\pm$	0.2	&	3727.6	&	1.93	$\pm$	0.16	\\
J092620.62+101734.8	&	3727.9	$\pm$	0.8	&	3727.1	&	0.92	$\pm$	0.35	\\
J105705.40+580437.4	&	3728.0	$\pm$	0.4	&	3727.3	&	1.49	$\pm$	0.15	\\
J113630.11+621902.4	&	3728.0	$\pm$	0.3	&	3726.4	&	1.86	$\pm$	0.15	\\
J113651.66+445016.4	&	3727.0	$\pm$	0.2	&	3727.1	&	1.33	$\pm$	0.09	\\
J123431.08+515629.2	&	3728.5	$\pm$	0.3	&	3728.0	&	1.44	$\pm$	0.17	\\
J125219.55+182036.0	&	3728.5	$\pm$	0.2	&	3727.2	&	1.50	$\pm$	0.17	\\
J134748.06+404632.6	&	3727.5	$\pm$	0.5	&	3725.9	&	0.75	$\pm$	0.28	\\
J134938.08+033543.8	&	3729.1	$\pm$	0.2	&	3728.0	&	0.94	$\pm$	0.17	\\
J151600.39+572415.7	&	3727.8	$\pm$	0.1	&	3726.6	&	1.25	$\pm$	0.07	\\
J155950.79+512504.1	&	3728.8	$\pm$	0.2	&	3727.3	&	1.32	$\pm$	0.18	\\
J161002.70+202108.5	&	3727.7	$\pm$	0.2	&	3726.7	&	1.27	$\pm$	0.11	\\
J162612.16+143029.0	&	3728.2	$\pm$	0.5	&	3726.8	&	0.76	$\pm$	0.18	\\
J170250.46+334409.6	&	3729.0	$\pm$	0.4	&	3728.3	&	0.95	$\pm$	0.25	\\
\hline		
\end{tabular}
  Notes: (1) SDSS ID of the object; (2) [OII]$\lambda\lambda$3726,3729 effective wavelength; 
  (3) [OII]$\lambda\lambda$3726,3729 effective wavelength corrected for SSP shift; 
  (4) R - ratio between intensity of [SII]$\lambda\lambda$6717,6731 lines.  }
 \end{table}

 \subsection{Relation between velocity dispersion of stellar and narrow-line components}

There is considerable interest in the correlation of the supermassive black holes (SMBHs)
masses \mbh, with the stellar velocity dispersion  of  the host galaxy bulge \citep{gebhardtetal00, ferraresemerritt00,kormendyho13} because of its important implications to the coevolution of galaxies  and their SMBHs. A problem affecting the definition of the \mbh-$\sigma_{\star}$ relation for AGNs is that a strong optical  continuum emission from the AGN accretion disk can make measuring $\sigma_{\star}$\ difficult. 
\citet{nelsonwhittle96} proposed 
using FWHM \oiii\ as a proxy for $\sigma_{\star} \times \ 2.355$ because  the \oiii\ lines are strong and easily observable. The problem is that \oiiiopt\  often display blue asymmetries, most often explained as an outflow component \cite{heckmanetal81}, that increase the scatter of the
\mbh -- $\sigma_{\star}$ relation. Measuring of $\sigma_{\star}$ is more complicated in the case of AGN Type 1 because of the high influence of broad emission lines and strong 
featureless non-stellar continuum. Besides, just in several cases of recent studies 
\citep{duetal16,sextonetal19} $\sigma_\mathrm{[OIII]}$ and $\sigma_{\star}$ has been measured simultaneously.  
\citet{sextonetal19} showed  that fitting the \oiii\ line with a single Gaussian or Gauss-Hermite polynomials 
overestimates $\sigma_{\star}$\ by more then 50\%. Moreover, they showed that even when they exclude line asymmetries 
from non-gravitational gas motion in a fit with two Gaussians, there is no correlation between narrow component of 
$\sigma_\mathrm{[OIII]}$ and $\sigma_{\star}$. The fact that these two parameters have the same range, average and 
standard deviation implies that they are under the same gravitational potential \citep{sextonetal19}. 
They suggest that the large scatter is probably caused by dependency between the line profiles and the light 
distribution and underlying kinematic field. Because of this \cite{sextonetal19} strongly caution that the [OIII] 
width can not be used as a proxy for $\sigma_{\star}$ on an individual basis. This confirms the results of 
\citet{bennertetal18} who showed that $\sigma_\mathrm{[OIII]}$ can be used as a surrogate for $\sigma_{\star}$ only in 
statistical studies.\\
\indent \citet{komossaxu07}  suggested to use \sii\ as a surrogate for $\sigma_{\star}$
since the sulfur lines have a lower ionization potential and do not suffer from significant asymmetries, but the 
scatter is comparable to that of the core of the [OIII] line.\\
\indent In this work we confirm results of \citep{sextonetal19} since we found no correlation between $\sigma_{\star}$  
and the velocity dispersion of \oiii \ narrow component. We instead found a high correlation between $\sigma_{\star}$ and 
the velocity dispersion of \hb\ narrow component ($r\approx$0.64, $P\approx1.93E-05$). There is an overall consistency 
between the  values of $\sigma_{\star}$ and both $\sigma_\mathrm{[OIII]}$ and
$\sigma_\mathrm{H\beta}$: the median values of the ratios involving the three parameters and their semi inter-quartile  ranges (SIQR) are:
$<\sigma_\mathrm{[OIII]}$/$\sigma_{\star}> \approx 0.98 \pm 0.23$, and
$\sigma_\mathrm{H\beta}$/$\sigma_{\star} \approx 1.03 \pm 0.19$, 
$\sigma_\mathrm{[OIII]}$/$\sigma_\mathrm{H\beta} \approx 0.93 \pm 0.22$. \\

\subsection{No strong outflows  diagnosed by the \oiii\ profile}

\begin{figure*}
	\begin{center}
		\includegraphics[width=0.4\textwidth]{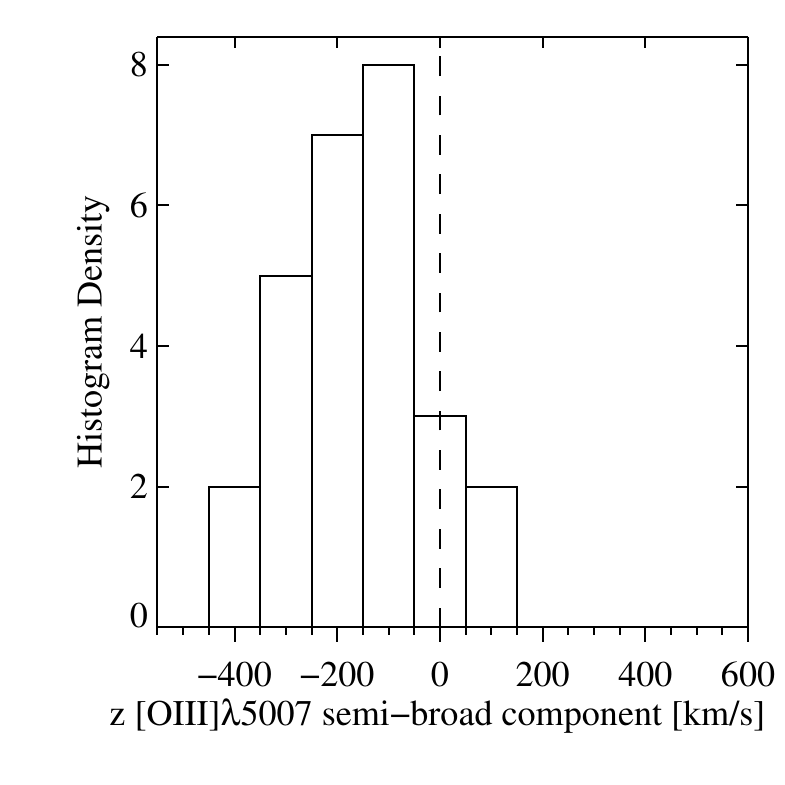}
		\includegraphics[width=0.4\textwidth]{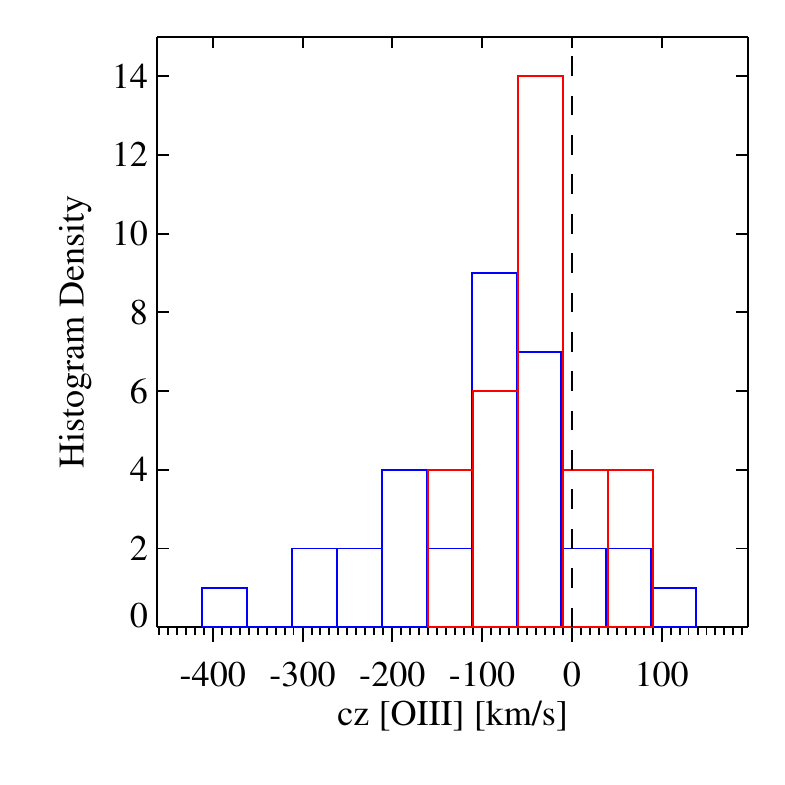}
		
		\caption{Left: Distribution of the shift of [OIII]$\lambda$5007 semi-broad component. 
		Right: Distribution of the \cqm\ \oiii\ (blue) and of the $c$(0.9) \oiii (red). }\label{OIIIw_shift}
	\end{center}
\end{figure*}

As mention above \oiii \ lines were fitted with two components - a narrower associated with the core of the
line, and a semi-broad component that corresponds to the radial motions
\citep[e.g.,][]{komossaetal08,zhangetal11}. The  spectral range  around \oiii \ lines is zoomed  on the
middle plot of the figure \ref{fits_plots}. Figure \ref{OIIIw_shift} shows the distribution of the shift
of the semi broad component in the HG sample.  {As for typical type 1 AGN, the distribution of the sources in our sample is
skewed to the blue, especially toward the line base. The amplitude of the blue-shifts is however modest, and as so, not 
as strong as in the real xA
sample}. Looking at the full \oiiiopt\ profiles we see again that no object qualifies as a blue outlier
following the definition of \citep{zamanovetal02}, by far. The highest amplitude blueshift at 0.9 peak
intensity is $\approx -150$ \kms. The distribution of \cqm\ values is skewed toward blueshifts, as observed
in most samples \citep[e.g.,][]{gauretal19, bertonetal16, zhangetal11}, in both Pop. A and Pop. B. The
conclusion is that, for most objects, we have no evidence of xA properties from the \oiiiopt\ profiles:
large shifts are common among xA sources, with a high frequency of blue-outliers
\citepalias{negreteetal18}.

\subsubsection{The H$\beta$ broad profile}

We expect a significant blueward asymmetry in the \hb\ broad profile of xA sources. If the profile is fit by a symmetric 
and unshifted Lorentzian function, a residual excess emission appears on the blue side of the Lorentzian profile (several 
examples are shown by \citealt{negreteetal18}). The blueshifted emission is associated with outflows that emit more 
prominently in high-ionization lines such as \civ\ \citep[see eg.,][for a systematic comparison]{marzianietal10}. Table 
\ref{classification} reports the centroid shifts of the \hbbc\  broad component.  We see one clear example of blue shifted 
\cqm\ in the source SDSS J105530.40+132117.7, which has the highest \rfe\ in our sample. Only this object appears to be a 
bona fide xA source.  However, sources with \cqm\ $\lesssim -300$ \kms\  (assumed as a typical uncertainty at 1/4 maximum) 
are rare, just 6 out of 33.  Most sources are symmetric or with the \cqm\ displaced to the red: more than one half (21/33)  
have a significant redward displacement. Prominent redward asymmetries are found among Pop. B sources, both radio quiet 
and radio-loud, with extreme cases in the 
radio-loud population \citep[e.g.,][]{punsly13}. The  redward excess is  associated with low Eddington ratio, although its 
origin is still not well-understood: tidally disrupted dusty clumps infalling toward the central black could be the cause 
of a net redshift \citep{wangetal17}, although other lines of evidence challenge this interpretation 
\citep[e.g.,][and references therein]{bonetal15}.



\subsection{Properties of the host galaxy}

In almost all objects we uncovered very high fraction of SSP spectra to the total flux (in the case of 17 objects even
higher then 40\%). 

\indent Restored mean stellar
velocity (cz) is between -50 \kms and 170 \kms. Stellar velocity dispersion is between 90\kms and 220 \kms.
On the figure \ref{chi2maps} 
we show $\chi^2$ maps in the space of SSP mean stellar velocity (cz) and SSP velocity dispersion. 
All SSP cz obtained from the single best fit 
are in a good agreement with values obtained from the $\chi^2$ maps, while SSP velocity dispersion obtained 
from the single fit are usually lower then those obtained from $\chi^2$ maps. \\
\indent We found mostly old SSP (older then 1 Gyr). The metallicities of SSPs in our 32 sample are mainly Solar like.
This property is at variance with the star formation
property expected for xA sources. The UV
spectral properties indicate extreme metal enrichment \citep{martinez-aldamaetal18}, most likely
associated with extreme star formation detected in the FIR \citep[] [in the most luminous cases, SFR$\sim 10^{3}$ M$_{\odot}$ yr$^{-1}$.]{sanietal10,gancietal19}.\\

 \begin{table*}[t]
 	\centering\tabcolsep=3pt
 	\caption{Basic physical properties of the HG sample.}\label{masses}
 	\begin{tabular}{ccccccr}
 		\hline\hline
 		\hspace{8mm}SDSS ID &  	 	 	 $\log L_{5100}$ 	&	 	$\log M_\mathrm{BH}$	&	$\log \frac{L}{L_\mathrm{Edd}}$  &  $\log M_\mathrm{BH} \rm \ corr$	&   $\log \frac{L}{L_\mathrm{Edd}} \rm \ corr$    \\
 		      &   [erg s$^{-1}$]	&	[\msol]	  	 &	&   [\msol]	 &    	\\
 		 \hline
J003657.17-100810.6	&	43.94	&	8.2	&	-1.42	&	8.1	&	-1.27	\\
J010933.91+152559.0	&	43.73	&	8.2	&	-1.54	&	7.7	&	-1.07	\\
J011807.98+150512.9	&	44.13	&	8.9	&	-1.92	&	8.5	&	-1.53	\\
J031715.10-073822.3	&	44.08	&	8.8	&	-1.85	&	8.5	&	-1.50	\\
J075059.82+352005.2	&	44.26	&	8.7	&	-1.56	&	8.3	&	-1.12	\\
J082205.19+584058.3	&	44.07	&	8.8	&	-1.81	&	8.3	&	-1.36	\\
J082205.24+455349.1	&	44.42	&	8.9	&	-1.57	&	8.2	&	-0.94	\\
J091017.07+060238.6	&	44.08	&	8.6	&	-1.68	&	8.3	&	-1.36	\\
J091020.11+312417.8	&	44.08	&	8.5	&	-1.51	&	8.5	&	-1.58	\\
J092620.62+101734.8	&	43.83	&	8.2	&	-1.49	&	7.7	&	-0.97	\\
J094249.40+593206.4	&	43.95	&	8.1	&	-1.22	&	7.3	&	-0.49	\\
J094305.88+535048.4	&	44.05	&	8.9	&	-1.96	&	8.6	&	-1.62	\\
J103021.24+170825.4	&	44.12	&	8.6	&	-1.58	&	8.1	&	-1.08	\\
J105530.40+132117.7	&	44.19	&	8.4	&	-1.36	&	7.4	&	-0.35	\\
J105705.40+580437.4	&	43.67	&	7.5	&	-0.95	&	7.5	&	-0.96	\\
J112930.76+431017.3	&	43.65	&	8.2	&	-1.63	&	7.6	&	-1.02	\\
J113630.11+621902.4	&	43.95	&	8.6	&	-1.73	&	8.0	&	-1.17	\\
J113651.66+445016.4	&	43.51	&	8.1	&	-1.69	&	7.6	&	-1.21	\\
J123431.08+515629.2	&	43.94	&	8.4	&	-1.56	&	7.9	&	-1.07	\\
J124533.87+534838.3	&	44.14	&	8.6	&	-1.58	&	8.2	&	-1.16	\\
J125219.55+182036.0	&	43.93	&	8.5	&	-1.65	&	7.9	&	-1.10	\\
J133612.29+094746.8	&	43.88	&	8.3	&	-1.52	&	8.1	&	-1.37	\\
J134748.06+404632.6	&	43.96	&	8.2	&	-1.33	&	8.0	&	-1.19	\\
J134938.08+033543.8	&	43.84	&	8.0	&	-1.26	&	7.9	&	-1.16	\\
J135008.55+233146.0	&	44.31	&	9.1	&	-1.88	&	8.5	&	-1.27	\\
J141131.86+442001.0	&	43.91	&	8.4	&	-1.66	&	8.3	&	-1.47	\\
J143651.50+343602.4	&	43.85	&	8.5	&	-1.79	&	8.3	&	-1.59	\\
J151600.39+572415.7	&	43.82	&	8.0	&	-1.27	&	7.9	&	-1.18	\\
J155950.79+512504.1	&	43.91	&	8.2	&	-1.42	&	7.8	&	-1.03	\\
J161002.70+202108.5	&	43.82	&	8.1	&	-1.41	&	7.6	&	-0.90	\\
J162612.16+143029.0	&	43.69	&	8.7	&	-2.09	&	8.4	&	-1.88	\\
J170250.46+334409.6	&	43.90	&	8.3	&	-1.57	&	7.8	&	-1.01	\\
\hline
 	\end{tabular}
 \end{table*}


\section{Discussion}
\label{disc}

\subsection{Interpretation in the Eigenvector 1 context}

\indent We notice consistent results between measurements of \cite{shenho14} and ours, albeit with a bias in favour of higher \rfe\ for \citet{shenho14}.  According to the position of the spectra on the MS diagram of Fig. \ref{fig:e1fl}, objects are mainly Pop. B, with the exception of 5   sources that are of the Pop. A class. 
The distribution of the quasar data points is  centered in Pop.B2, with 22 sources objects that belong or are likely to belong to Pop. B2,  including borderline.  

Apart from the location along the MS of Fig. \ref{fig:e1fl}, the conclusion that the wide majority of the HG sample sources are not xAs (only one source (J105530.40) meets in full the criterion of \citet[][\rfe $\gtrsim 1.2$]{negreteetal18} to qualify as an xA source) is reinforced by several  lines of evidence: (a) \oiiiopt\ profile without large blueshifts and \oiiiopt\ consistent with the rest frame; (b) \hbbc\ profile symmetric or redward asymmetric; (c) HG component predominantly dominated by an old stellar population; (d) conventional estimates of the \lledd\ $\ll$1. On this last line, we discuss in \S \ref{ph} a discrepancy between \lledd\ estimates based on scaling laws and the new approach of the accreting black hole fundamental plane \citep{duetal16}.

The previous analysis points toward a sample showing relatively low luminosity, and ``milder" signs of nuclear activity 
with respect to the extreme radiators of xA. This does not mean that a similar phenomenology concerning nuclear outflows is not occurring, but its detectability is limited to some particular manifestations, such  us the blueshift of \oiiiopt\ close to the line base.

\subsection{Basic physical properties}
\label{ph}

\begin{figure*}
	\begin{center}
		\includegraphics[width=0.45\textwidth]{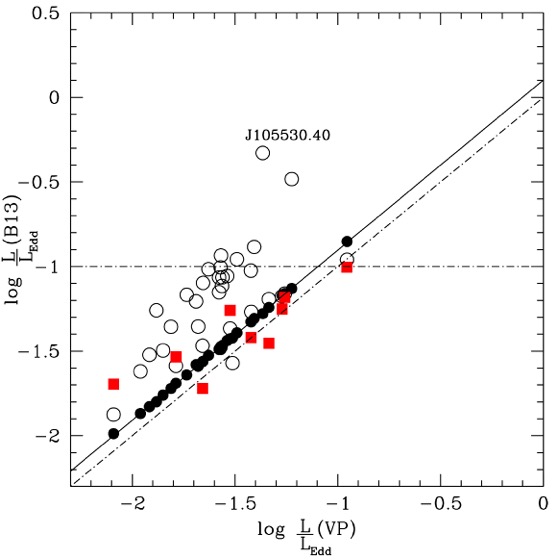}\vspace{6mm}
			\includegraphics[width=0.45\textwidth]{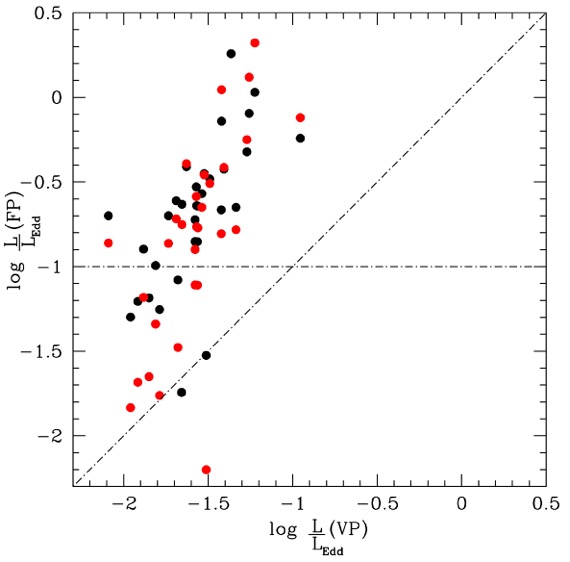}
	\caption{Left: relation between the Eddington ratio with \mbh\ 
	computed from the scaling law of VP and with the $r_\mathrm{BLR}$\ 
	from \citet{bentzetal13} employing an uncorrected FWHM value (black circles). 
	The open circles represents estimates using $f^\star_2$ (Eq. \ref{eq:f}) and $r_\mathrm{BLR}$\ corrected following 
	Eq. \ref{eq:drb}. Red squares represents \lledd\ estimates from the \citet{mcconnelletal11} scaling law, 
	with the restriction of $\sigma_\star \gtrsim 150$\kms. The oblique dot-dashed line is the equality line; 
	the horizontal one marks the conventional limit separating Pop. A and B. The filled line is the result of an 
	unweighted lsq fit between the VP and the uncorrected  \citet{bentzetal13} \mbh\ 
	estimates.    Right: relation between the Eddington ratio with \mbh\ computed from the scaling law of
	VP and the fundamental plane of \citet{duetal16a} (black circles), 
	and after the re-fit of this paper (red circles).   \label{fig:ml}}
	\end{center}
\end{figure*}

{Black hole masses estimates using scaling laws for large samples of AGN are subject to a large uncertainty, 
due to both systematic and random errors \citep[exhaustive reviews are given in][]{marzianisulentic12,shen13}. However, 
in the case of the low-$z$ sample of the present work, we can count on the \hb\ line width that is considered a reliable 
``virial broadening estimator'' \citep[][ with the caveats of \citealt{marzianietal13a,marzianietal19}]{trakhtenbrotnetzer12}.
Table \ref{masses} lists basic physical properties of the AGN - the log of the 5100 $\AA$ AGN luminosity scaled by the AGN 
power-law continuum fraction to the total flux, the black hole mass \mbh\ computed following  the prescription of 
\citet[][hereafter, VP]{vestergaardpeterson06}, and \lledd\ (assuming the bolometric luminosity to be 10 times the luminosity 
at 5100 \AA; \citealt{richardsetal06}). {In this table we present also the corrected \mbh and \lledd\ according to the prescription of \citet{martinez-aldamaetal19} (see \ref{orient})}. The \mbh\ values following VP indicate a population of quasars of relatively modest \mbh, $\sim 10^8$ \msol.  Accepted at a face value,  \lledd\ is typical of Pop. B, with some objects close to the boundary between A and B but formally on the side of Pop. B, if  \lledd $\approx 0.1 - 0.2$ is assumed as the \lledd\ threshold for Pop. A sources. 

We can   write the expression of the virial mass as  follows:

\begin{equation}\label{eq:m}
M_\mathrm{BH} = f  \frac{r_\mathrm{BLR} (\delta v)^{2}}{G} =  f_{1}(\dot{m}, a) f_{2}(\theta\, |\, \dot{m})  \frac{r_\mathrm{BLR} (\delta v)^{2}}{G},
\end{equation}

where $r_\mathrm{BLR}$ is the BLR radius, $a$\ is the spin parameter of a black hole, and we considered as estimator of the virial broadening velocity spread $\delta v = $FWHM, the FWHM of the broad component of \hb.  We have written the structure or form factor $f$ as the product of two terms, one depending on accretion rate and black hole spin, and one depending on orientation.  
The dependence of $f_{1}$ on dimensionless accretion rate has been emphasized  by the $r_\mathrm{BLR}$\ dependence on 
luminosity \citep{duetal16}  which, for xA sources, is not consistent with the general AGN population. The dependence of 
$f_{1}$ on the spin parameter is unknown, but  is expected since the spin in influencing the temperature of the accretion disk 
and hence the SED of the ionizing continuum \citep[e.g.,][]{wangetal14}. To complicate the issue, the orientation effects are 
also expected to be dependent on $\dot{m}$ \citep{wangetal14a}, as a 
geometrically thin optically thick disk may be considered as a Lambertian radiator (with some limb-darkening effects at high inclination \citealt{netzer13}), but free of the 
self-shadowing effects expected for a geometrically thick disk.  
Keeping for the moment with the simplest approach, we can compute the  \mbh\ by using the \citet[][hereafter, B13]{bentzetal13} correlation between  $r_\mathrm{BLR}$  and optical luminosity,  assuming $f=1$. The results are tightly correlated with the mass estimate obtained from the   VP relation (Fig. \ref{fig:ml}). The extremely tight correlation is expected as the  VP assumes the same virial relation and only a slightly different value of the zero point and of the $r_\mathrm{BLR}$ - $L$\ correlation. The small bias between the two relations is understood in terms of a constant difference in the $f$  factor, since VP assumed $f = 0.75$. In both cases no orientation effects are considered. 
Typical uncertainties in the \mbh\ are expected to be $\approx 0.3$ dex at 1 $\sigma$ 
\citep{vestergaardpeterson06,marzianietal19}, most likely because of differences in $f$ \ associated with different 
structure \citep[$f \lesssim$ 1 and $\approx 2$ were derived for Pop. B and A, respectively][]{collinetal06}, and with 
the effect of orientation. The $r_\mathrm{BLR}$ - $L$\ is also known to be dependent on $\dot{m}$\ \citep[][and references 
therein]{duwang19}.  The main source of uncertainty in luminosity estimates at 5100 \AA\ is the continuum placement and the 
error associated with the decomposition of between the AGN and host continua. Even if formal errors are low, it is 
unlikely that the uncertainty is less than $\approx$ 10\%, which we assume as an indicative value. The computation of the 
bolometric luminosity suffers from the additional scatter associated with the diversity in the AGN SEDs; scatter at 
1$\sigma$\ could be assumed $\sim $ 20\%\ \citep{elvisetal94,richardsetal06}. We expect a dependence of the bolometric 
correction along the main sequence; more recent estimates suggest a dependence on luminosity, spin and dimensionless 
accretion rate \citep[e.g.,][]{runnoeetal13,netzer19}, but they are relatively untested and were sparsely considered in 
past work. We assume  bolometric correction 10. 

\subsubsection{\lledd\ estimate using the Fundamental Plane}
\label{fp}

A second method to estimate \lledd\ can be based on the fundamental plane (FP) of accreting black holes described by \citet{duetal16a}. \citet{duetal16a} introduced the notion of the fundamental plane of  SEAMBHs defined by a bivariate correlation between the parameter $\dot{{\mathcal M}} = \frac{ \dot{M_{\rm BH}}c^{2}}{L_\mathrm{Edd}}$\ i.e., the dimensionless accretion rate $\dot{m} = \frac{\eta \dot{M}c^{2}}{L_\mathrm{Edd}}$\   for $\eta = 1$ \citep{duetal15}, the Eddington ratio, and the   observational parameters \rfe\ and D parameter (ratio $FWHM/\sigma$ of H$\beta$, where $\sigma$ is the velocity dispersion of the broad component of \hb). The FP can then be written as two linear relations   between $\log  \dot{\cal M}$\ and $ L_{\rm L}/L_{\rm Edd}$ versus  $\approx a  + b \frac{\rm FWHM}{\sigma}+c R_{\rm FeII},$ where $a, b, c$ are  reported by \citet{duetal16a}. The identification criteria included in the fundamental plane are consistent with the ones derived from the E1 approach (\lledd\ and $\dot{{\mathcal M}}$\ increase as the profiles become Lorentzian-like, and \rfe\ becomes higher).

To investigate the origin of this disagreement, we considered that the fit provided by \citet{duetal16a} is very good for high Eddington radiators but is biased if low \lledd\ data are considered. The upper panel of Fig. \ref{fig:fp} shows that there is a significant residual between data and fit values that is dependent on \lledd: at low Eddington ratio, $\log $ \lledd\ $\sim -2$, the FP plane fit reported by \citet{duetal16a} predicts a value of \lledd\ almost one order of magnitude systematically higher with respect to the one inferred by the distribution of the data points. The residuals can be fit by a linear function ($\delta = \log$\lledd $-\log$\lledd(FP) that zeroes the trend in Fig. \ref{fig:fp} (red dots), with a post-correction  best fitting line consistent with $\delta$(\lledd) $\equiv 0$. Applying the correction to the residuals to obtain new values of \lledd\ we obtain this slightly modified equation for the fundamental plane $\log$ \lledd = $\alpha + \beta D + \gamma$\rfe $\approx 0.774 - 1.33 D + 1.30$\rfe.  The estimates with this new law, although lower at the low \lledd\ do not solve the disagreement between the VP conventional estimates and the FP estimates  (Fig \ref{fig:ml}). The right panel of Fig. \ref{fig:ml} shows that the FP \lledd\ estimates are in large disagreement with respect to the VP and B13 estimates with both the old and new equation for FP : VP estimates are below by more than one order of magnitude the ones based on the FP. The disagreement is so serious that the highest radiating source with \lledd $\sim $ 2 according to the FP has \lledd $\approx 0.04$ following VP, and that it leads to inconsistencies between the MS interpretation and spectral type assignment: 
the same source would qualify as a Pop. B source (VP) and as an xA (FP). Using the modified VP with the parameters reported above, the only effect is to bring in agreement only 6-7 points at the low-\lledd\ end. The bulk of the data point remains above the VP estimates by $\approx$1 dex. 

To further investigate the issue, we computed   \mbh\   from the stellar velocity dispersion $\sigma_\star$\ of the host bulge,
using the scaling law \mbh\ $\approx 1.95 \cdot 10^8 \left(\sigma_\star/200\right)^{5.12}$ \msol\ \citep{mcconnelletal11},  which is an updated formulation of the original scaling law of \citet{ferraresemerritt00}.  Fig. \ref{fig:mhost} shows that  the VP \mbh\  and \mbh\ from host  show systematic differences that are strongly 
correlated with $\sigma_\star$, increasing with decreasing $\sigma_\star$. The unweighted  least squares fitting line shown in Fig. \ref{fig:mhost} represent a highly significant but likely spurious correlation.  {When physical velocity dispersion is of the same order or smaller than the instrumental velocity dispersion, it is advisable 
\citet{Kol09} to inject line spread 
function (LSF) of the spectrograph in the model of SSP, in order to adjust the resolution of the spectra and the model.  We re-fitted the spectra where $\sigma_\star$ was below 150 \kms, with injected LSF in the SSP model, but restored $\sigma_\star$\ was just slightly higher then the first estimation of $\sigma_\star$, and still within the error bars of the first estimation. Therefore, we concluded that LSF injection would not solve the problem of discrepancies between two estimations of the masses.  There is a possibility, discussed in \S\ \ref{orient} }that $\sigma_\star$\ is associated 
with systems observed face-on, and that are therefore also affected by orientation effects. 

The FP estimates are based on two parameters that do not include information on  line broadening. 
The parameter $D$\ is somewhat redundant as the shape of the \hb\ profile is known to be a MS correlate: 
the profiles are Gaussian-like ($D \approx 2.3$) in Population B, while become Lorentzian-like in Pop. A 
(spectral type 1) and are consistent with Lorentzian-like up to the highest \rfe\ value, albeit with a 
blueshifted excess interpreted as Balmer emission from a high-ionization wind, more easily detected in 
high-ionisation lines such as \civ\ \citep[e.g.,][]{richardsetal11}.  Therefore the behaviour of the 
parameter D is not expected to be monotonic along the sequence: it should increase from extreme Pop. B toward A1, 
where the most Lorentzian-like profiles are observed, and decrease again where a blueshift excess provides 
a significant deviation from a Lorentzian profile (ST A3 and A4). In addition, what would be the prediction 
passing from A2 to B2 and  from A3 to B3 according to the fundamental plane? The xA sources of A3 in 
the sample of \citet{duetal16} show a typical $D \approx 1.5$; in B3 the profiles are more Gaussian-like, 
and we can assume a conservative $D \approx 2$; for the same average \rfe = 1.25, the change in \lledd\ would 
be more than a factor $2.5$. These consideration focus the  issue on the nature of Pop. B2 and B3. Populations B2 and B3 are 
rare at low $z$ (B2 are $\lesssim$3\%\ in the sample of \citealt{marzianietal13a}; B3 is not even detected, 
which implies a prevalence $\lesssim 0.2$\%), and represent a poorly understood classes. There is a degeneracy 
between effects of orientation and \mbh\ in the optical plane of the MS; for a fixed \mbh, A2 sources seen at 
higher inclination may be displaced into B2 \citep{pandaetal19}. At the same time we cannot exclude that higher \mbh\ 
sources are located within B2. In both cases, for a fixed luminosity, we expect a significant decrease in \lledd\ 
passing from B2 to A2. 

Besides, the object could appear as
B2 type, due to different response of H$\beta$ and \feii  \ flux to the variability of ionizing
continuum. 
Higher  \rfe \  could be caused by two variability effects: (1) a
faster response of H$\beta$ flux to the variability of ionizing
continuum; (2) a larger amplitude of H$\beta$ flux
variations compared to the amplitude of \feii  \ flux variations
\citep[see e.g.][]{huetal15,barthetal13}.
In case of observing single epoch spectra, depending on variability state
both effects, together with the line width response to the flux variations,  
could contribute to estimates of 
mass and \lledd\ .  
Also, these effects could produce the
trend of \lledd\  decreasing with  \rfe \ \citep{bonetal18},
which is opposite to the trend along EV1 where \lledd\  increase with \rfe\ \citep{marzianietal13}.

\begin{figure}
	\centering
	\includegraphics[width=\columnwidth]{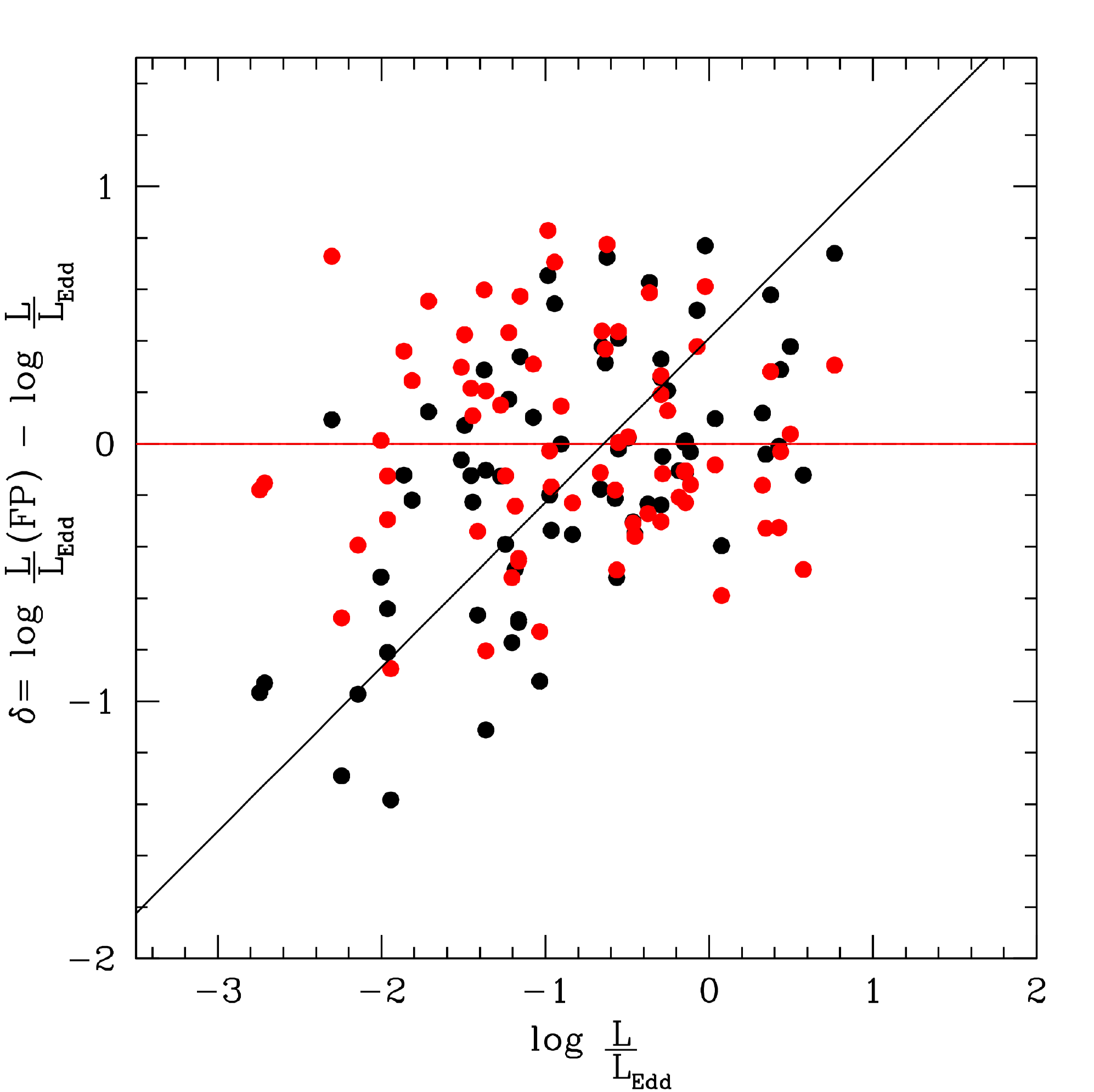}
	\includegraphics[width=\columnwidth]{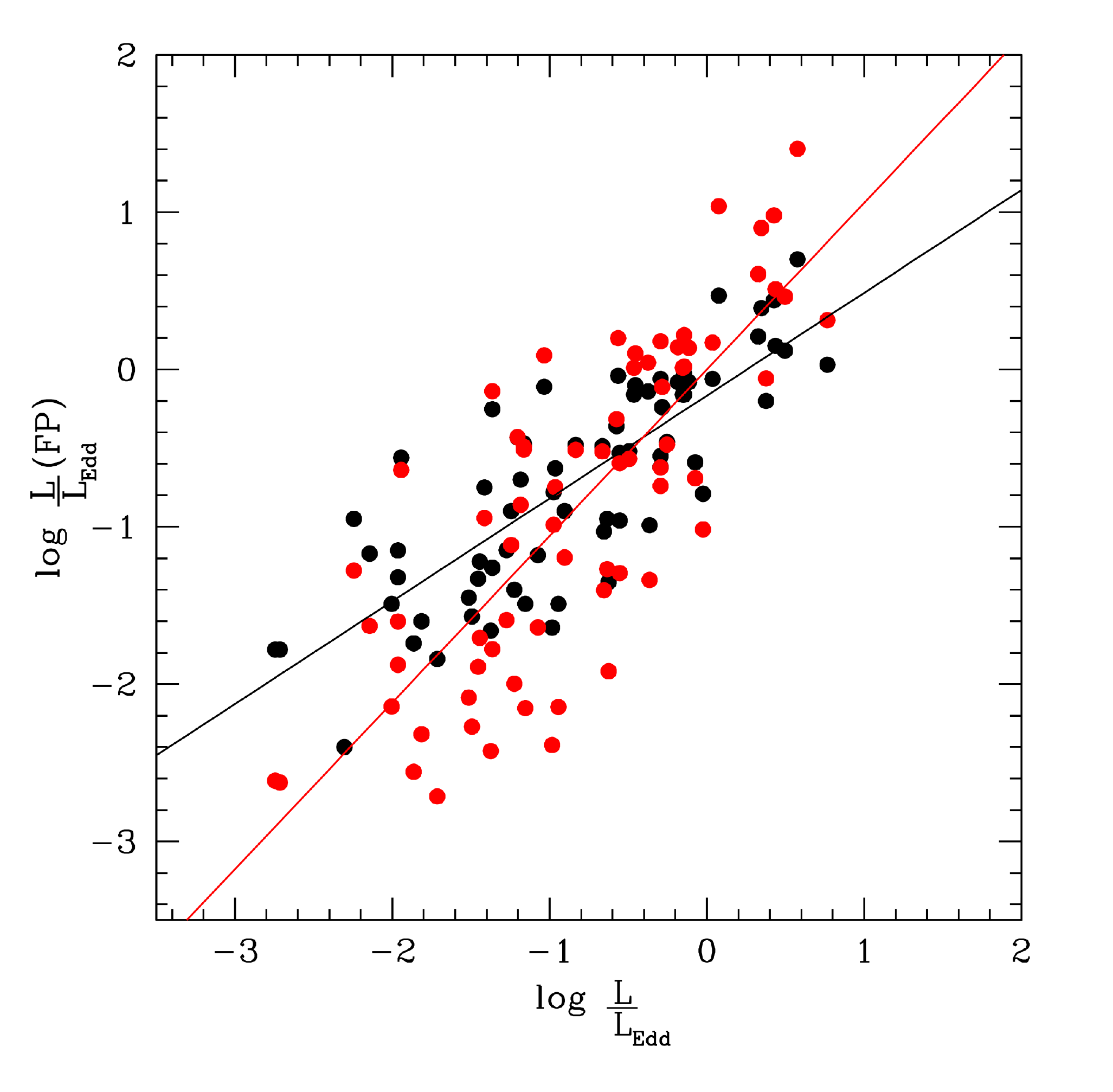} 
	\caption{Improvement on the fundamental plane of accreting black holes. Top panel shows residuals between the original FP fitting equation of 
	\citet{duetal16a} and the data (black circles). There is a significant linear trend; the black line traces the 
	unweighted lsq best fit. The red circles show the residuals with the modified FP (see \S\ \ref{fp}); 
	there is no trend, and the best fitting (red) line is consistent with zero slope (black line). 
	The bottom panel shows the  data points with \lledd\ estimated with the original (black) and revised (red) FP equation. \label{fig:fp}	}
\end{figure}

\begin{figure}
	\begin{center}
	\includegraphics[width=0.45\textwidth]{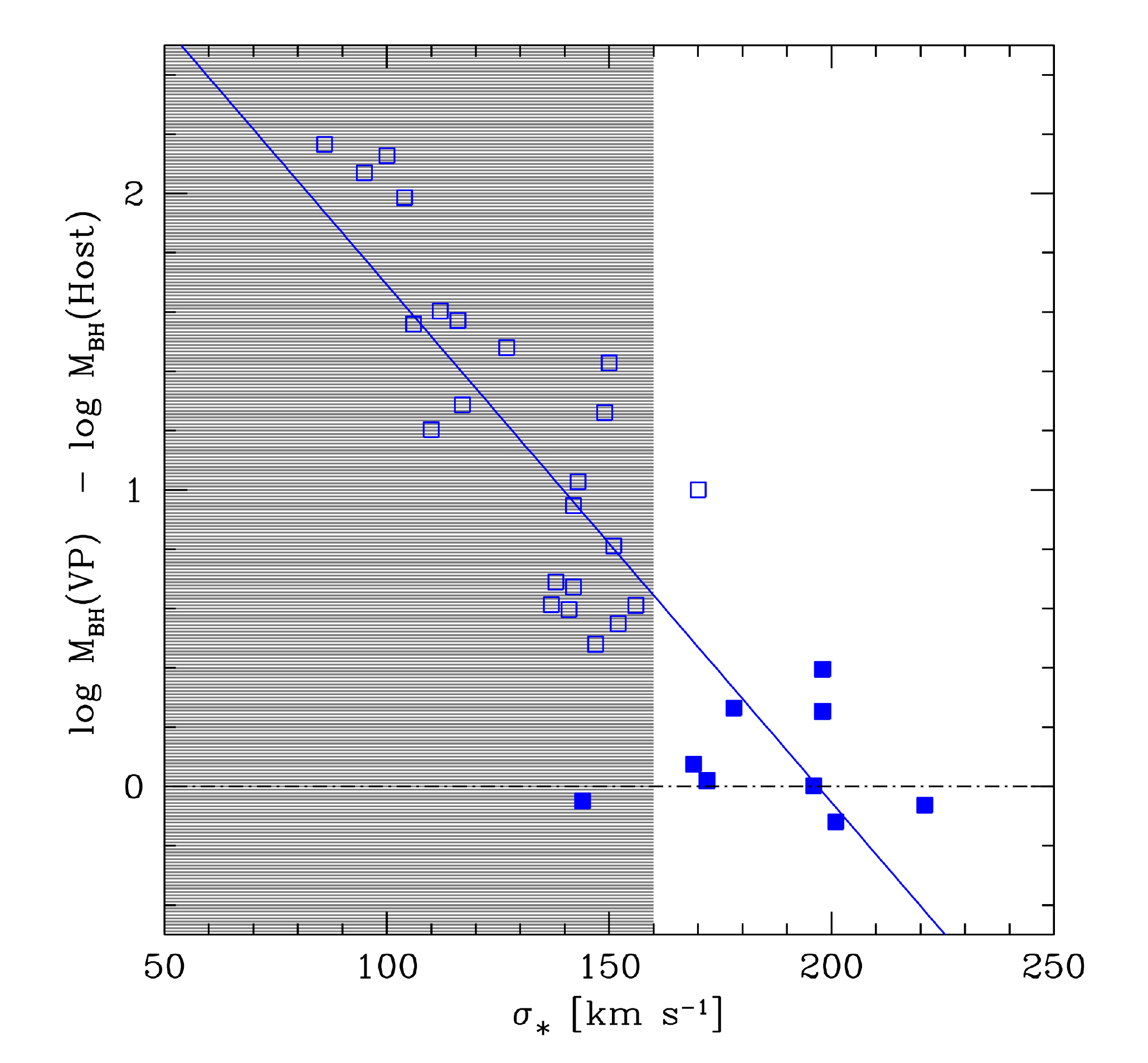}
	\includegraphics[width=0.45\textwidth]{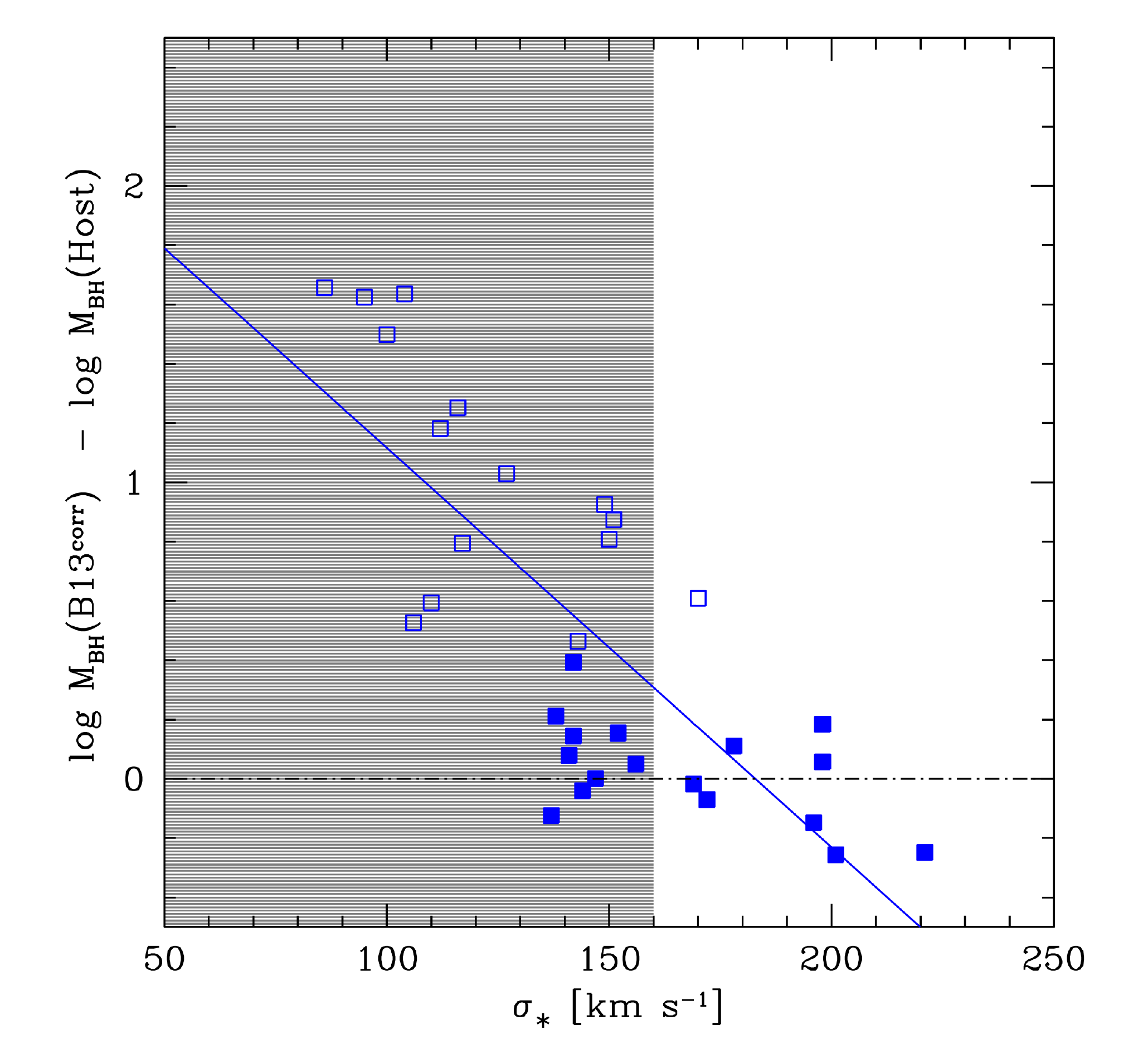}
	\caption{Top: Relation between velocity dispersion $\sigma_\star$\ and difference between \mbh\ estimates from $\sigma_\star$\ using the scaling law of \citet{mcconnelletal11} and the scaling law from VP. The shaded area identifies the range of  $\sigma_\star \le 160 $\ \kms. All data point save one  with \citet{mcconnelletal11} \mbh\ values yielding  large systematic differences ($>$ 0.4, represented with open squares) with respect  to the VP \mbh\  are within the shaded area. Bottom: same as in the top panel but with the \mbh\ values computed with the  correction described in \S \ref{orient}. \label{fig:mhost}}
	\end{center}
\end{figure}

\subsection{Orientation and physical parameter estimates}
\label{orient}

The previous analysis ignored the effect of orientation on  the \mbh\ computation.  However, growing evidence suggests that the low-ionization lines-emitting BLR is highly flattened \citep[e.g.,][and references therein]{mejia-restrepoetal18}.  If this is the case, the observed velocity can be parameterized as  $  \delta v^{2}_\mathrm{obs} = \delta v_\mathrm{iso}^{2}/3 + v_\mathrm{Kepl}^{2} \sin^{2}\theta$, and if $\delta v_\mathrm{iso}/\delta v_\mathrm{Kepl}\approx 0.1$, where  $\delta v_\mathrm{iso}$\ is an isotropic velocity component,  and $\delta v_\mathrm{Kepl}$\ the Keplerian velocity.  For a  geometrically thin disk, it implies   $\delta v_\mathrm{obs} \approx \delta v_\mathrm{Kepl}/\sin\theta$ {(if the FWHM is taken as the $\delta v_{\rm obs}$, and $\delta v_{\rm Kepl} = 0$\ i.e., in the case of isotropic velocity dispersion,  $f_{2}  = \frac{3}{4}$)}.  If the VBE estimates are not corrected beforehand for orientation,  the structure (or form) factor is $f_{2}  \propto 1/\sin^{2}\theta$\ \citep[e.g.,][]{mcluredunlop01,jarvismclure06,decarlietal11}, and more precisely (we assume $f_{1} \equiv 1$): 
 
 \begin{equation}
f_{\rm 2}  = \frac{1}{4 \left[ \frac{1}{3}\left(\frac{\delta v_{\rm iso}}{\delta v_{\rm K}}\right)^{2} + \sin^{2} \theta  \right]   }. \label{eq:fs}
\end{equation}

We attempt to consider the effect of the viewing angle on the \hb\ line width by considering   that the virial factor is anti-correlated with the FWHM of broad emission line. For the \hb\ line the relation is given by  
\begin{equation} \label{eq:f}
f_2^\star = \left(\frac{{\mathrm{FWHM}}}{4550}\right)^{ -1.17},
\end{equation}
 \citep{mejia-restrepoetal18}. This implies that sources with FWHM \hb\ {\em narrower} than $4550$\kms\ should have their mass increased by a factor that can be as large as $\approx 5$ in the case of the narrowest \hb\ profiles observed in NLSy1s. The effect is milder than the one predicted by Eq. \ref{eq:fs}, and may be better suited for the general population of quasars encompassing both typical Pop. A and B sources.   In addition, \citet{martinez-aldamaetal19} suggested a correction to the B13 $r_\mathrm{BLR} $\ estimate, following the reverberation mapping campaign of highly accreting quasars  \citep[][and references therein]{duetal18}, $\delta r_\mathrm{BLR}=\mathrm{log}\,\left( {r_\mathrm{BLR} }/{r_\mathrm{{BLR,B13}}}\right)$. According to \citet{martinez-aldamaetal19}, with the $f^\star_{2}$ dependence on  FWHM, the correction to $r_\mathrm{BLR}$\  is:

\begin{equation} \label{eq:drb}
\delta r_\mathrm{BLR} =\left(-0.271\pm0.030\right) \mathrm{log} \, \frac{L\mathrm{_{bol}}}{L\mathrm{_{Edd}}^\star} \, +   \left(-0.396\pm0.032\right),
 \end{equation}
 
where ${L\mathrm{_{Edd}}^\star}$\ means that the Eddington luminosity has been computed with virial mass relation assuming $f^\star_2$\ (Eq. \ref{eq:f}). The \lledd\ values computed with this approach deviate significantly from the VP and B13 \lledd, yielding a higher value of \lledd\ (Fig. \ref{fig:ml}), because most of the sources have FWHM \hb$ \gtrsim 4500 $, and hence $f^\star  < 1$\ which implies a lower \mbh.  The effect visible  in Fig. \ref{fig:ml} comes mostly from the $f^\star_2$: the $\delta r_\mathrm{BLR}$\ is small since \lledd\ is low ($\log $ \lledd $\sim -1. -1.5$).  The agreement between the modified VP and the FP remains poor. After the corrections, however, the xA candidate \object{SDSSJ105530.40+132117.7} is recognised as the highest radiator, with \lledd\ $\approx -0.35$, close to the conventional lower limit for xA sources. We can conclude that the objects in our sample are safely not xA (apart from the one source mentioned right above, and perhaps a couple of borderline cases). Are they of Pop. A? Following the FP all save 7 should be of Pop. A, with 5 xA candidates (Fig. \ref{fig:ml}). Following the modified expression of Eq. \ref{eq:f}, about one half of the sources has $\log$ \lledd $\gtrsim -1$, the conventional boundary between Pop. A and B.  

If we consider the modified viral mass as described in the previous paragraph, any correlation with the \citet{mcconnelletal11} remains fairly weak ($r \approx 0.385$, significant at just $2\sigma$ confidence level). However, the bottom panel of Fig. \ref{fig:mhost} shows that now more than half of the sample has masses in reasonable agreement with ones from the host. {For the remaining sources with small $\sigma_\star$, we consider that  bulges seen face-on should be considerably affected by orientation in their measured velocity dispersion, as they are   rotationally supported \citep{kormendyillingworth82}. A test of this possibility goes beyond the goals of the present paper. }

The existence of a relation between \rfe\ and \lledd\  is a robust result and rests on several lines of evidence (\citealt{grupeetal99,kuraszkiewiczetal04,dongetal11,marzianietal13,sunshen15}, in addition to the analysis of \citet{duetal16,duwang19}). For instance \citet{marzianietal13} showed (their Fig. 6) that there is a \lledd\ systemic trend along Pop. A, with \lledd\ increasing with \rfe.  Independent evidence is provided by \citet{sunshen15}: for the luminosity in a fixed range, the $\sigma_\star$\ (a proxy of \mbh) decreases systematically with increasing \rfe\ up to \rfe $\sim 1$. Indirect evidence is also provided by one of the correlations of 4DE1: the highest correlation coefficient is between \rfe \ and \civ\ blueshift amplitude \citep{sulenticetal00c,sulenticetal07,sulenticetal17}. The balance between radiative and gravitational forces is able to account for a large part of the quasars optical/UV phenomenology along the quasar MS \citep{ferlandetal09}.  Therefore the key factor in the discrepancy of the \lledd\ estimates around A2/B2 is most likely to be orientation, as the viewing angle $\theta$\ affects FWHM \hb\ linearly, and \mbh\ quadratically.  By the same token a large fraction of the  B2 sources could be very well intrinsically Pop. A, especially the ones with \rfe\ significantly above 0.5 and FWHM borderline. 

Great care should be used in the computation of the \mbh\ and \lledd\ if no orientation correction is possible. As a test not related to the present sample, we considered the sources of the \citet{duetal16} sample that are xA i.e., satisfying the criterion \rfe $>1.2$ following \citet{negreteetal18}. For these 6 sources the \lledd\ average value applying the original FP is $\approx 0.45$, implying that they should be considered true ``super-Eddington"  accreting massive black holes \citep[SEAMBHs][]{wangetal14b}.  For these xA sources, it is possible to derive an estimate of the viewing angle $\theta$\ \citep{negreteetal18}: the difference between the virial luminosity estimate and the concordance luminosity (assumed to be the correct luminosity) is expected to be mainly dependent on the viewing angle, which is assumed to strongly affect the FWHM according to Eq. \ref{eq:fs}. For these sources,  $\theta \sim$0.2 rad: their emitting regions are expected to be seen almost face-on. If   the \mbh\ are recalculated  following Eq. \ref{eq:m} and  Eq. \ref{eq:fs} with the $\theta$\ estimated following  \citet{negreteetal18}, the \lledd\ is lowered to an average $\approx -0.147$, consistent with the estimates using the conventional approach \citep{marzianisulentic14}. The  \citet{negreteetal18} approach can not be extended to the sample of the present paper, but it is a strong indication that \lledd\ actual values depend on a normalization factor that is in turn dependent on the viewing angle, via the dependence on the viewing angle of \mbh.  The xA sources apparently radiate at a limiting Eddington ratio along the MS.  However, it is unclear whether the xA sources are truly SEAMBHs. Even if there is a consistency in the selection criteria, their Eddington ratios and $\dot{{\mathcal M}}$\ are too much affected by uncertainties in the \mbh\ to be a safe discriminant.

\section{Conclusions} \label{sec:conclusions}

This paper has analysed a spin-off of the \citet{negreteetal18} sample, 
and precisely spectra that were preliminary selected as xA candidates and afterwards found to suffer 
strong contamination by the spectrum of the host galaxy. Main results encompass 

\begin{enumerate}

\item  A proper identification of xA sources requires a careful simultaneous multicomponent fit in order to 
retrieve information on the stellar continuum and on the FeII emission, especially if the AGN is of low luminosity, and the data
are from optical fiber with a relatively large angular coverage.
Inclusion of the spurious xA sources should be avoided, as in cosmological studies it may dramatically increase the dispersion in the 
Hubble diagram of quasars obtained from virial luminosity estimates. 
\item Objects of our sample with strong host galaxy contamination show properties that suggest more modest activity, in
comparison to xA sources. Modest activity means that they lack extreme outflows, strong
starburst activity, and high accretion rate, typical of xA sources.
\item We found a high fraction of host galaxy spectrum (in half of the sample even higher then 40\%). We conclude that 
when absorption lines are prominent, and the fraction of the host galaxy
is high, SSP is mimicking FeII, and that this can
lead to mistaken identification of FeII spectral features. We have identified several stellar absorption lines that, 
along with the continuum shape, may lead to an 
overestimate of \rfe, and therefore to the misclassification of sources as xA sources. 
Our results lead support to the results of 
\citet{Sniegowska2018} who found that only six sources 
out of a sample of 23 could be classified as xA after
 a careful decomposition of all spectral components involving also the spectrum of the host galaxy. 
\item We have studied the 32 sources with high host galaxy contamination as an independent sample which has an interest 
of its own. {We have used a host galaxy shift as a reference frame in order to study shifts of emission lines 
more precisely. Unlike xA sources, there is a very good agreement between the shift of absorption spectrum 
and the shifts of \hbnc, \oii, and the \oiii. } 
The good agreement between  \oiii\ and the narrow low-ionisation lines \hbnc\ and \oii\ has  important consequences
 for the systemic redshift estimates in case no host absorptions could be detected.  
\item We have considered the effect of the density on the effective wavelength of the \oii\ doublet, providing a relation 
linking $n_\mathrm{H}$\ and $\lambda_{eff}$. The dependence on density introduces a significant error in \oii-based redshift 
estimates. However, the effective wavelength density dependence could be in principle provide a diagnostics of the relative 
importance of the AGN NLR and of circum-nuclear star formation producing low-density \hii\ regions.   
\item We found mostly old SSP (older then 1 Gyr) for the HG. The metallicities of SSPs in our 32 sample are mainly solar.
\item The HG sources cluster around spectral type B2. Considering the spectral type correlation with \lledd, a large 
fraction of them should be considered inclined Pop. A sources. 
\item Computations of \mbh\ are problematic, especially if  small samples of heterogeneous sources involving a 
broad range of\ \hb\ FWHM are considered \citep[see e.g.,][]{shen13}.  We have discussed estimates of \mbh\ and  \lledd, 
and emphasized the effect of orientation that should be considered if a meaningful comparison of \lledd\ values between 
sources of widely different width has to be done. In principle, if viewing angles were known for each source, the \mbh\ 
and \lledd\ values could be normalised to a standard $\theta$, to let physical trends emerge more clearly. Individual 
$\theta$\ estimates are still unavailable for the general population of quasars outside of the MS extremes such 
as the ones considered in this study, although   $\theta$\ computations may become widespread in the coming years. Spectropolarimetric measurements, even if demanding in terms of telescope time, have provided individual $\theta$\ values for sources in different spectral types along the MS \citep{afanasievpopovic15}. 
Other techniques, based on the SED also show promising possibilities \citep{capellupoetal15,mejia-restrepoetal18}.
\item At this point, one has to consider that the VP scaling law, perfectly consistent with the use of the more recent 
\citet{bentzetal13}\ $r_\mathrm{BLR}$\ scaling law is biased in favor of broader sources (i.e., the many Population 
B sources that were targets of early reverberation mapping campaigns). On the converse, the \lledd\ estimates from the 
\citet{duetal16a} fundamental planes are apparently biased toward narrow sources (i.e., the many NLSy1s and Population 
A sources  that are included in the \citealt{duetal16a} sample). In both cases, a small FWHM is taken as a synonym of 
small \mbh, and a broad FWHM as synonym of large \mbh. Presently, we know this is not the case because of the degeneration 
between mass and orientation effects. The present work let emerge how VP overestimate the \mbh\ and underestimate \lledd, 
and how \citet{duetal16a} underestimate the \mbh, and overestimate \lledd. These problems are likely to be overcome by the next-generation SDSS-V panoptic spectroscopy \citep{kollmeieretal17}, multi-epoch spectroscopic survey of over six million objects that plans reverberation-mapping quality monitoring for thousands of quasars.
 \end{enumerate}

\begin{acknowledgements}
This research is part of the projects 176001 ''Astrophysical spectroscopy
of extragalactic objects`` and 176003 ''Gravitation and the large scale structure of the Universe", 
funded by Ministry of Education, Science and Technological Development of the Republic of Serbia.
PM and MDO acknowledge funding from the INAF PRIN-SKA 2017 program 1.05.01.88.04. PM also acknowledges the Programa de Estancias de Investigaci\'on (PREI) No. DGAP/DFA/2192/2018 of UNAM. 
A.d.O. acknowledges financial support from the Spanish Ministry of Economy
and Competitiveness through grant AYA2016-76682-C3-1-P and from the
State Agency for Research of the Spanish MCIU through the
``Center of Excellence Severo Ochoa`` award for the Instituto de Astrof\'{\i}sica
de Andaluc\'{\i}a (SEV-2017-0709).
ML. M. A. acknowledges financial 
support of National Science Centre, Poland, grant No. 2017/26/A/ST9/00756 (Maestro 9). DD and AN 
acknowledge support from grants 
PAPIIT, UNAM 113719, and CONACyT221398, Instituto de Astronomia, UNAM, CDMX 04510,
Mexico.  We thank to Pu Du for his help and constructive comments.


Funding for the SDSS and SDSS-II has been provided by the Alfred P. Sloan Foundation, the Participating Institutions, the National Science Foundation, the U.S. Department of Energy, the National Aeronautics and Space Administration, the Japanese Monbukagakusho, the Max Planck Society, and the Higher Education Funding Council for England. The SDSS Web Site is http://www.sdss.org/. The SDSS is managed by the Astrophysical Research Consortium for the Participating Institutions. The Participating Institutions are the American Museum of Natural History, Astrophysical Institute Potsdam, University of Basel, University of Cambridge, Case Western Reserve University, University of Chicago, Drexel University, Fermilab, the Institute for Advanced Study, the Japan Participation Group, Johns Hopkins University, the Joint Institute for Nuclear Astrophysics, the Kavli Institute for Particle Astrophysics and Cosmology, the Korean Scientist Group, the Chinese Academy of Sciences (LAMOST), Los Alamos National Laboratory, the Max-Planck-Institute for Astronomy (MPIA), the Max-Planck-Institute for Astrophysics (MPA), New Mexico State University, Ohio State University, University of Pittsburgh, University of Portsmouth, Princeton University, the United States Naval Observatory, and the University of Washington.

\end{acknowledgements}

\vfill\eject\eject\newpage\pagebreak
\newpage\pagebreak
\clearpage

\begin{appendix}			
\section{SSP analysis atlas}		
\label{sap}

Fig. \ref{fits_plots} shows the appearance of the spectrum in the range 4000 -- 5500 \AA\ where the fit was carried out,  
a zoom around \hb\ and \oiiiopt, and the cleaned \hb\ profile for the sources of the HG sample.  
The follow-up measurements of the individual spectra obtained after the fitting with ULySS are reported in 
Table \ref{longtable}.   Although in the case of few spectra the narrow and semi-broad component of \oiii\ lines have 
switched places, we carefully disentangle the two components according to their widths. 


 \begin{figure*}
	\begin{center}
		\includegraphics[width=0.3\textwidth]{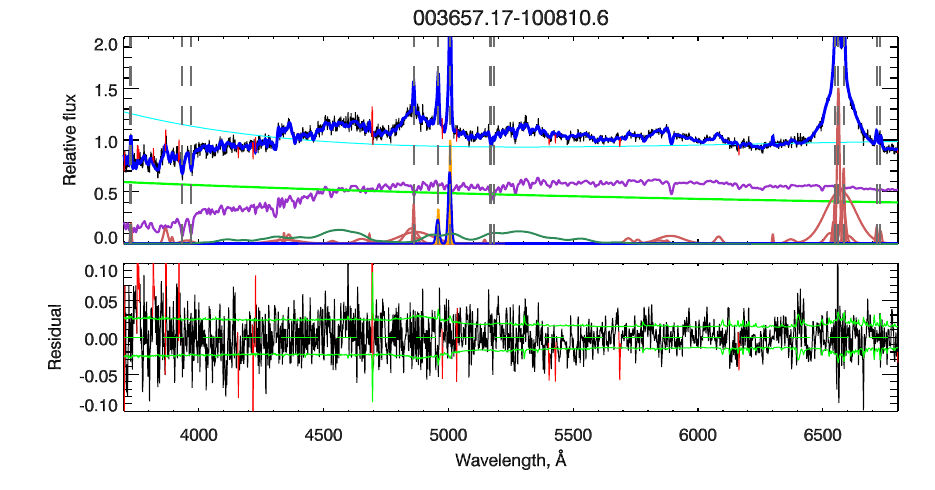}
	        \includegraphics[width=0.3\textwidth]{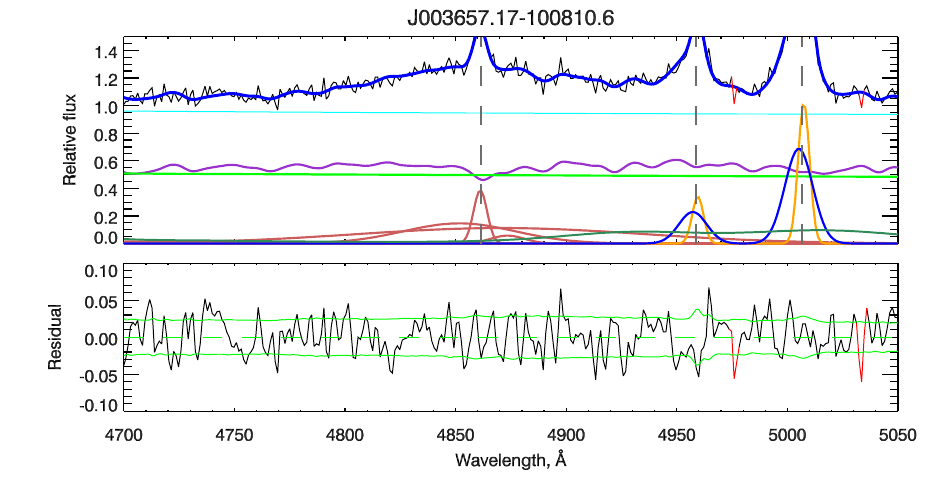}
		\includegraphics[width=0.16\textwidth]{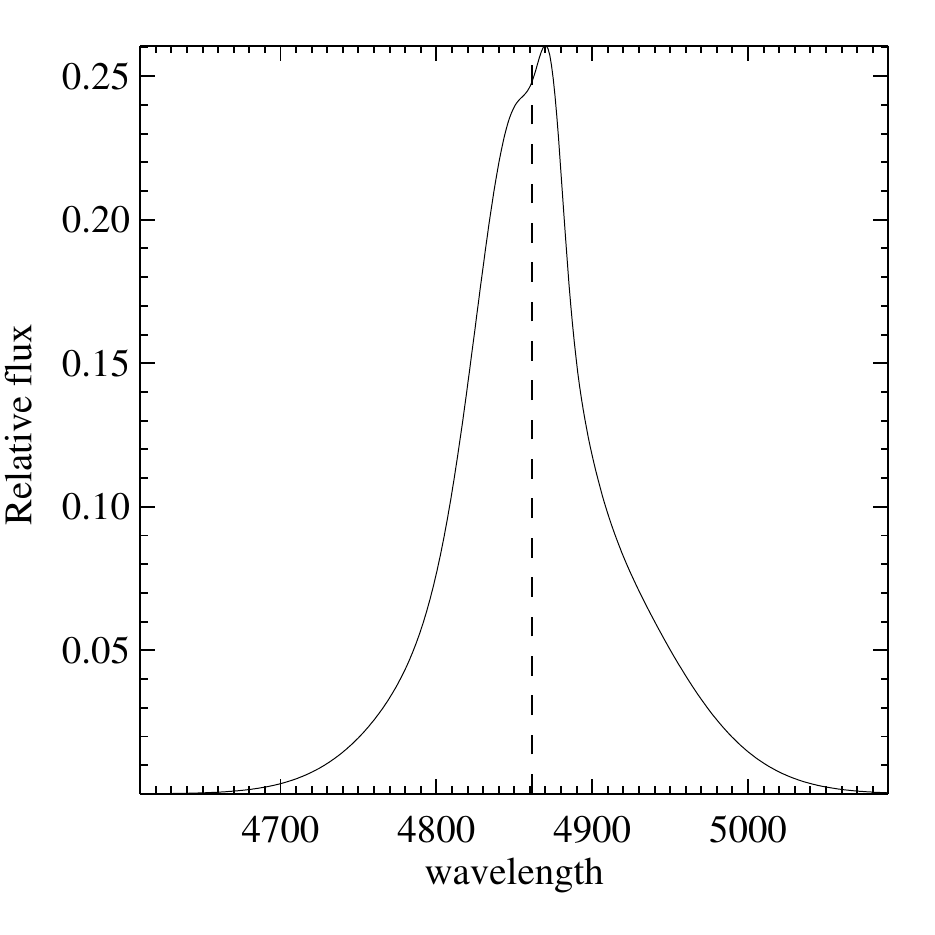}
		\includegraphics[width=0.3\textwidth]{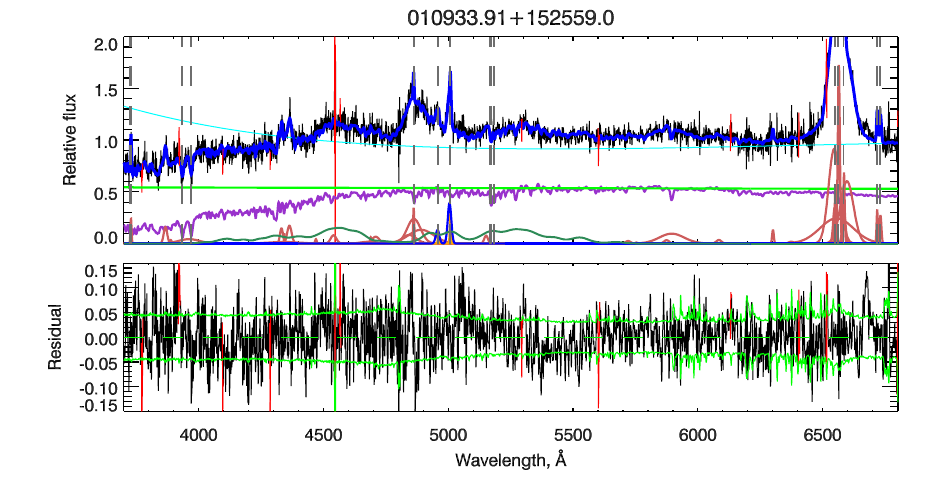}
		\includegraphics[width=0.3\textwidth]{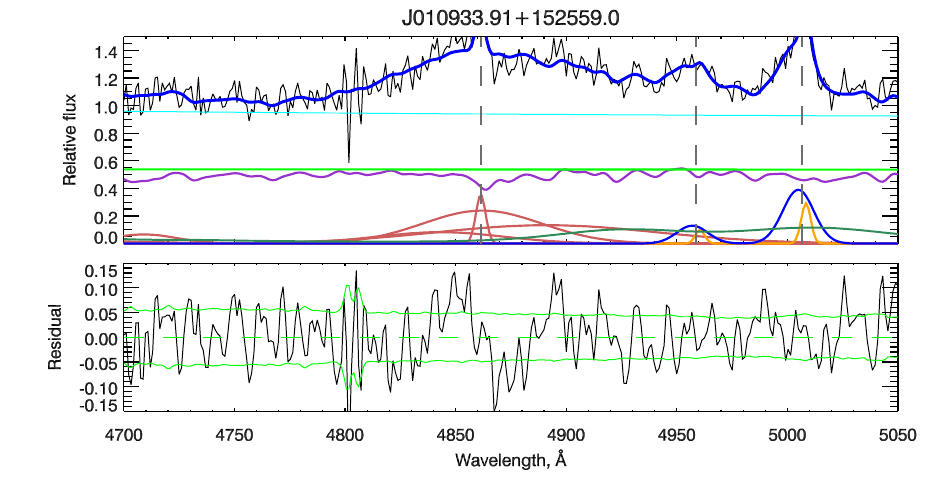}
                \includegraphics[width=0.16\textwidth]{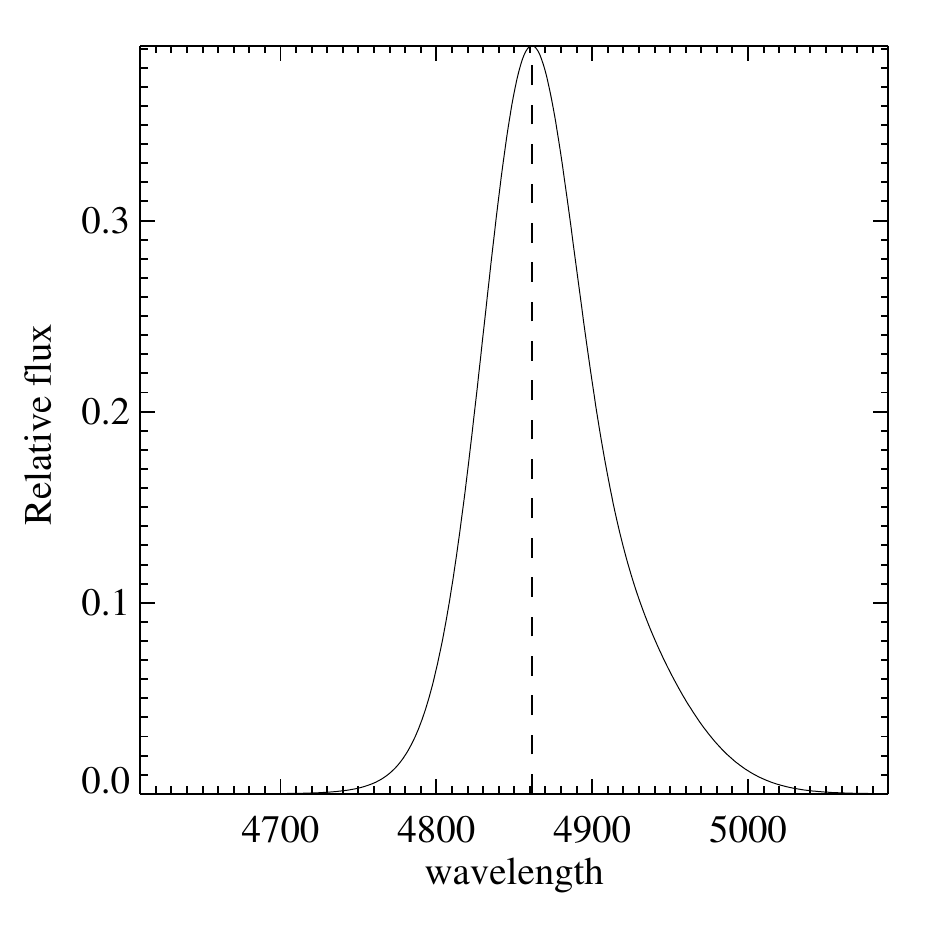}
		\includegraphics[width=0.3\textwidth]{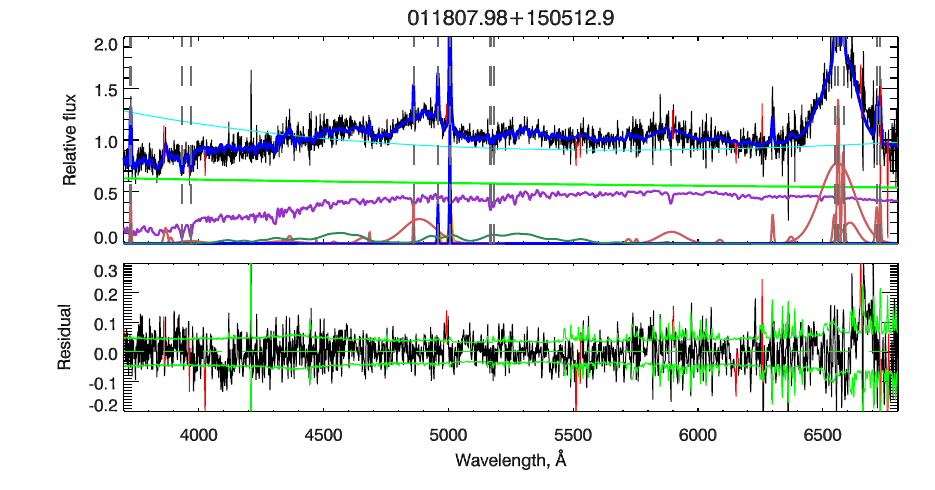}
		\includegraphics[width=0.3\textwidth]{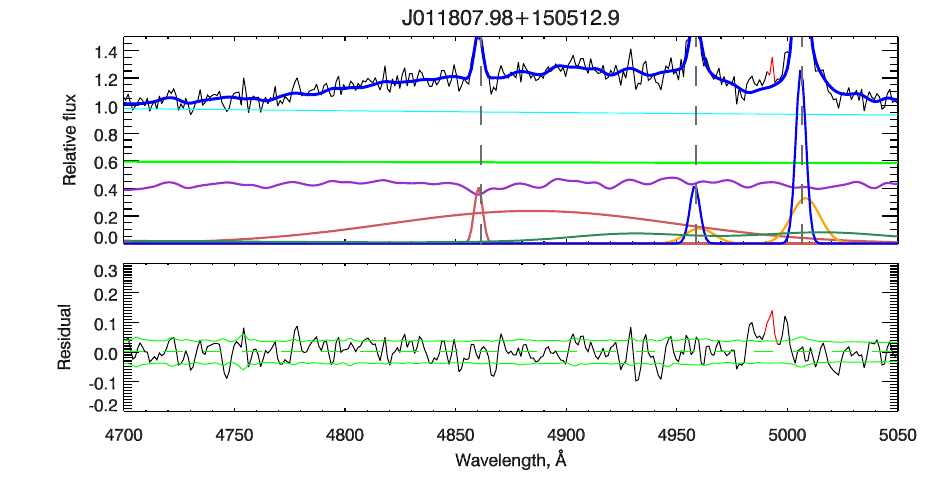}
                \includegraphics[width=0.16\textwidth]{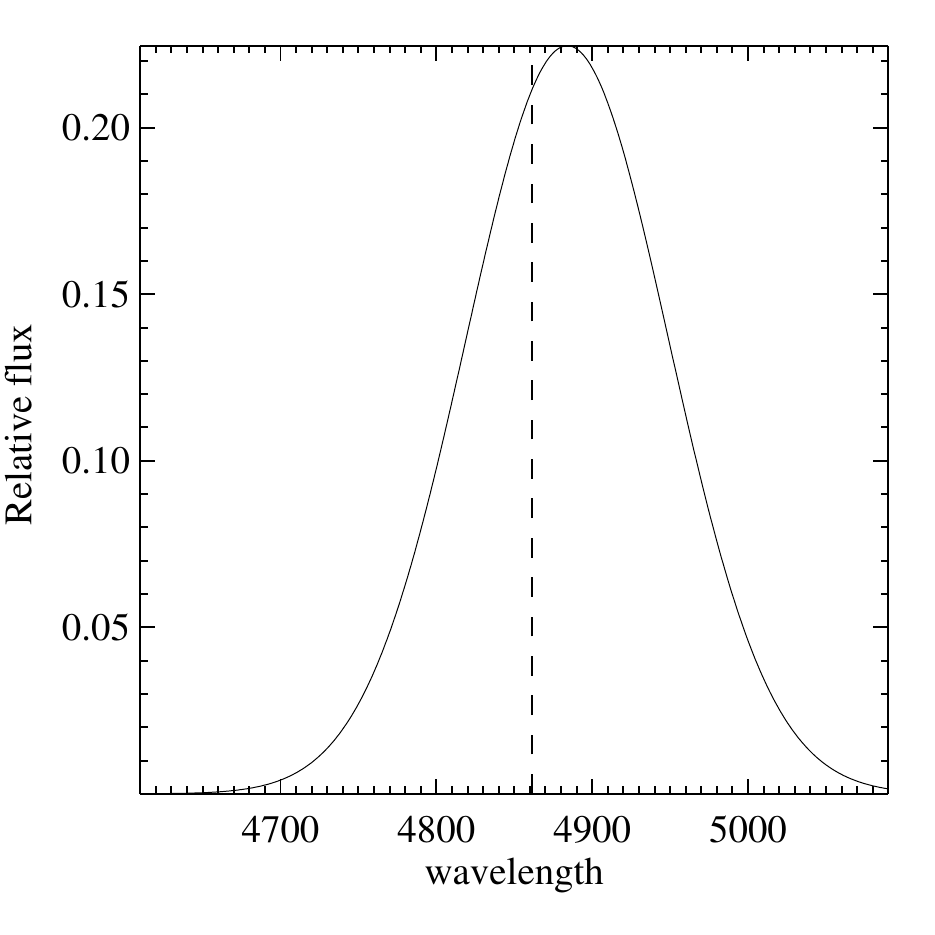}
		\includegraphics[width=0.3\textwidth]{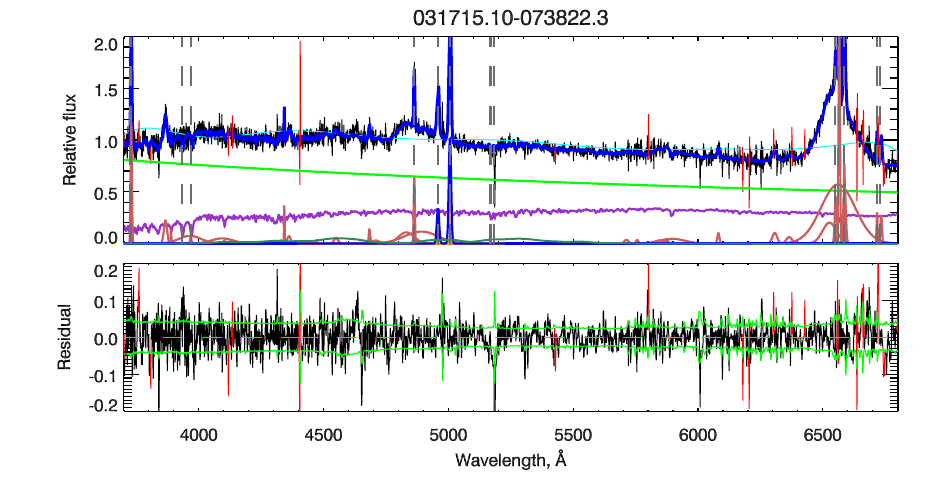}
		\includegraphics[width=0.3\textwidth]{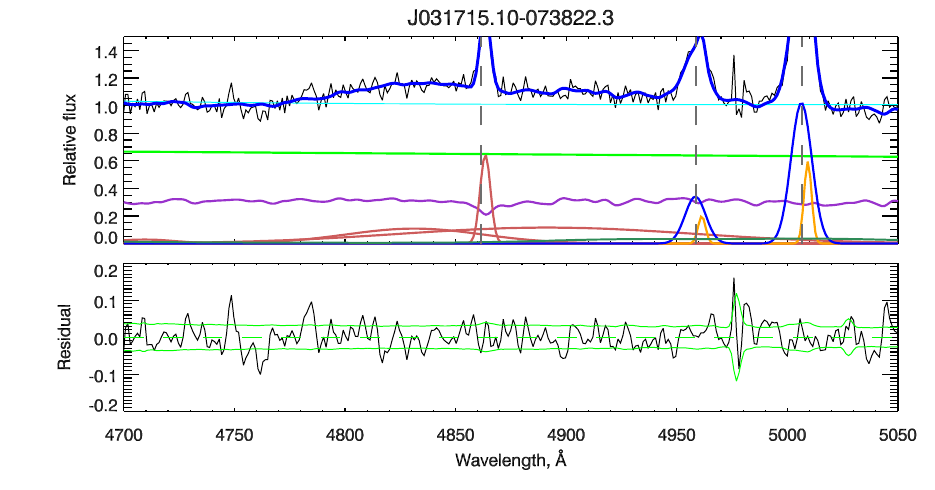}
		\includegraphics[width=0.16\textwidth]{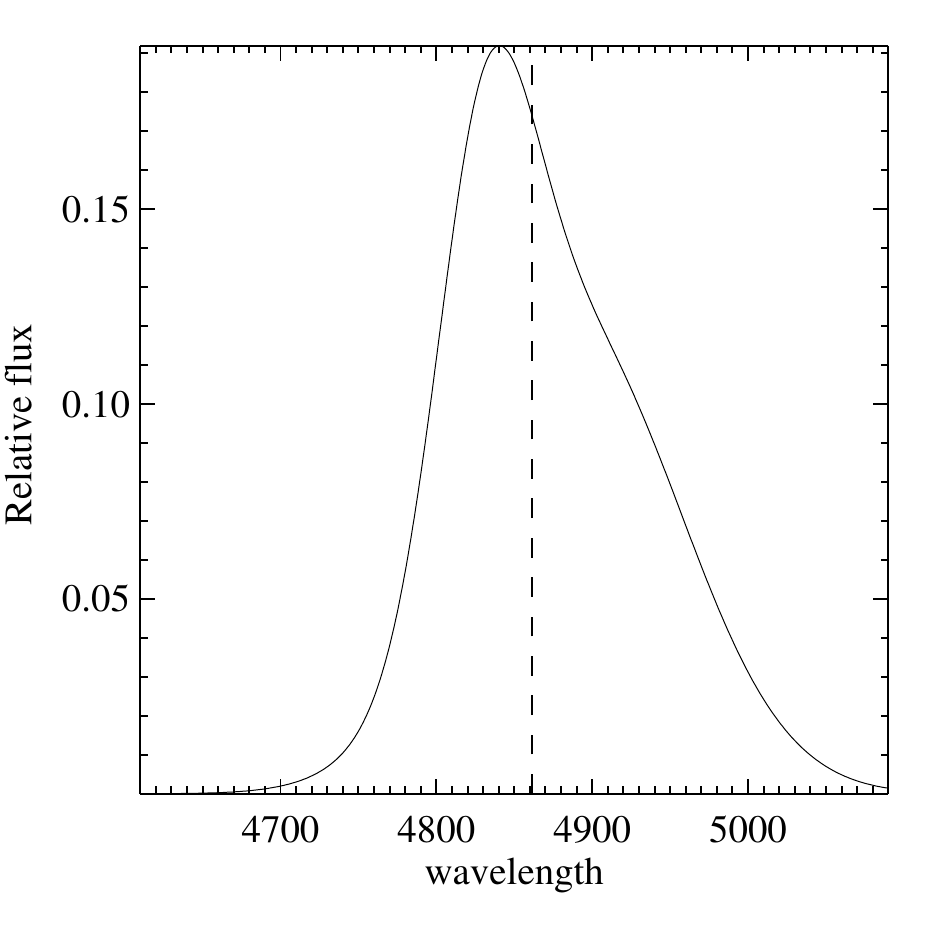}
		\includegraphics[width=0.3\textwidth]{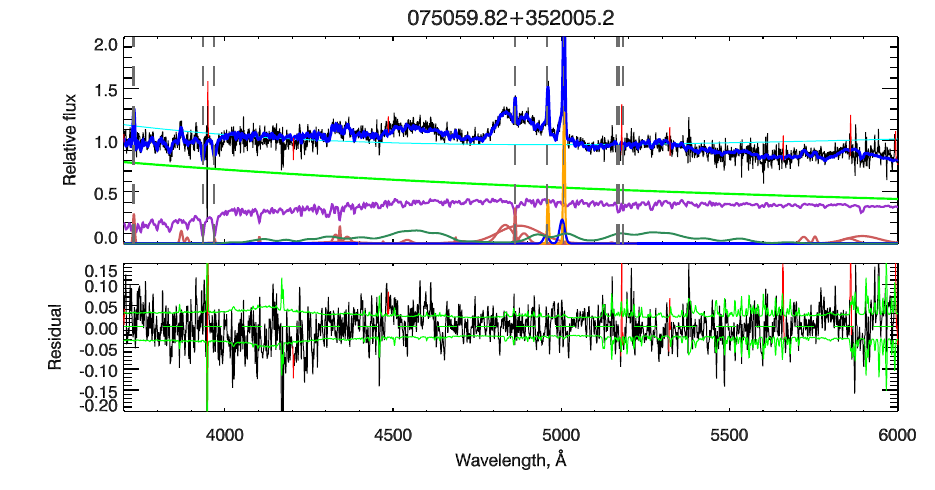}
		\includegraphics[width=0.3\textwidth]{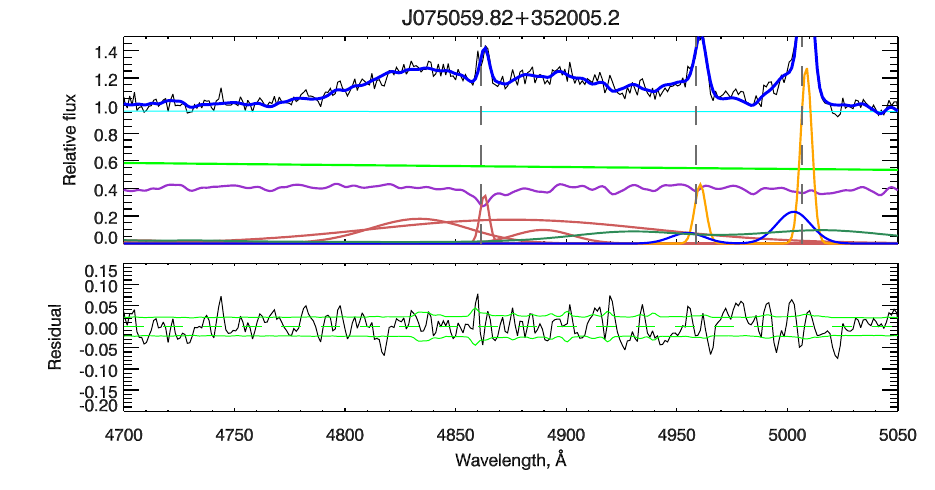}
		\includegraphics[width=0.16\textwidth]{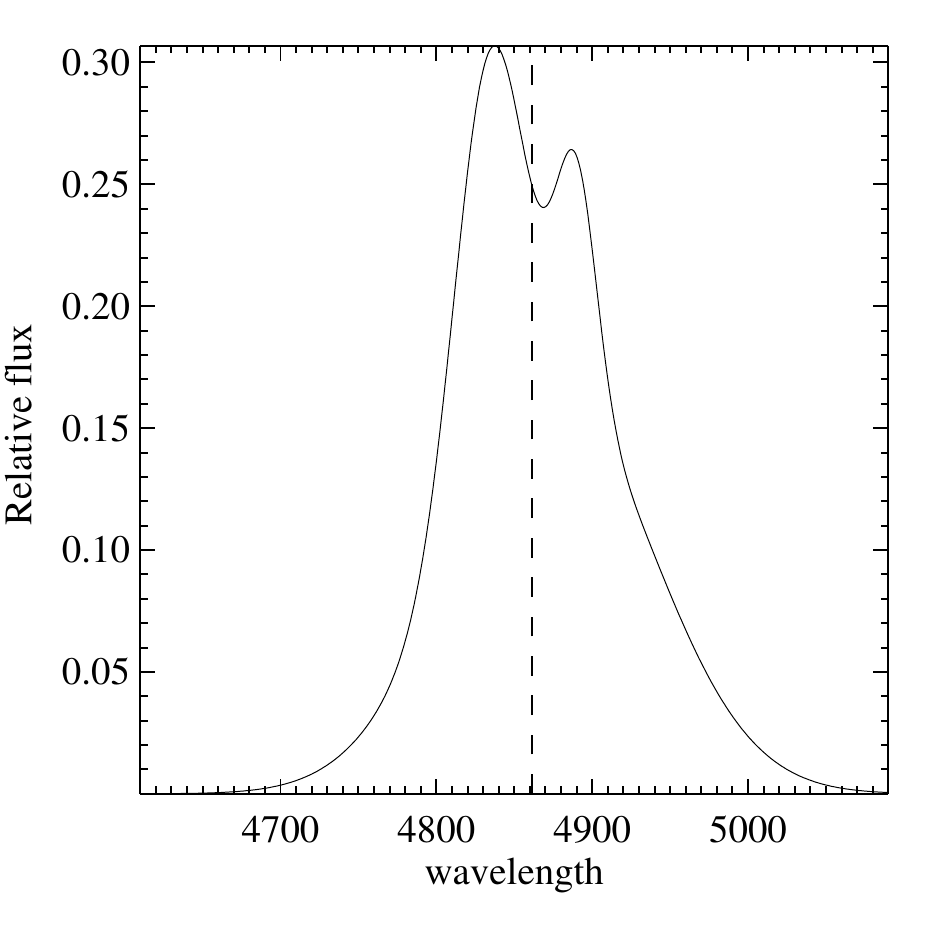}
		\includegraphics[width=0.3\textwidth]{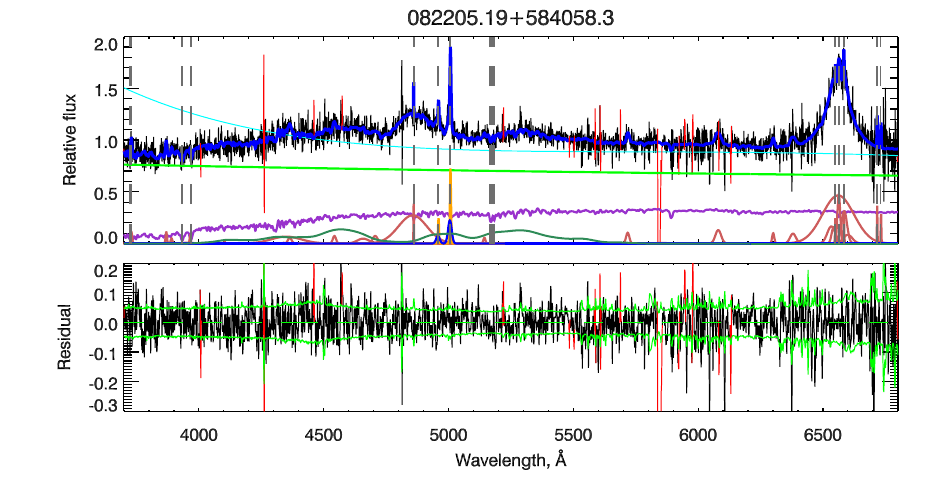}
		\includegraphics[width=0.3\textwidth]{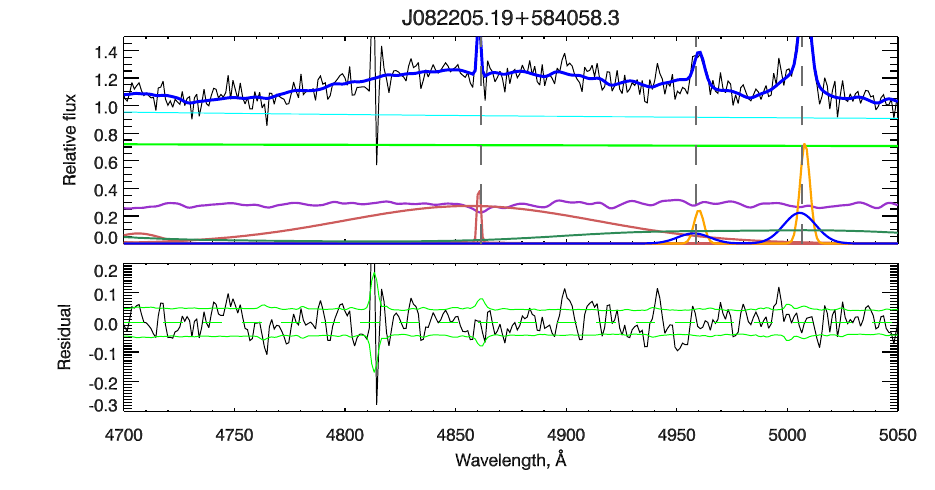}
		\includegraphics[width=0.16\textwidth]{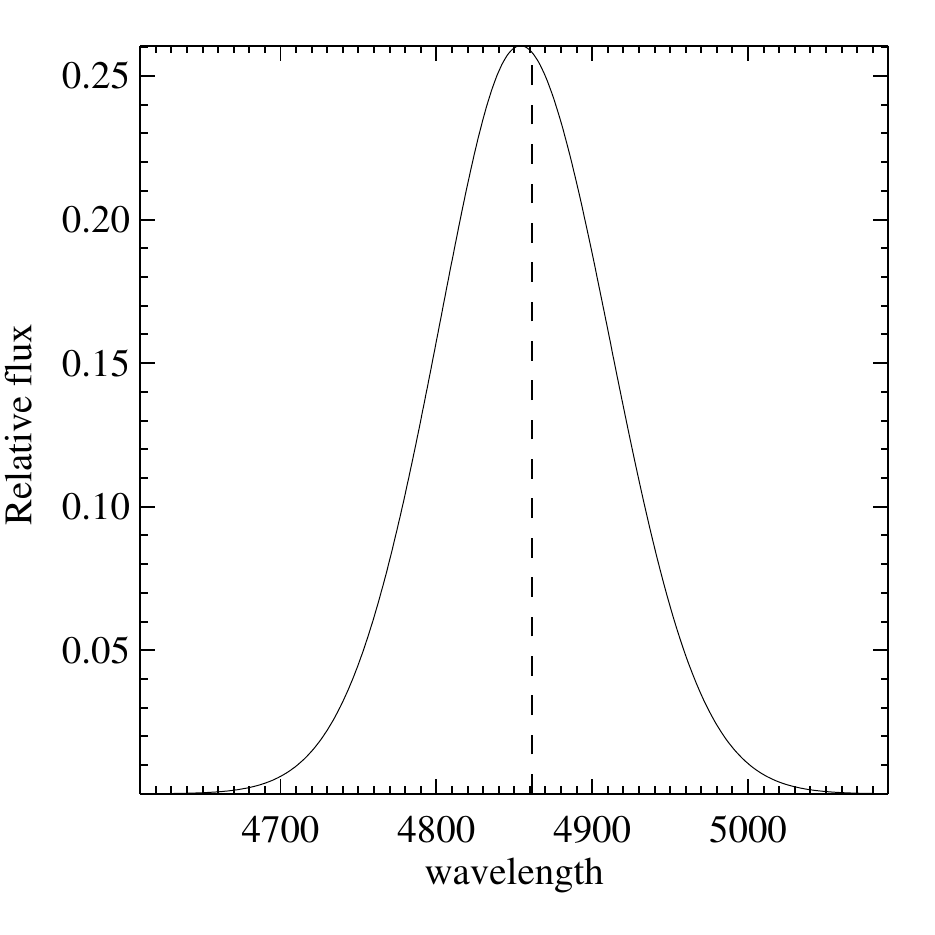}
		\includegraphics[width=0.3\textwidth]{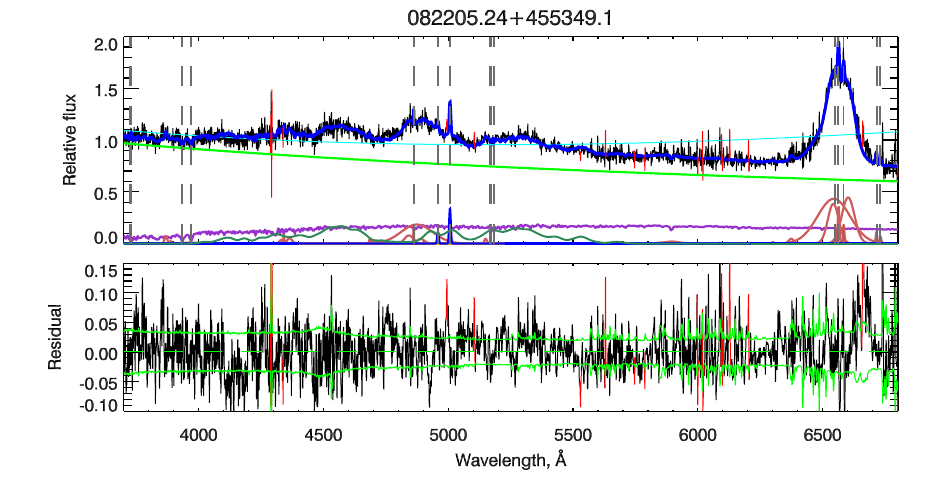}
		\includegraphics[width=0.3\textwidth]{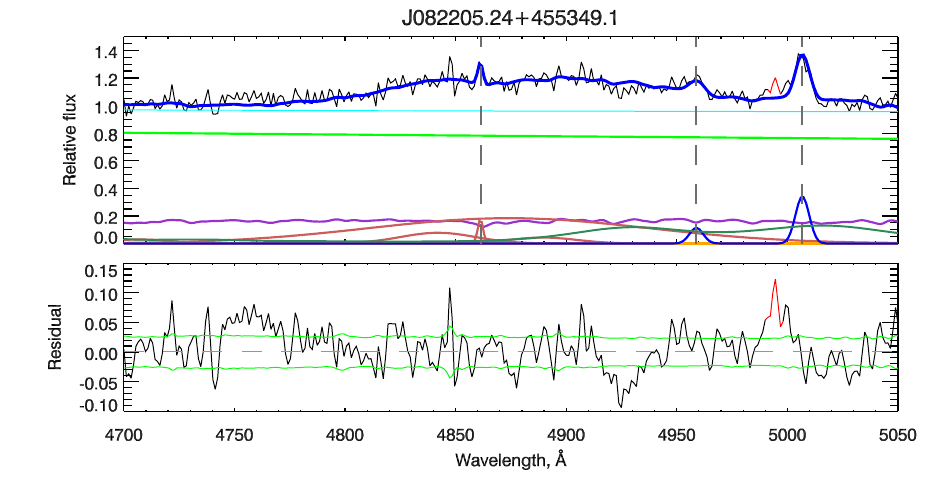}
		\includegraphics[width=0.16\textwidth]{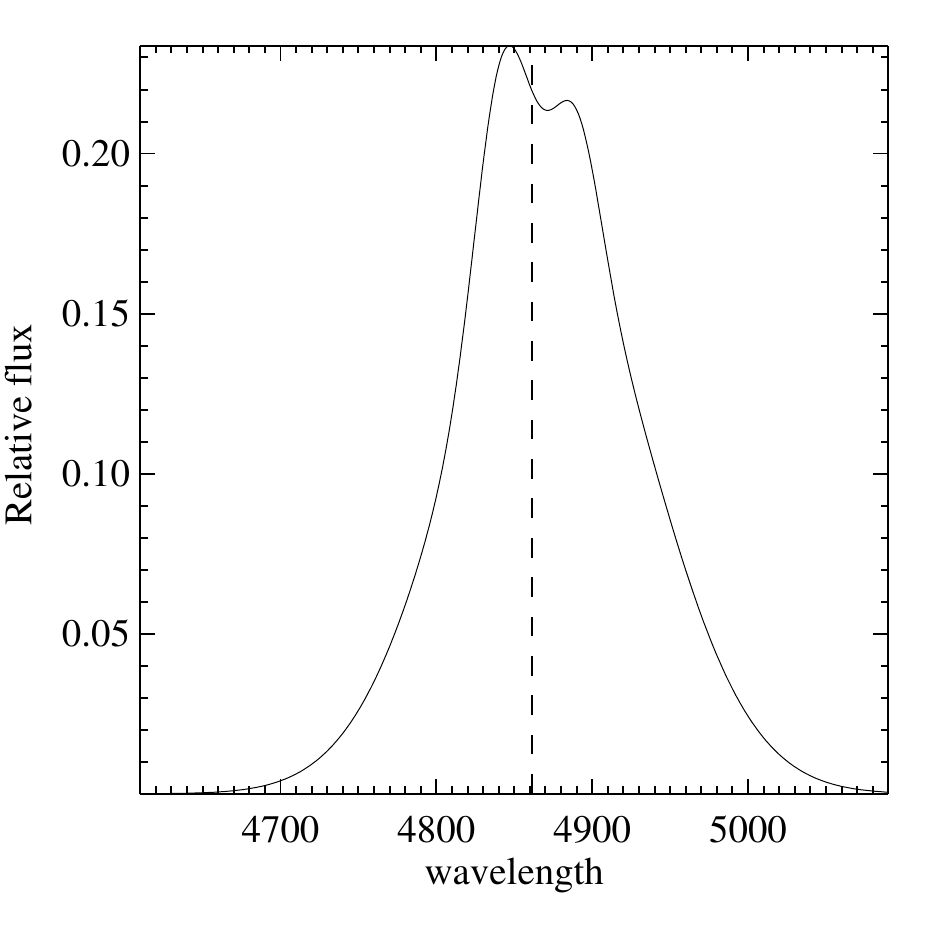}
		\includegraphics[width=0.3\textwidth]{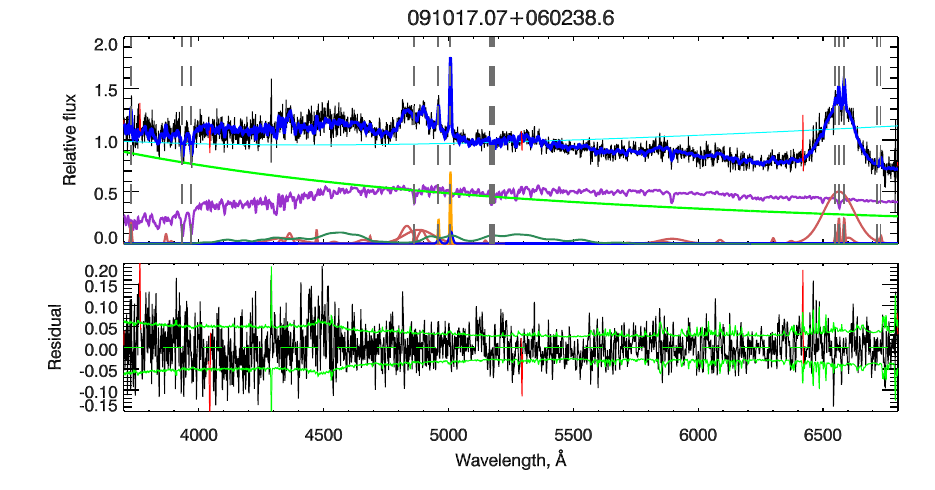}
		\includegraphics[width=0.3\textwidth]{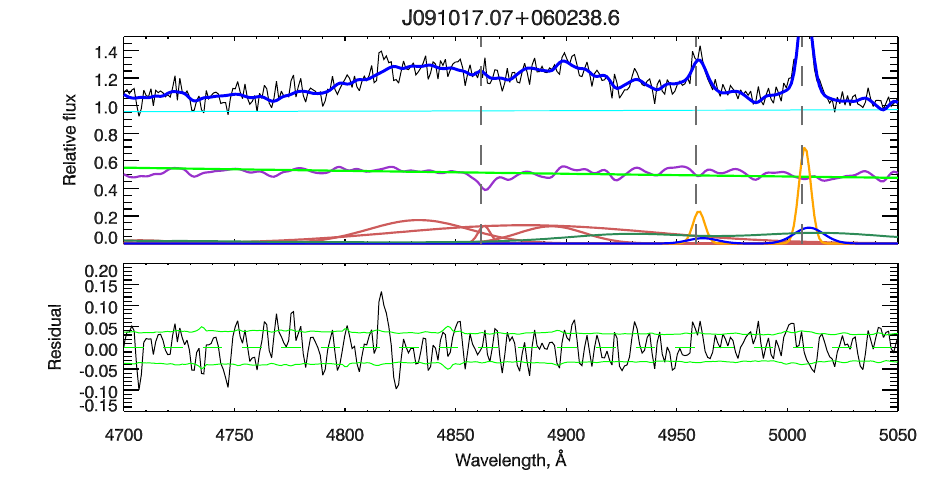}
		\includegraphics[width=0.16\textwidth]{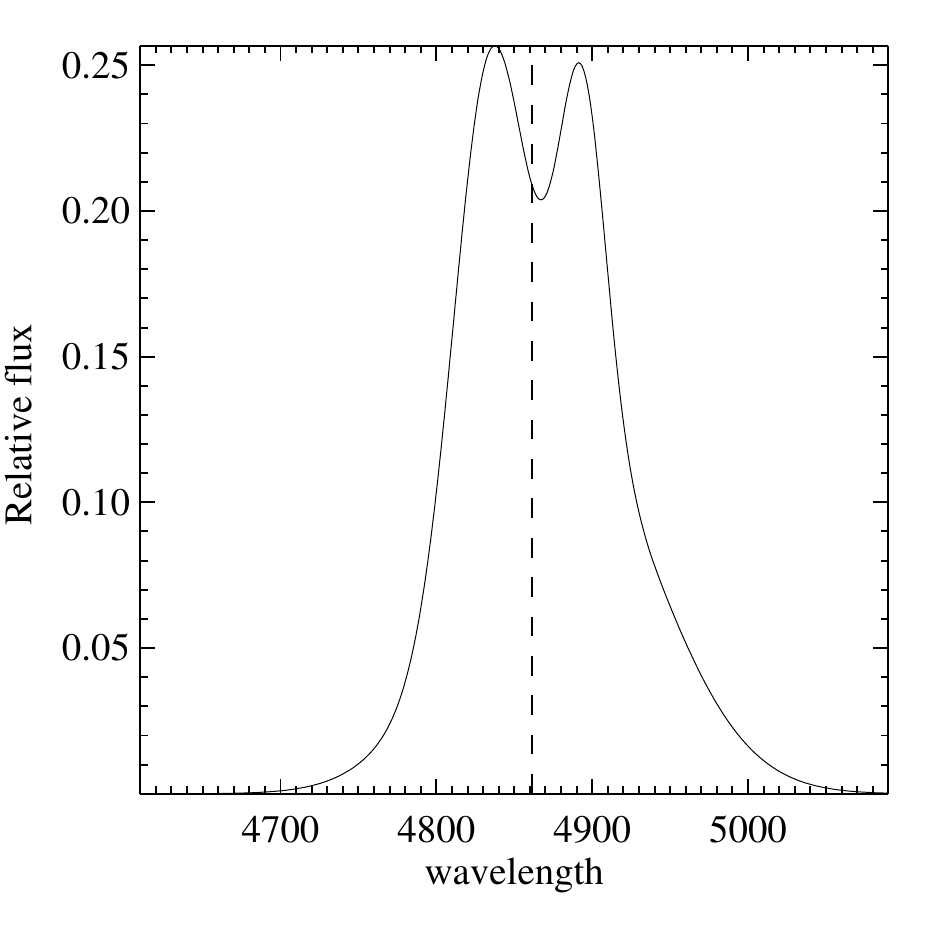}
		
\caption{Ulyss fits to the HG sample. In the upper left panel, the black line represents
the observed spectrum, the blue one the best fit, the red line the multiplicative polynomial, while the
green, light red, and violet lines represent components of the best fit: violet -- stellar population, red
-- emission lines, and green --  AGN continuum. The bottom left panel shows the residuals of the best fit
(black line). The green solid line shows the level of the noise, and the dashed line is the zero-axis.
Middle panels zoom the domain around H$\beta$ and \oiii\ lines, while right hand panels show the model
broad H$\beta$.  }\label{fits_plots}
	\end{center}
\end{figure*}

\begin{figure*}
	\begin{center}
		\ContinuedFloat*
                \includegraphics[width=0.3\textwidth]{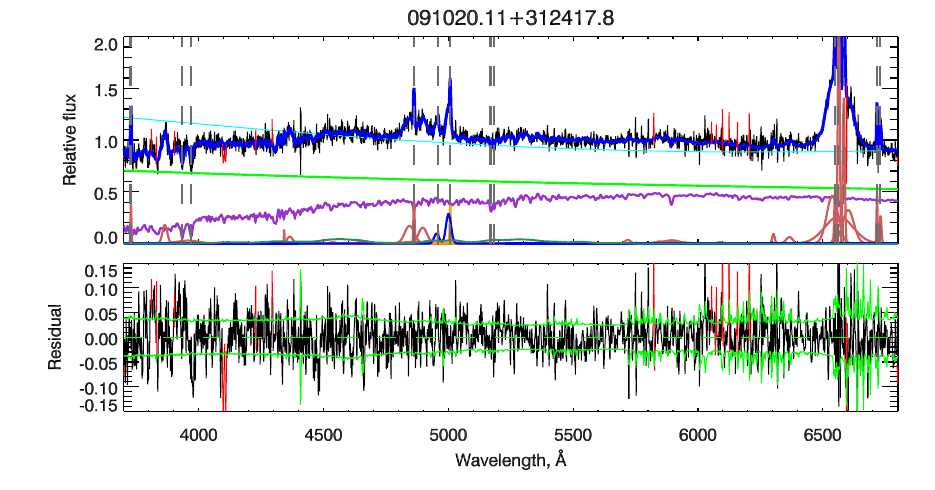}
		\includegraphics[width=0.3\textwidth]{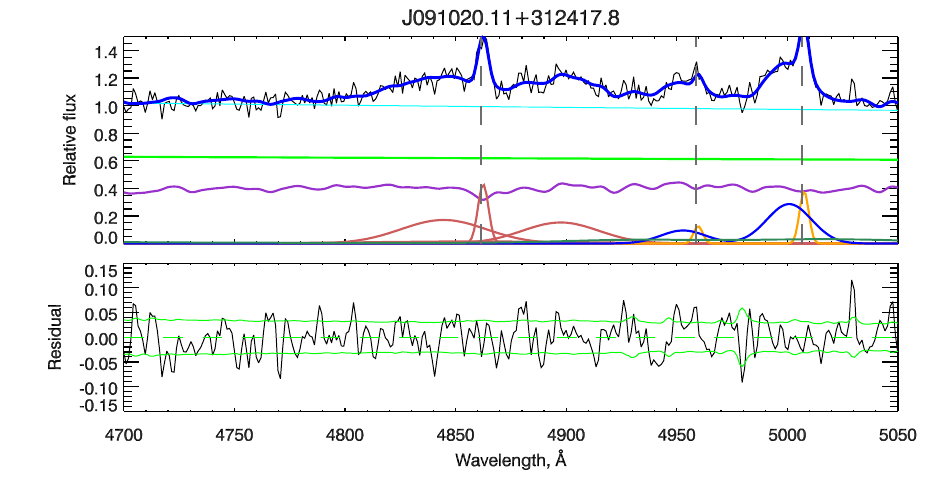}
		\includegraphics[width=0.16\textwidth]{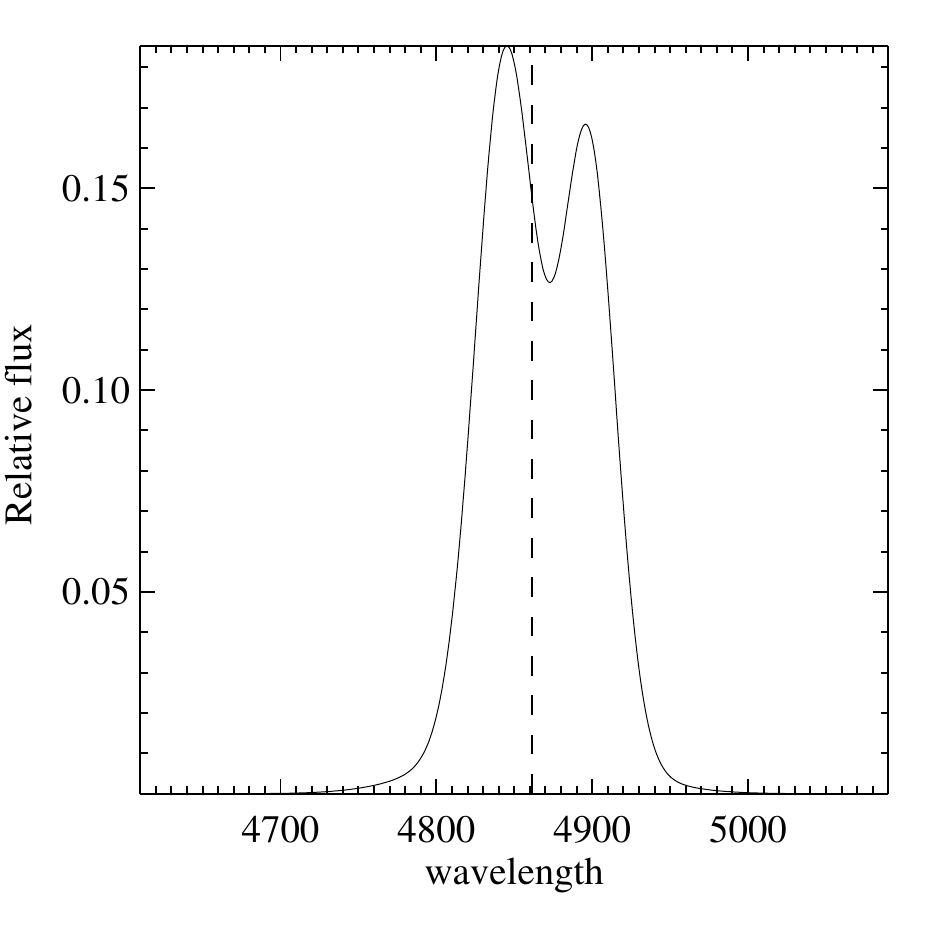}
	        \includegraphics[width=0.3\textwidth]{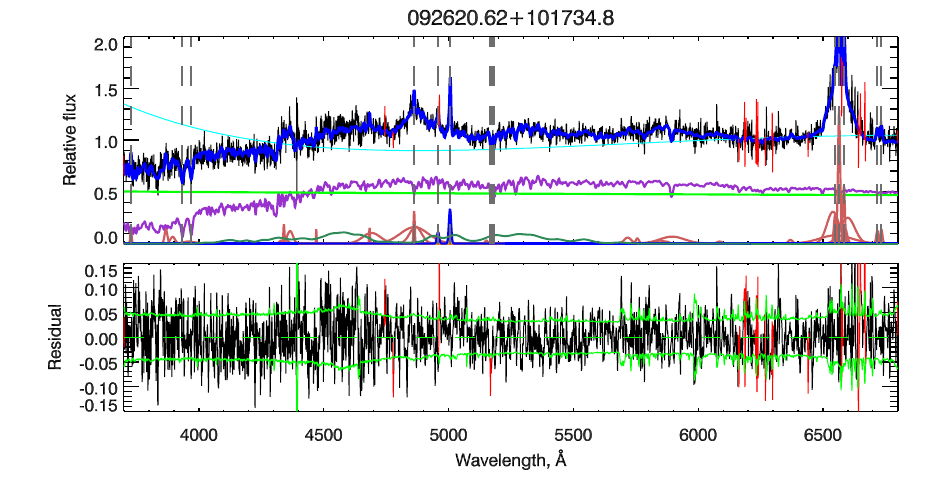}
		\includegraphics[width=0.3\textwidth]{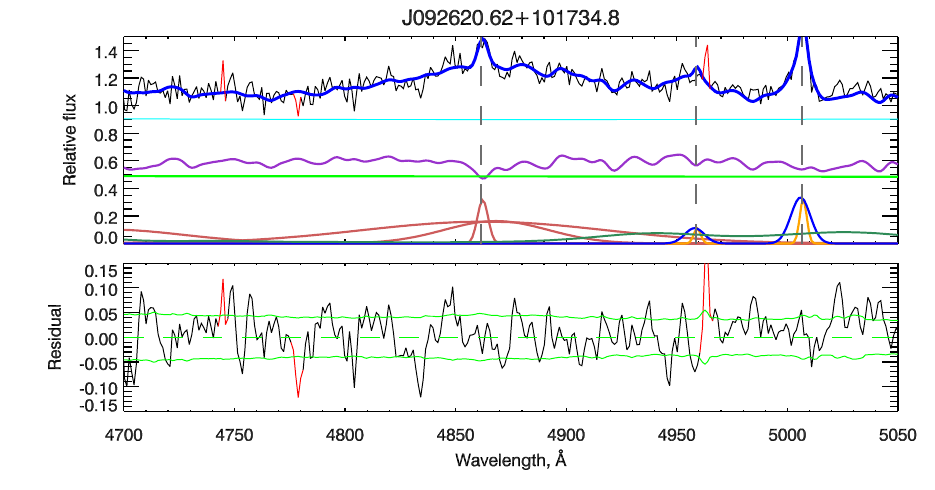}
		\includegraphics[width=0.16\textwidth]{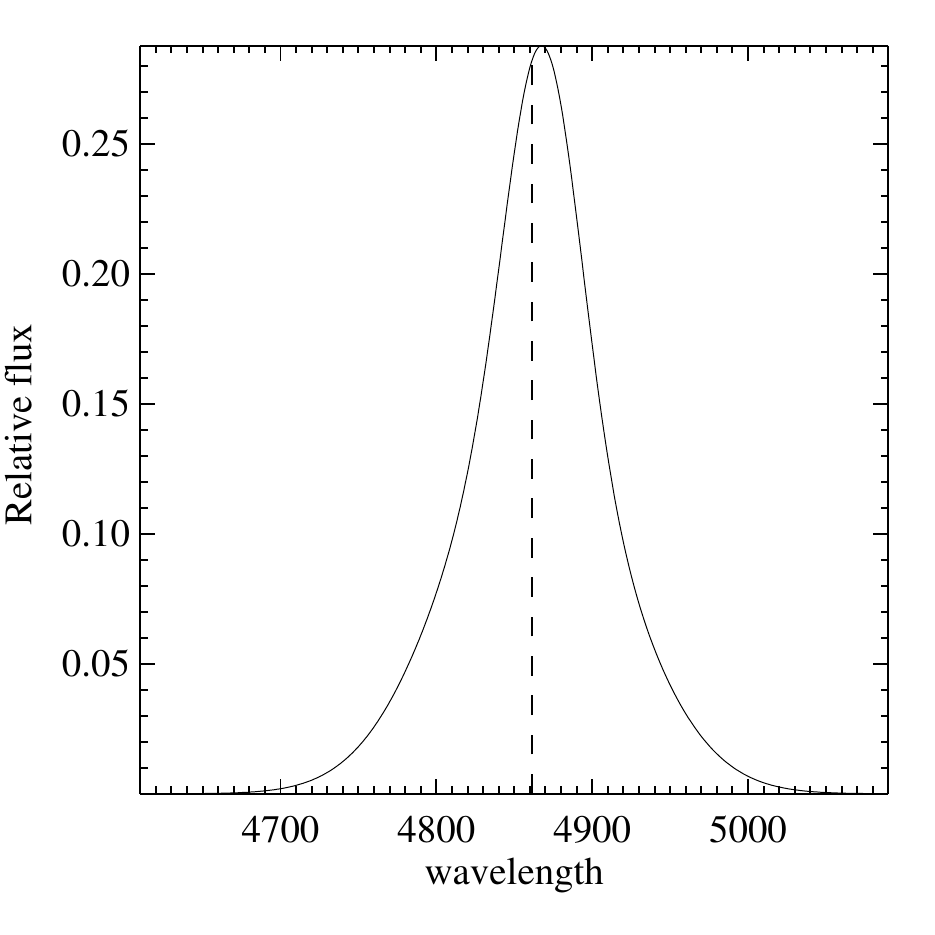}
		\includegraphics[width=0.3\textwidth]{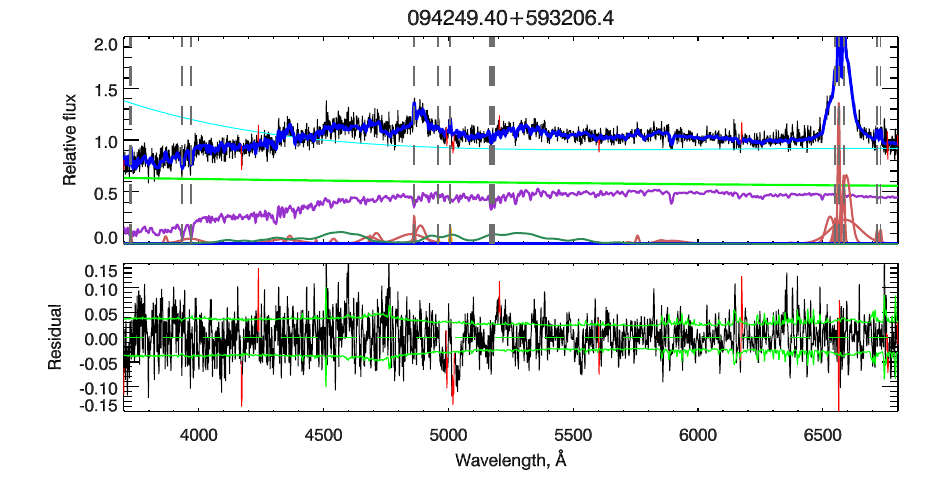}
		\includegraphics[width=0.3\textwidth]{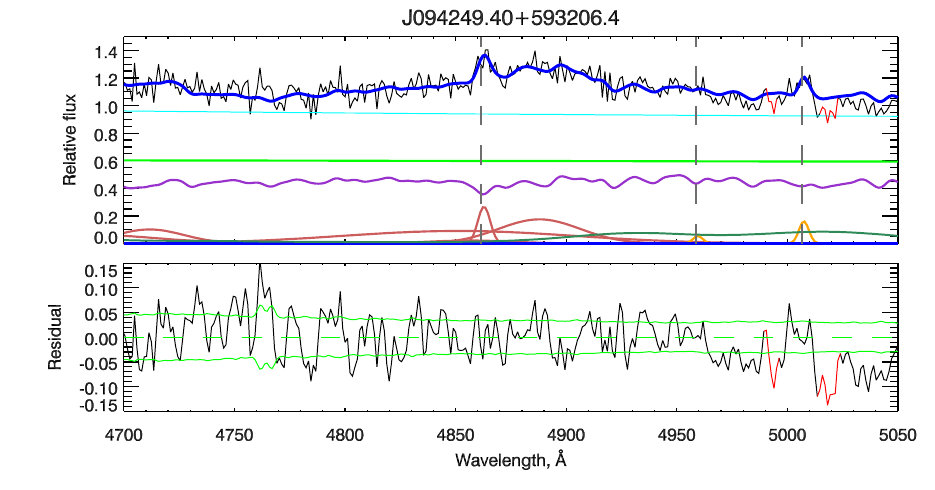}
		\includegraphics[width=0.16\textwidth]{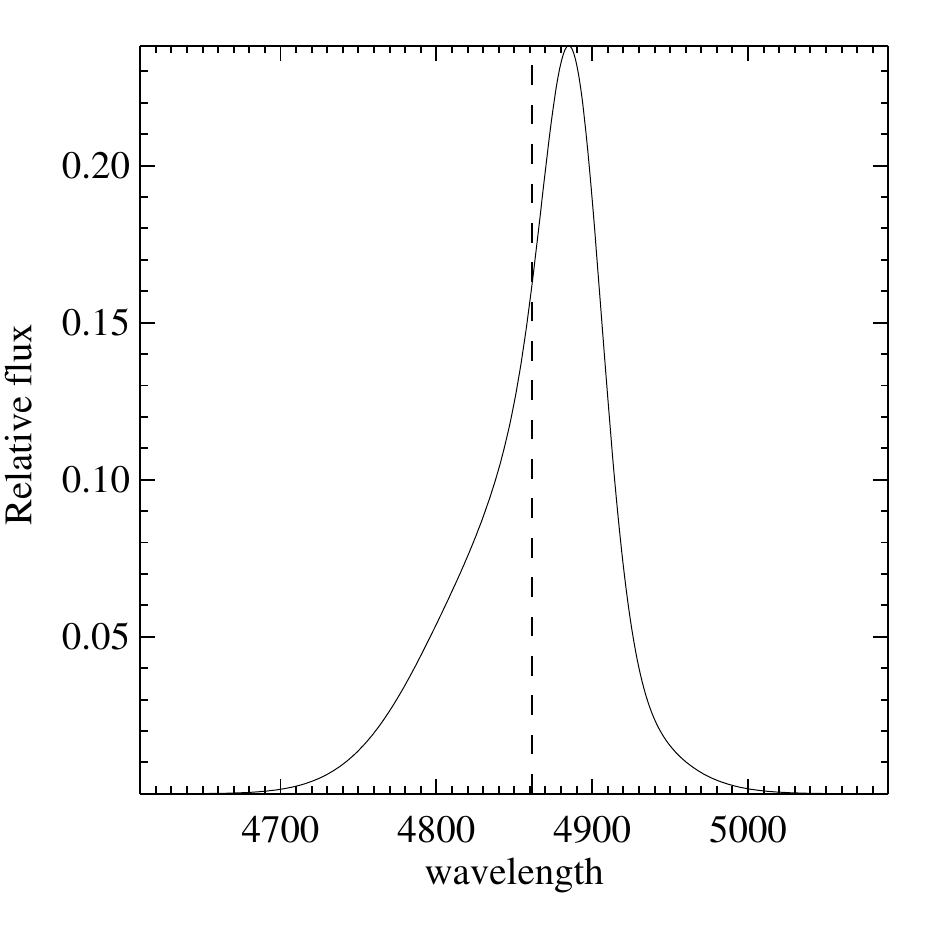}
		\includegraphics[width=0.3\textwidth]{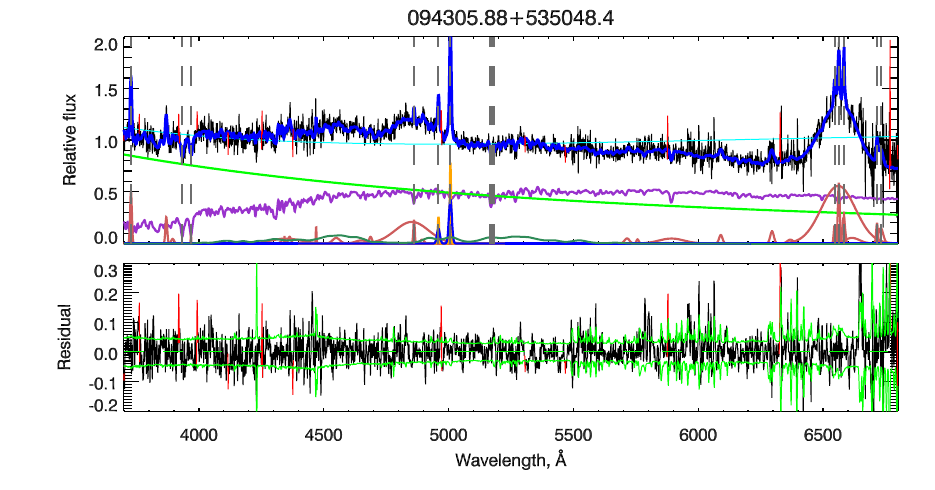}
		\includegraphics[width=0.3\textwidth]{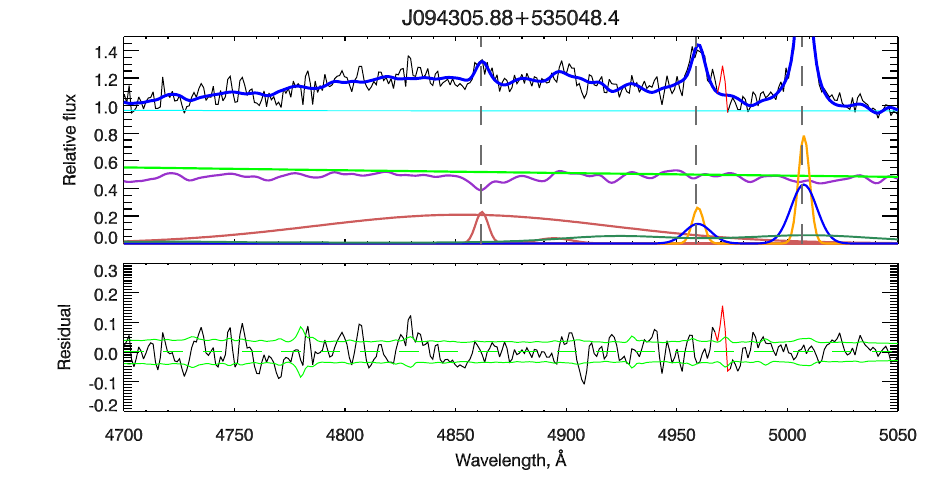}
		\includegraphics[width=0.16\textwidth]{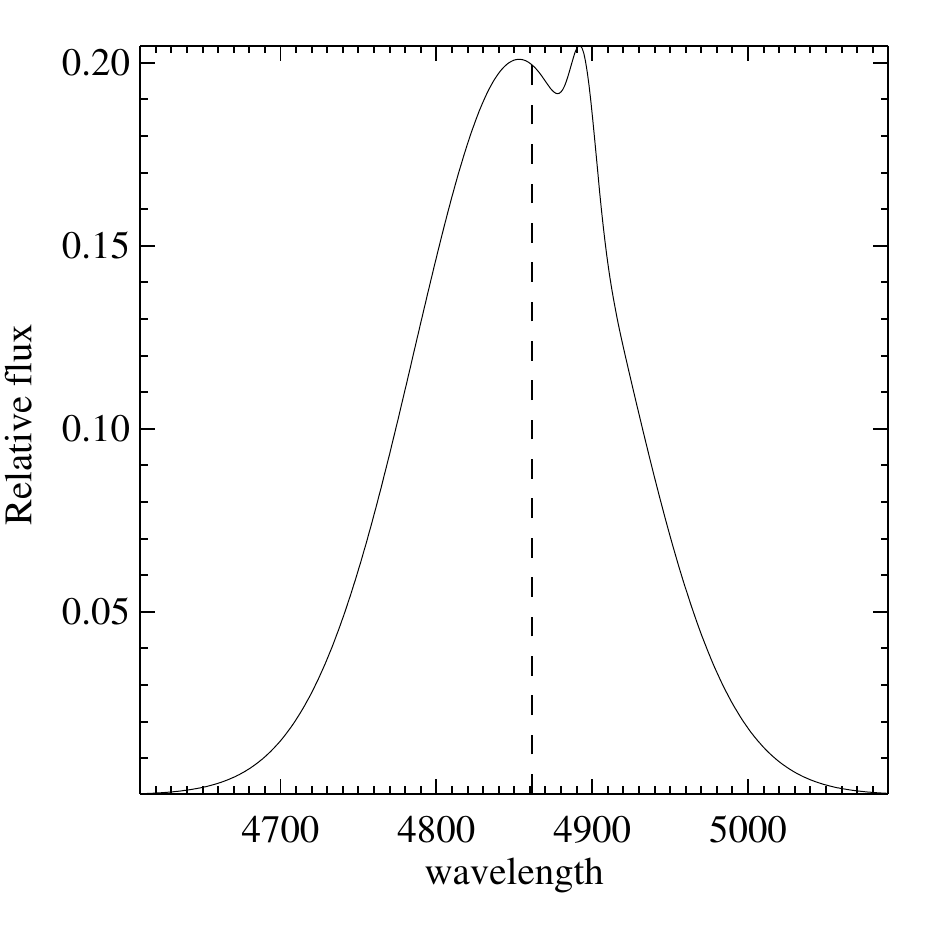}
		\includegraphics[width=0.3\textwidth]{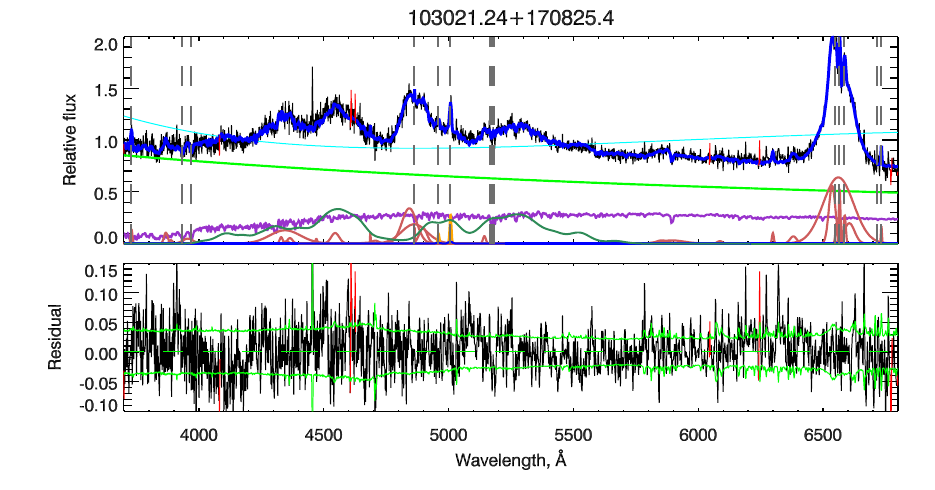}
		\includegraphics[width=0.3\textwidth]{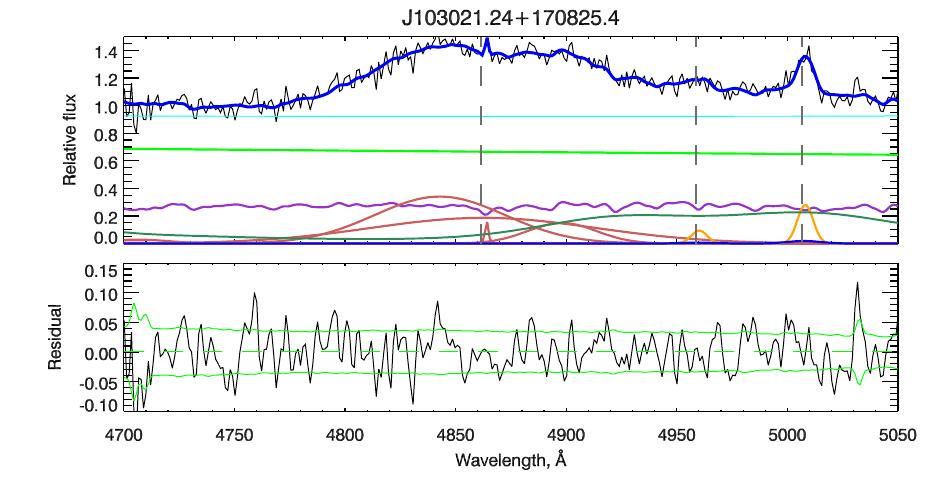}
		\includegraphics[width=0.16\textwidth]{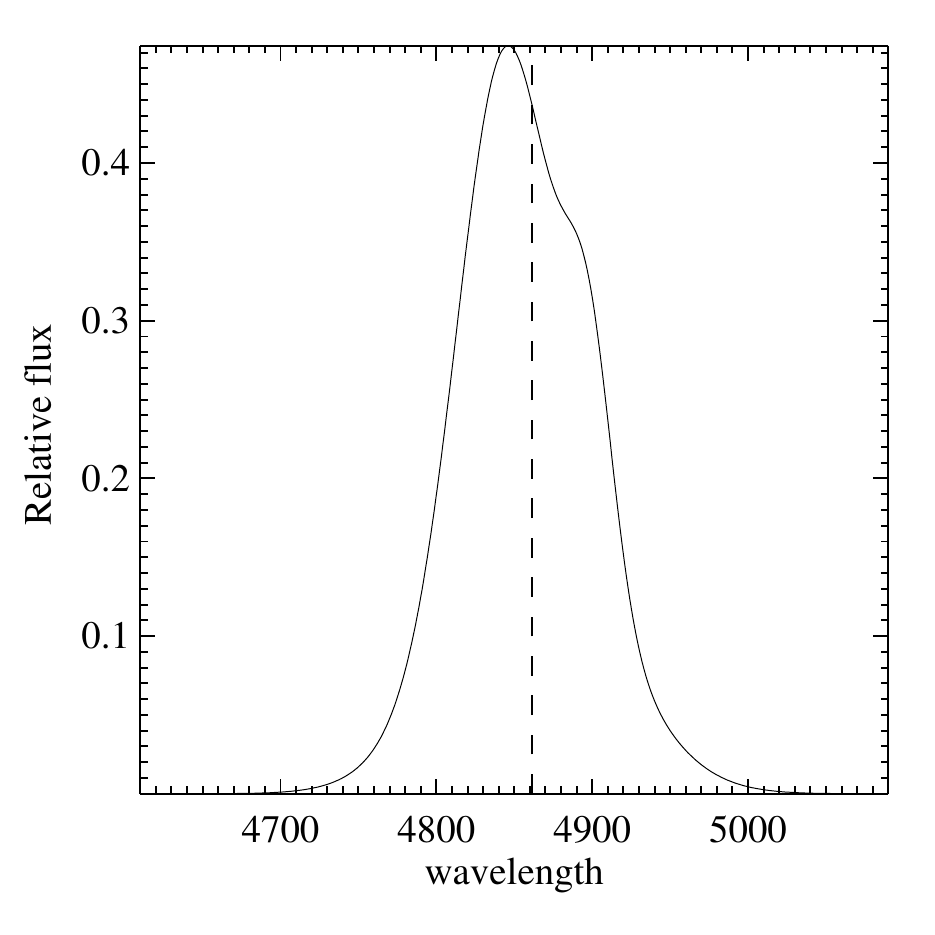}
		\includegraphics[width=0.3\textwidth]{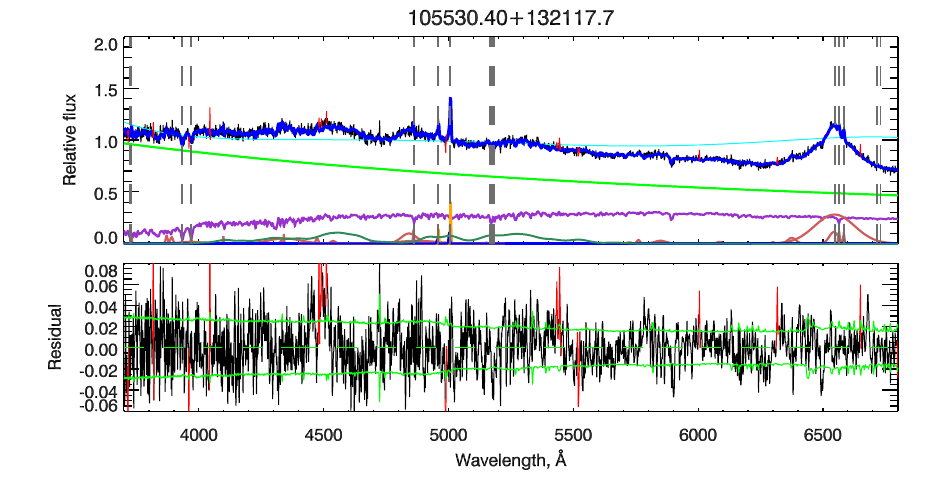}
		\includegraphics[width=0.3\textwidth]{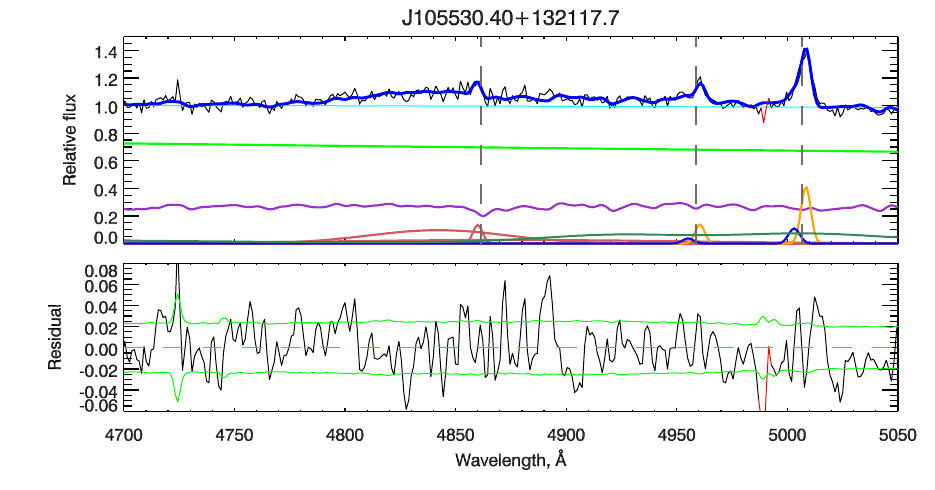}
		\includegraphics[width=0.16\textwidth]{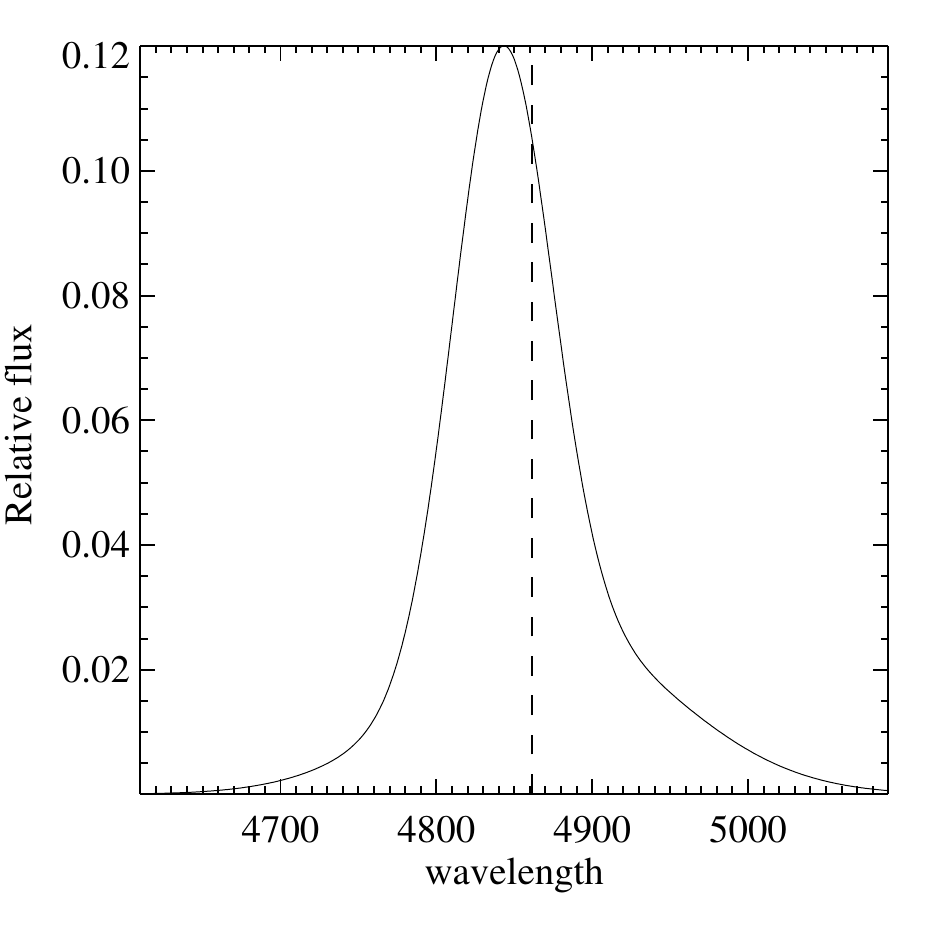}
		\includegraphics[width=0.3\textwidth]{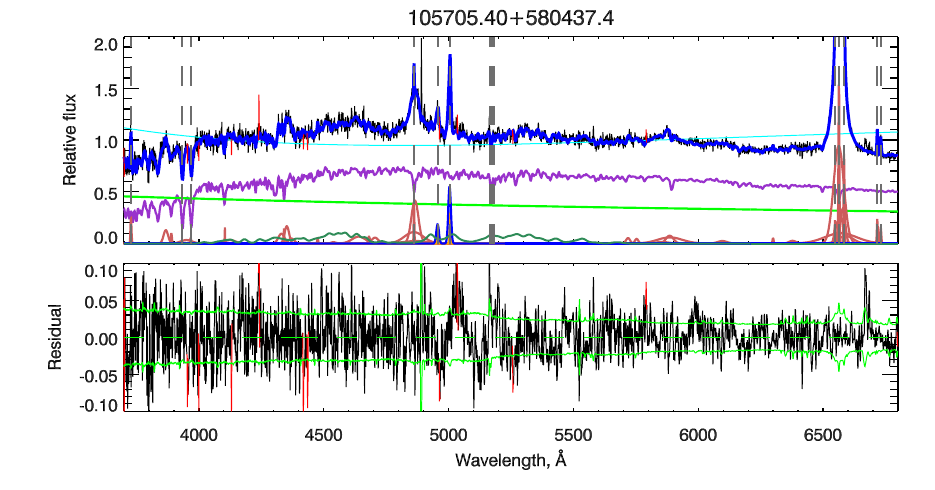}
		\includegraphics[width=0.3\textwidth]{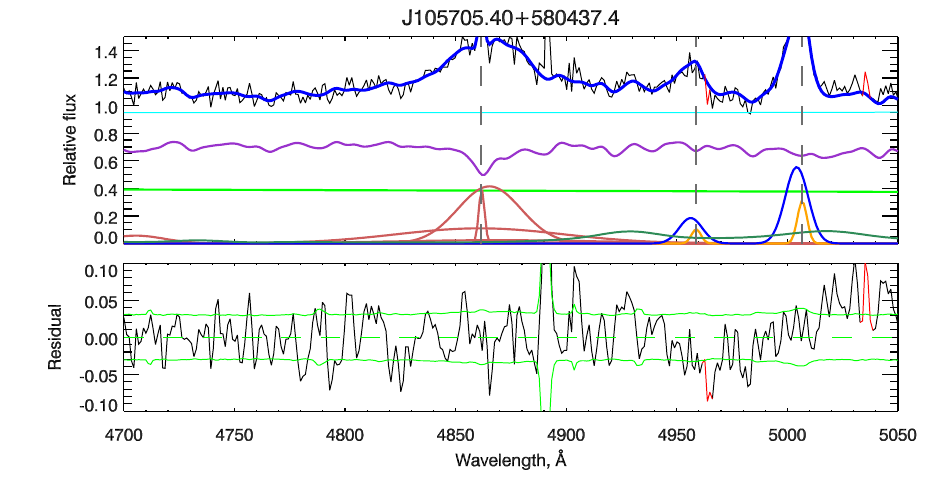}
		\includegraphics[width=0.16\textwidth]{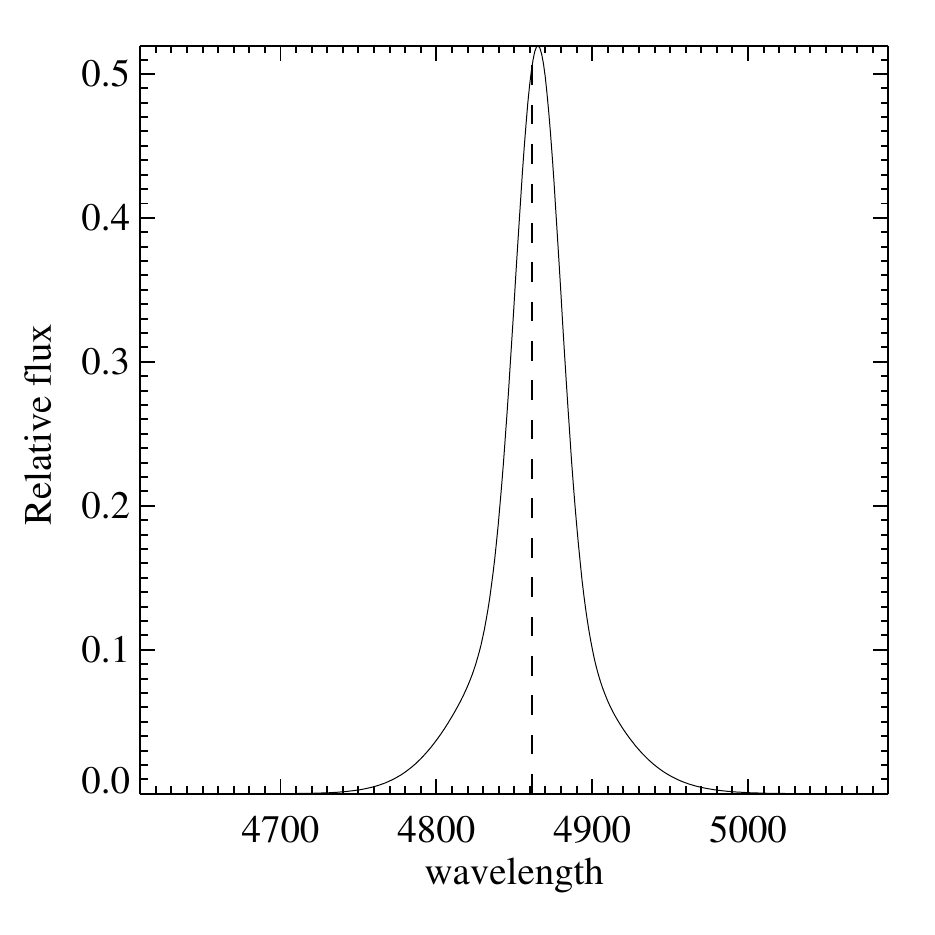}
		\includegraphics[width=0.3\textwidth]{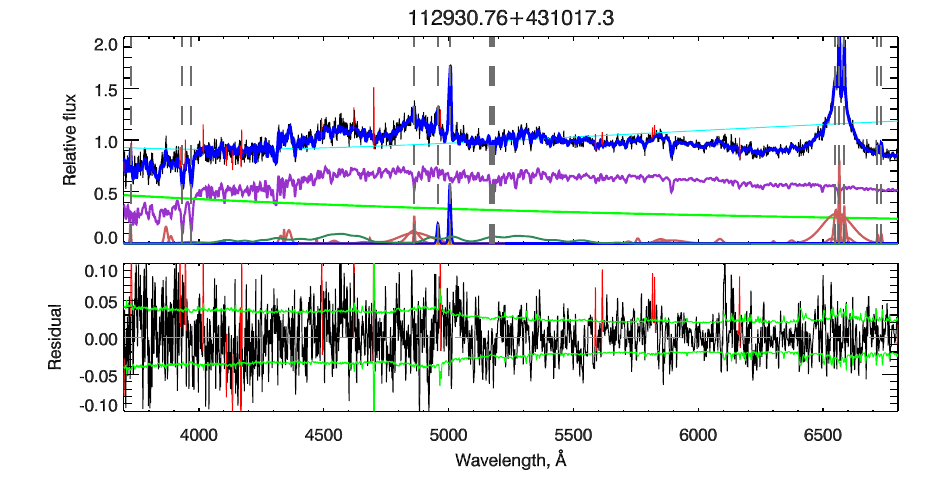}
		\includegraphics[width=0.3\textwidth]{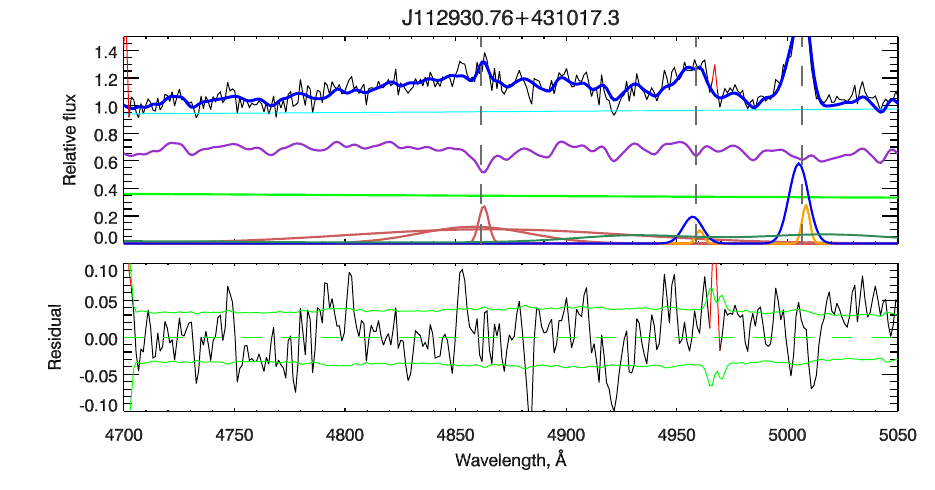}
		\includegraphics[width=0.16\textwidth]{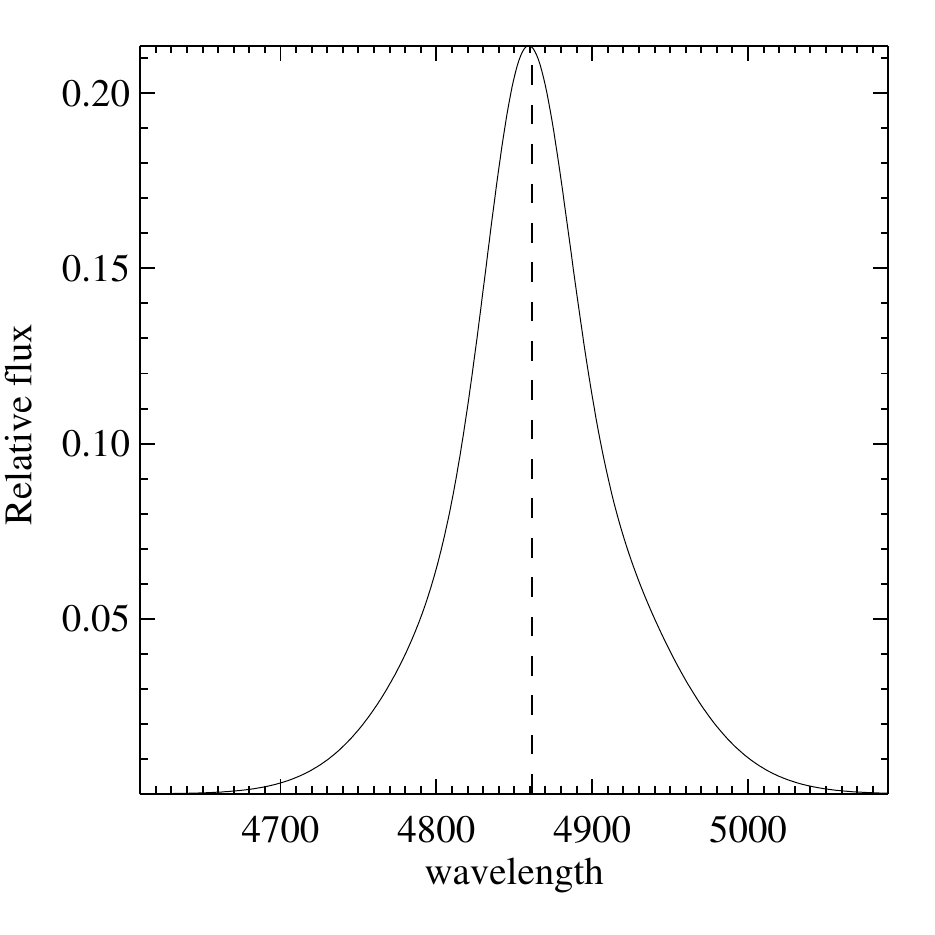}
		\includegraphics[width=0.3\textwidth]{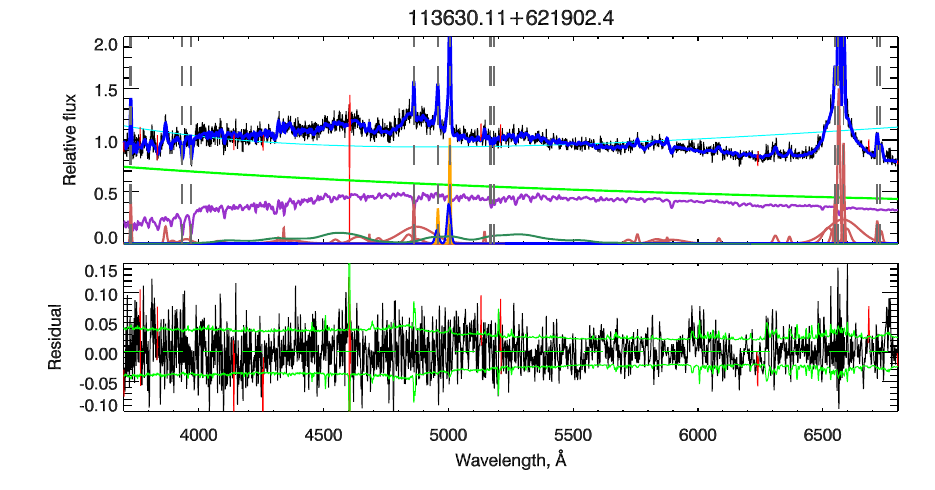}
		\includegraphics[width=0.3\textwidth]{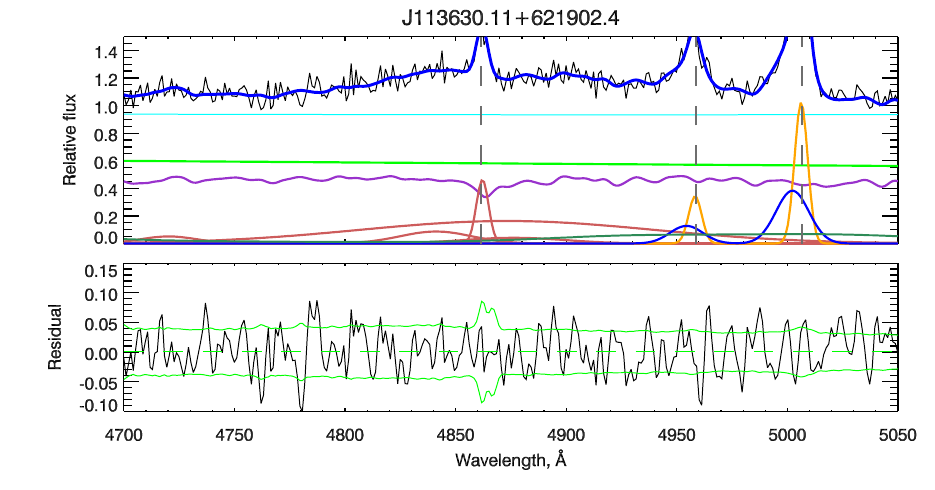}
		\includegraphics[width=0.16\textwidth]{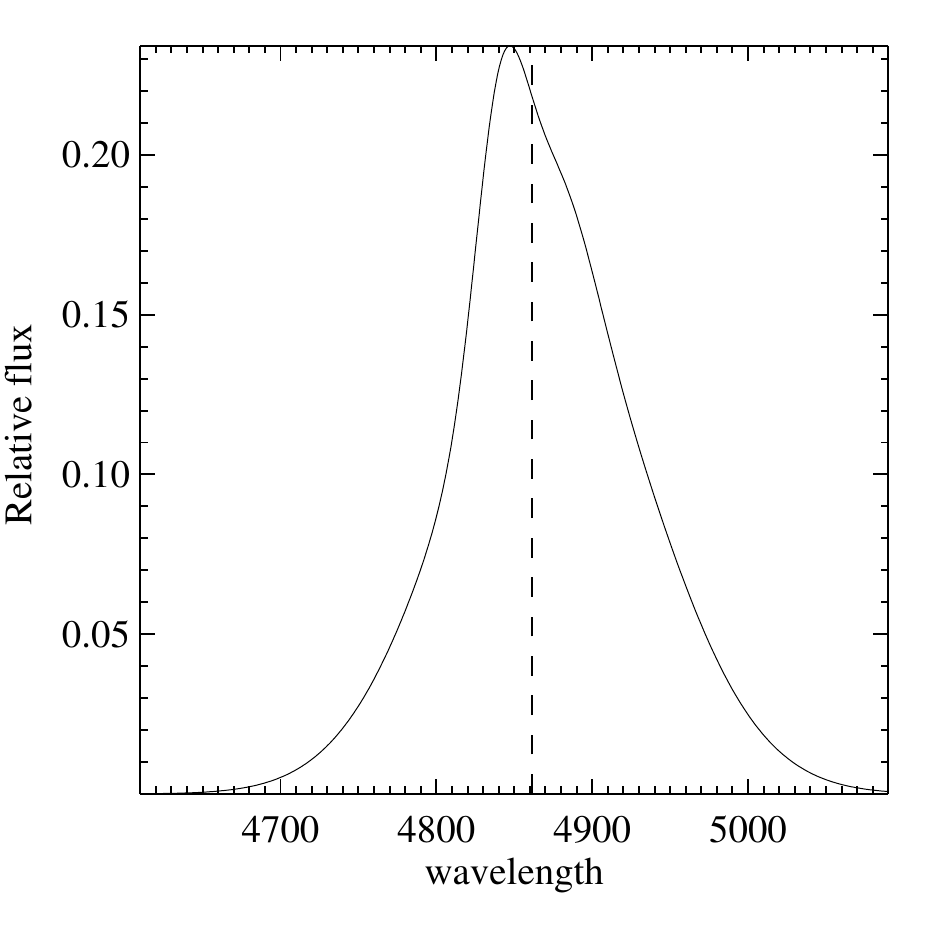}
		
		\caption{\textit{ -- continued}}
	\end{center}
\end{figure*}


\begin{figure*}
	\begin{center}
		\ContinuedFloat*
                \includegraphics[width=0.3\textwidth]{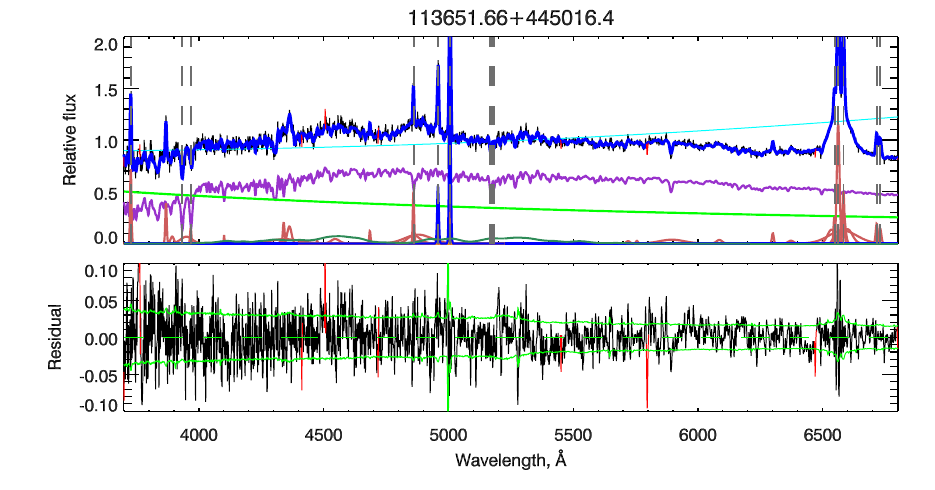}
		\includegraphics[width=0.3\textwidth]{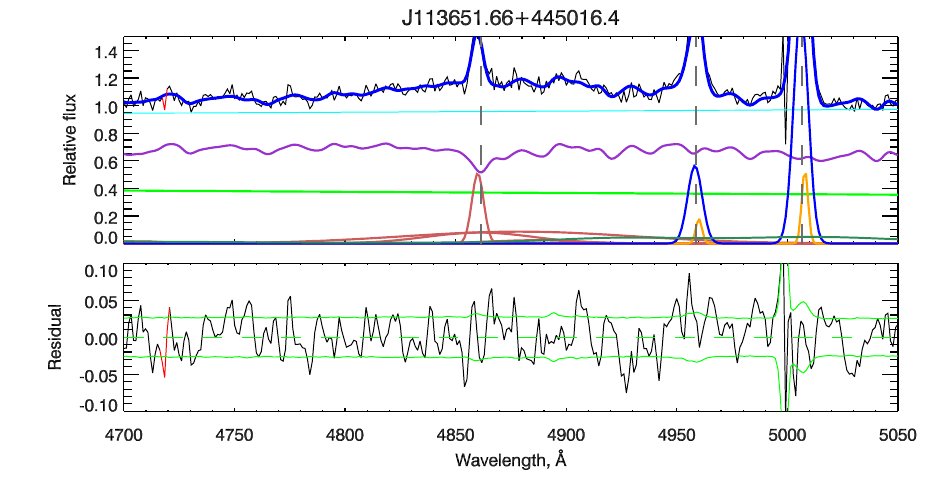}
		\includegraphics[width=0.16\textwidth]{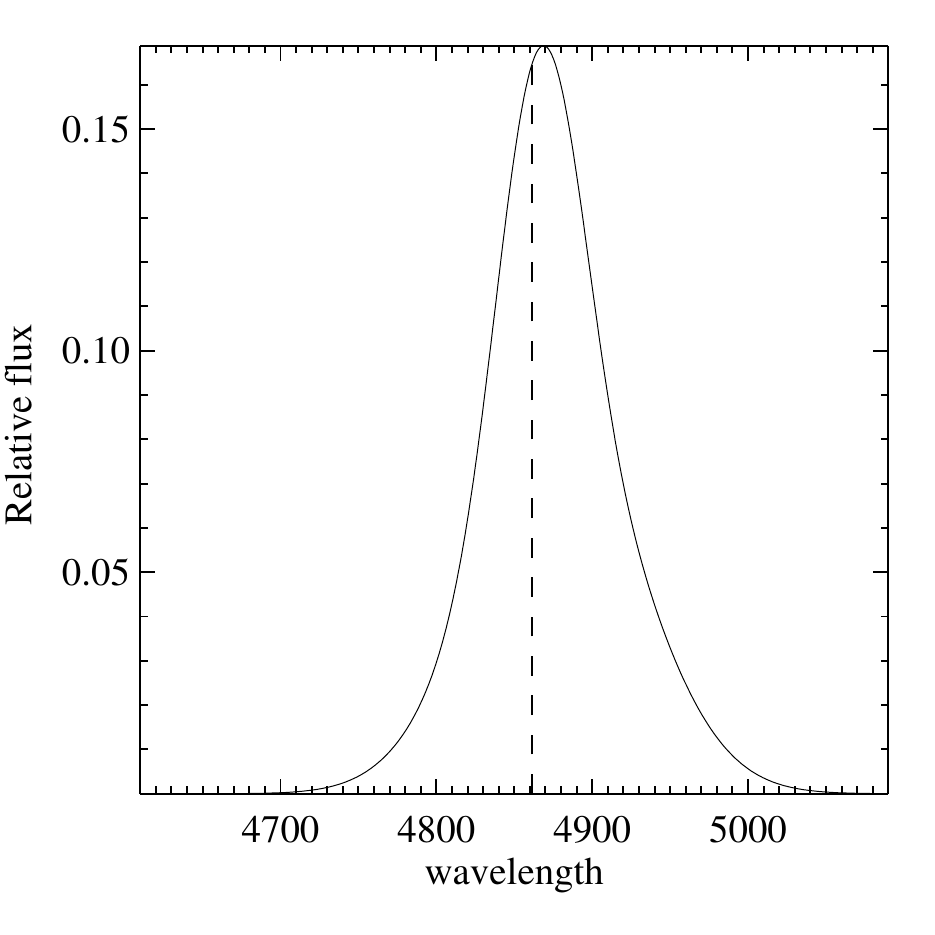}
		\includegraphics[width=0.3\textwidth]{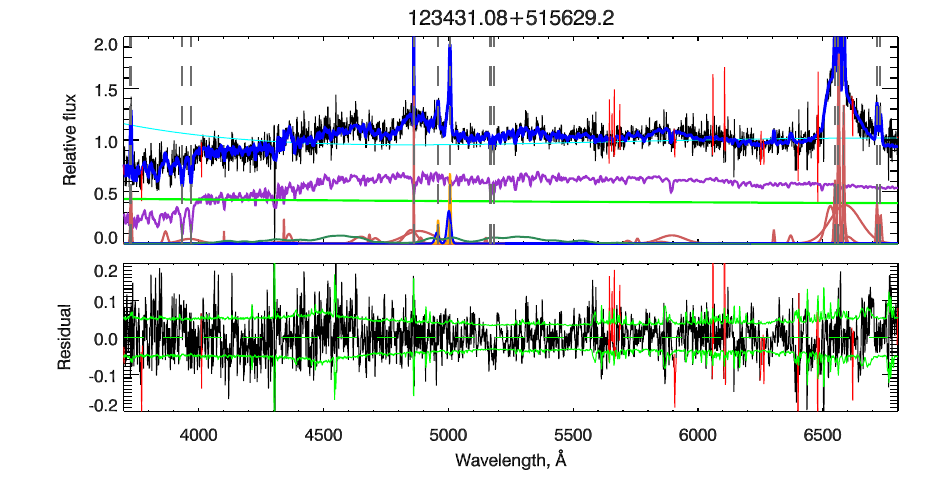}
		\includegraphics[width=0.3\textwidth]{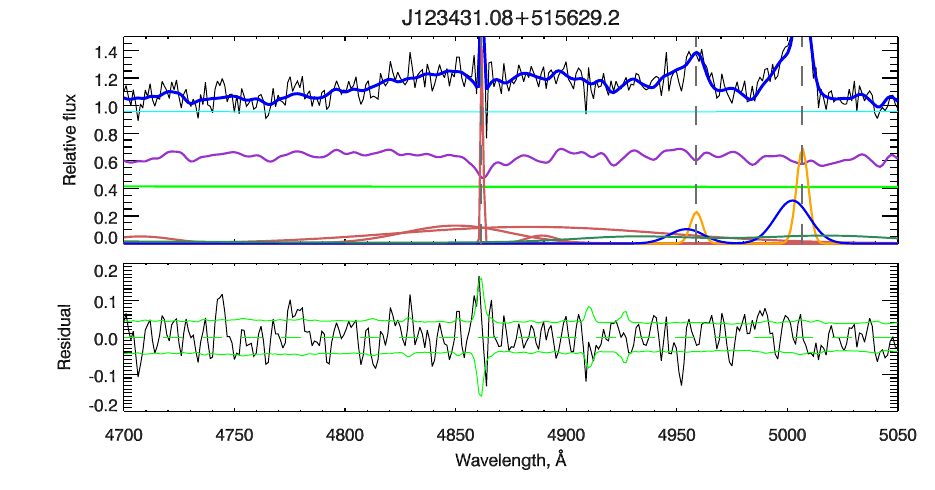}
		\includegraphics[width=0.16\textwidth]{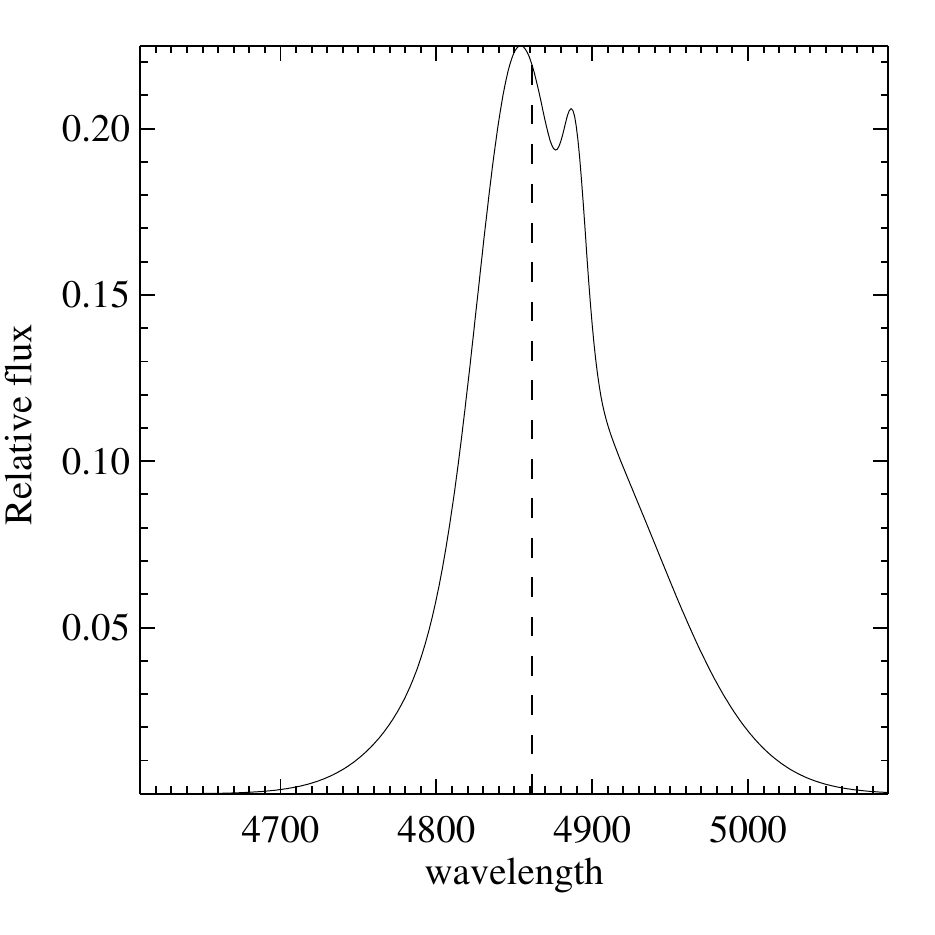}
		\includegraphics[width=0.3\textwidth]{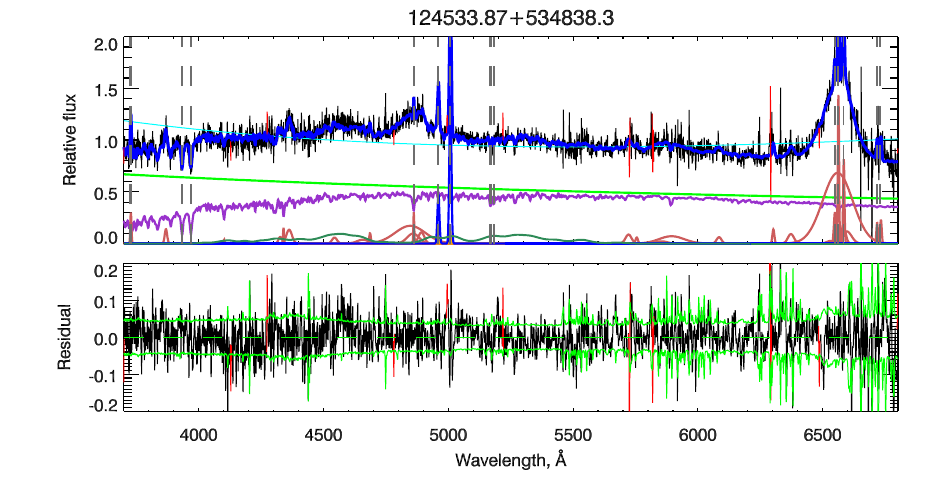}
		\includegraphics[width=0.3\textwidth]{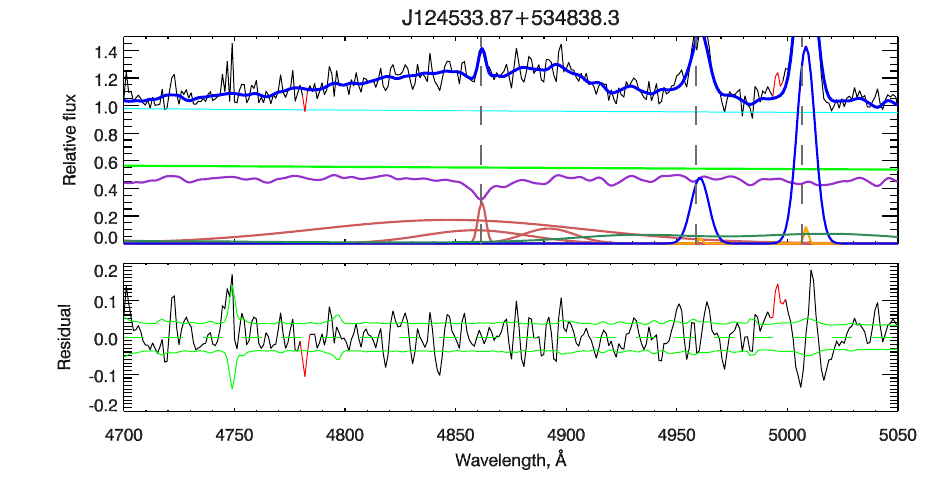}
		\includegraphics[width=0.16\textwidth]{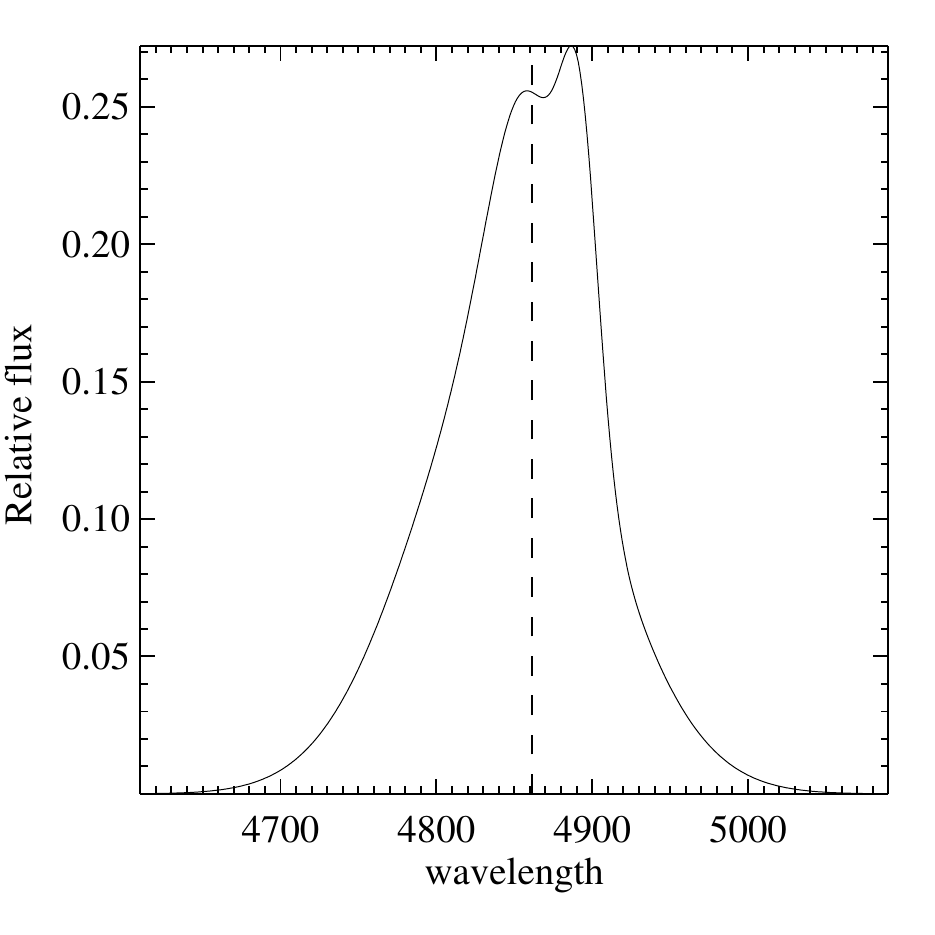}
		\includegraphics[width=0.3\textwidth]{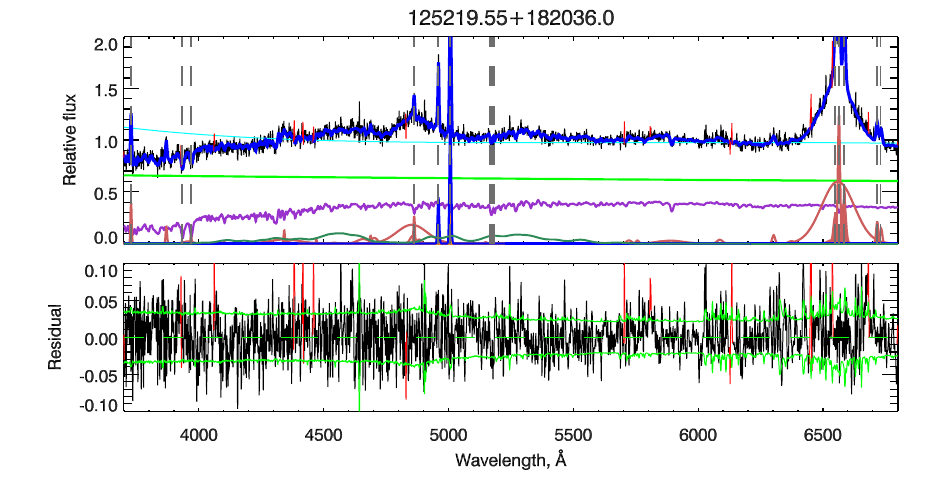}
		\includegraphics[width=0.3\textwidth]{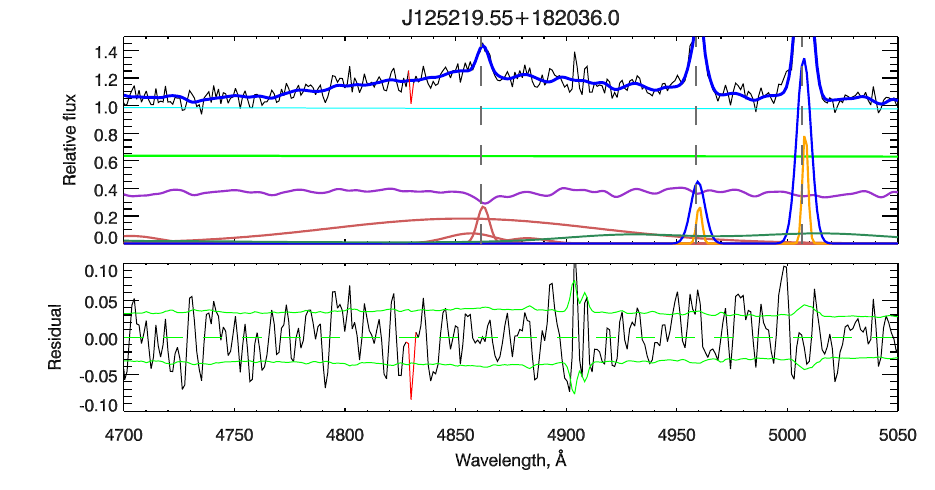}
		\includegraphics[width=0.16\textwidth]{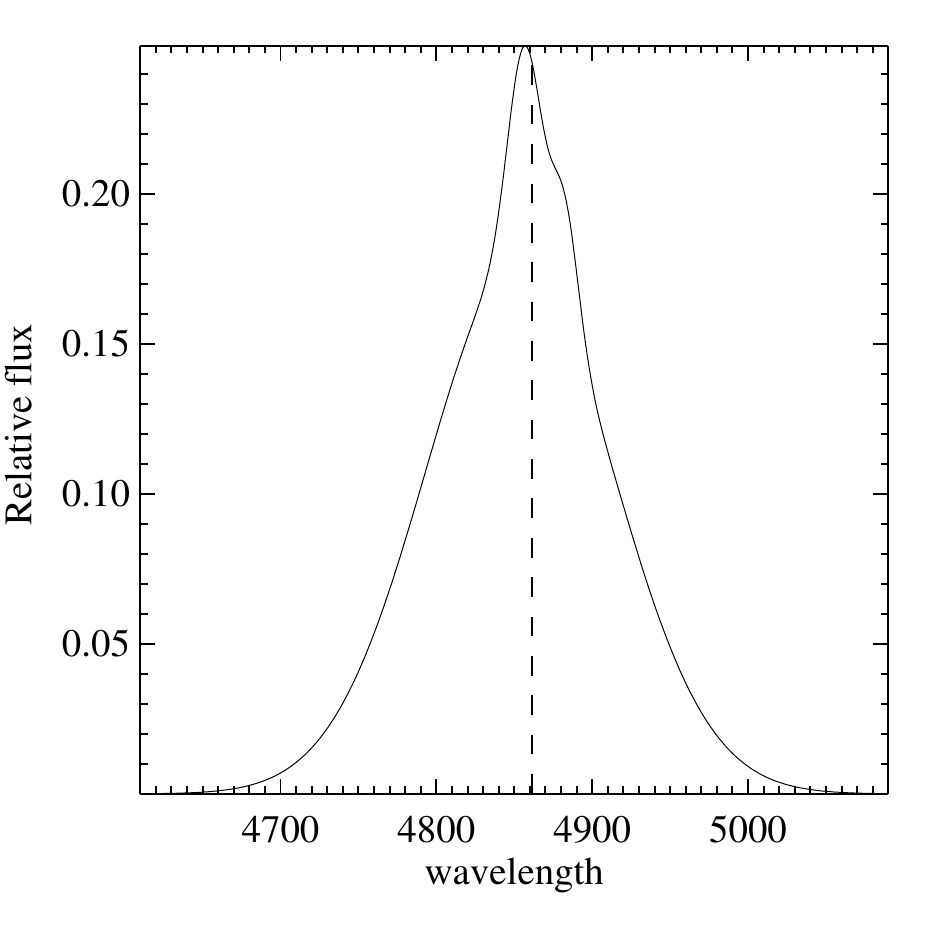}
		\includegraphics[width=0.3\textwidth]{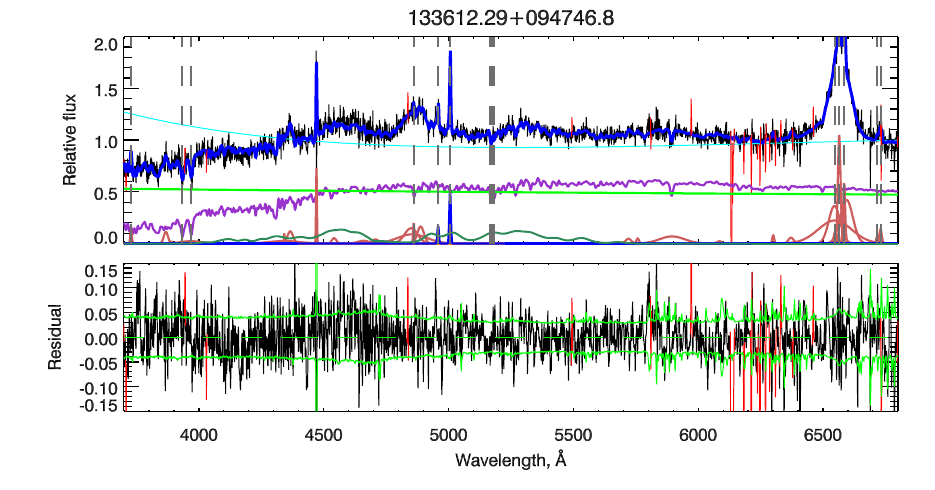}
		\includegraphics[width=0.3\textwidth]{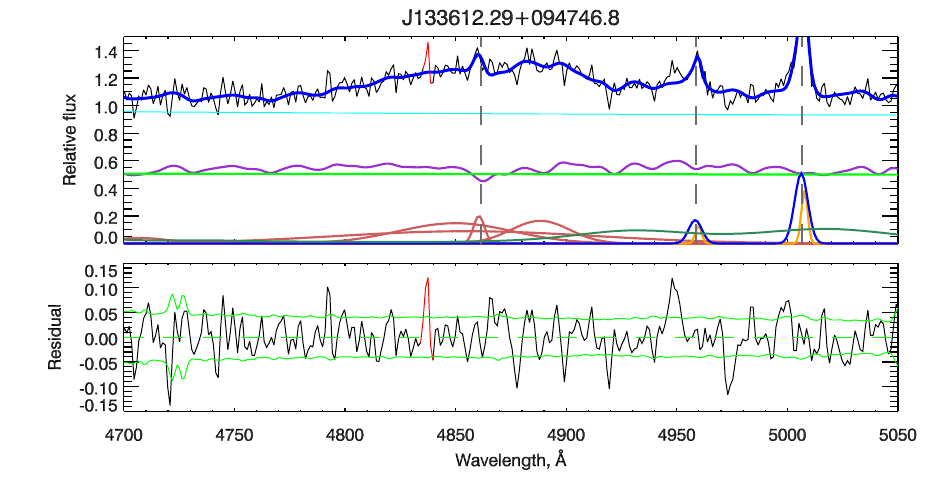}
		\includegraphics[width=0.16\textwidth]{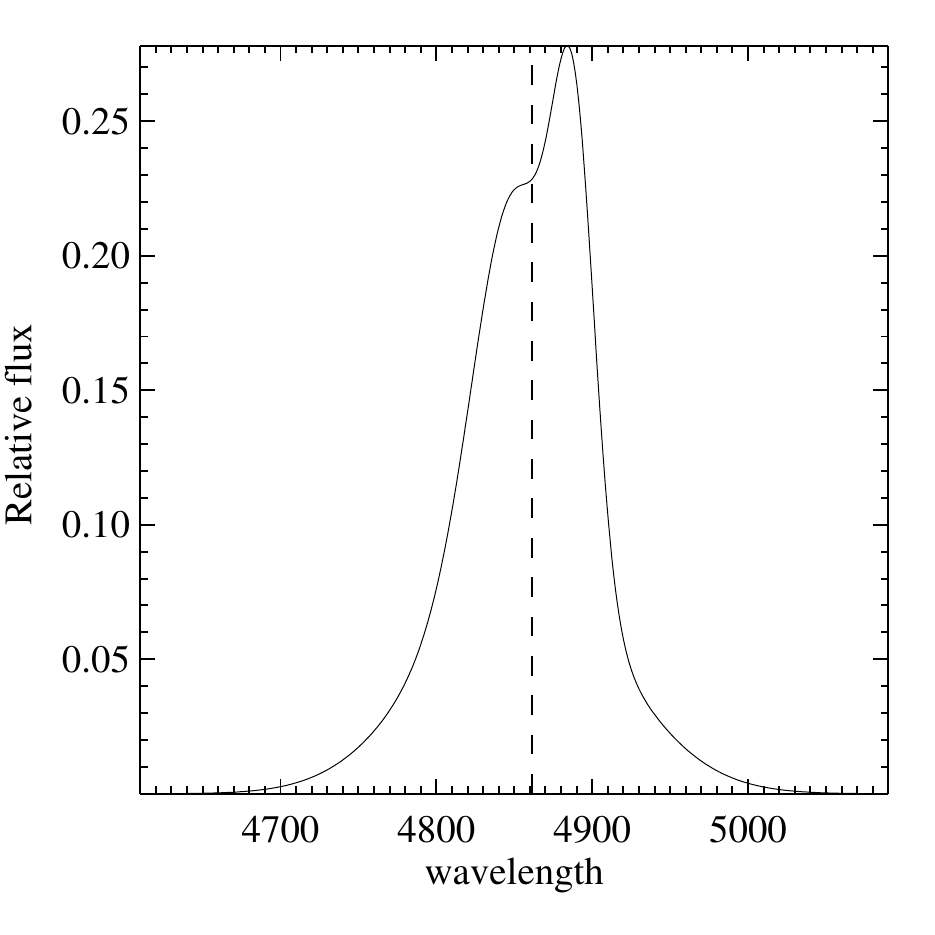}
		\includegraphics[width=0.3\textwidth]{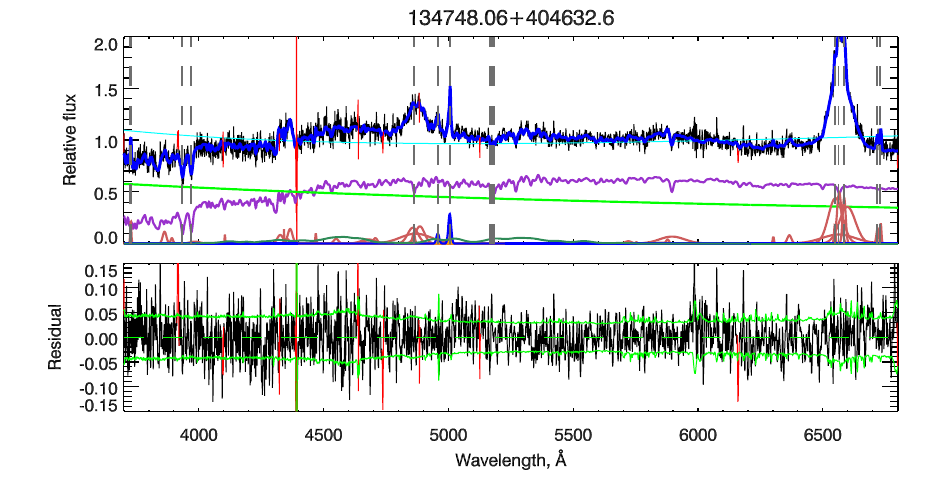}
	        \includegraphics[width=0.3\textwidth]{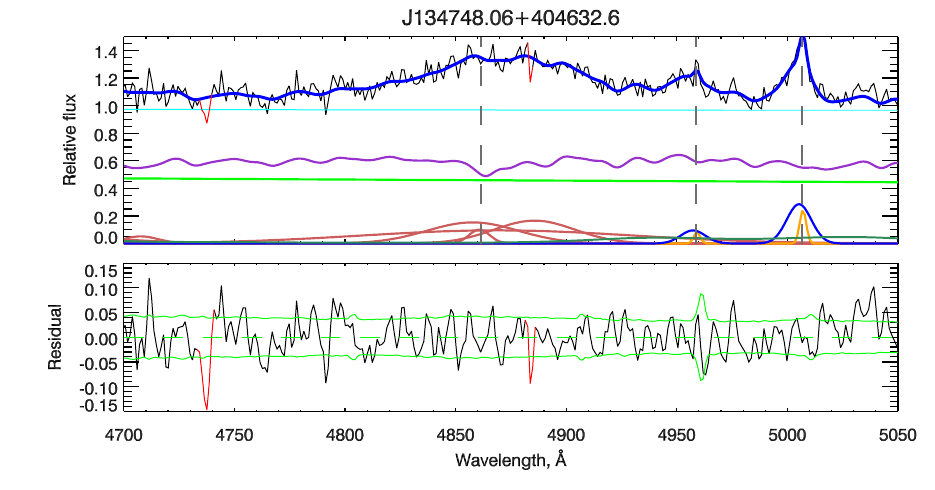}
		\includegraphics[width=0.16\textwidth]{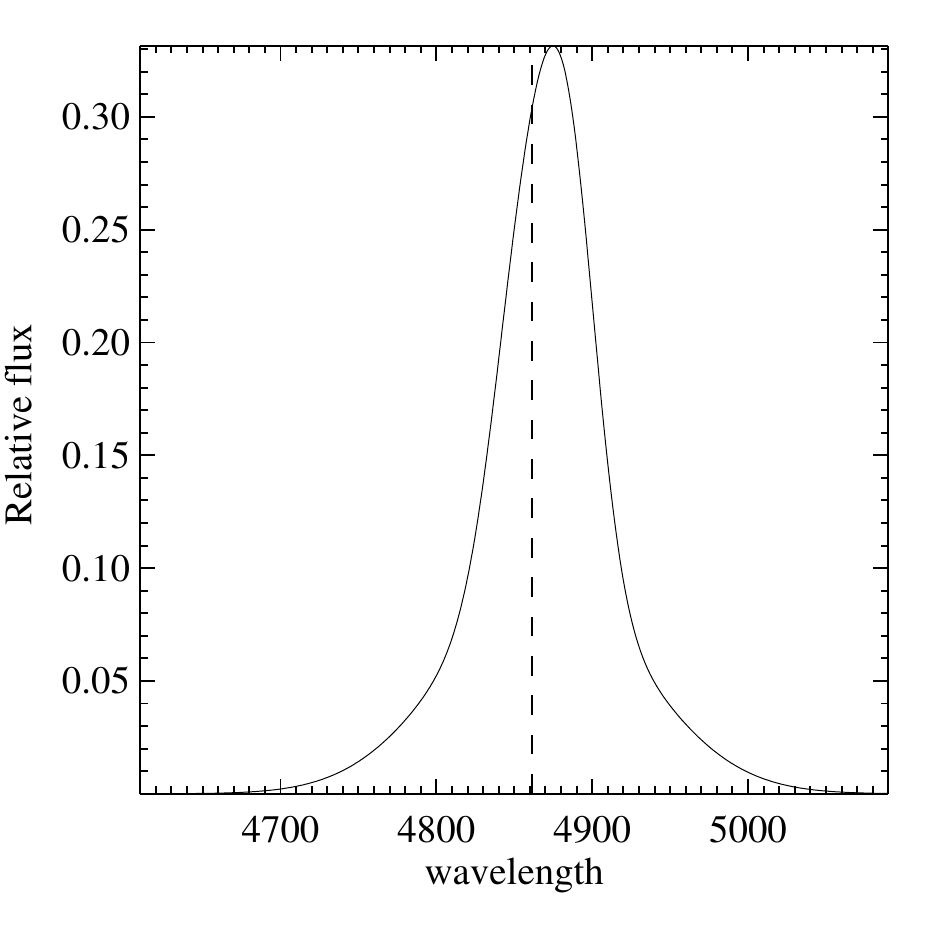}
		\includegraphics[width=0.3\textwidth]{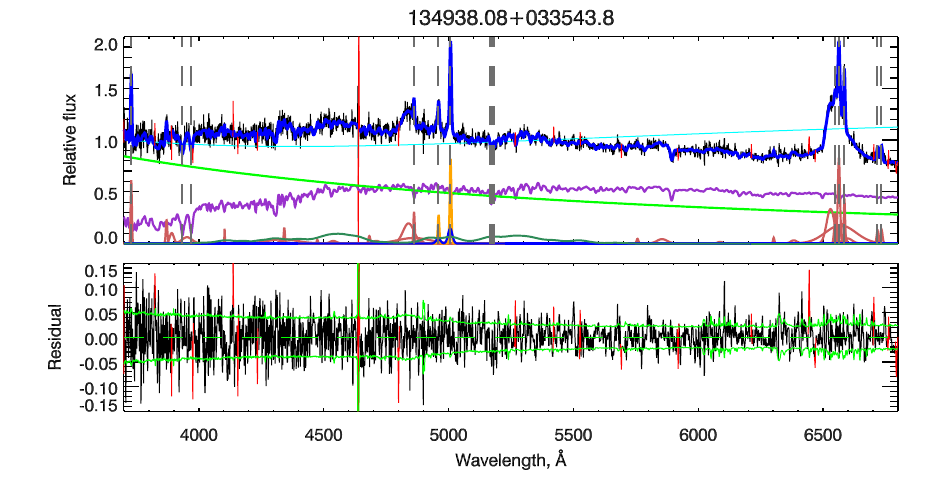}
		\includegraphics[width=0.3\textwidth]{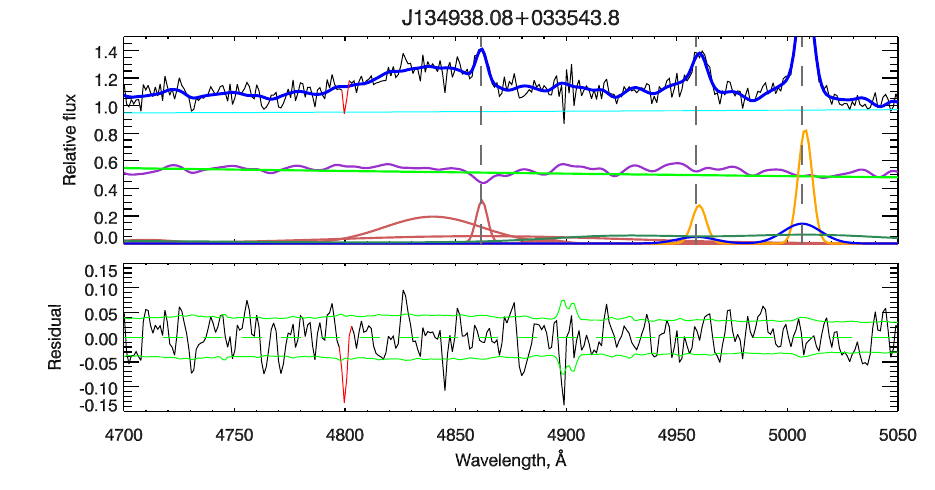}
		\includegraphics[width=0.16\textwidth]{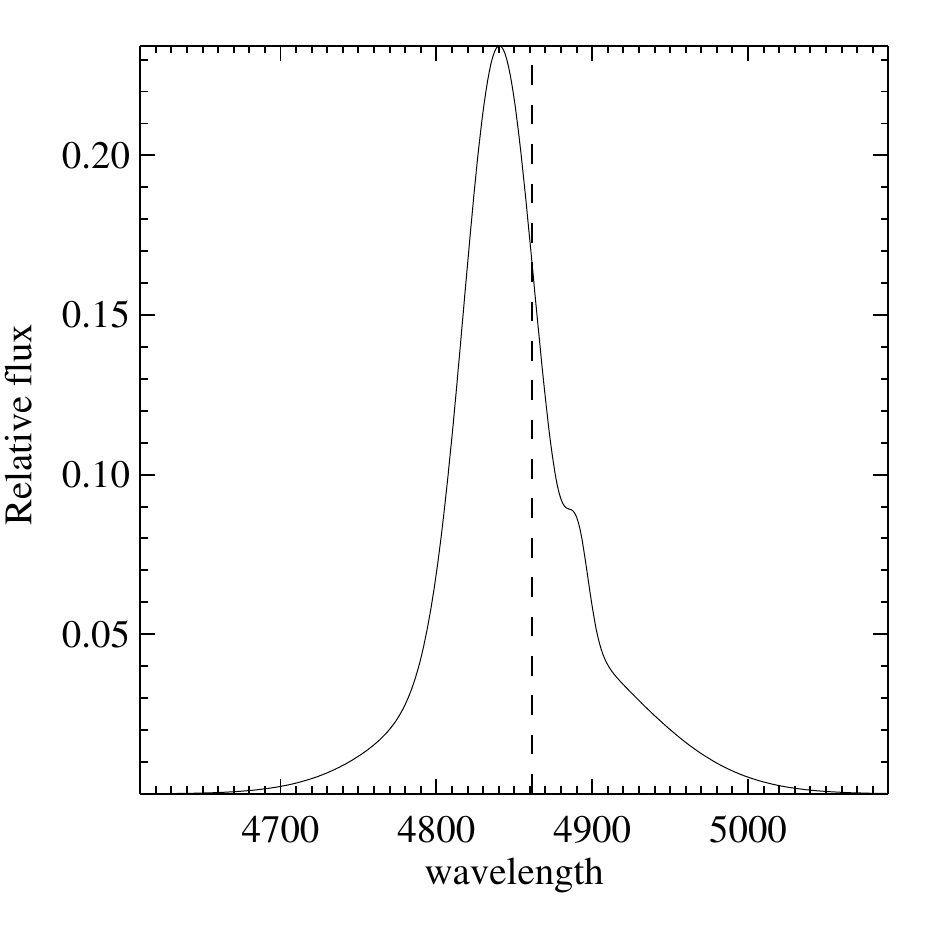}
		\includegraphics[width=0.3\textwidth]{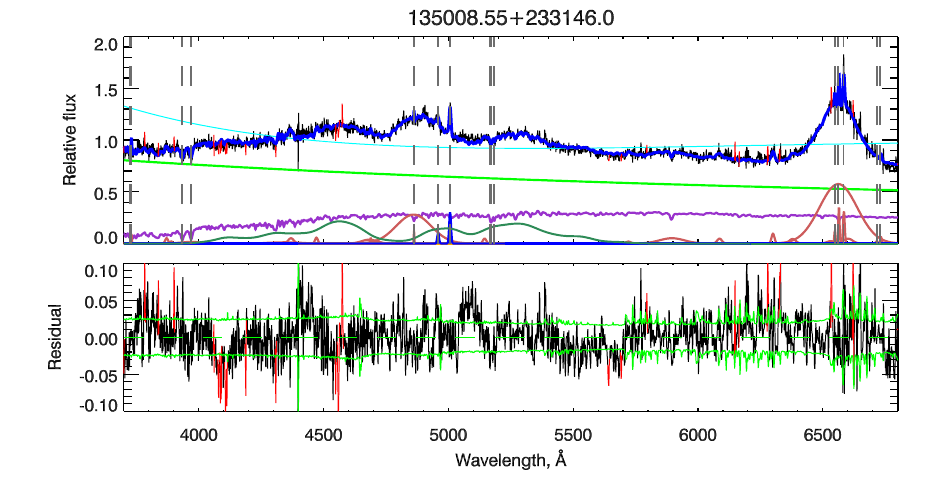}
		\includegraphics[width=0.3\textwidth]{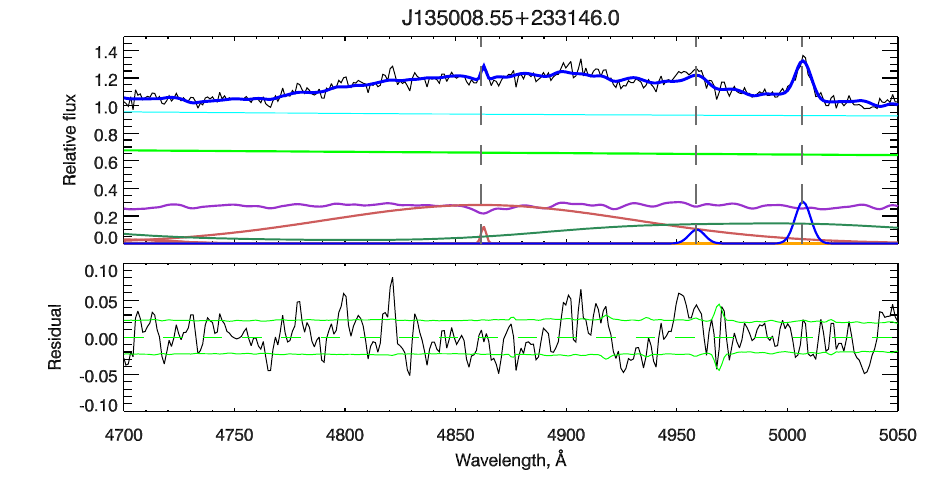}
		\includegraphics[width=0.16\textwidth]{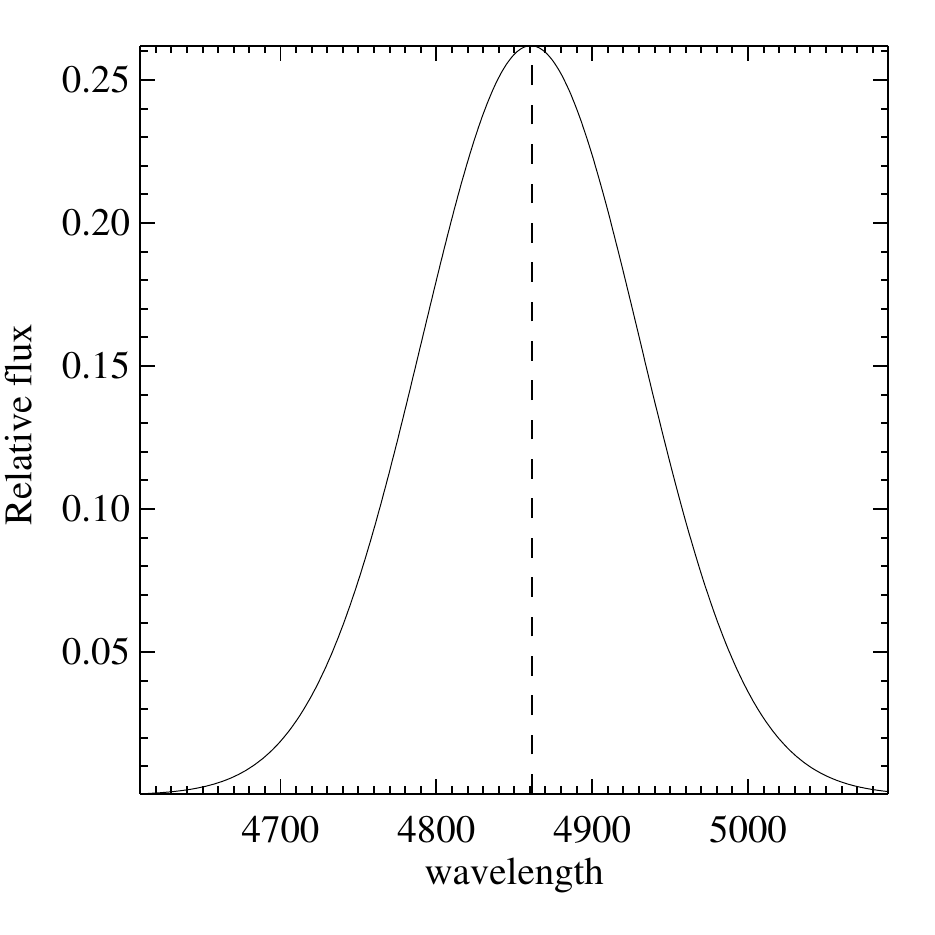}
		\includegraphics[width=0.3\textwidth]{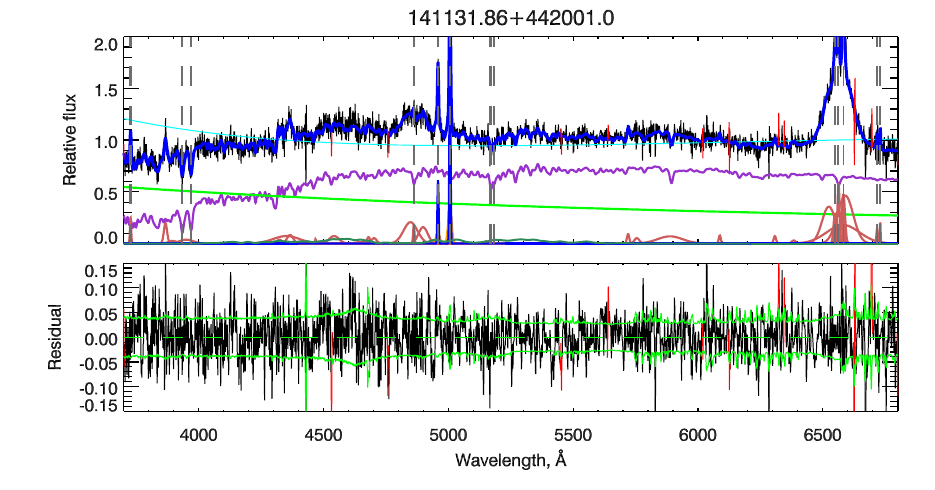}
		\includegraphics[width=0.3\textwidth]{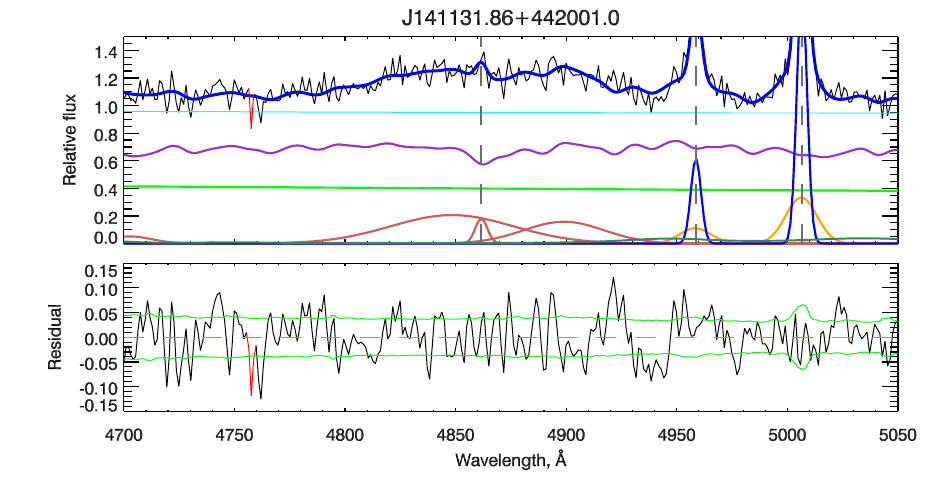}
		\includegraphics[width=0.16\textwidth]{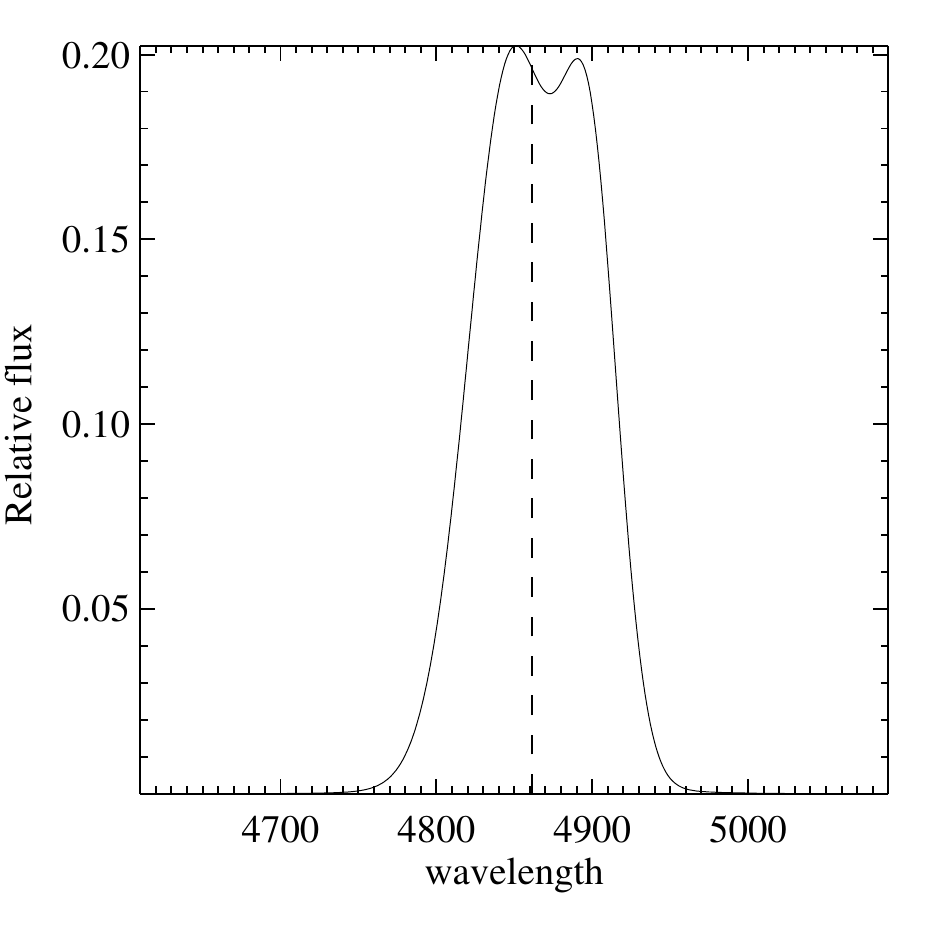}
		
				\caption{\textit{ -- continued}}
	\end{center}
\end{figure*}


\begin{figure*}
	\begin{center}
		\ContinuedFloat*
                \includegraphics[width=0.3\textwidth]{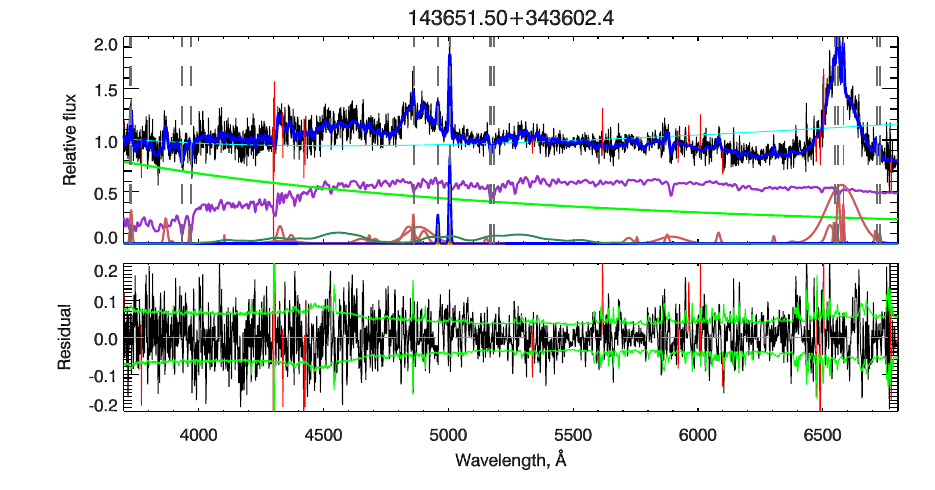}
		\includegraphics[width=0.3\textwidth]{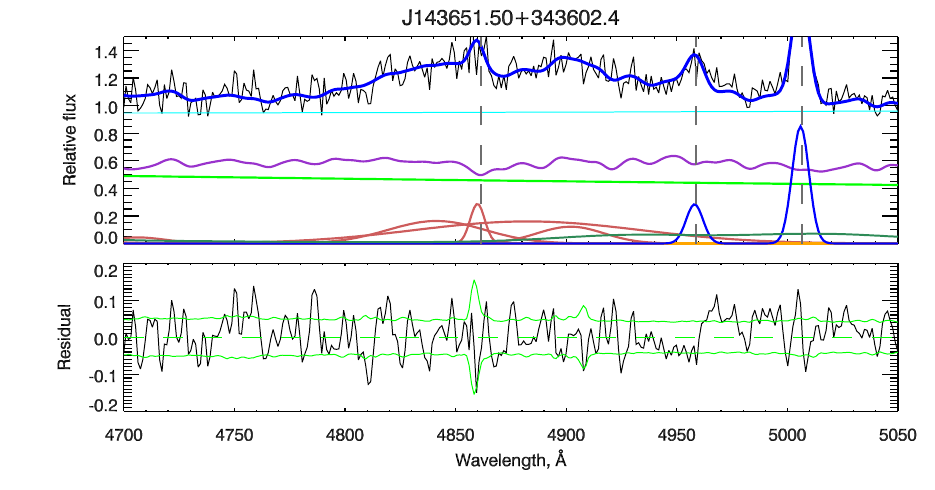}
		\includegraphics[width=0.16\textwidth]{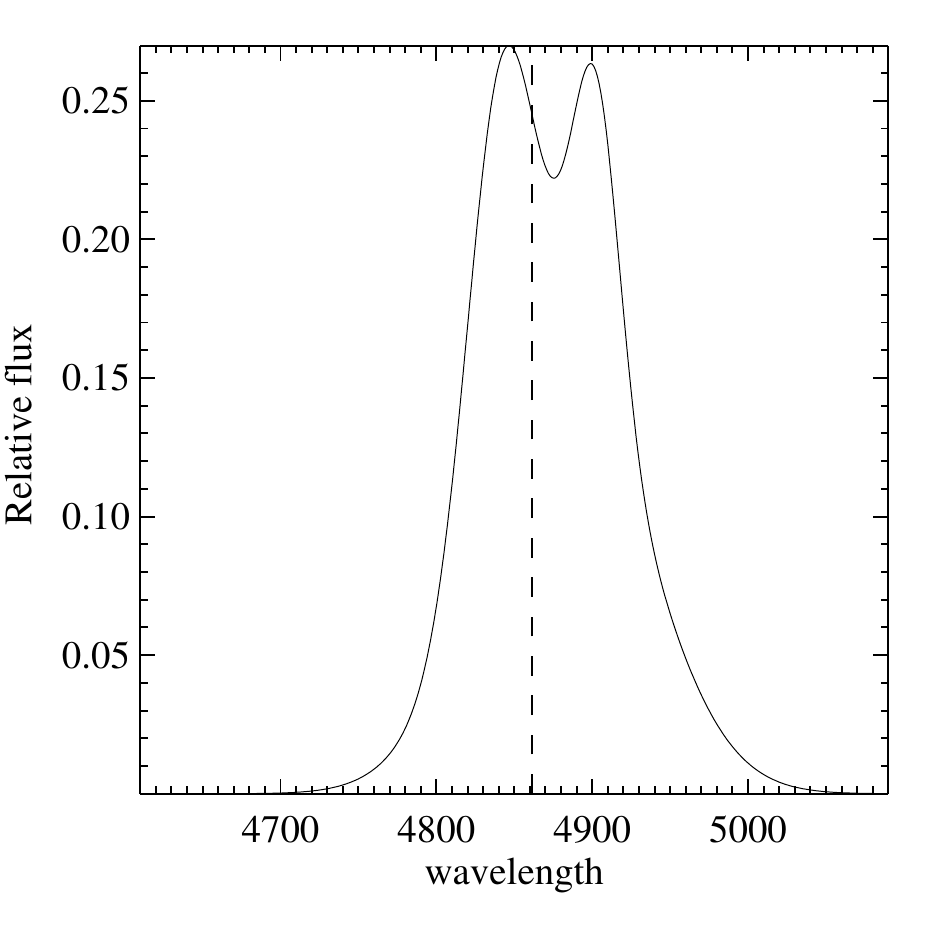}
		\includegraphics[width=0.3\textwidth]{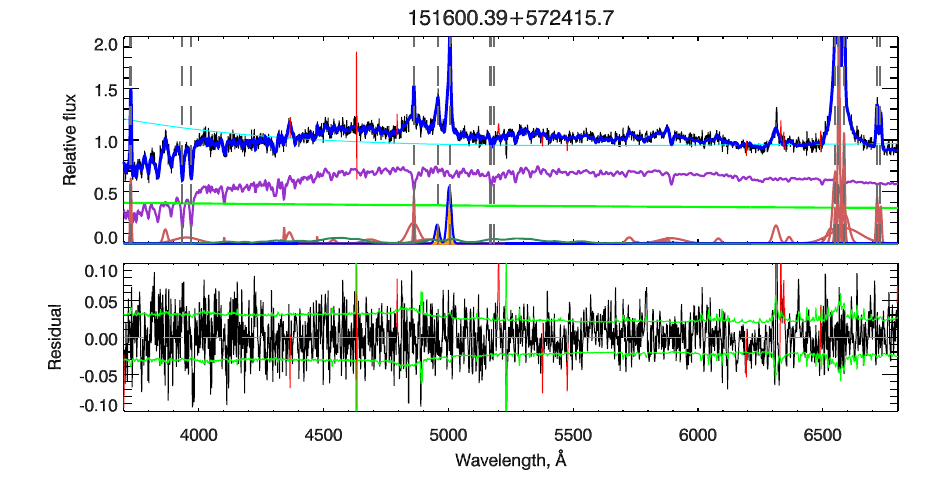}
		\includegraphics[width=0.3\textwidth]{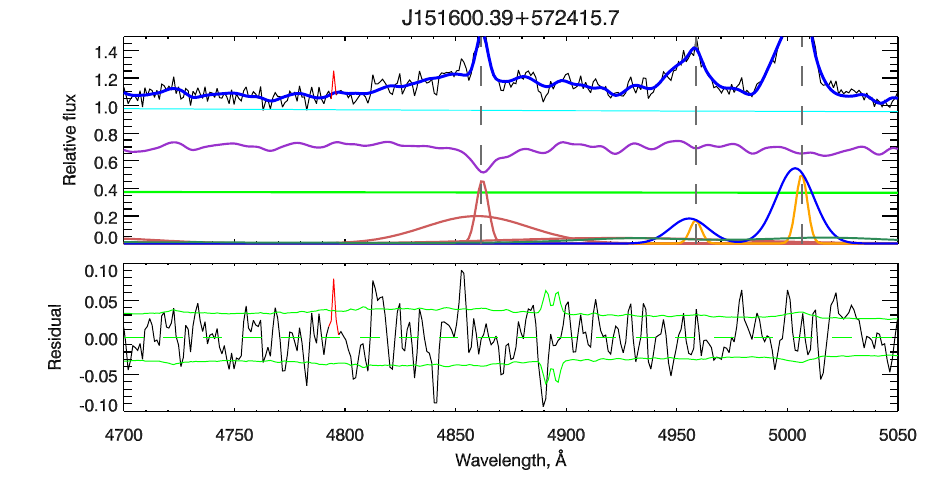}
		\includegraphics[width=0.16\textwidth]{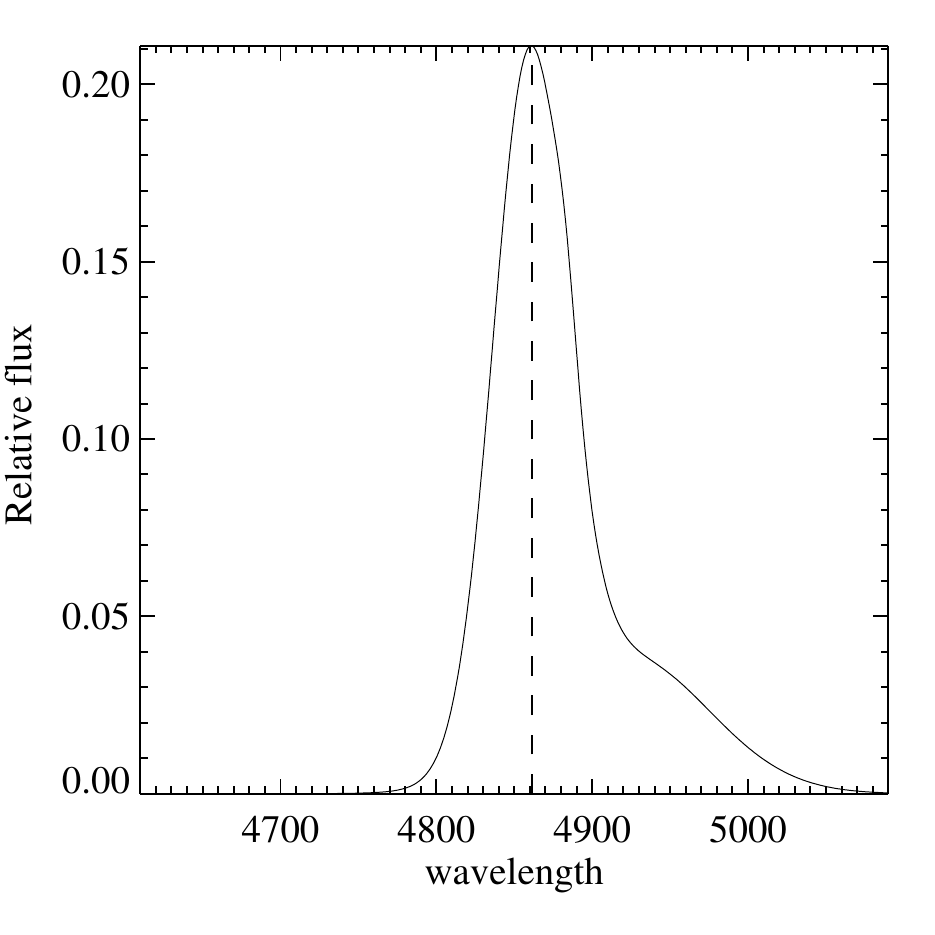}
		\includegraphics[width=0.3\textwidth]{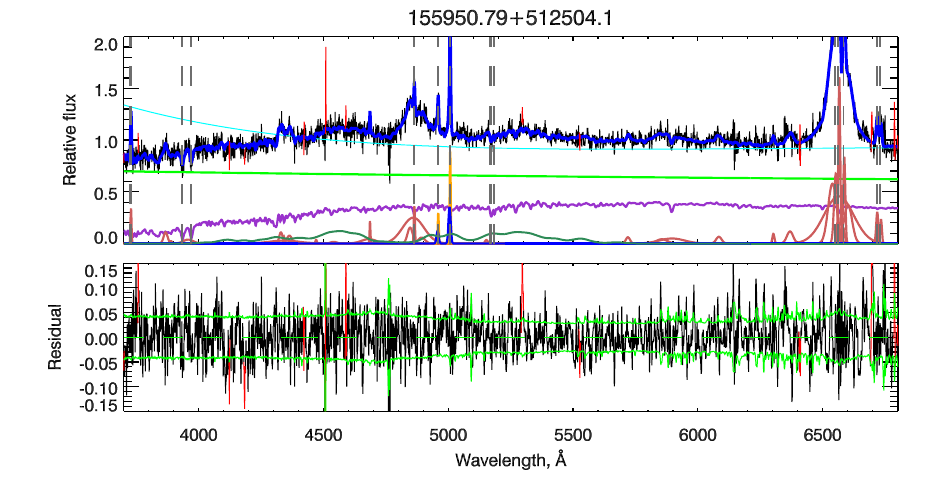}
		\includegraphics[width=0.3\textwidth]{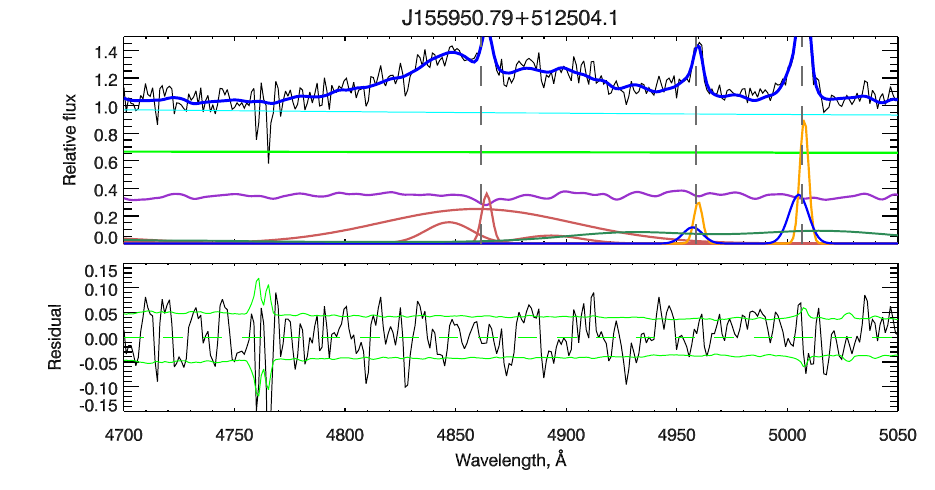}
		\includegraphics[width=0.16\textwidth]{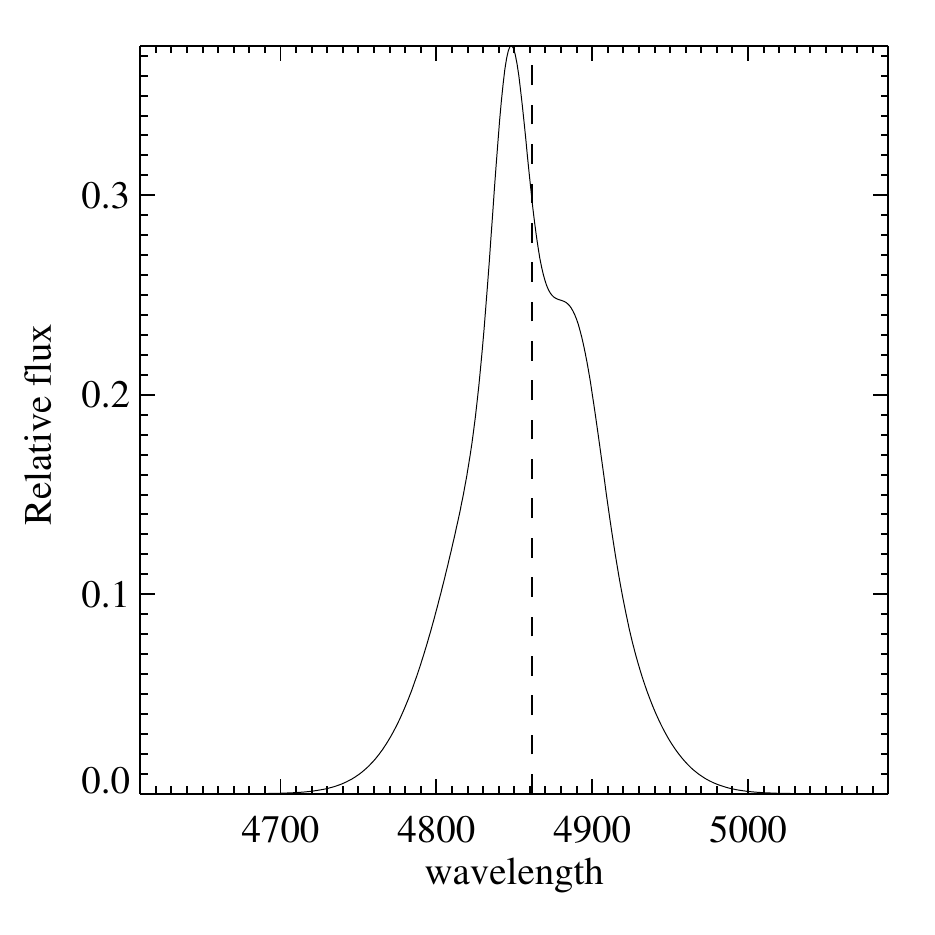}
		\includegraphics[width=0.3\textwidth]{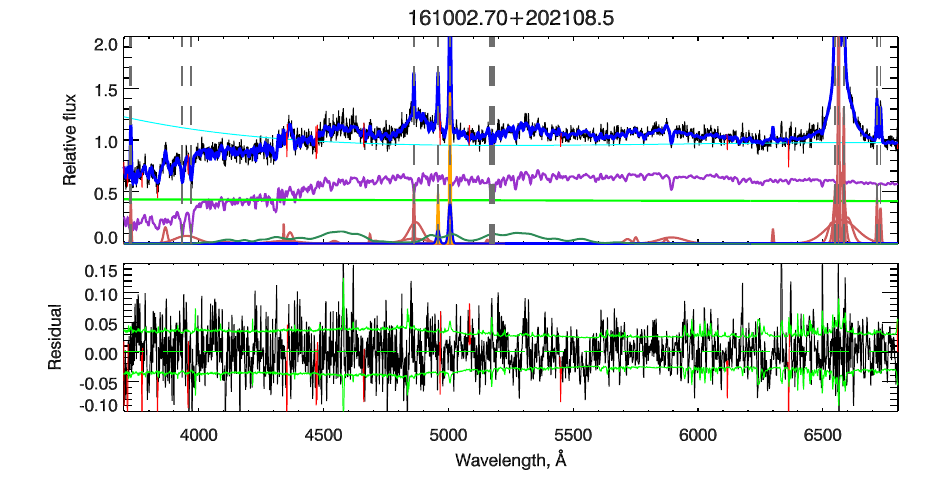}
		\includegraphics[width=0.3\textwidth]{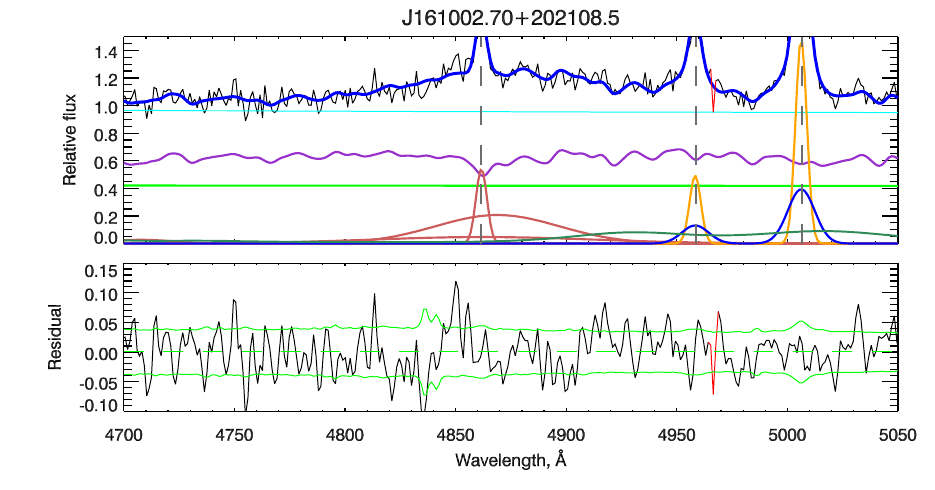}
		\includegraphics[width=0.16\textwidth]{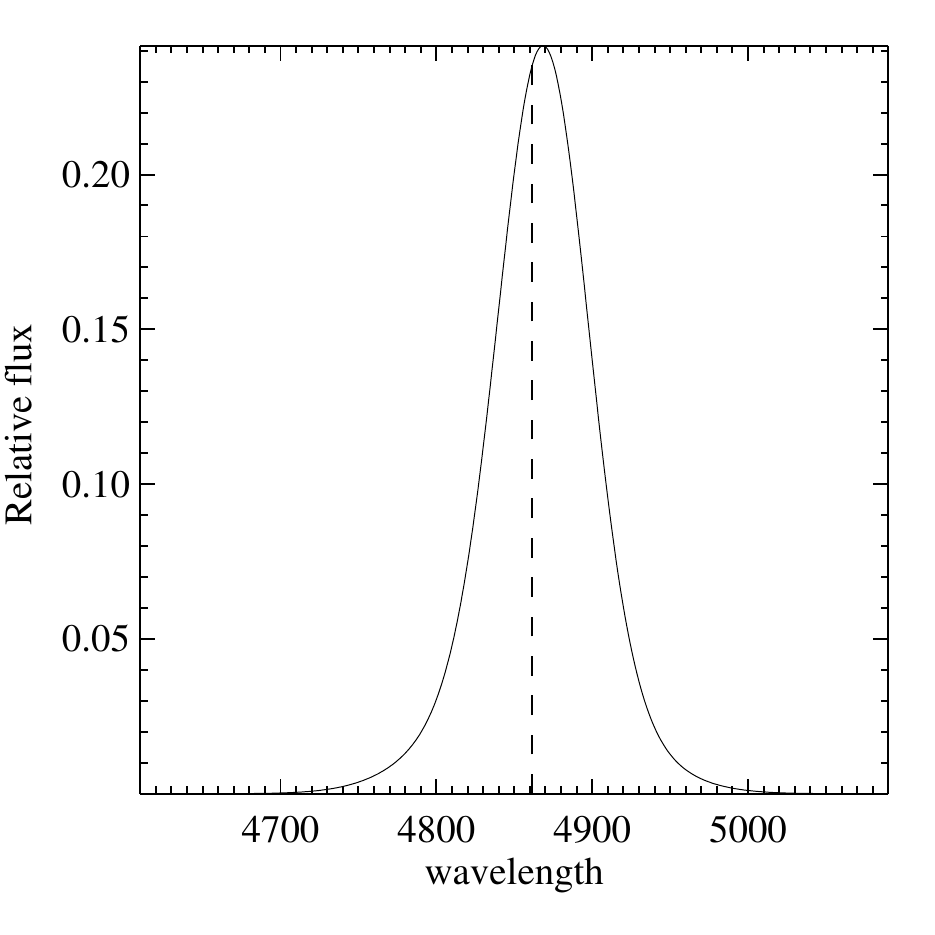}
		\includegraphics[width=0.3\textwidth]{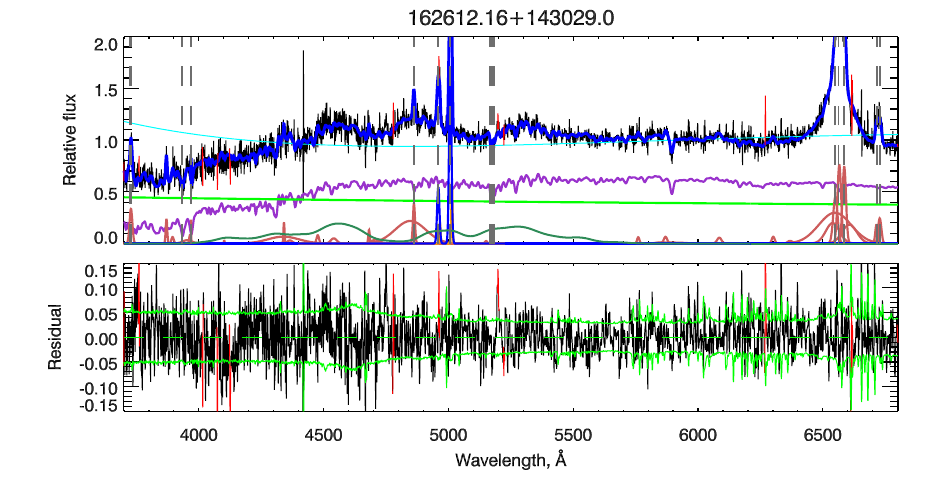}
		\includegraphics[width=0.3\textwidth]{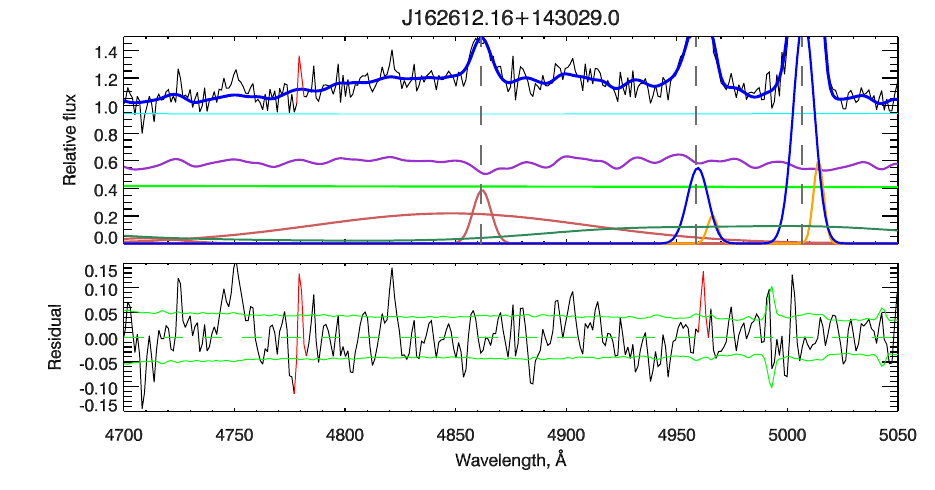}
		\includegraphics[width=0.16\textwidth]{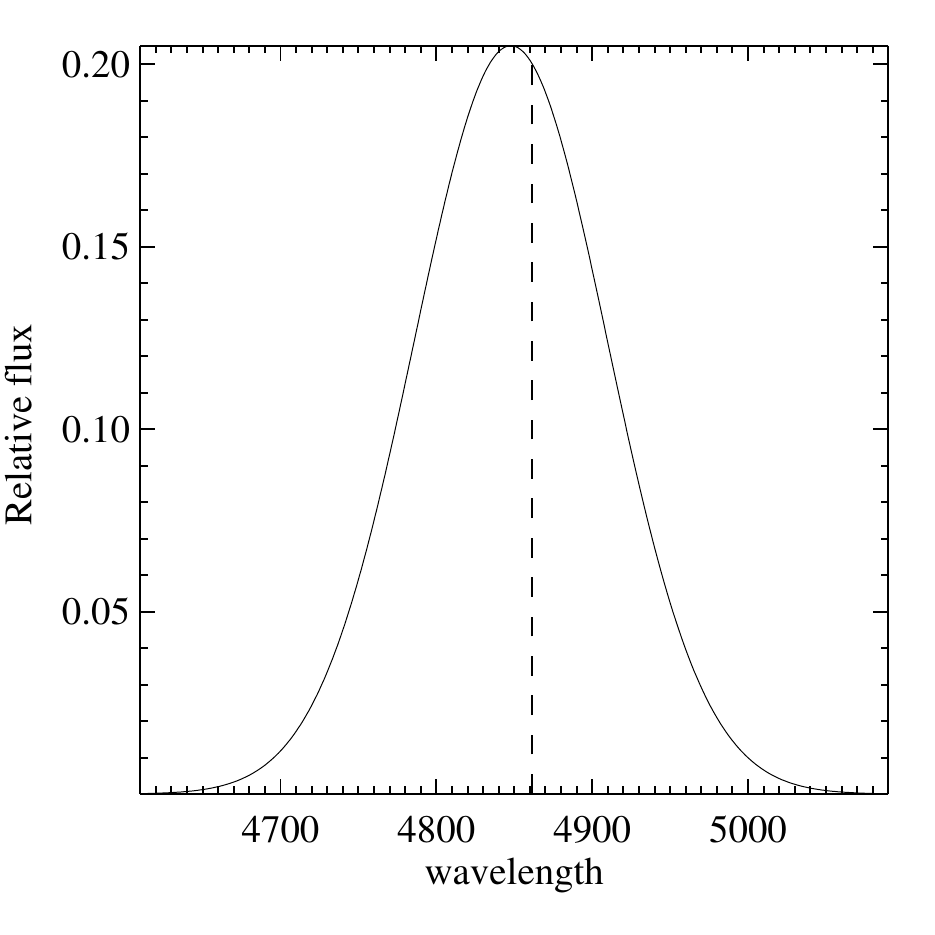}
		\includegraphics[width=0.3\textwidth]{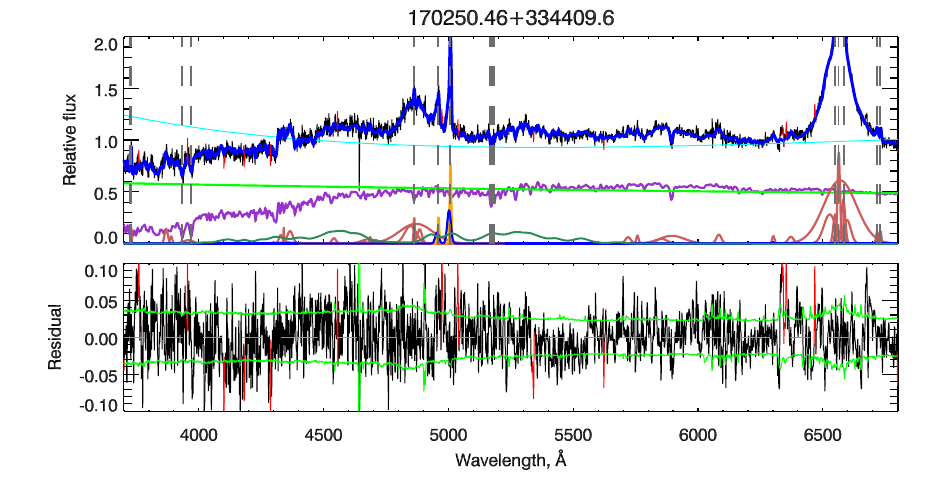}
		\includegraphics[width=0.3\textwidth]{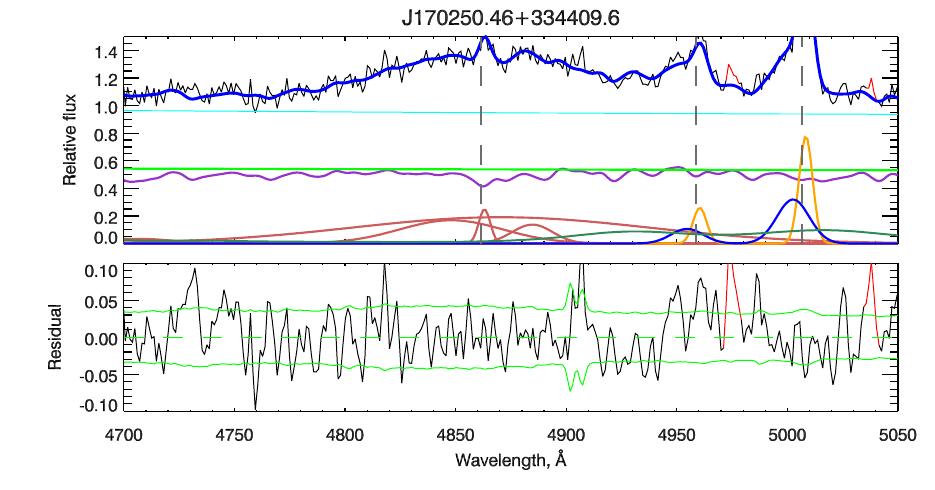}
		\includegraphics[width=0.16\textwidth]{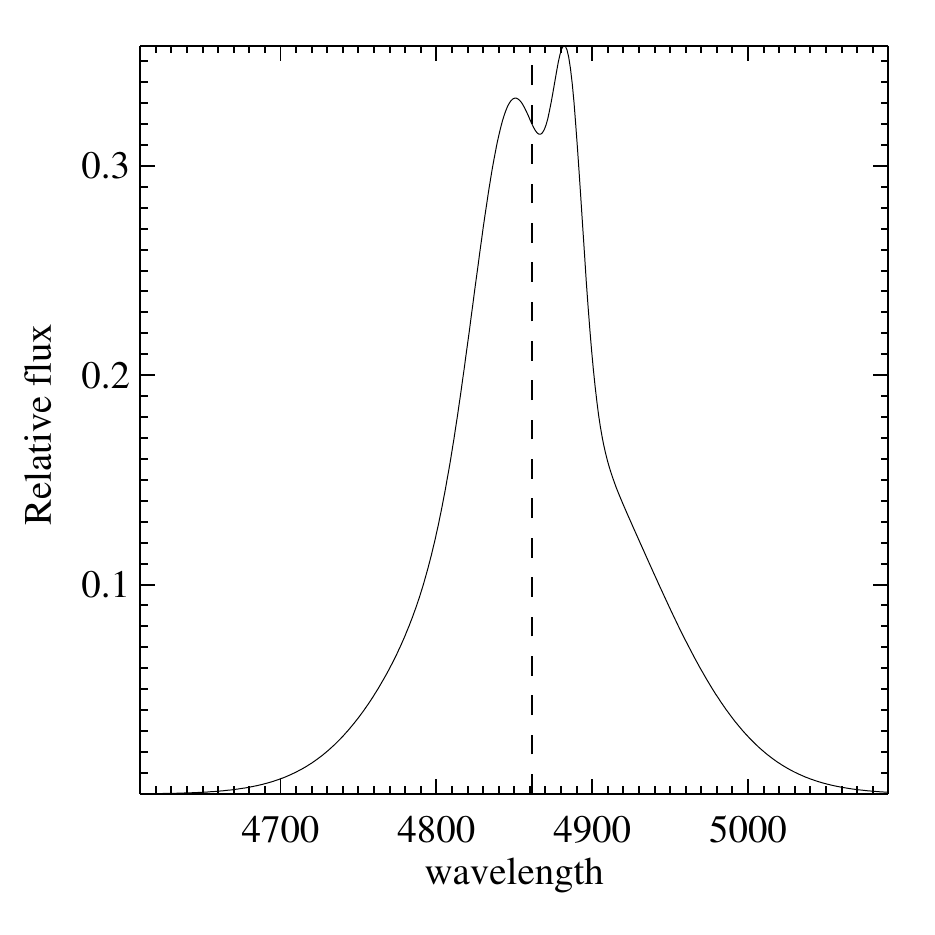}
		\caption{\textit{ -- continued}}
	\end{center}
\end{figure*}

Figure \ref{chi2maps} represents the $\chi^2$ maps for the whole sample in the parameter space of the stellar velocity dispersion 
and the mean stellar velocity.

\begin{figure*}
	\begin{center}
	  \includegraphics[width=0.225\textwidth]{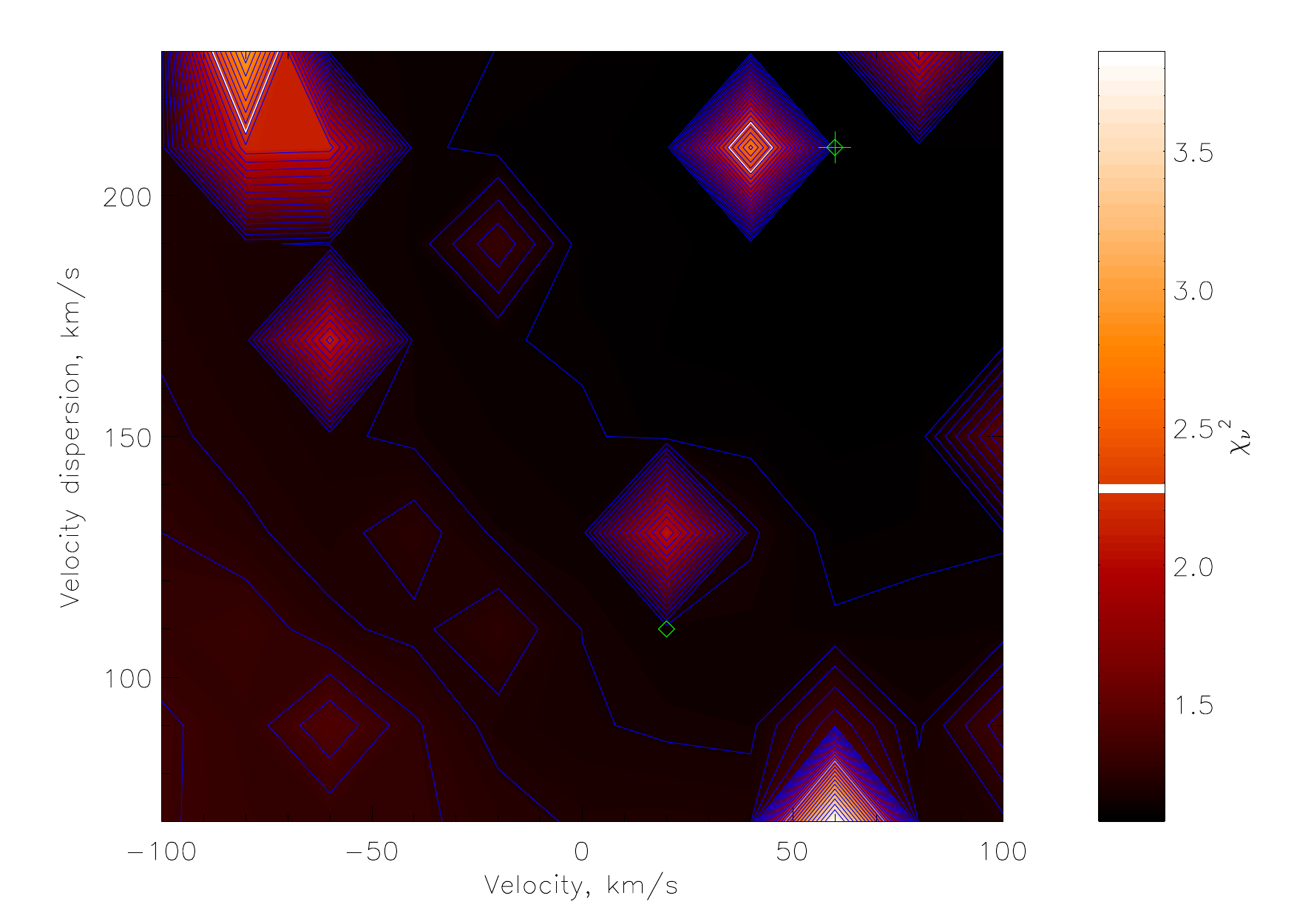}
	  \includegraphics[width=0.225\textwidth]{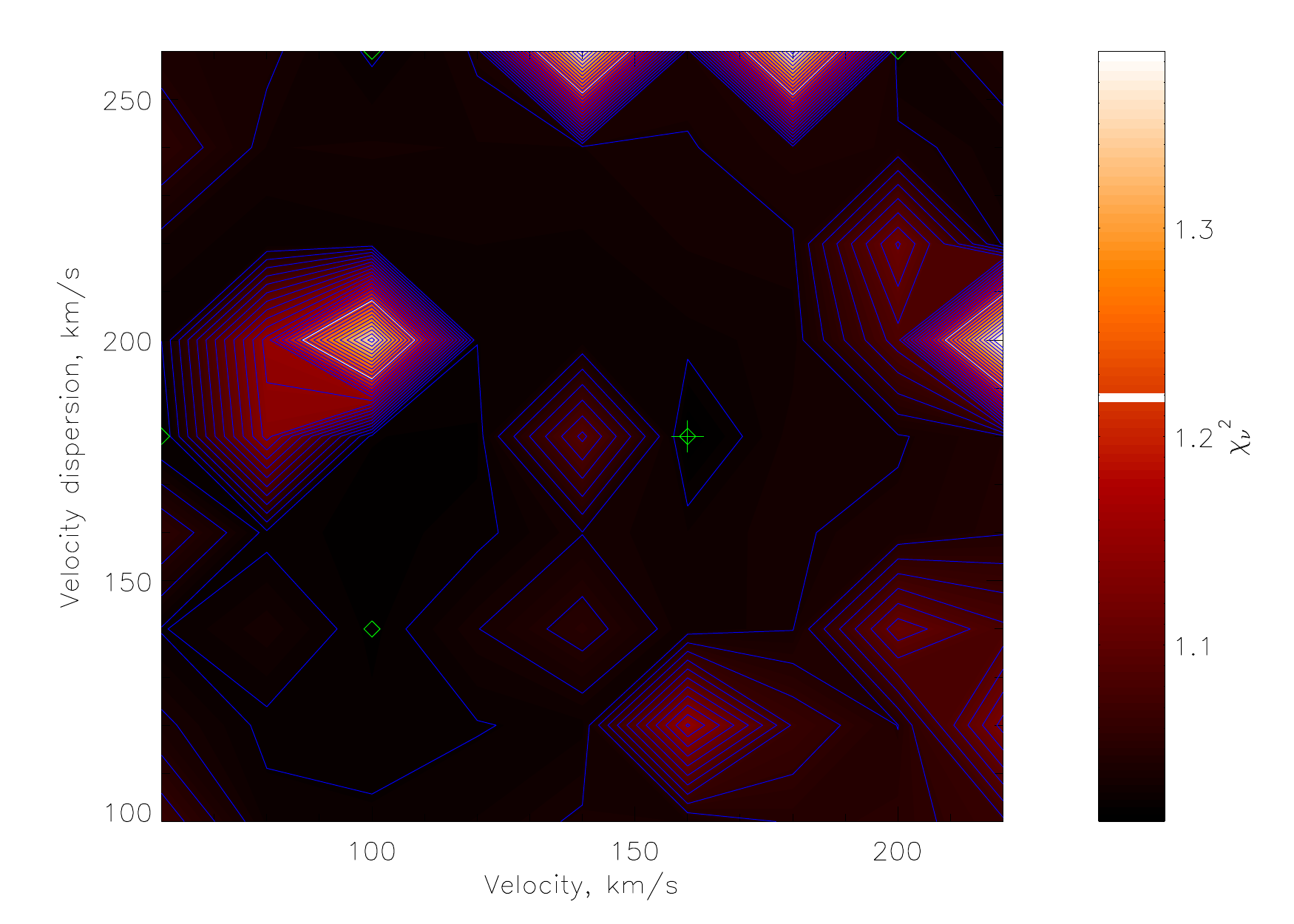}
	  \includegraphics[width=0.225\textwidth]{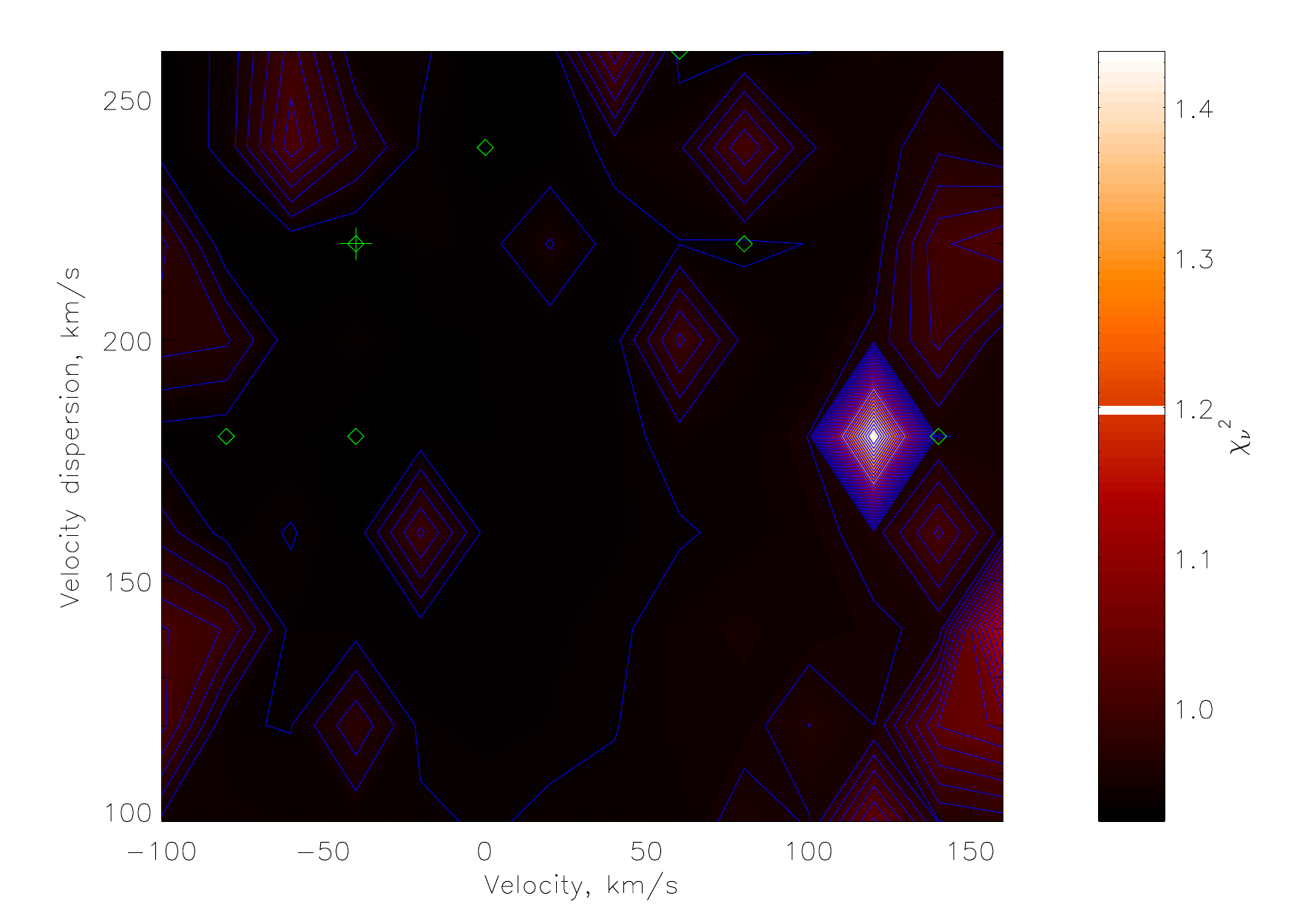}
	  \includegraphics[width=0.225\textwidth]{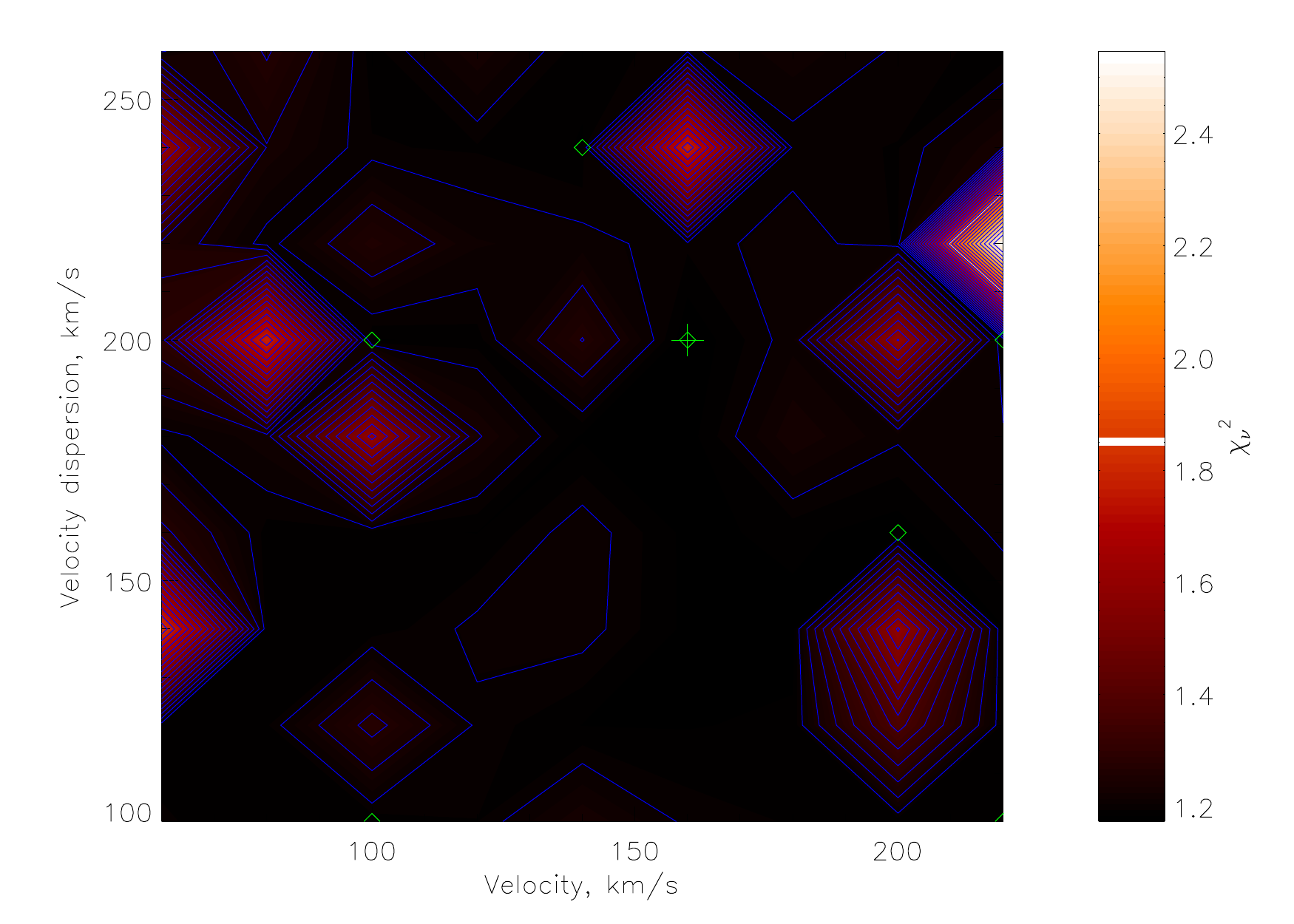}
	  \includegraphics[width=0.225\textwidth]{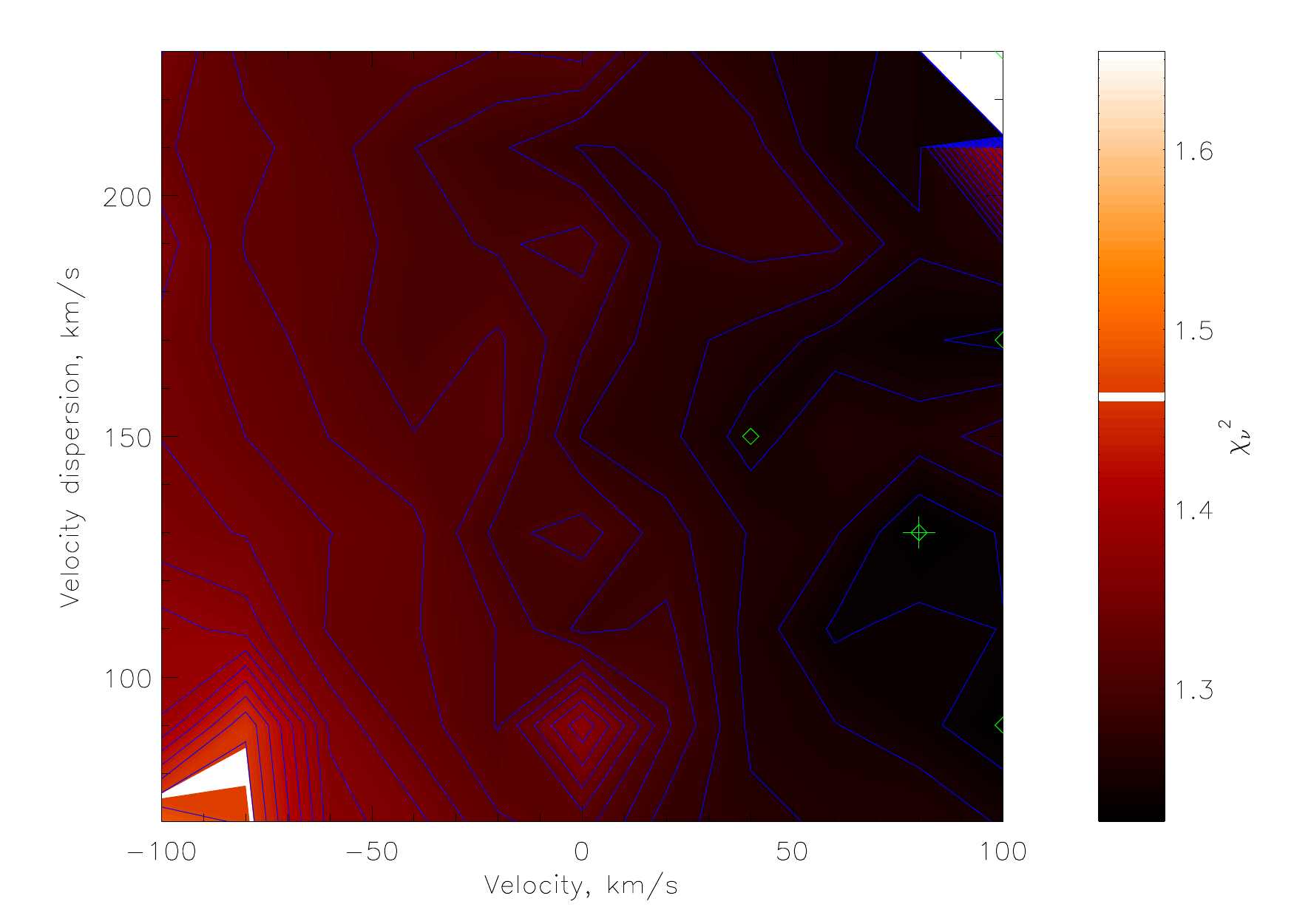}
	  \includegraphics[width=0.225\textwidth]{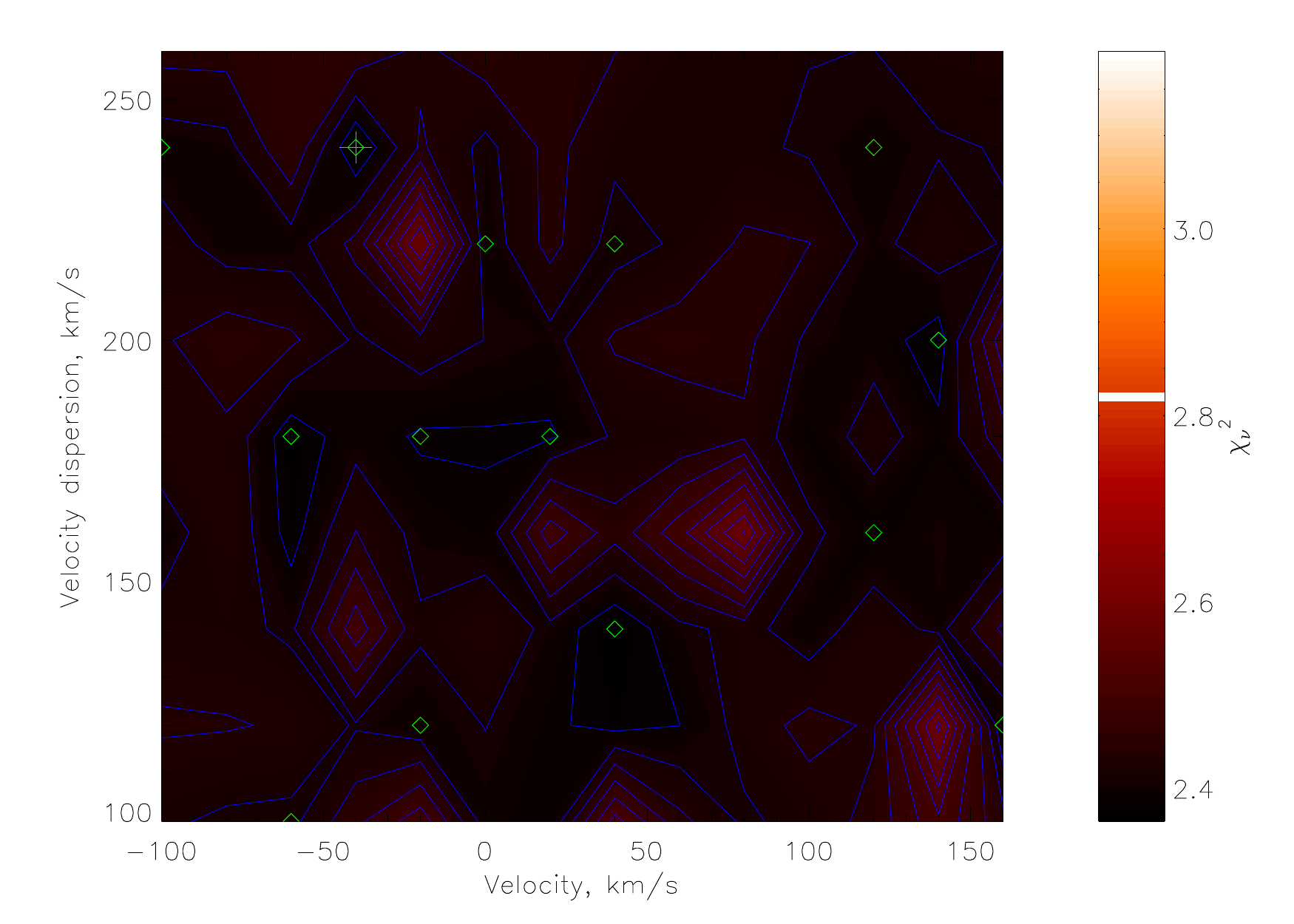}
	  \includegraphics[width=0.225\textwidth]{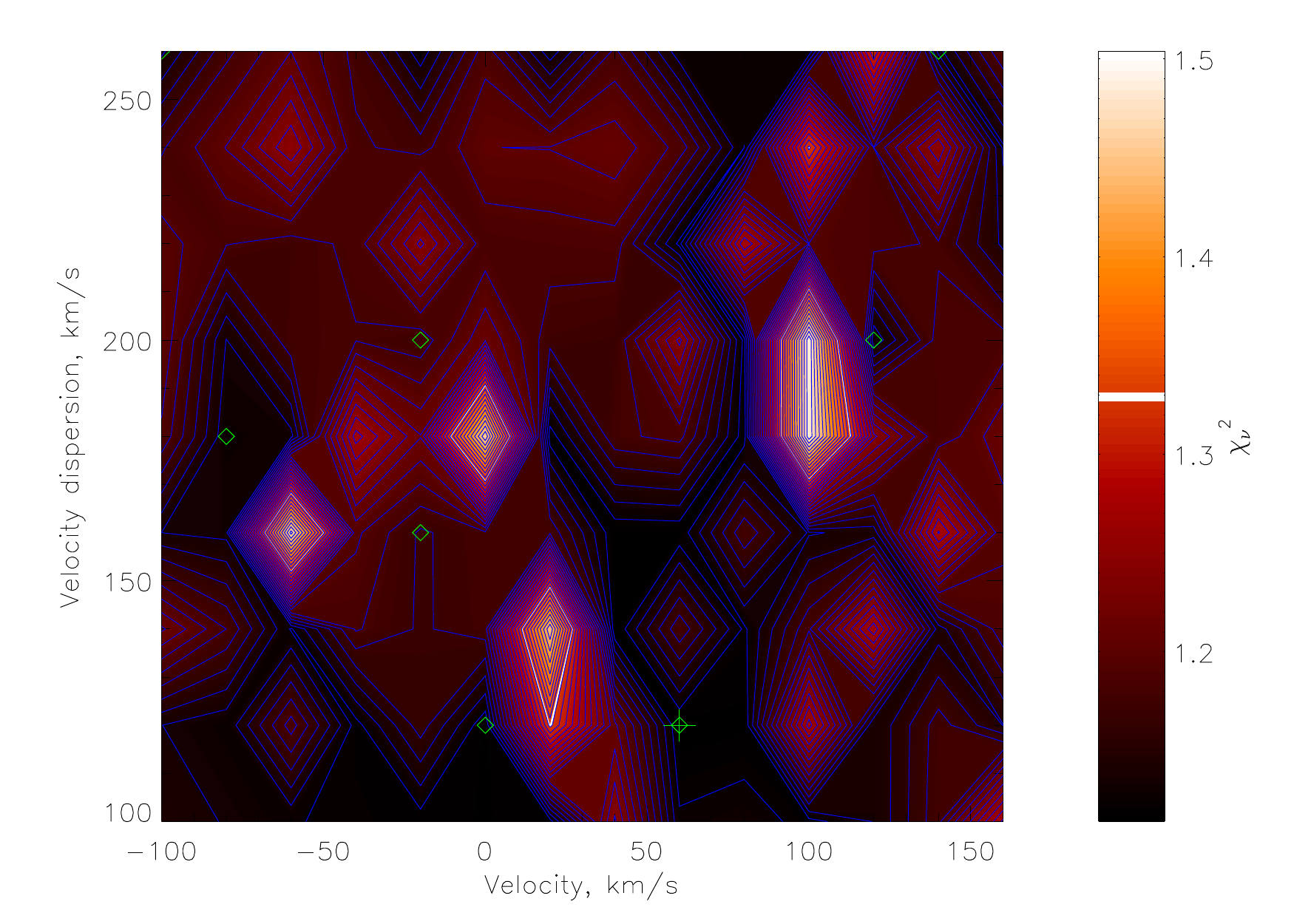}
	  \includegraphics[width=0.225\textwidth]{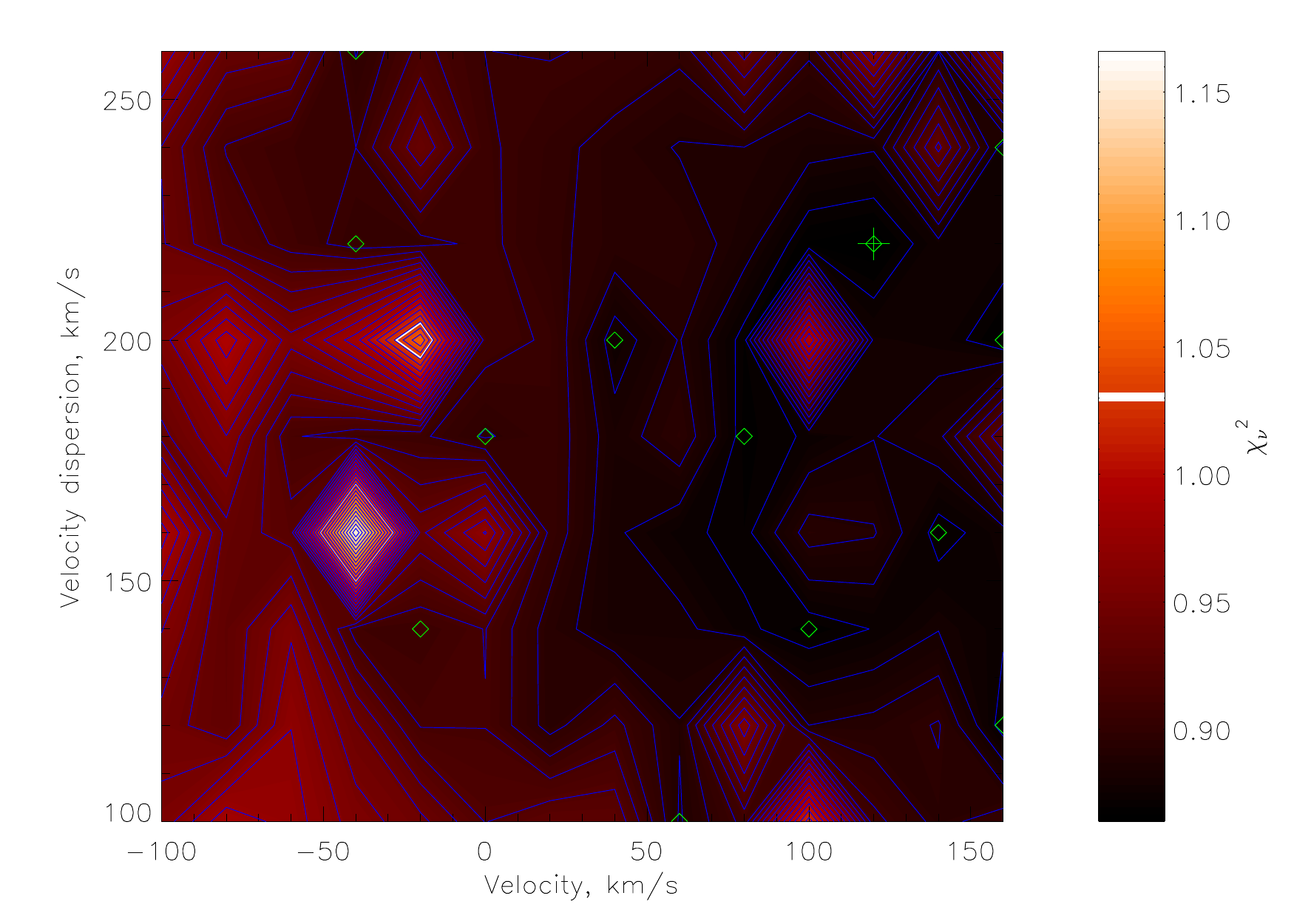}
	  \includegraphics[width=0.225\textwidth]{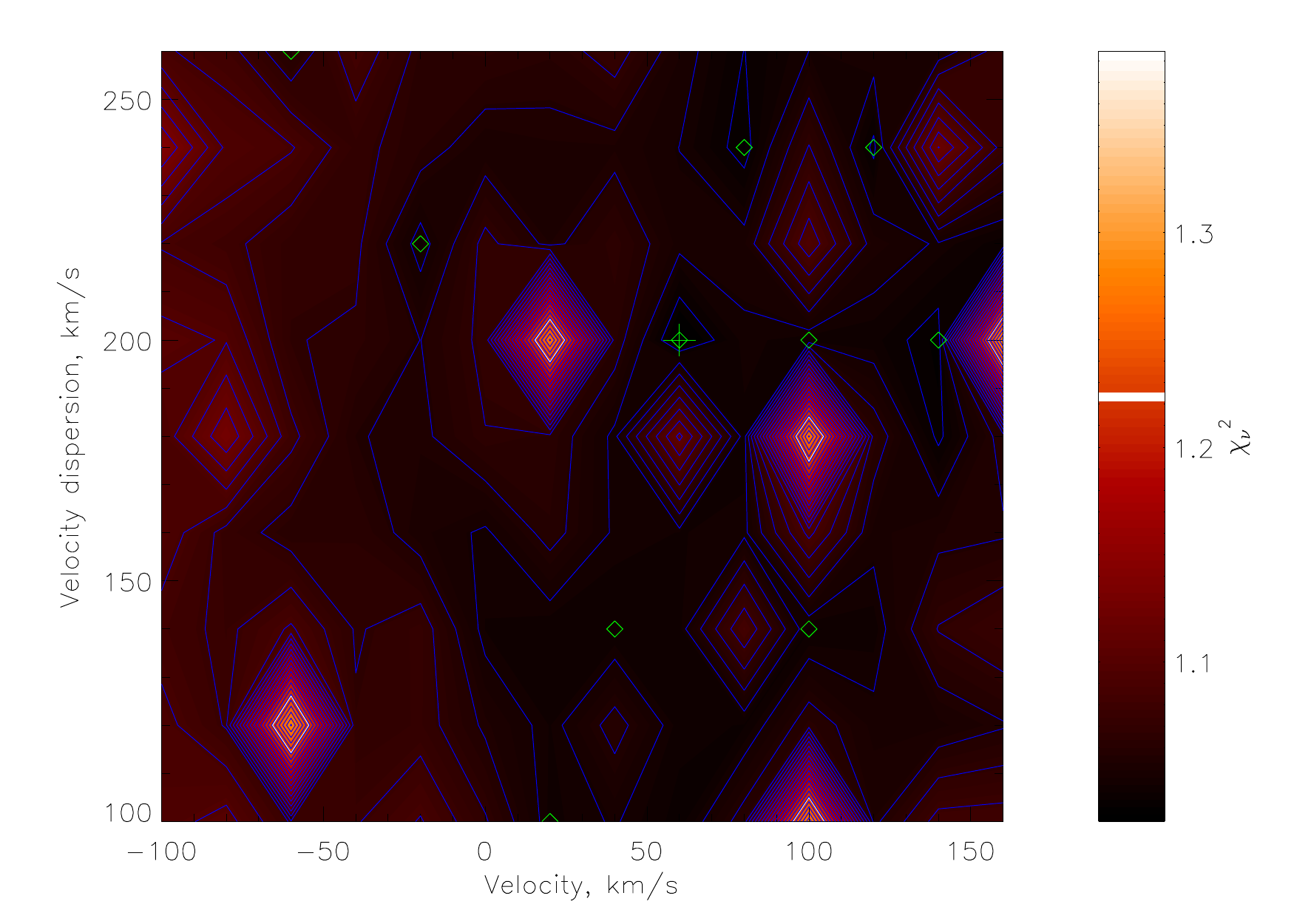}
	  \includegraphics[width=0.225\textwidth]{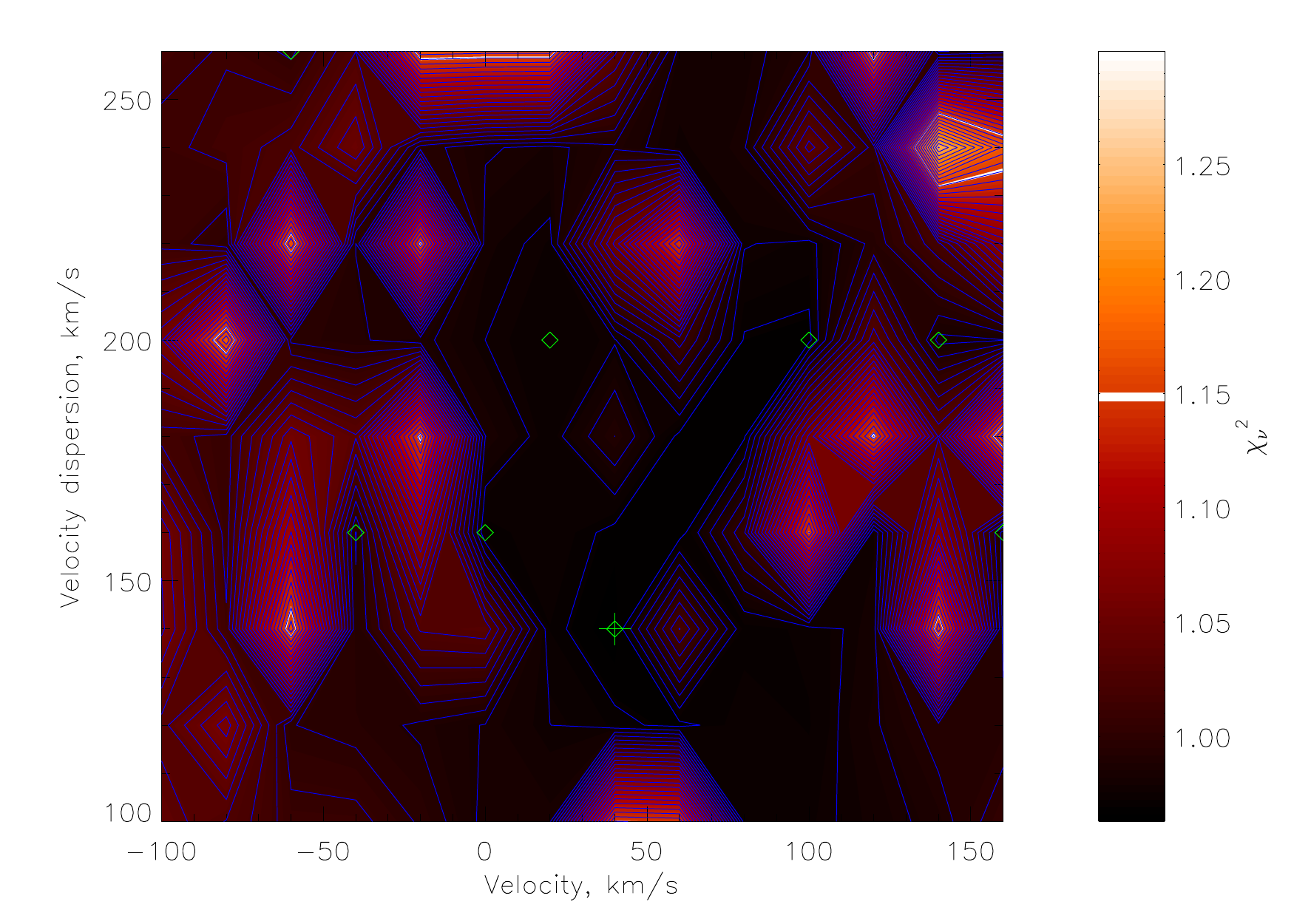}
	  \includegraphics[width=0.225\textwidth]{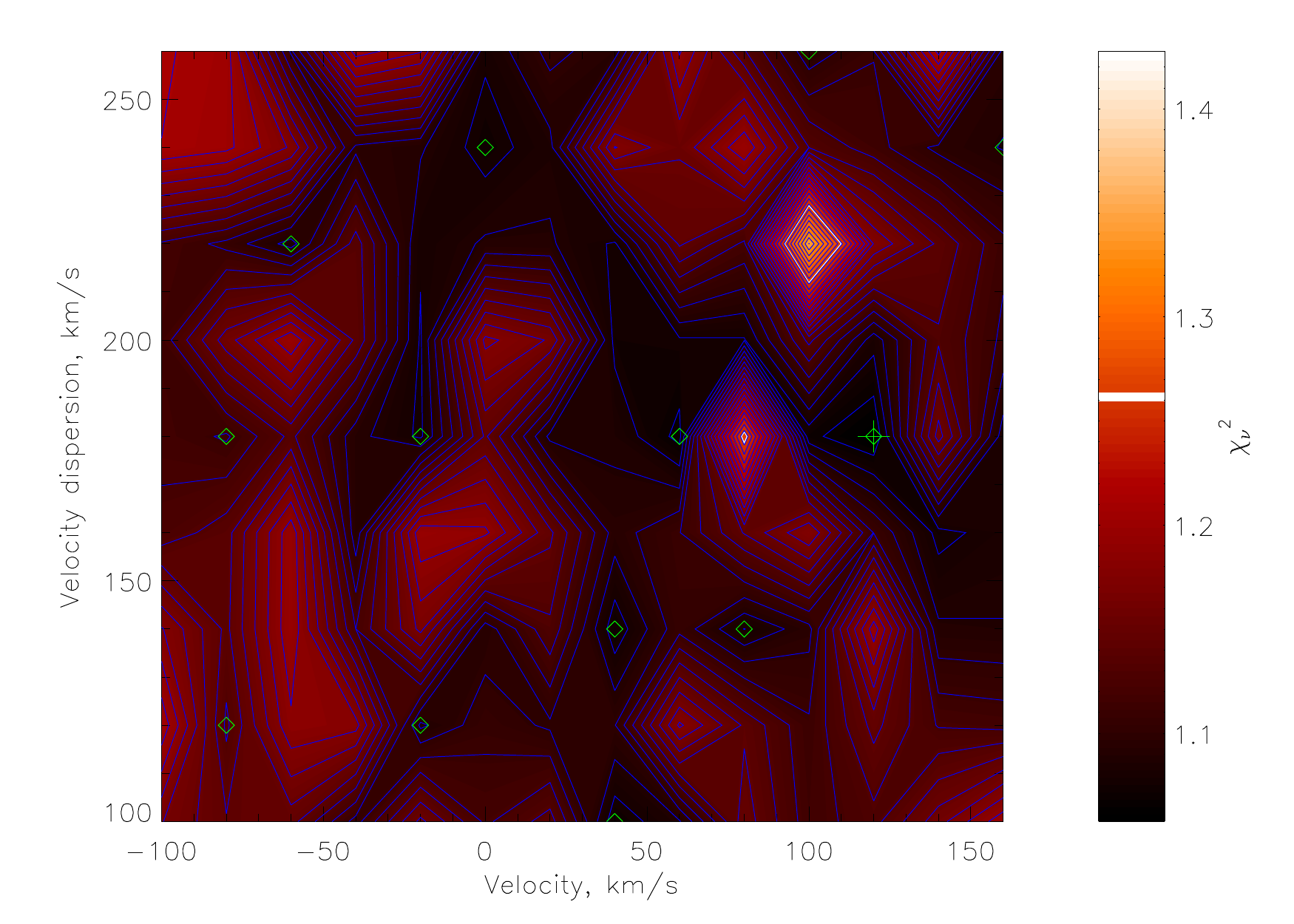}
	  \includegraphics[width=0.225\textwidth]{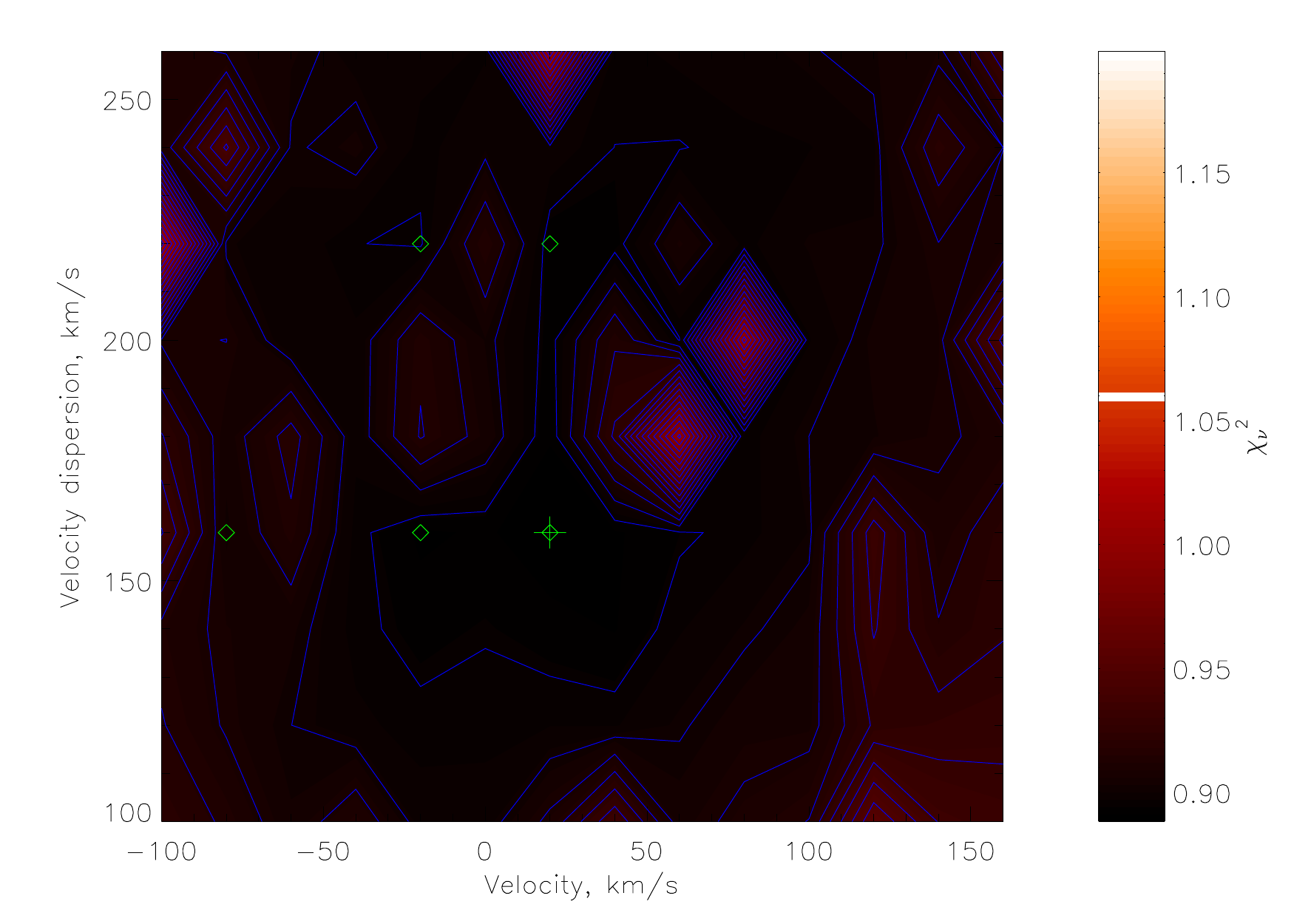}
	  \includegraphics[width=0.225\textwidth]{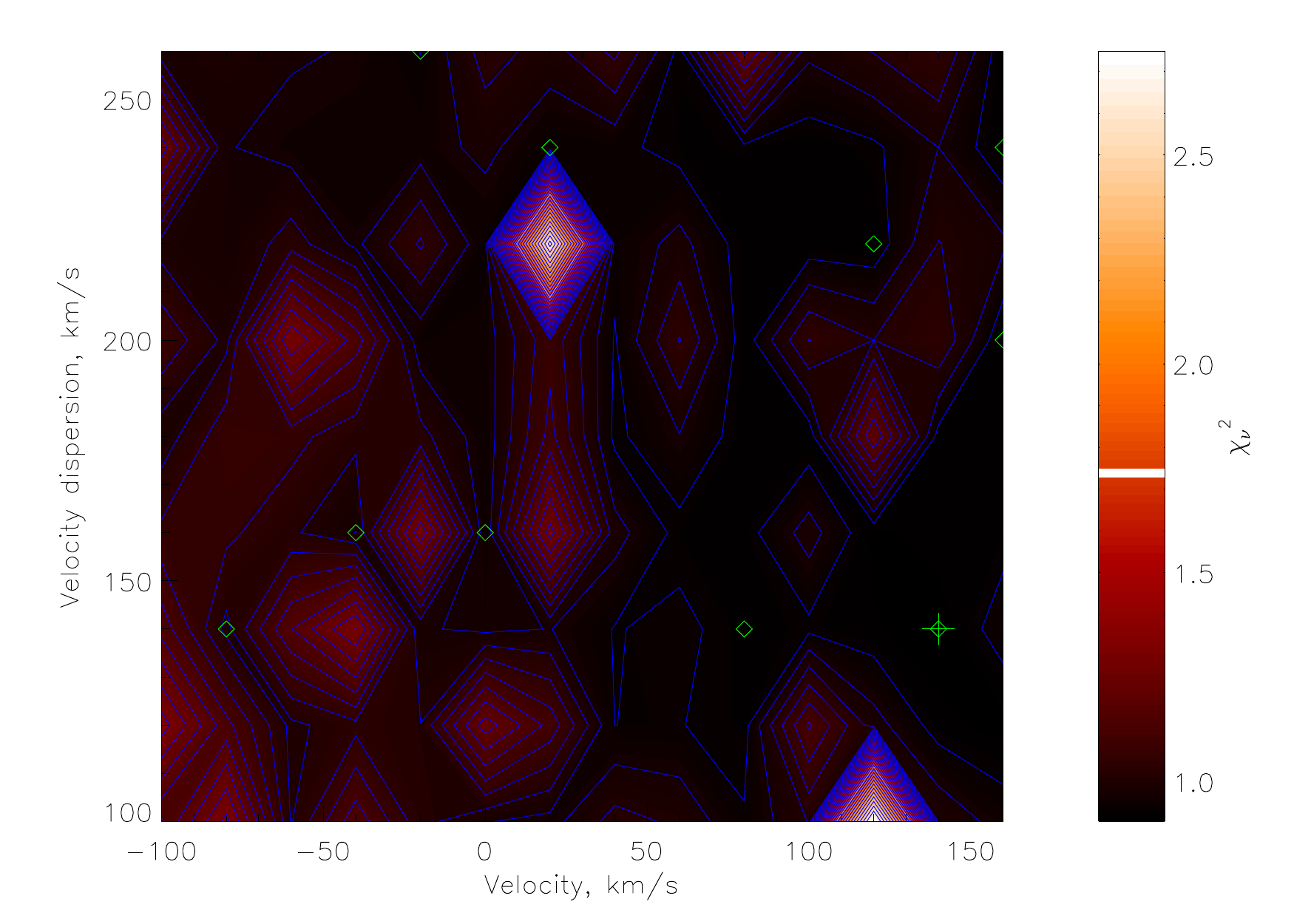}
	  \includegraphics[width=0.225\textwidth]{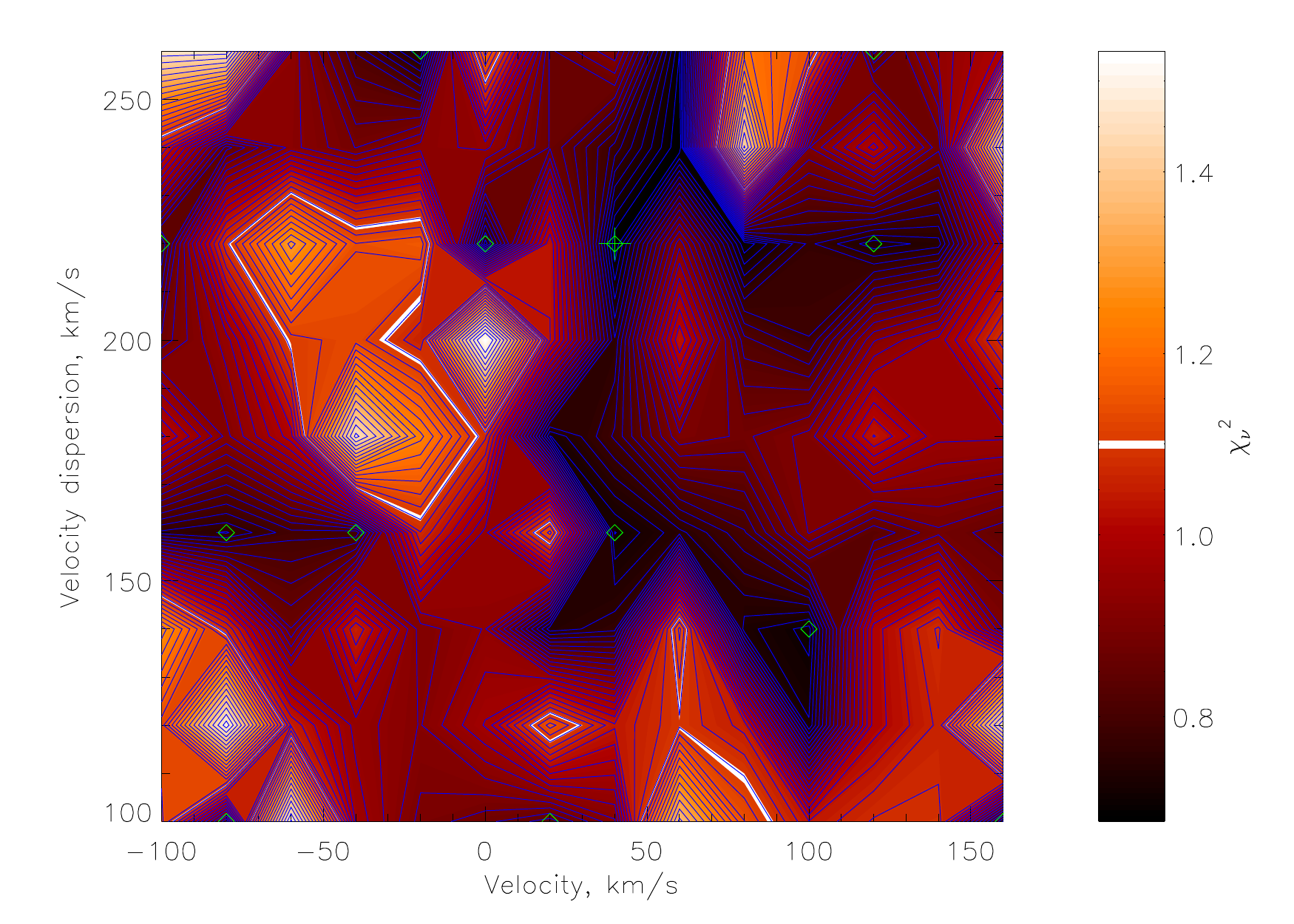}
	  \includegraphics[width=0.225\textwidth]{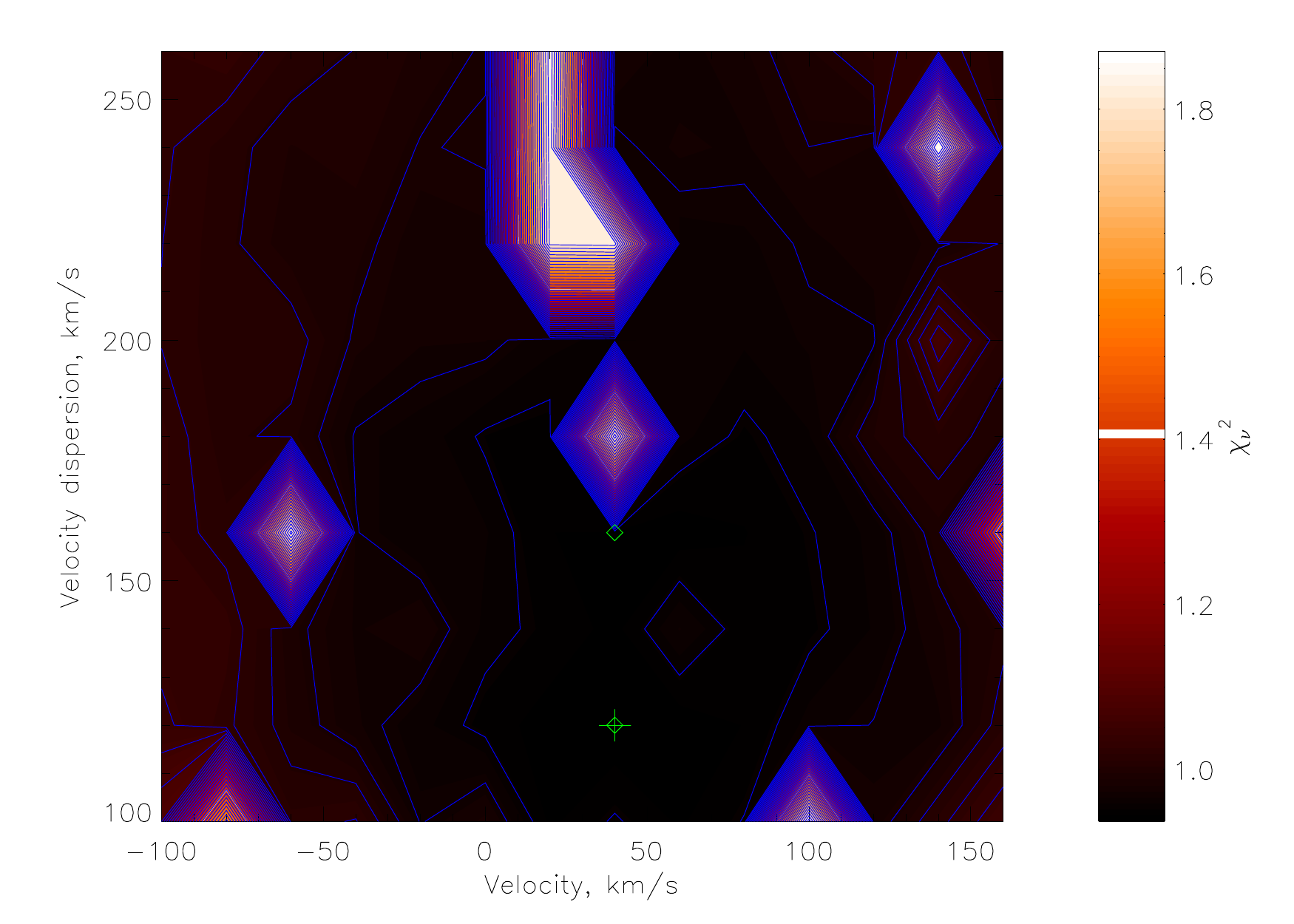}
	  \includegraphics[width=0.225\textwidth]{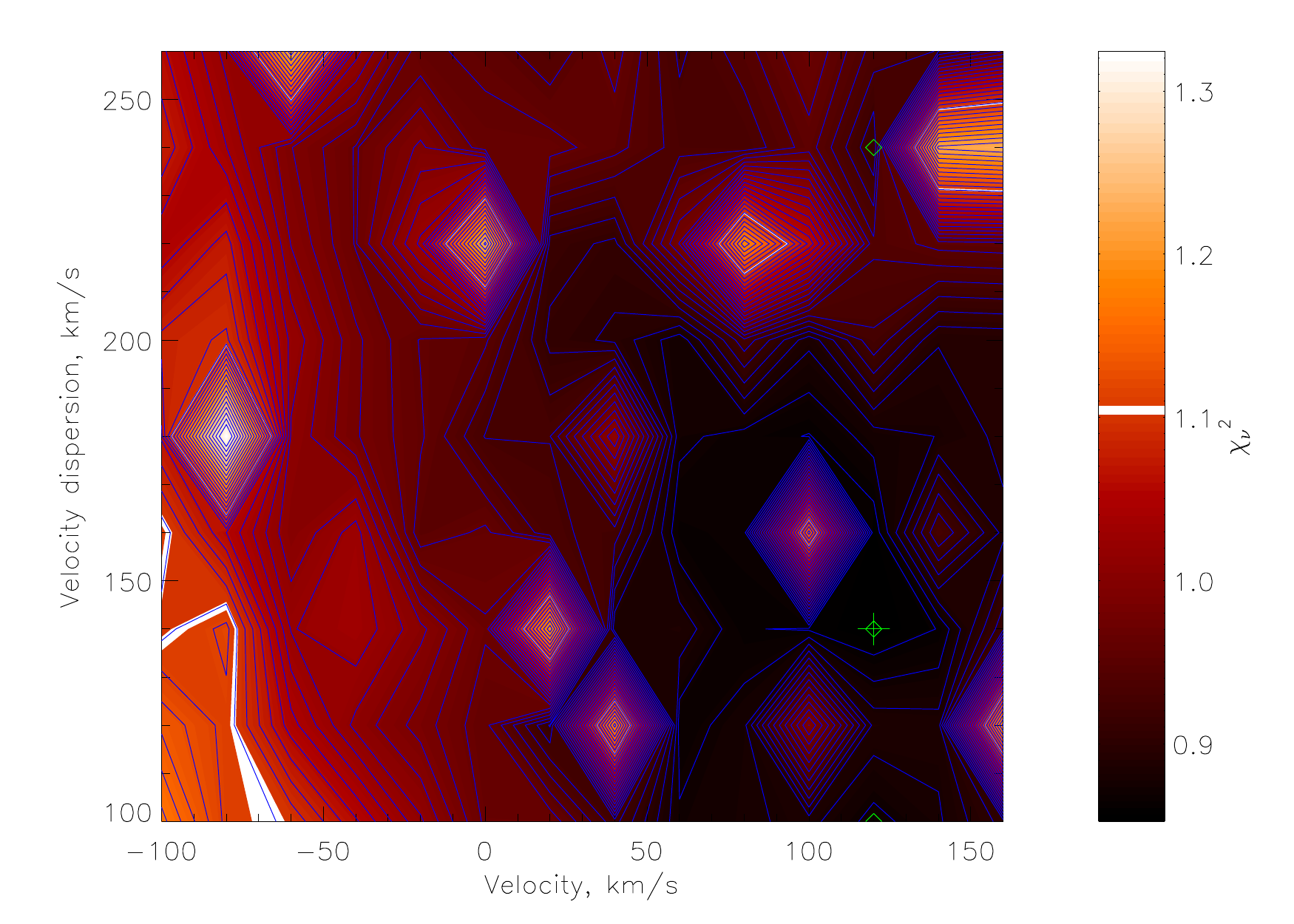}
	  \includegraphics[width=0.225\textwidth]{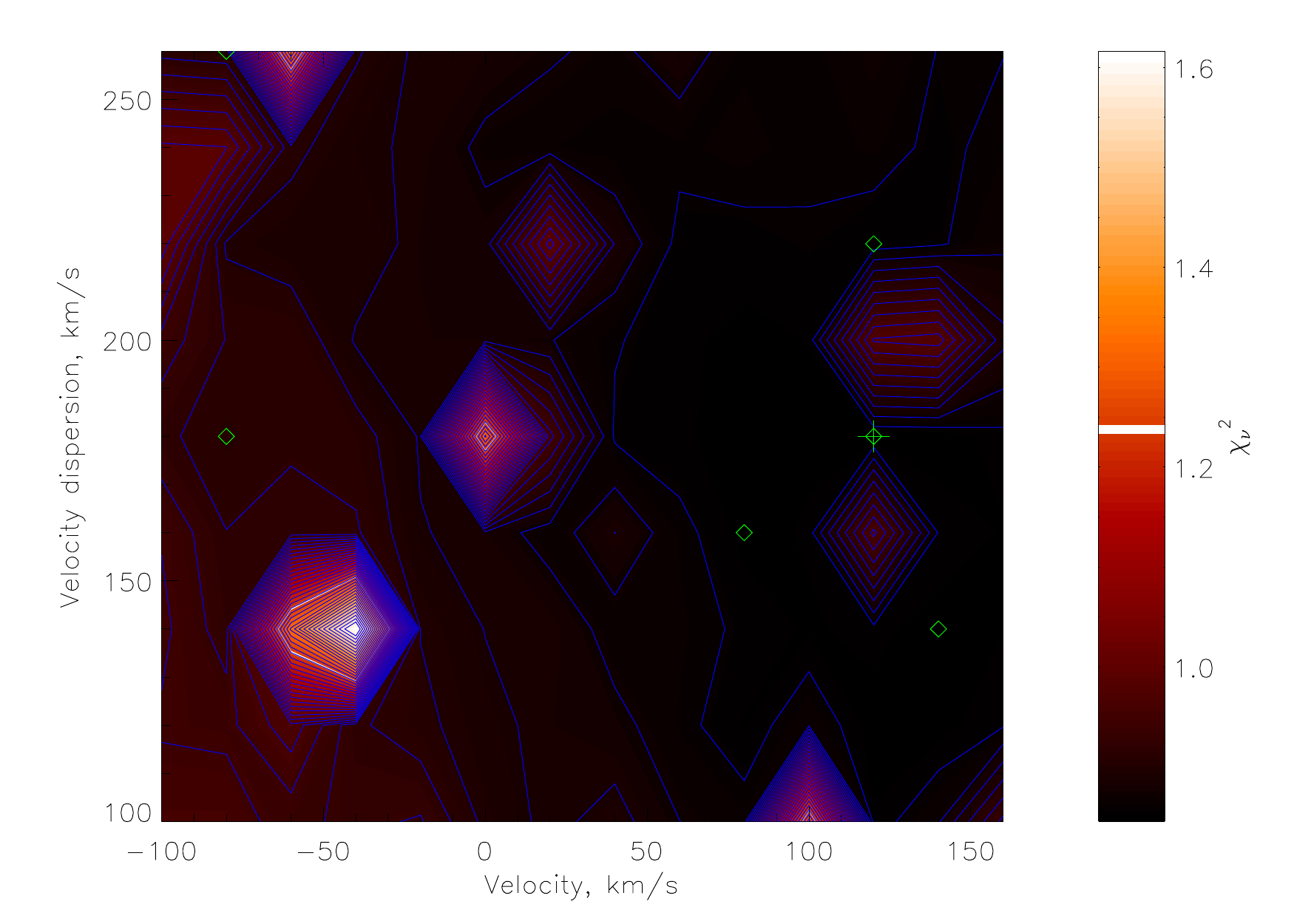}
	  \includegraphics[width=0.225\textwidth]{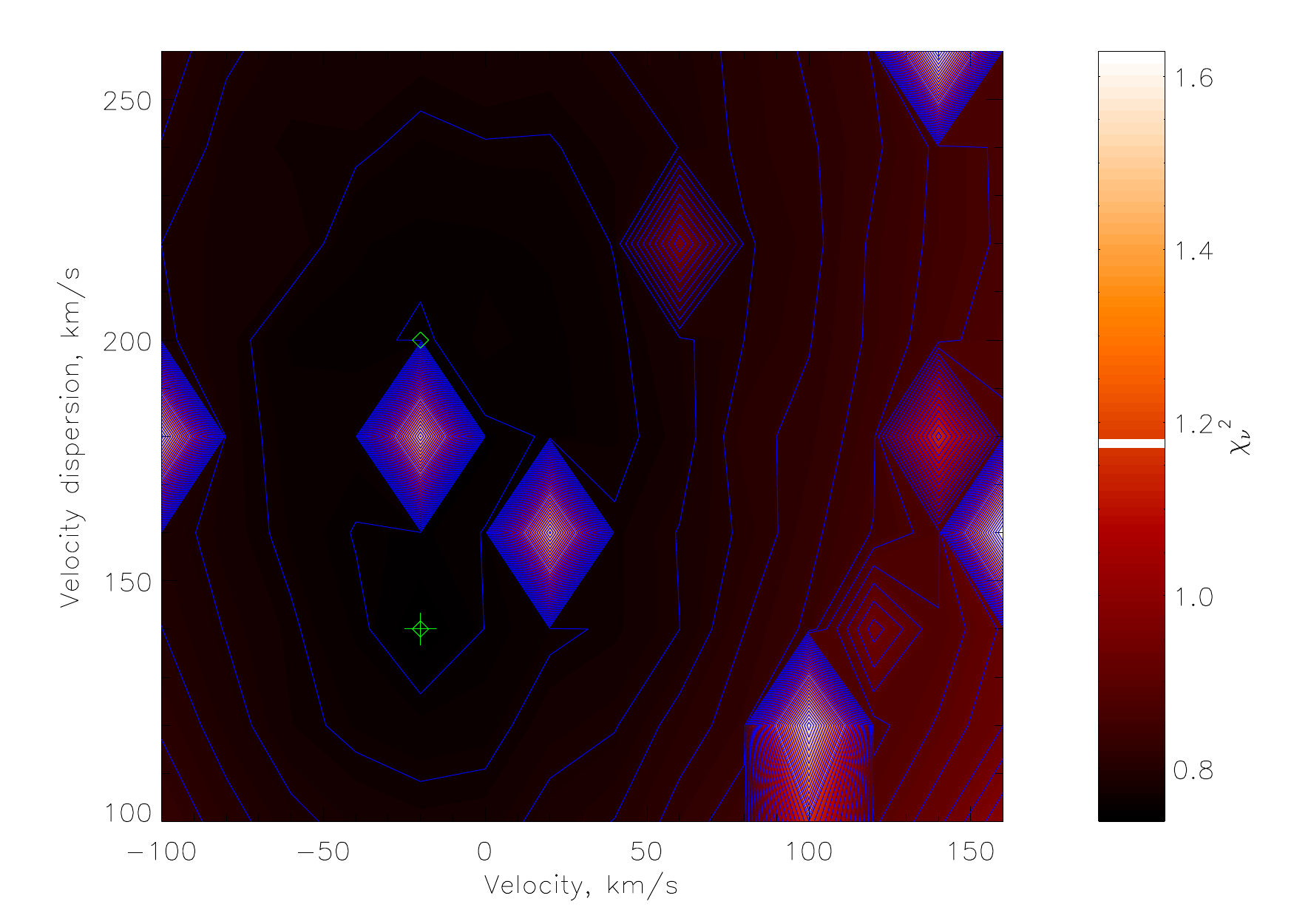}
	  \includegraphics[width=0.225\textwidth]{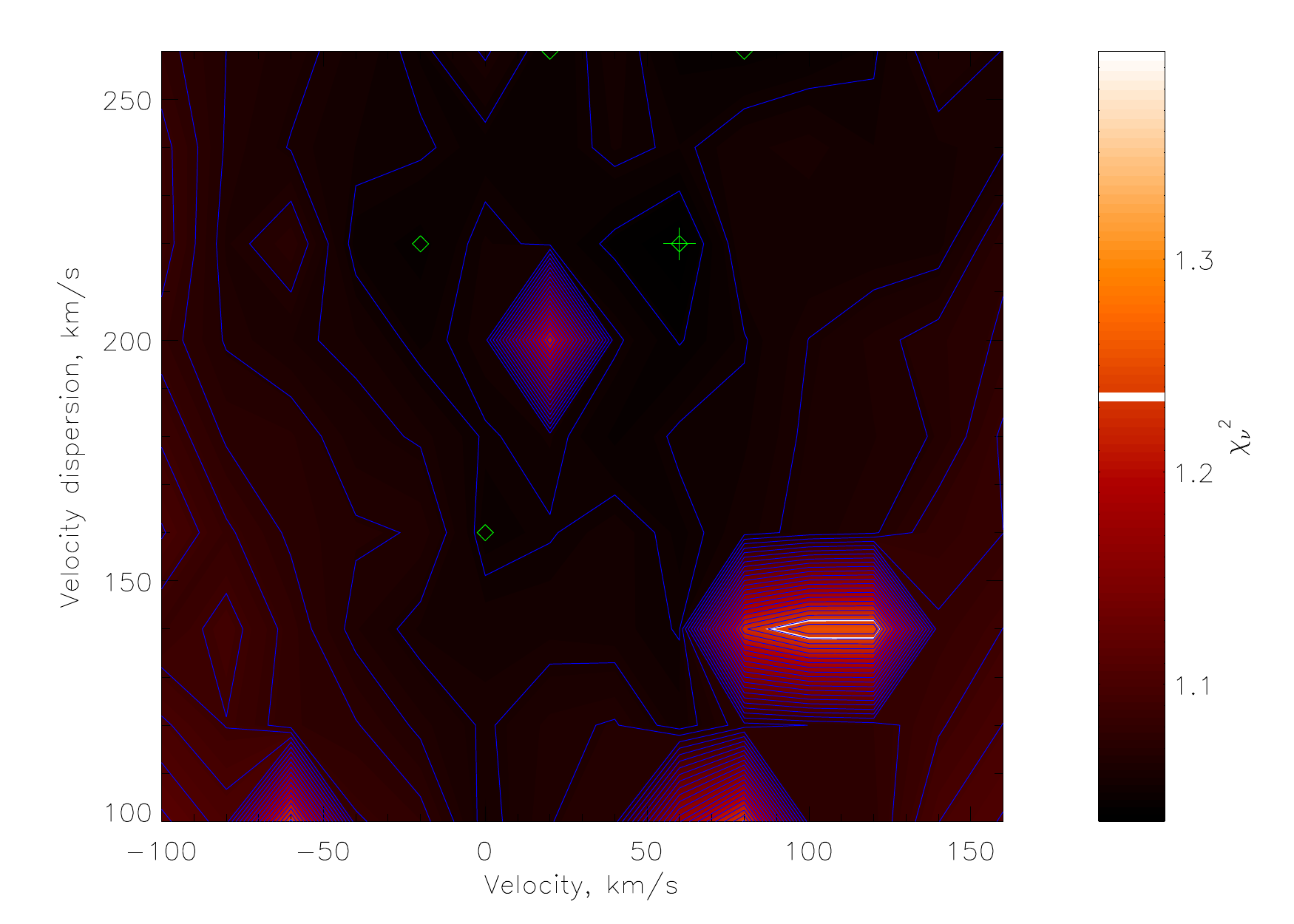}
	  \includegraphics[width=0.225\textwidth]{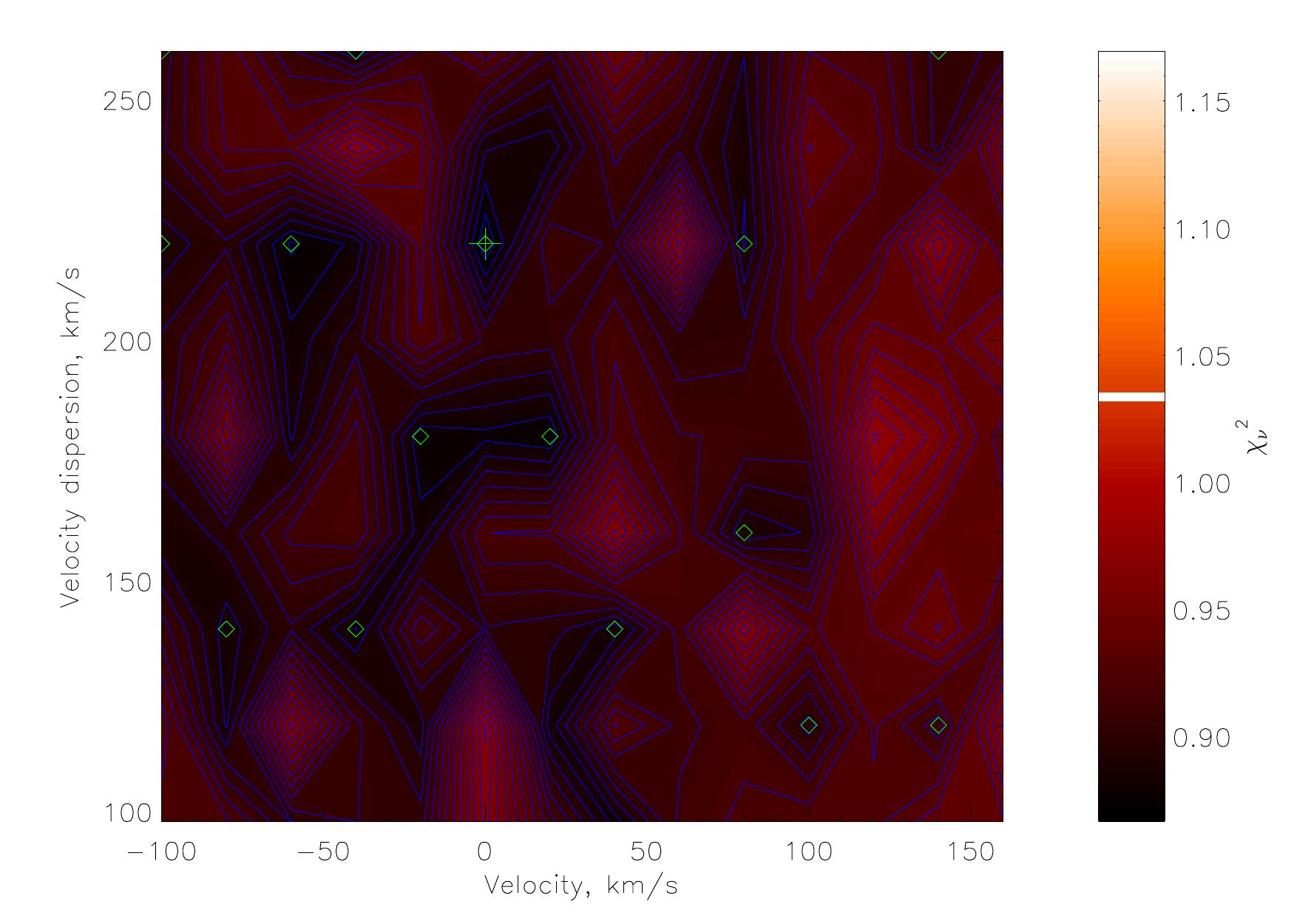}
	  \includegraphics[width=0.225\textwidth]{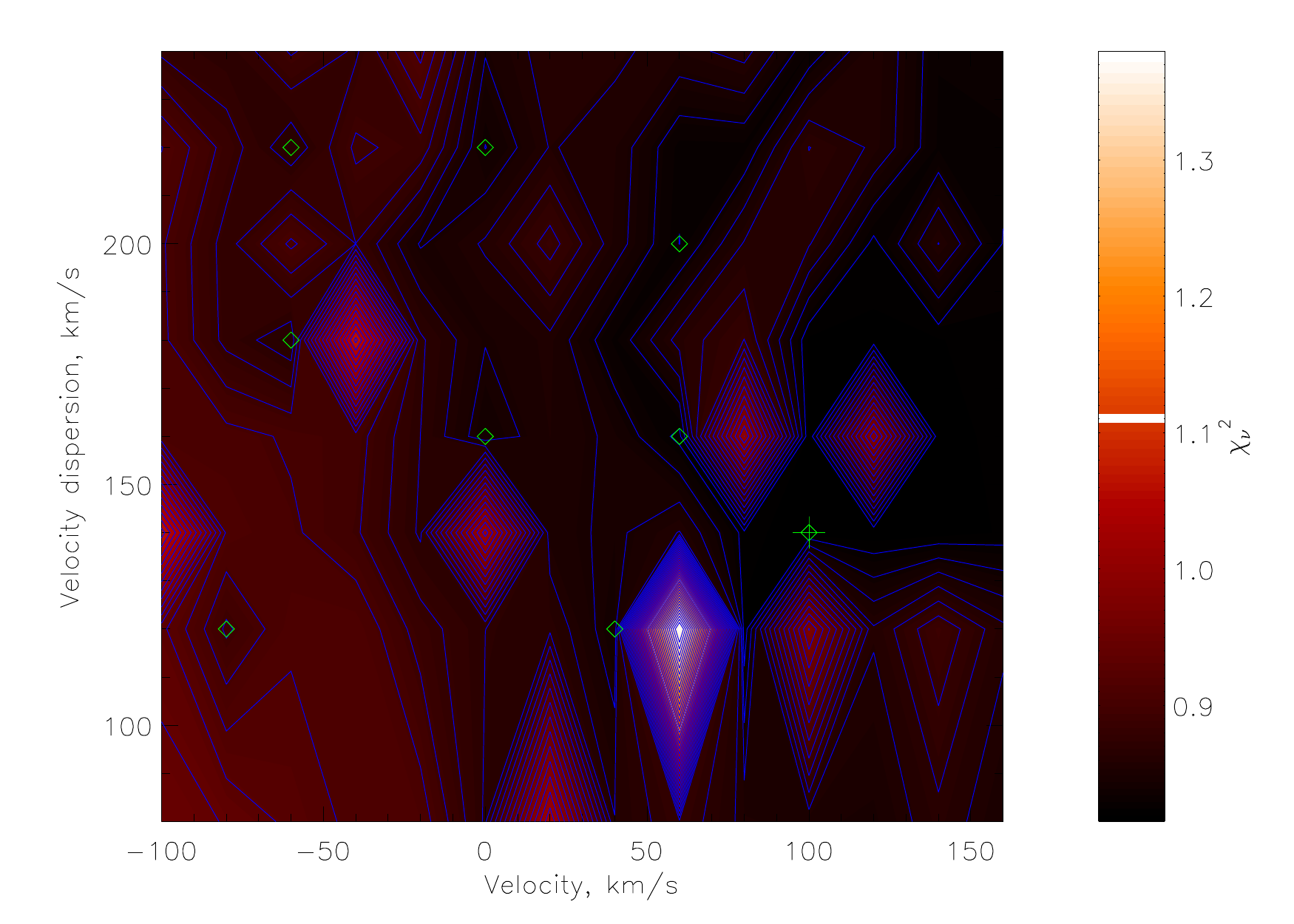}
	  \includegraphics[width=0.225\textwidth]{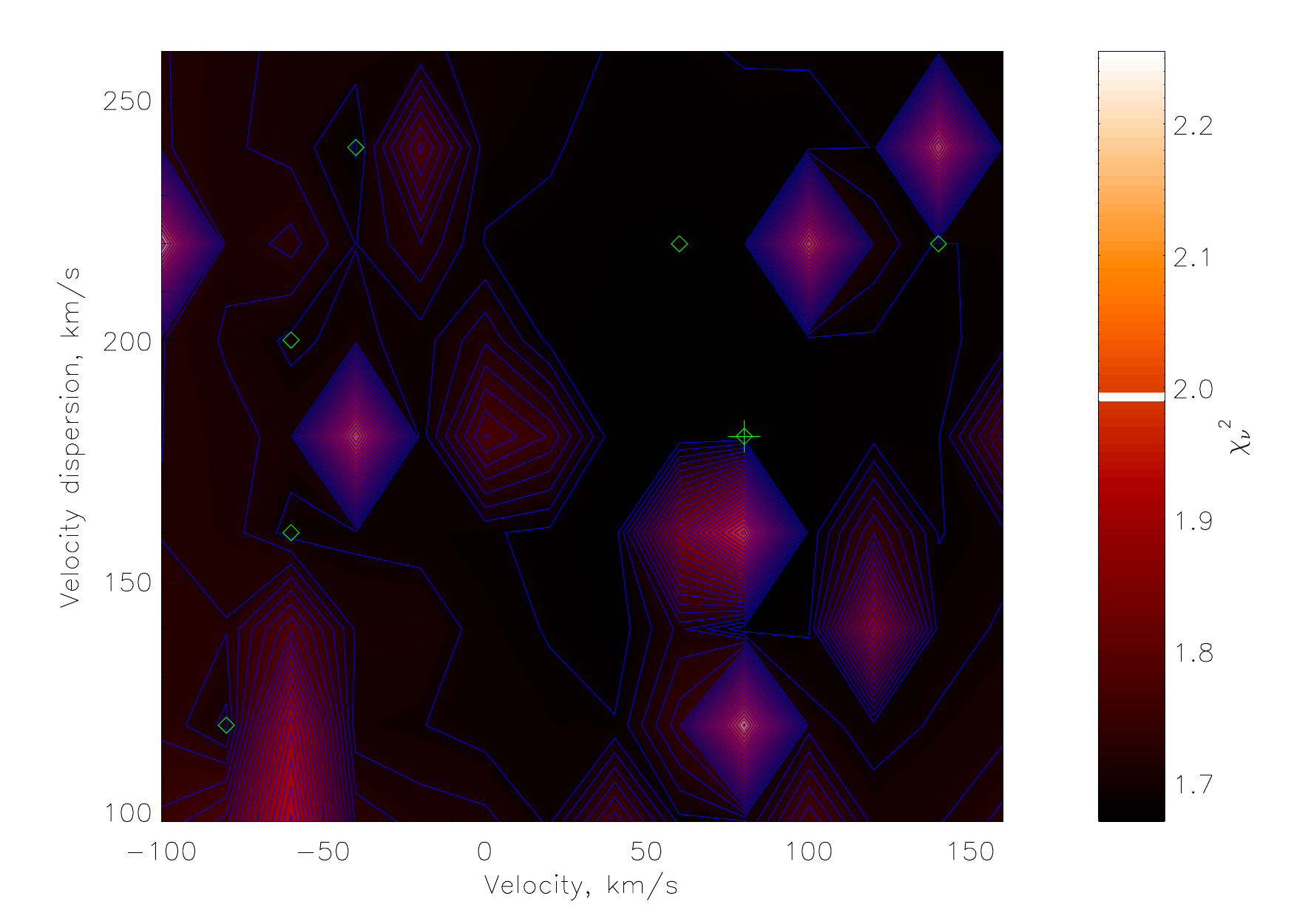}
	  \includegraphics[width=0.225\textwidth]{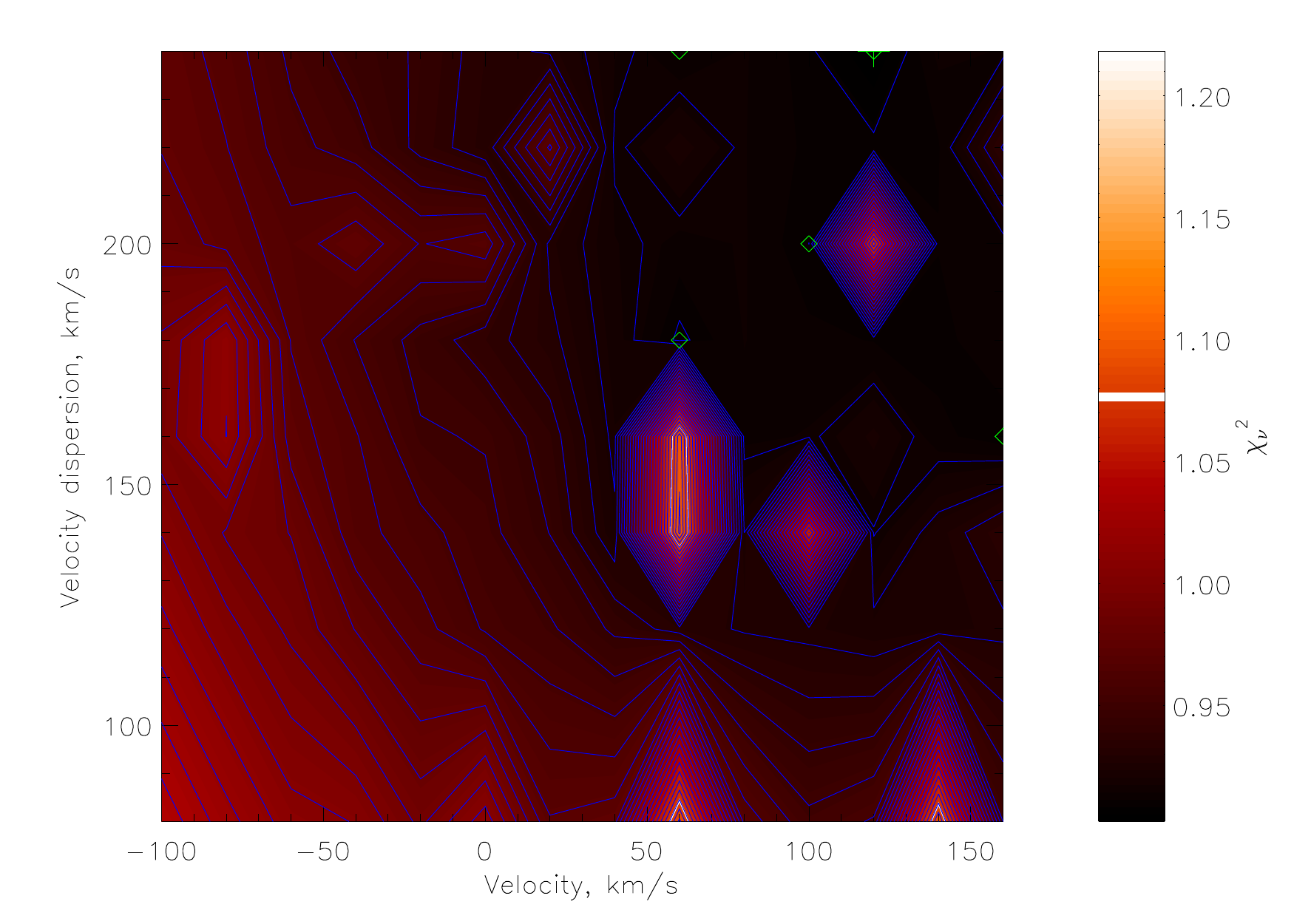}
	  \includegraphics[width=0.225\textwidth]{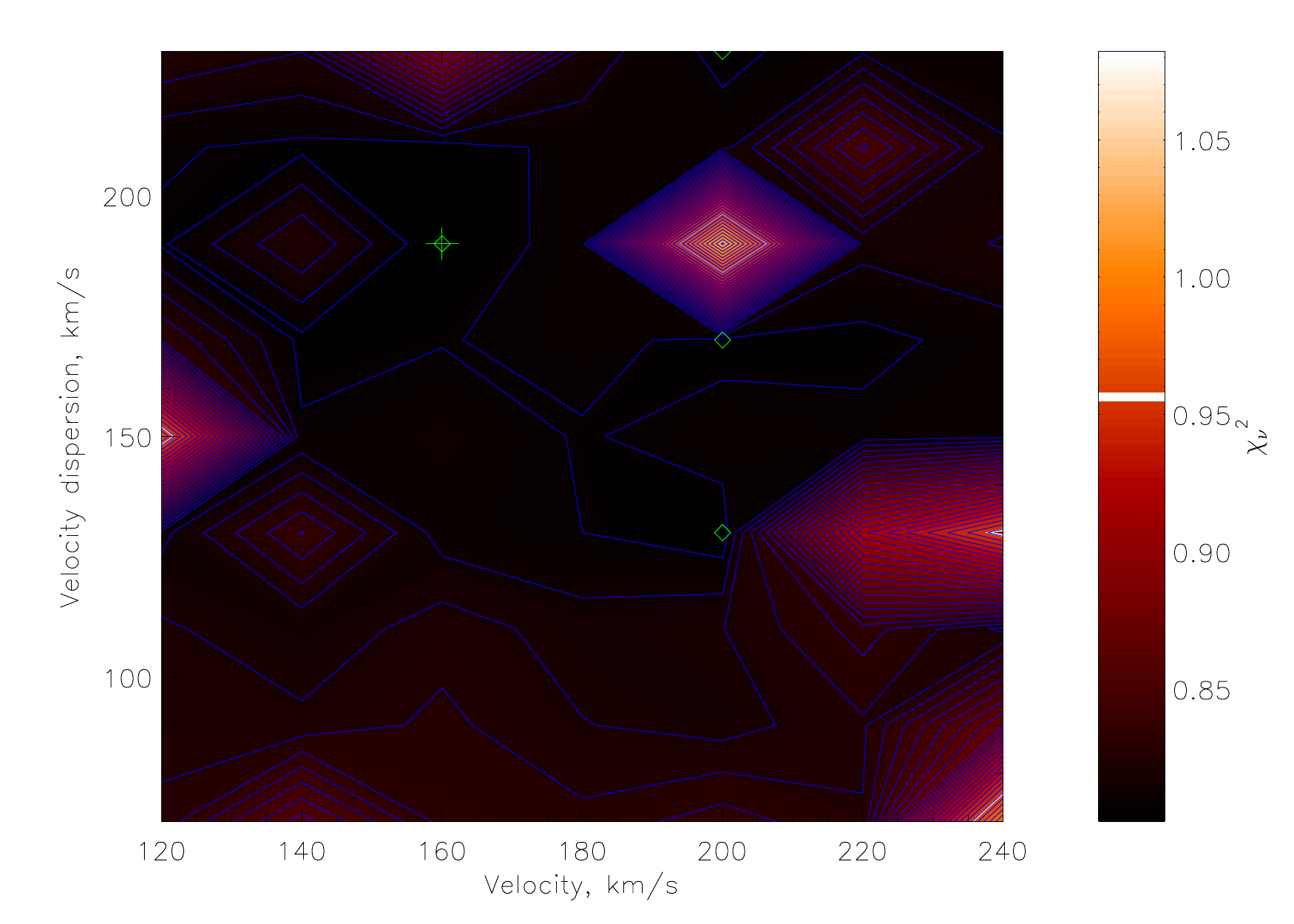}
	  \includegraphics[width=0.225\textwidth]{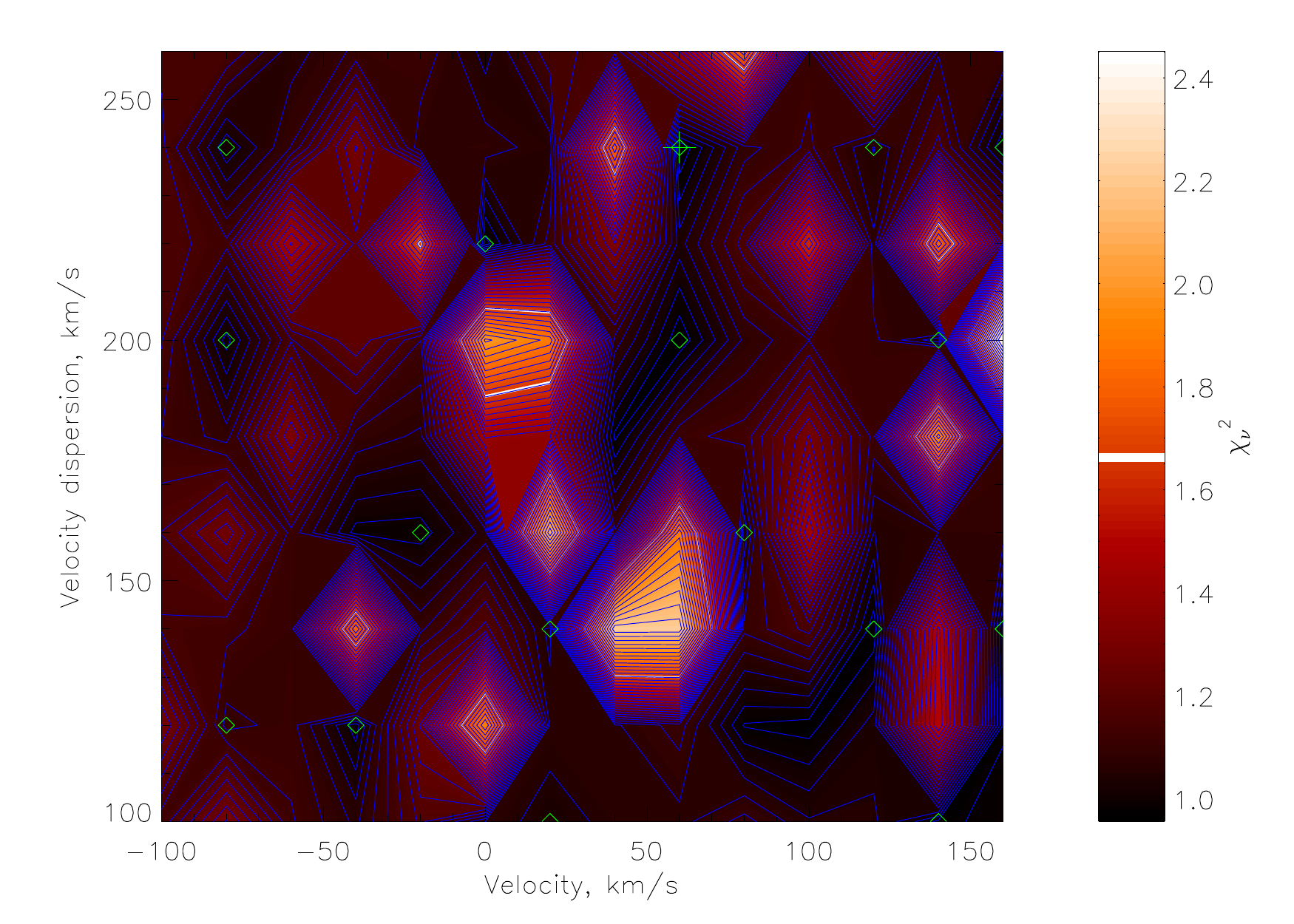}
	  \includegraphics[width=0.225\textwidth]{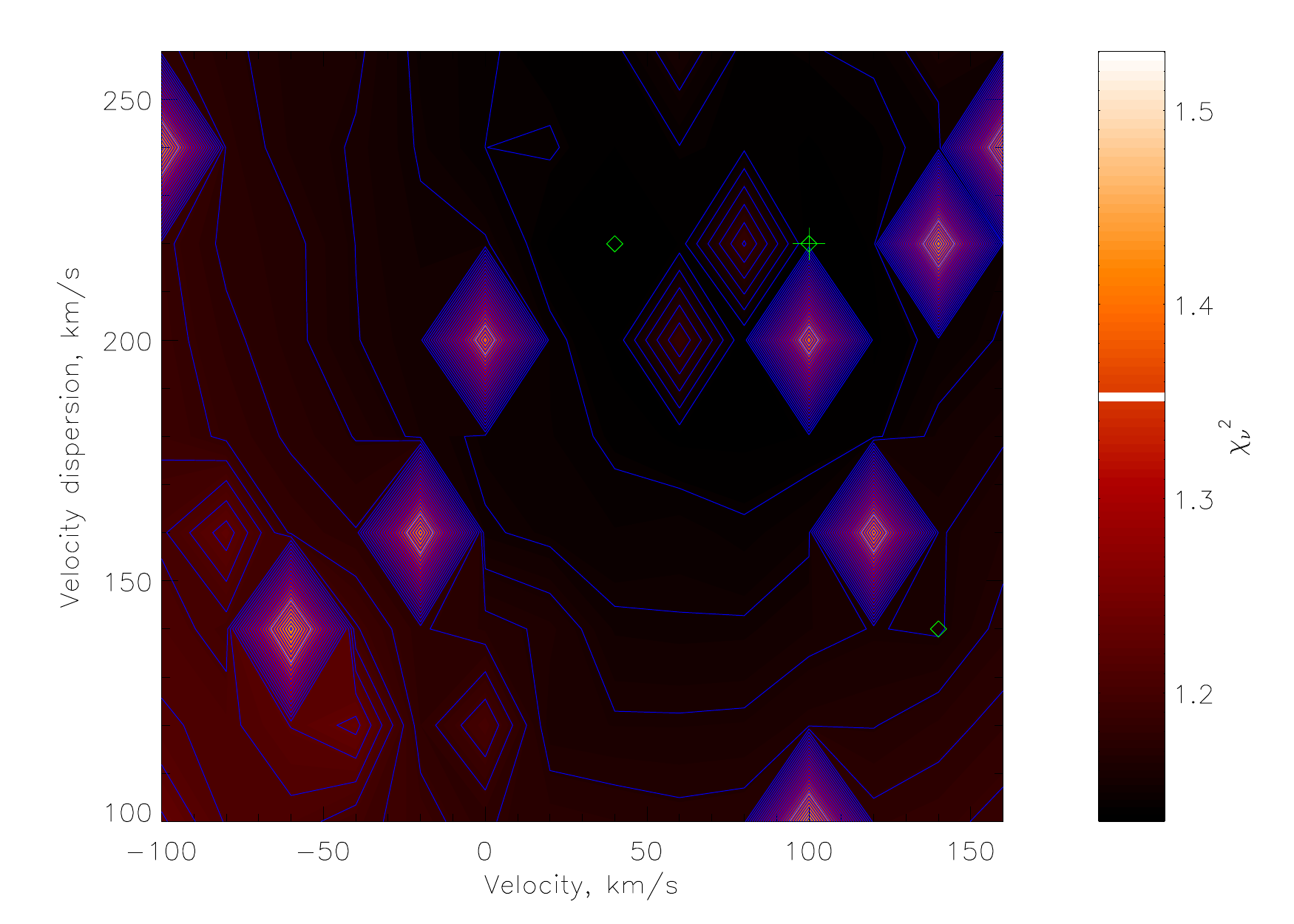}
	  \includegraphics[width=0.225\textwidth]{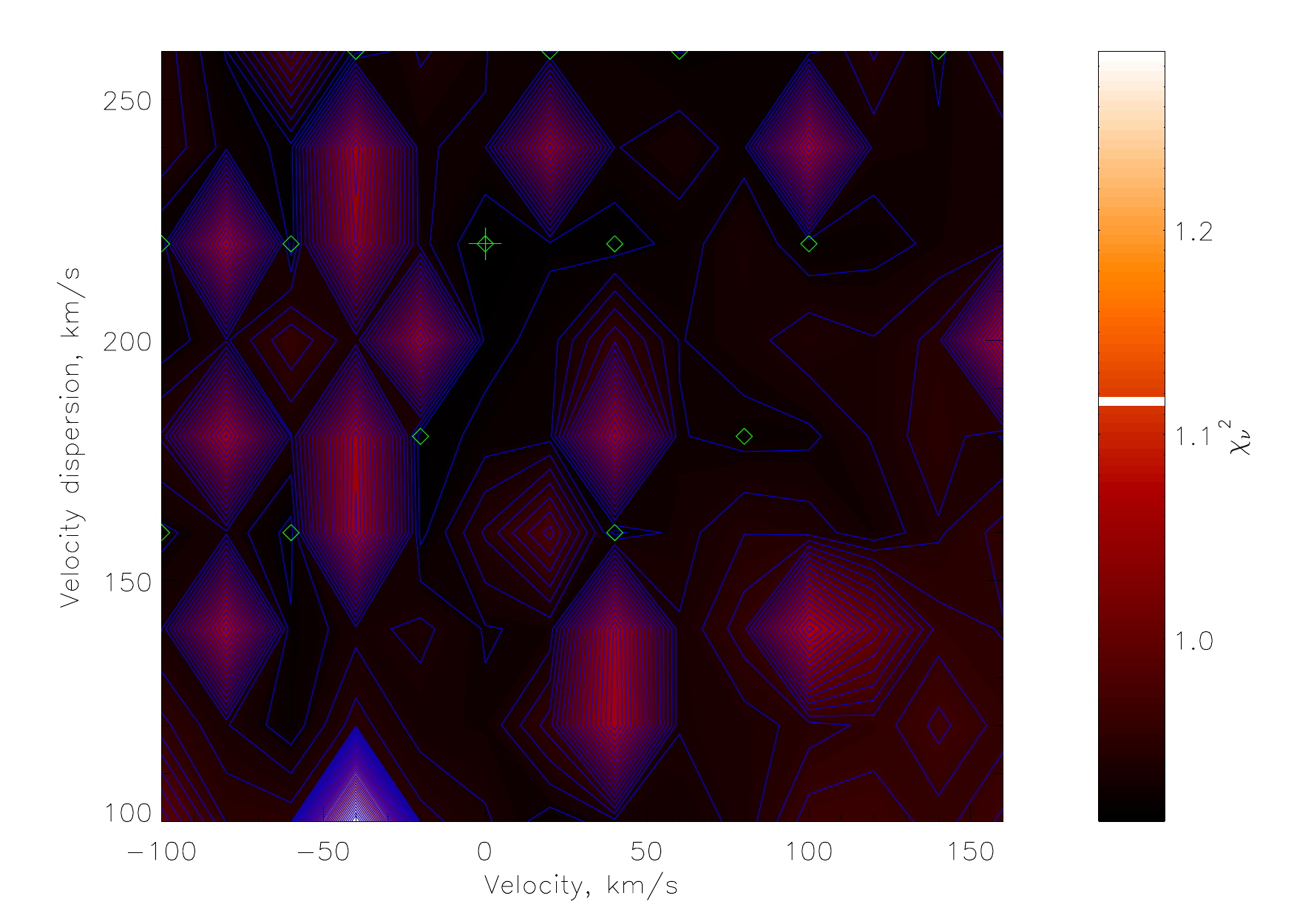}
	  \includegraphics[width=0.225\textwidth]{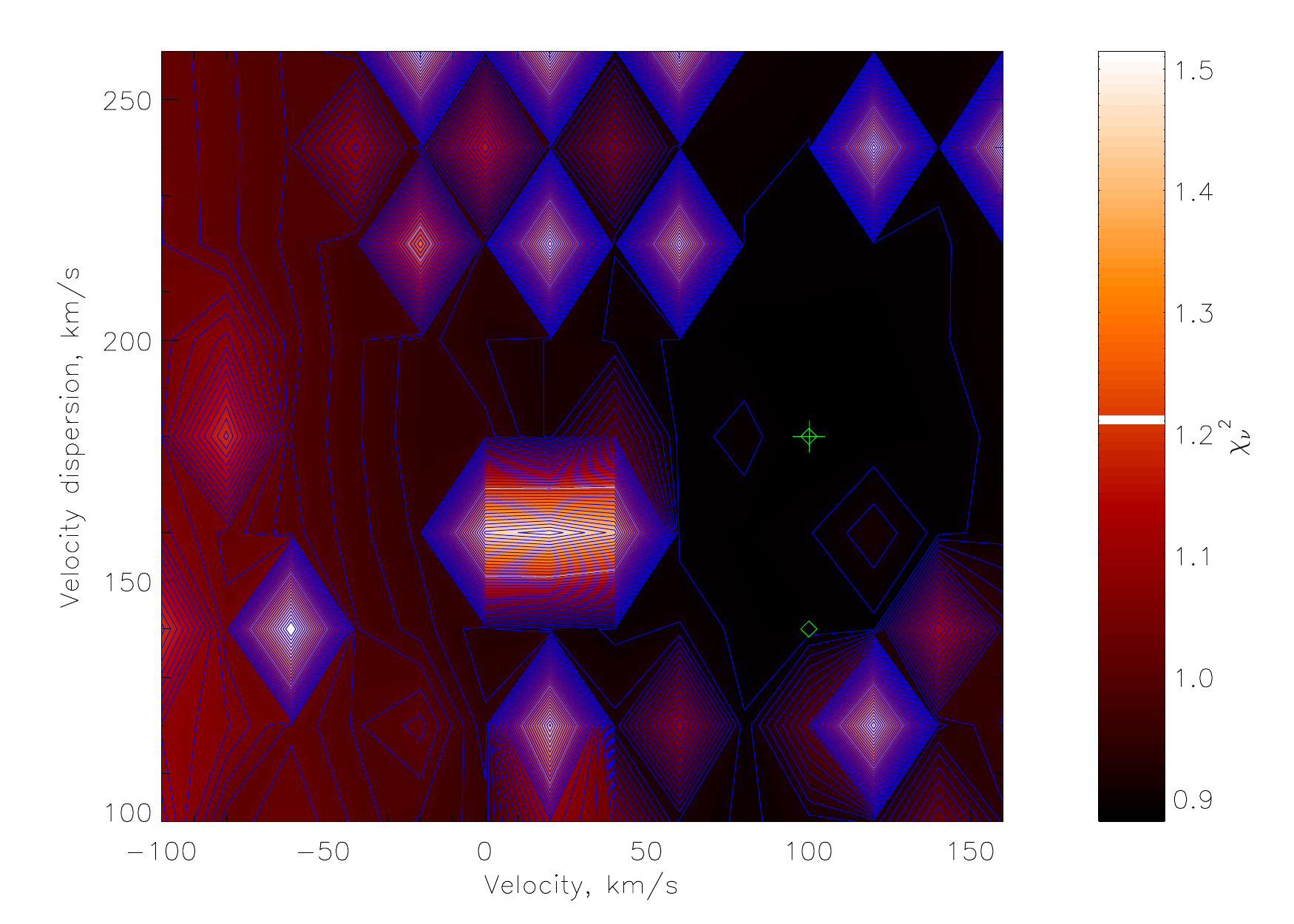}
	  \includegraphics[width=0.225\textwidth]{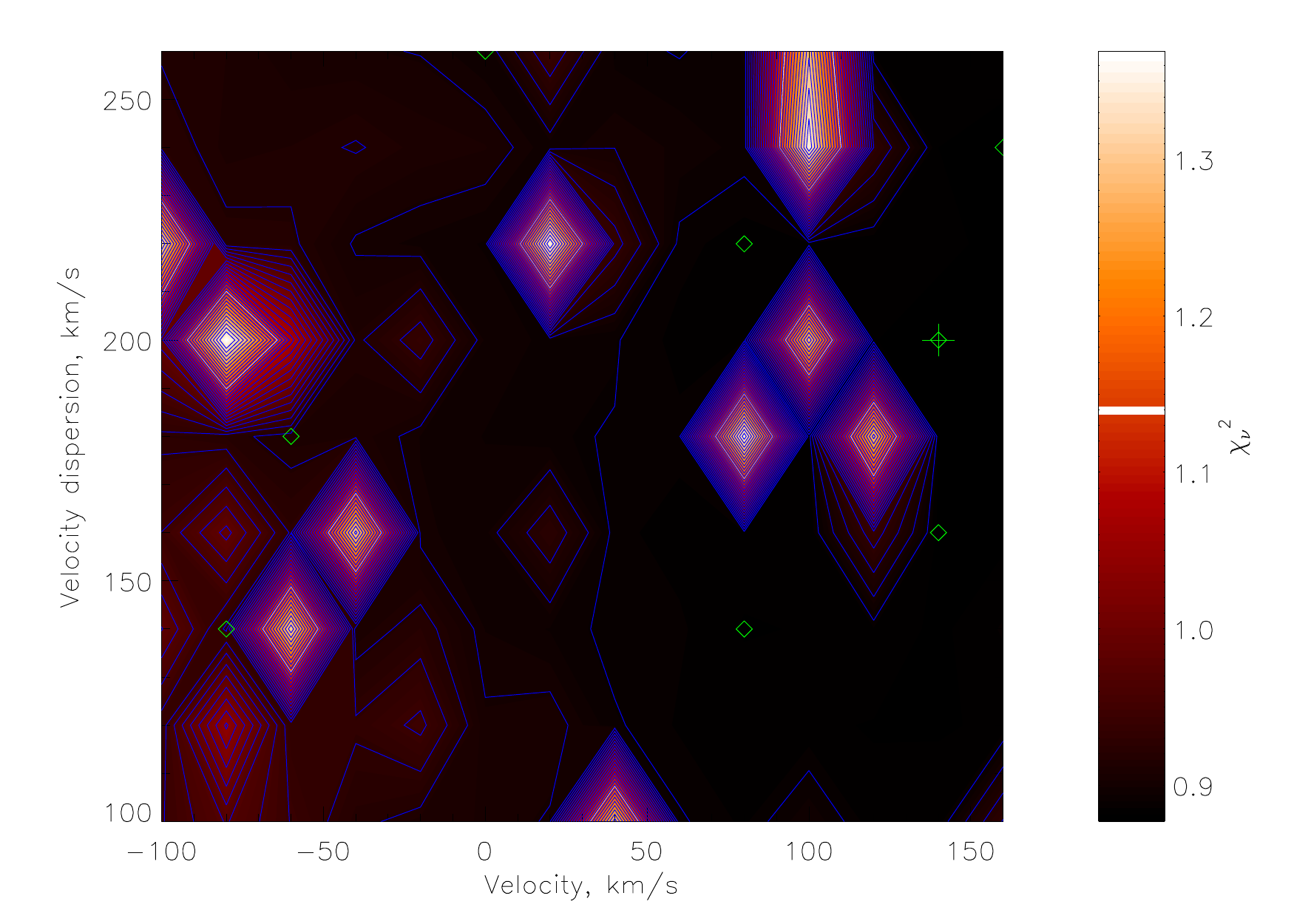}
	  \includegraphics[width=0.225\textwidth]{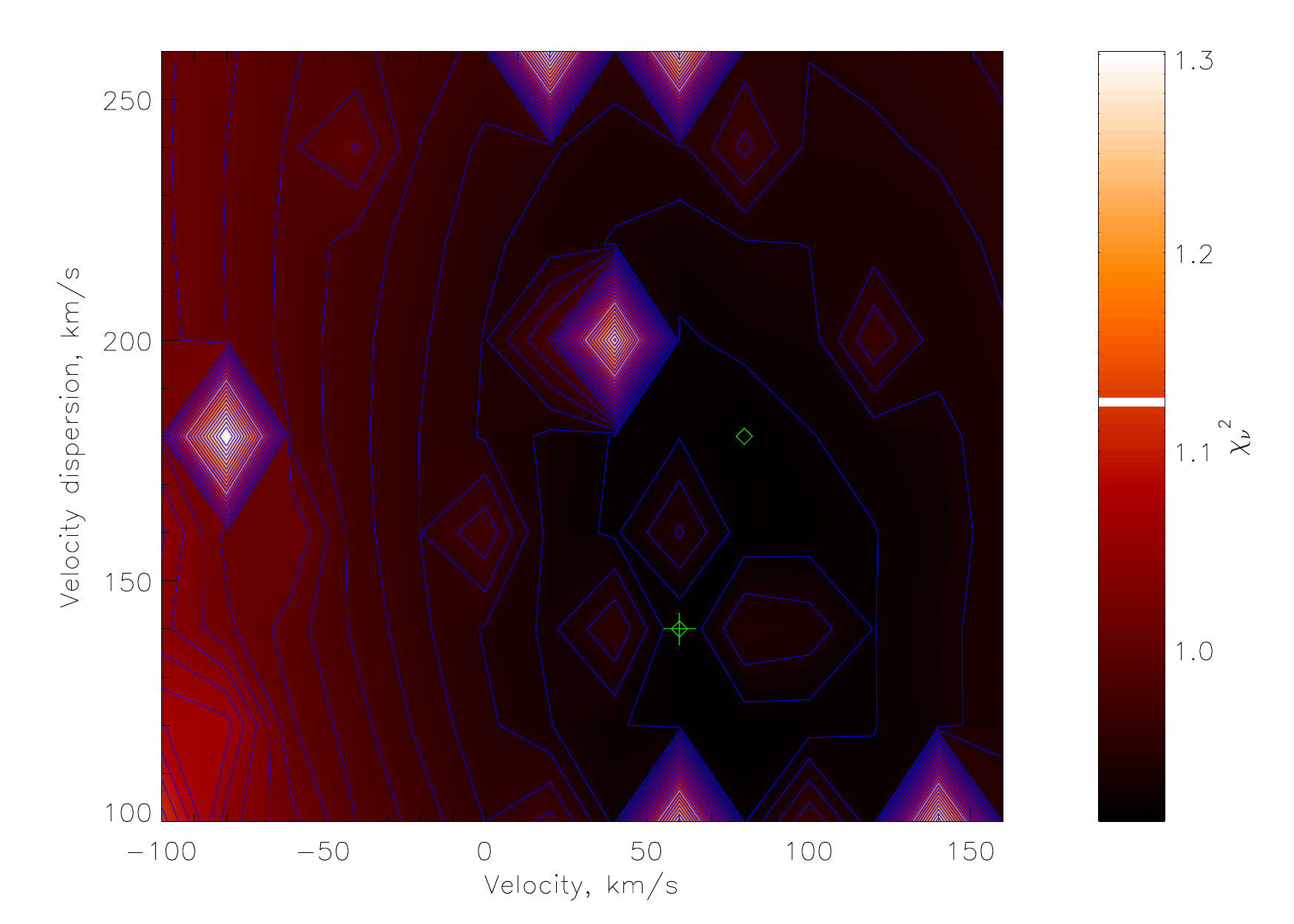}
	  \includegraphics[width=0.225\textwidth]{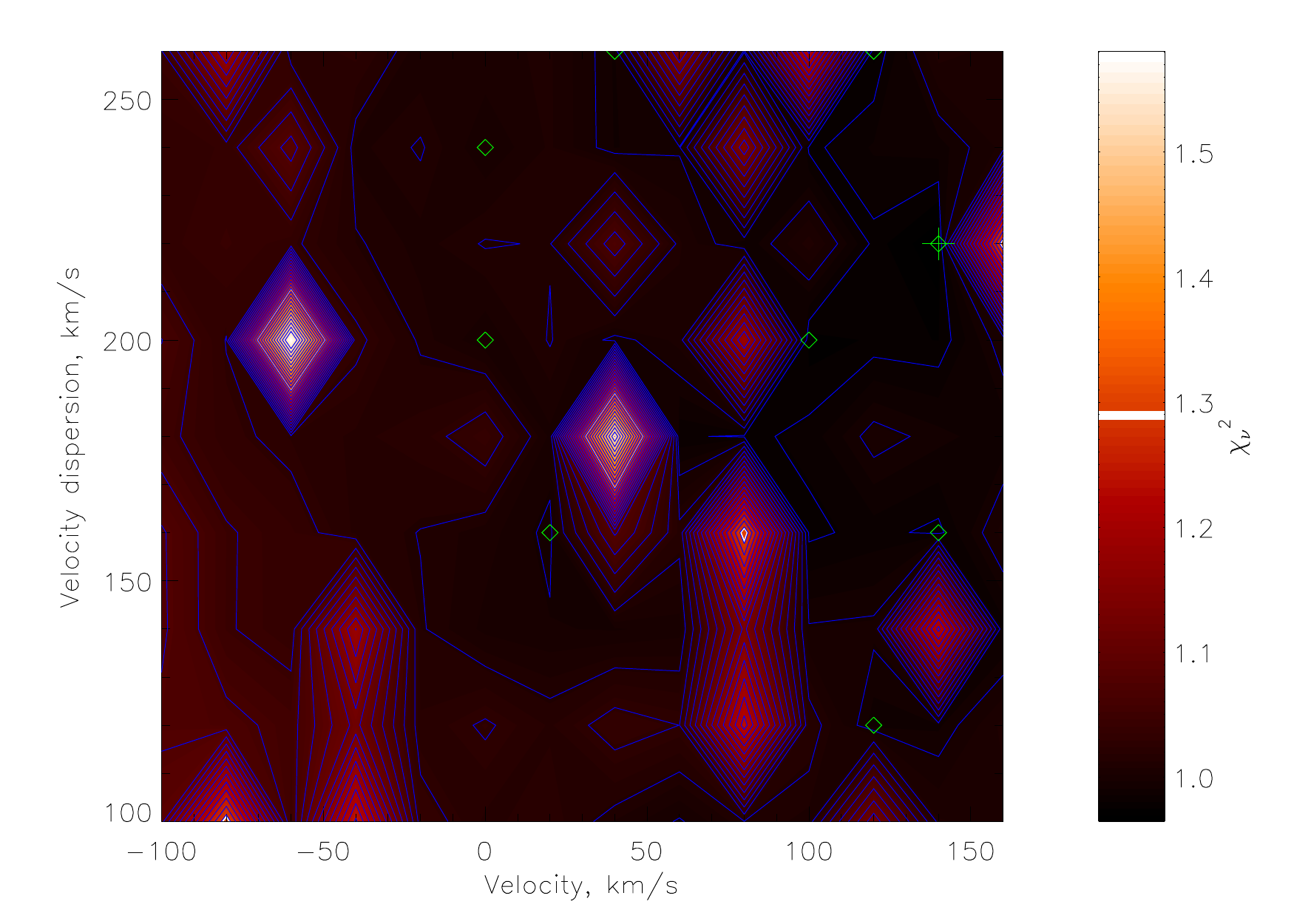}
	  \includegraphics[width=0.225\textwidth]{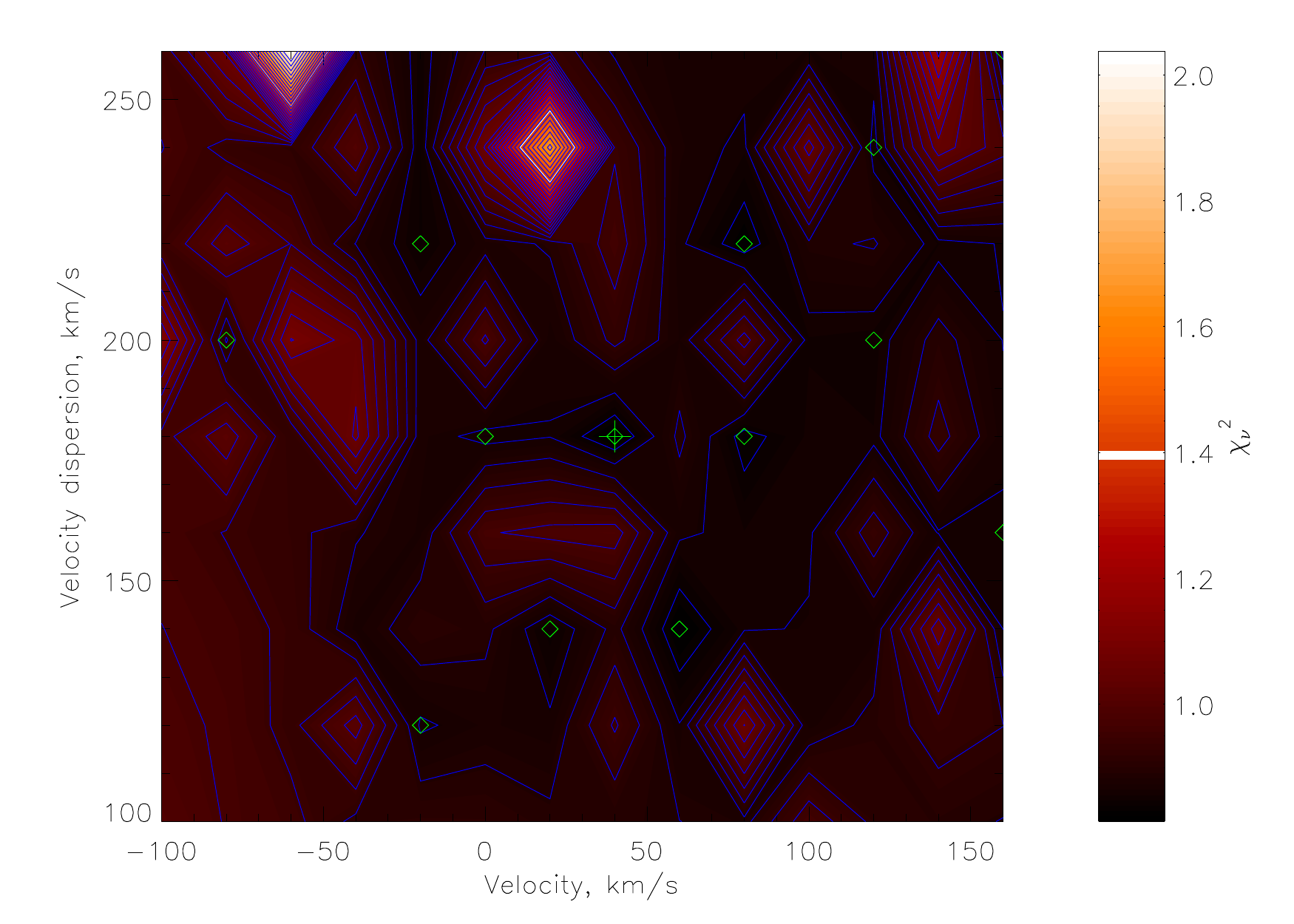}		
		\caption{$\chi^2$ maps in the space of SSP mean stellar velocity and SSP velocity dispersion. 
		Color-bar on the right hand side of $\chi^2$ maps show $\chi^2$ value normalized for the signal-to-noise value of the spectrum, measured by the SDSS in the $g$-band.}\label{chi2maps}
	\end{center}
\end{figure*}

\end{appendix}	

\clearpage

\vfill\eject\eject\newpage
\newpage\pagebreak

\begin{appendix}

\section{A method to derive emissivity weighted $n_{e}$\ from $\lambda_\mathrm{eff}$\ of \oii} \label{consistency_cz_hb_oii}

Figure \ref{fig:doublets} shows the behavior of $\mathcal{R}$=$I$([O\textsc{ii}]$\lambda$3729)/\-$I($[O\textsc{ii}]$\lambda$3726) as a function of [O\textsc{ii}]\-$\lambda\lambda$\-3726,3729 doublet effective wavelength ($\lambda_\mathrm{eff}$)\ for an unresolved mock doublet of 4 and 5 \AA. The $\lambda_\mathrm{eff}$\ has been measured for mock profiles built for 11 values of $R$\ with a step $\delta \mathcal{R}=0.1$. A cubic or quadratic fit reproduces the value of $R$:
\begin{equation} \label{eq:rl}
\mathcal{R}(n_\mathrm{e}) = k_{3}  (\lambda_\mathrm{eff}-3727)^{3} + k_{2}  (\lambda_\mathrm{eff}-3727)^{2} + k_{1} (\lambda_\mathrm{eff} - 3727) + k_{0}
\end{equation}
with the coefficients given in Table \ref{tab:bestfit}. The three width cases suggest a monotonic behavior of  $\mathcal{R}$  as a function of $\lambda_\mathrm{eff}$, with $\lambda_\mathrm{eff}$\ changing by $1.5$\AA\ from 3727.5 \AA\ and 3729.0 \AA. The measurements carried out from 
half peak intensity (filled circles in Fig. \ref{fig:doublets}) are more sensitive to the centroid differences and should 
be preferred in practice to fits from the base of the line.

 The best-fit parameters of $\mathcal{R}$\ as a function of $n_\mathrm{e}$\ using up-to-date atomic data are shown in Table~\ref{tab:bestfit1} for [O\textsc{ii}] from \citet{sandersetal16}.  Coefficients refer  to   a function of the form
\begin{equation}\label{eq:frat}
\mathcal{R}(n_\mathrm{e})=a \frac{b + n_\mathrm{e}}{c + n_\mathrm{e}}.
\end{equation}

\begin{figure}
 \centering
\includegraphics[width=\columnwidth]{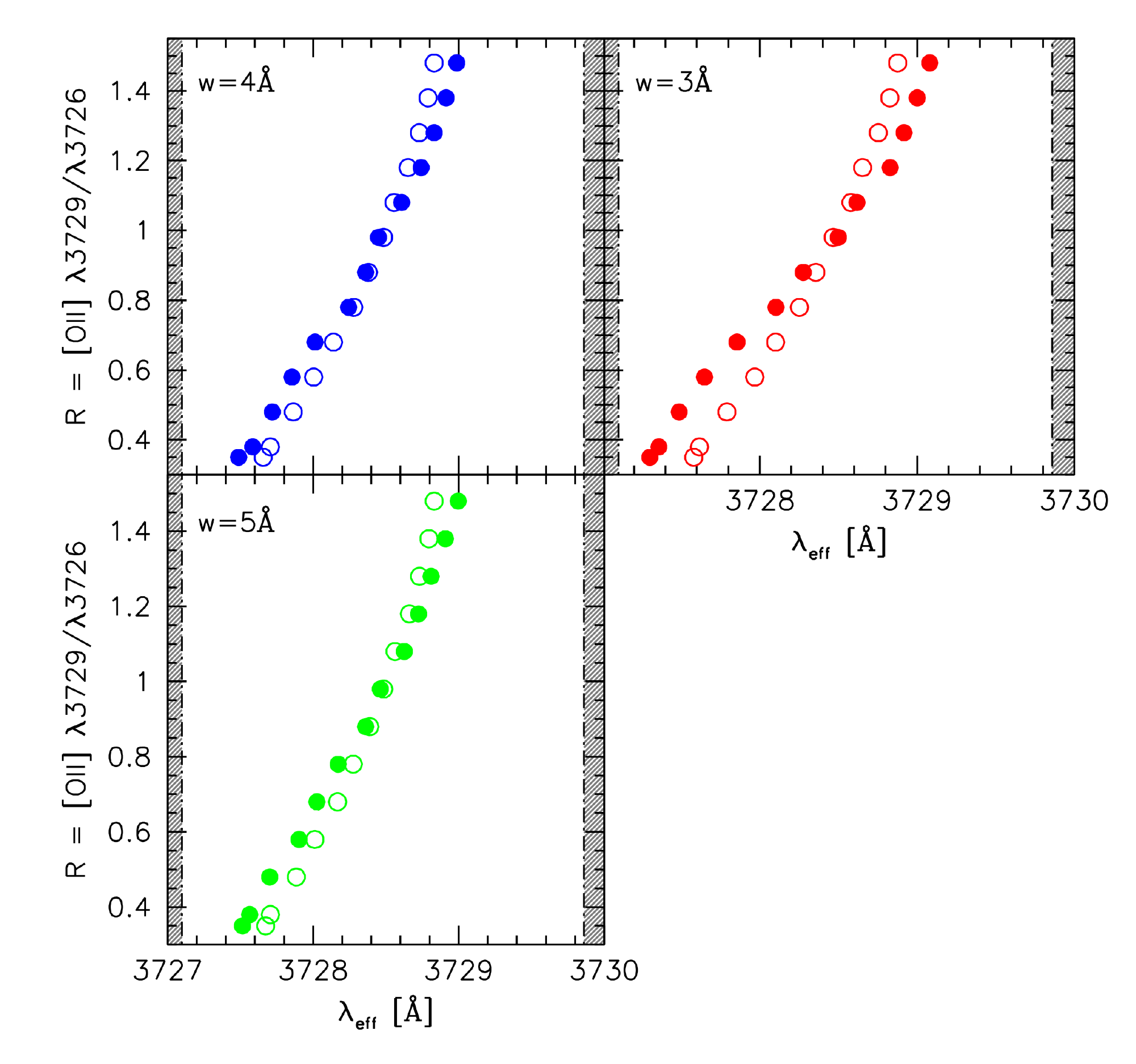} 
 \caption{$\mathcal{R}$[OII] as a function of [O\textsc{ii}]$\lambda\lambda$3726,3729 doublet effective wavelength $\lambda_\mathrm{eff}$ for an unresolved mock doublet of FWHM 3, 4 and 5 \AA. In each panel the open circles refer to Gaussian fits from the line base, filled circles from half peak intensity. The vertical dot-dashed lines mark the position of the individual component of the [O\textsc{ii}] doublet.}\label{fig:doublets}
\end{figure}

 \begin{table}[t]
 \centering
 \caption{Coefficients   in equation~\ref{eq:rl}}\label{tab:bestfit}
 \begin{tabular}{ c  c c   c c  c c}
   \hline\hline
      $w$\ & Fit  & $k_{3}$  & $k_{2}$ & $k_{1}$ &  $k_{0}$  \\ \hline
    3  & half   & 0.241 & - 0.7672 & + 1.2428& + 0.0302\\
    3  & full   &   0   & 	0.3353 & + 0.0105& + 0.2473\\
    4  & half   &   0   & 	0.1831 & + 0.2726& + 0.1818\\
    4  & full   &   0   & 	0.4356 & - 0.1684& + 0.2879\\
    5  & half   &   0   & 	0.1713 & + 0.3065& + 0.1628\\
    5  & full   &   0   & 	0.4492 & - 0.2008& + 0.2982\\
   \hline
 \end{tabular}
 \end{table}
 
\begin{table}[t]
 \centering
 \caption{Coefficients and limiting line ratios for [O~\textsc{ii}]  in equation~\ref{eq:frat}}\label{tab:bestfit1}
 \begin{tabular}{ c | c c c | c c}
   \hline\hline
   $\mathcal{R}$ & $a$ & $b$ & $c$ & $\mathcal{R}_{\rm min}^{a}$& $\mathcal{R}_{\rm max}^{b}$ \\ \hline
   {[O\textsc{ii}]$\lambda$3729/$\lambda$3726} & 0.3771 & 2,468 & 638.4 & 0.3839 & 1.4558 \\
   \hline
 \end{tabular}
 \begin{flushleft}
 {$^{a}$}{Theoretical minimum line ratio calculated in the high-density limit of 100,000~cm$^{-3}$}\\
 {$^{b}$}{Theoretical maximum line ratio calculated in the low-density limit of 1~cm$^{-3}$}\\
 \end{flushleft}

\end{table}

So the $\lambda_\mathrm{eff}$ of [O~\textsc{ii}] can be related to the density as follows:

\begin{equation}\label{eq:lambdaeff}
\mathcal{R}(n_\mathrm{e}) = k_{2}  \lambda_\mathrm{eff}^{2} + k_{1} \lambda_\mathrm{eff} + k_{0}   = a \frac{b + n_\mathrm{e}}{c + n_\mathrm{e}}
\end{equation}
\vspace{0.01cm}

Making this relation explicit for density:
\begin{eqnarray}
\label{eq:edens}
n_\mathrm{e}(\mathcal{R})& = & \frac{c\mathcal{R} - ab}{a - \mathcal{R}}\\
n_\mathrm{e}(\lambda_\mathrm{eff})&=  & \frac{c(k_{2}  \lambda_\mathrm{eff}^{2} + k_{1} \lambda_\mathrm{eff} + k_{0})-ab}{a-
(k_{2}  \lambda_\mathrm{eff}^{2} + k_{1} \lambda_\mathrm{eff} + k_{0} )}
\end{eqnarray}

\begin{figure}
	\centering
	\includegraphics[width=\columnwidth]{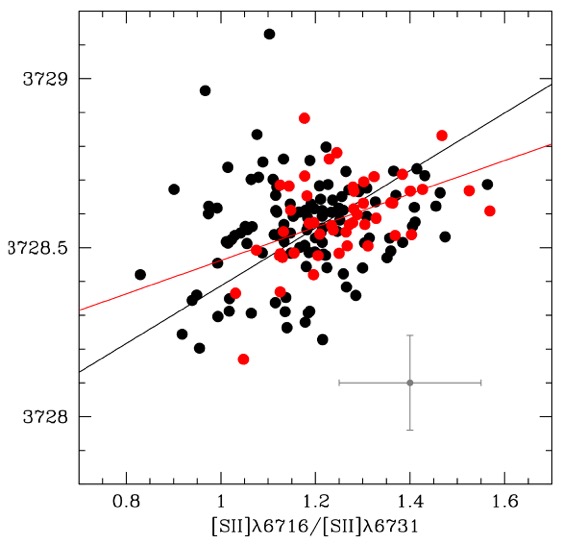} 
	\caption{Relation between the \oii\ effective wavelength 
	$\lambda_\mathrm{eff}$ and the  $R$[SII] 
	intensity ratio for a sample of \hii\ regions. Red circles refer to a vetted subsample in which 
	the uncertainty in the \sii\ doublet ratio $R$\ is less than 10\%. Best fitting lines are obtained with 
	the bisector method (full sample, black line) and with an unweighted least square fit (vetted sample, blue line). 
	Typical errors are $\delta \lambda_\mathrm{eff} \approx \pm 0.15 $ \AA\ for the effective wavelength and 
	$\delta R$[SII]$\approx \pm 0.15$ for the \sii\ ratio.  
	}\label{fig:oiisii}
\end{figure}

An application of the method to the sample of \hii\ region from the SDSS DR1 by \citet{kniazevetal04} is shown in Fig. \ref{fig:oiisii}.  The $\lambda_\mathrm{eff}$ and ${R}$[SII]\ are correlated. The scatter is relatively large. Typical errors ($\delta {R}$[SII] $\approx$ 0.15) suggest that measurement uncertainties account for most or all of it. If a restriction of $\delta R  $[SII] $\lesssim$ 0.10 is applied, the correlation is better defined, with a Pearson correlation coefficient $r \approx 0.46$, implying a significance $\gtrsim 4\sigma$. The average value of $ {R}$ [SII] and  $\lambda_\mathrm{eff}$\ \oii\  are 1.21 and 3728.56 \AA, respectively. The $R$ [SII] value implies $n_\mathrm{e} \approx 10^{2.4}$ cm$^{-3}$; $\lambda_\mathrm{eff}$\ \oii\ implies $\mathcal{R}$ [OII] $\approx 1.1$, which in turn yields $n_\mathrm{e} \approx 10^{2.5}$ cm$^{-3}$. The estimators are therefore consistent on average. This result indicates that, especially in some ideal cases, the $\lambda_\mathrm{eff}$\ can be considered as a proxy of the \oii\ doublet component ratio $\mathcal{R}$\ and hence an appropriate \nelec\ estimator. 

\end{appendix}

\clearpage

\bibliographystyle{aa}
\bibliography{biblioletter2a}

\vfill
\newpage
\pagebreak

\end{document}